\documentclass{aa}
\usepackage{color}
\usepackage{graphicx}
\usepackage{txfonts}
\usepackage{amssymb}

\usepackage{lscape}
\bibpunct{(}{)}{;}{a}{}{,}

\usepackage{color}

\begin{document} 

\title{Chemical evolution in the early phases of massive star formation \sc I\thanks{Based on observations carried out with the IRAM 30m Telescope. IRAM is supported by INSU/CNRS (France), MPG (Germany) and IGN (Spain). Reduced spectra as FITS files are available at the CDS via anonymous ftp to \url{cdsarc.u-strasbg.fr} (130.79.128.5)
or via \url{http://cdsarc.u-strasbg.fr/viz-bin/qcat?J/A+A/vol/page}
}}

\author{T. Gerner\inst{1}, H. Beuther\inst{1}, D. Semenov\inst{1},
H. Linz\inst{1}, T. Vasyunina\inst{2,3}, S. Bihr\inst{1}, Y. L. Shirley\inst{4}, \and Th. Henning\inst{1}}
\institute{Max-Planck-Institut f\"ur Astronomie, K\"onigstuhl 17, D-69117 Heidelberg, Germany \and Department of Chemistry, University of Virginia, Charlottesville, VA 22904, USA \and Current affiliation: Max-Planck-Institut f\"ur Radioastronomie, Auf dem H\"ugel 69, D-53121, Bonn, Germany \and Steward Observatory, University of Arizona, Tucson, AZ 85721, USA\\ \email{gerner@mpia.de}}

\titlerunning{Chemical evolution in early phases of massive star formation \sc I}
\authorrunning{Gerner et al.}

\abstract
{Understanding the chemical evolution of young (high-mass) star-forming regions is a central topic in star formation research. Chemistry 
is employed as a unique tool 1) to investigate the underlying physical processes and 2) to characterize the evolution of the chemical composition. With these aims in mind, we observed a sample of 59 high-mass star-forming regions at different evolutionary stages varying from the early starless phase of infrared dark clouds to high-mass protostellar objects to hot molecular cores and, finally, ultra-compact H{\sc ii} regions at 1\,mm and 3\,mm with the IRAM 30\,m telescope. We determined their large-scale chemical abundances and found that the chemical composition evolves along with the evolutionary stages. On average, the molecular abundances increase with time.
We modeled the chemical evolution, using a 1D physical model where density and temperature vary from stage to stage coupled with an advanced gas-grain chemical model and derived the best-fit $\chi^2$ values of all relevant parameters. A satisfying overall agreement between observed and modeled column densities for most of the molecules was obtained. With the best-fit model we also derived a chemical age for each stage, which gives the timescales for the transformation between two consecutive stages. The best-fit chemical ages are $\sim10\,000$~years for the IRDC stage, $\sim60\,000$~years for the HMPO stage, $\sim40\,000$~years for the HMC stage, and $\sim10\,000$~years for the UCH{\sc ii} stage. Thus, the total chemical timescale for the entire evolutionary sequence of the high-mass star formation process is on the order of $10^5$~years, which is consistent with theoretical estimates. Furthermore, based on the approach of a multiple-line survey of unresolved data, we were able to constrain an intuitive and 
reasonable physical and chemical model. The results of this study can be used as chemical templates for the different evolutionary stages in high-mass star formation.}

\keywords{Stars: formation -- Stars: early-type -- ISM: molecules --  (ISM:) evolution}

\maketitle

\section{Introduction}\label{sec:introduction}
The chemical evolution of star-forming regions is an important topic from various perspectives. From an astrochemist's point of view, one wishes to understand the chemical properties of the gas to be able to characterize the chemical composition and evolution of the interstellar medium (ISM) in general. This starts with the chemistry associated with diffuse clouds and the formation of relatively simple molecules like H$_2$ or CO, and continues to the formation of complex molecules (even pre-biotic ones) in hot molecular cores or protostellar accretion disks. Complementing this approach, astrophysicists need to have a good understanding of the chemical constituents of the ISM to use different molecules as tools to probe physical conditions such as densities, temperatures, or kinematical properties. These two approaches are interdependent.

Chemical inventories of (high-mass) star-forming regions have so far largely concentrated either on complete line-surveys toward selected regions (e.g., Orion-KL or NGC6334I \&I(N), \citealt{sutton1985,schilke1997b,schilke2001,walsh2010}) or have targeted small spectroscopic setups toward specific subsamples (e.g., hot molecular cores, HMCs, or infrared dark clouds, IRDCs, \citealt{hatchell1998b,bisschop2007,vasyunina2011}). In addition
to these line surveys, a few interferometric high-spatial resolution studies of limited samples of high-mass star-forming regions exist (e.g., \citealt{blake1996,beuther2009a}).

Many studies suffer either from too limited sample sizes or spectral coverages to be able to characterize the chemistry of high-mass star-forming regions at various evolutionary stages in a statistical sense. A number of more recent studies aimed at a deeper understanding of a broader chemistry in the various evolutionary stages of high-mass star formation. \citet{vasyunina2011} compared abundances in the IRDCs with the data for more evolved sources taken from the literature, and the main problem was systematic, where different spectral setups, beam sizes, and integration times were used in different samples. \citet{fontani2005} studied a larger sample of protostellar candidates from the IRAS Point Source Catalog on the basis of their IR color and observed them in CS and C$^{17}$O and the 1.2\,mm continuum. Based on these observations and results from the literature, they were able to classify the evolutionary stages as very early and probably previous to the formation of an UCH{\sc ii} region. \citet{
zinchenko2009} 
observed five high-mass star-forming regions in a handful of simple molecular species to estimate their physical and molecular parameters. They found systematic differences in the distributions and the abundances of various molecules and discussed the importance of HCO$^+$, HNC and, especially, N$_2$H$^+$ as potentially valuable indicators of massive protostars. \citet{pirogov2007} mapped twelve high-mass star-forming regions in CS, N$_2$H$^+$ and 1.2\,mm dust continuum and derived the physical parameters, the density and chemical structure of the associated dense cores within.  \citet{reiter2011} also studied 27 high-mass clumps in a larger set of molecules and amongst others discussed the dependence of various physical clump properties on chemical properties. They found that molecular column densities are only weakly correlated with any of the derived physical properties of the clumps. Another still ongoing study is the MALT90 survey, which aims to chemically characterize dense molecular clumps \citep[][]{
foster2011,sanhueza2012,hoq2013,jackson2013}.

To continue the chemical characterization of high-mass star-forming regions, it is important to study all evolutionary stages in a systematic and consistent way for all sources. Furthermore, it is desirable to include more molecular species as well as be able to understand the complete chemistry along the evolutionary sequence with models.
In this work we divide the early evolution of the high-mass star-forming regions into four different stages that are observationally motivated and guided by the evolutionary sequence shown in \citet{beuther2006b} and also in \citet{zinnecker2007} who divided the different stages based on their physical conditions. In the first, earliest considered phase, quiescent infrared dark clouds (IRDCs) are formed. They consist of cold and dense gas and dust and emit mainly at (sub-)millimeter wavelengths. Their physical conditions are close to isothermal. In our sample this group consists of starless IRDCs as well as IRDCs already starting to harbor point sources at $\mu$m-wavelengths.

Previous theoretical works, for instance by \citet{krumholz2007,narayanan2008,heitsch2008}, suggested that there might be a long-lived pre-IRDC massive clump stage. This stage was also seen in the observations by \citet{barnes2011} and supported by extragalactic works, for example, by \citet{koda2011} who suggested a long-lived pre-MSF stage in M51. The existence of a long-lived pre-IRDC phase as a precursor is not doubted in our approach. But from a modeling point of view, in
our opinion $\sim20\,000$~years are enough to convert almost all initially atomic gas in regions with densities $\geq 10^5$~cm$^{-3}$ into molecular-rich gas and thus reach the molecular IRDC stage. Thus, based on chemistry, we define the year zero in our evolutionary sequence when the objects reach densities of $\gtrsim 10^4$~cm$^{-3}$ and become detectable as cold dense molecular clouds. The evolution prior to this stage is not considered in this work and took already an unknown amount of time.

In the second phase, the so-called high-mass protostellar objects (HMPOs) form, hosting an actively accreting protostar(s) with $>8$~M$_{\odot}$, which already shows an internal emission source(s) at mid-infrared wavelengths. The central temperature rises and the temperature profile starts to  deviate from the isothermal case. In this work we split the physically homogeneous HMPO phase into two subsequent phases with the HMPO being followed by the much warmer hot molecular core phase (HMC). This phase is distinguished from the earlier HMPO phase from a chemical point of view.
In the HMC stage the central source(s) heats the surrounding environment, evaporating molecular-rich ices and giving rise to molecular complexity in the gas. Finally, the UV-radiation from the embedded protostar(s) ionizes the surrounding gas and an ultra-compact H{\sc ii} (UCH{\sc ii}) region is formed (fourth stage). In these objects many of the previously formed complex molecules are no longer detected because they are most likely destroyed by the ionizing radiation. Physically, many UCH{\sc ii} regions have probably stopped accretion at this point, but this class is not entirely homogeneous either and accretion may still continue in some UCH{\sc ii} regions.

Because the evolution in high-mass star formation takes place on rather short timescales and in clustered environments, the transitions from one into the next stage are smooth and not always clearly distinct. There might be overlaps among the HMPO, HMC and even UCH{\sc ii} region stage. Some regions are already surrounded by an ionized medium but still show the rich hot core chemistry. Due to the large beam, regions with different evolutionary stages in their proximity can also be observed as one object showing the characteristics of the different stages. In this work we wish to disentangle this from an observational point of view to be able to characterize the evolution in every considered stage statistically and also to model the evolution on a consistent timeline. On the one hand, we follow the evolution based on physical quantities, especially the temperature, which is rising from IRDCs to HMPOs to UCH{\sc ii} regions and on the other hand, based on the chemistry which splits the HMPOs into early, 
chemical poorer HMPOs, and HMCs with a molecular richer chemistry.

To set the observational results into context, an extensive multidimensional modeling and fitting of the observationally derived column densities is performed. The modeling of each individual evolutionary stage is based on the iterative fitting of the data with a set of 1D power-law physical models coupled to a pseudo-time-dependent gas-grain chemical model, calculated over $1\,$~Myr. The cloud density and temperature structures of the environment, as well as its chemical age, are the variable parameters. The chemical characterization of the presumed evolutionary sequence in a statistical sense, together with the unique approach to fit the observed chemical evolution in high-mass star formation directly with a model,  enables us to set the results into a broader context.

We describe the source sample in Section~\ref{sec:sourcesample} followed by the observations in Section~\ref{sec:observations}. The derivation of the chemical abundances and a chemical characterization of the four different subsamples based on the observational data are given in Section~\ref{sec:results}. In Section~\ref{sec:discussion} we describe our model and the fitting process and interpret the observed values in combination with the fitting results in more detail and compare the results with other studies. We conclude with a summary in Section~\ref{sec:conclusion}.

\section{Source sample}\label{sec:sourcesample}

\begin{table*}
\small
\caption{Source list showing the position, the distance, and the evolutionary stage of all high-mass star-forming regions.}             
\label{tbl:sourcelist}      
\centering                          
\begin{tabular}{lccccccccr}        
\hline\hline                 
source & $\alpha$ & $\delta$ & galactic $l$ & galactic $b$ & distance\tablefootmark{a} & type & 24~$\mu$m & 70~$\mu$m & continuum data \\
 & (J2000.0) & (J2000.0) & $[^{\circ}]$ & $[^{\circ}]$ & $[\rm {kpc}]$ &  &  &  &  \\
\hline                        
IRDC011.1    &   18:10:28.4 &    -19:22:34 & 11.108 & -0.115 & 3.6 &  IRDC & y & y  & ATLASGAL    \\
IRDC028.1    &   18:42:50.3 &    -04:03:20 & 28.343 & 0.060 &  4.8 &  IRDC & y & y  & ATLASGAL    \\
IRDC028.2    &   18:42:52.1 &    -03:59:54 & 28.397 & 0.080 &  4.8 &  IRDC & y & y  & ATLASGAL    \\
IRDC048.6    &   19:21:44.4 &    +13:49:24 & 48.657 & -0.285 &  2.5 &  IRDC & n & n  & ATLASGAL    \\
IRDC079.1    &   20:32:22.0 &    +40:20:10 & 79.338 & 0.341 &  1.0 &  IRDC & - & y & SCUBA   \\
\vspace{0.1cm}
IRDC079.3    &   20:31:57.7 &    +40:18:26 & 79.269 & 0.386 &  1.0 &  IRDC & - & y & SCUBA   \\
IRDC18151    &   18:17:50.3 &    -12:07:54 & 18.319 & 1.792 &  3.0 &  IRDC & - & y & Mambo       \\
IRDC18182    &   18:21:15.0 &    -14:33:03 & 16.578 & -0.081 &  3.6 &  IRDC & y & y & Mambo       \\
IRDC18223    &   18:25:08.3 &    -12:45:27 & 18.605 & -0.075 &  3.7 &  IRDC & y & y & Mambo       \\
IRDC18306    &   18:33:32.1 &    -08:32:28 & 23.297 & 0.0550 &  3.8 &  IRDC & n & n & Mambo       \\
\vspace{0.1cm}
IRDC18308    &   18:33:34.3 &    -08:38:42 & 23.209 & -0.001 &  4.9 &  IRDC & y & y & Mambo       \\
IRDC18310    &   18:33:39.5 &    -08:21:10 & 23.478 & 0.115 &  5.2 &  IRDC & n & n & Mambo       \\
IRDC18337    &   18:36:18.2 &    -07:41:00 & 24.374 & -0.158 &  4.0 &  IRDC & y & y & Mambo       \\
IRDC18385    &   18:41:17.4 &    -05:09:56 & 27.179 & -0.104 &  3.3 &  IRDC & y & y\tablefootmark{b} & Mambo       \\
IRDC18437    &   18:46:21.8 &    -02:12:21 & 30.390 & 0.123 &  (6.2) 7.3\tablefootmark{c} & IRDC & y & y & Mambo \\
\vspace{0.1cm}
IRDC18454.1    &   18:48:02.1 &    -01:53:56 & 30.854 & -0.109 &  (3.5) 6.4\tablefootmark{d} & IRDC & n & n & Mambo \\
IRDC18454.3    &   18:47:55.8 &    -01:53:34 & 30.848 & -0.083 &  6.0 (6.4)\tablefootmark{e} & IRDC & n & n & Mambo \\
IRDC19175    &   19:19:50.7 &    +14:01:23 & 48.617 & 0.214 &  1.1 &  IRDC & n & n & Mambo       \\
IRDC20081    &   20:10:13.0 &    +27:28:18 & 66.145 & -3.197 &  0.7 &  IRDC & - & n\tablefootmark{f} & Mambo       \\
HMPO18089    &   18:11:51.6 &    -17:31:29 & 12.889 & 0.489 &  3.6 &  HMPO & & & Mambo       \\
\vspace{0.1cm}
HMPO18102    &   18:13:11.3 &    -18:00:03 & 12.623 & -0.017 &  2.7 &  HMPO & & & Mambo       \\
HMPO18151    &   18:17:58.1 &    -12:07:26 & 18.341 & 1.768 &  3.0 &  HMPO & & & Mambo       \\
HMPO18182    &   18:21:09.2 &    -14:31:50 & 16.585 & -0.051 &  4.5 (11.8)\tablefootmark{g} &  HMPO & &  & Mambo       \\
HMPO18247    &   18:27:31.7 &    -11:45:56 & 19.755 & -0.129 &  6.7 &  HMPO & &  & Mambo       \\
HMPO18264    &   18:29:14.6 &    -11:50:22 & 19.884 & -0.535 &  3.5 (12.5)\tablefootmark{g} &  HMPO & &  & Mambo       \\
\vspace{0.1cm}
HMPO18310    &   18:33:48.1 &    -08:23:50 & 23.455 & 0.063 &  5.2 (10.4)\tablefootmark{g} &  HMPO & &  & Mambo       \\
HMPO18488    &   18:51:25.6 &    +00:04:07 & 32.991 & 0.034 &  5.4 (8.9)\tablefootmark{g} &  HMPO & &  & Mambo       \\
HMPO18517    &   18:54:14.4 &    +04:41:40 & 37.430 & 1.517 &  2.9 &  HMPO & &  & Mambo       \\
HMPO18566    &   18:59:10.1 &    +04:12:14 & 37.554 & 0.200 &  6.7 &  HMPO & &  & Mambo       \\
HMPO19217    &   19:23:58.8 &    +16:57:44 & 51.679 & 0.720 & 10.5 &  HMPO & &  & Mambo       \\
\vspace{0.1cm}
HMPO19410    &   19:43:11.0 &    +23:44:10 & 59.784 & 0.066 &  2.1 &  HMPO & &  & Mambo       \\
HMPO20126    &   20:14:26.0 &    +41:13:32 & 78.122 & 3.633 &  1.7 &  HMPO & &  & Mambo       \\
HMPO20216    &   20:23:23.8 &    +41:17:40 & 79.127 & 2.279 &  1.7 &  HMPO & &  & Mambo       \\
HMPO20293    &   20:31:12.9 &    +40:03:20 & 78.982 & 0.352 &  1.3 (2.0)\tablefootmark{g} &  HMPO & &  & Mambo       \\
HMPO22134    &   22:15:09.1 &    +58:49:09 & 103.876 & 1.856 &  2.6 &  HMPO & &  & Mambo       \\
\vspace{0.1cm}
HMPO23033    &   23:05:25.7 &    +60:08:08 & 110.093 & -0.067 &  3.5 &  HMPO & &  & Mambo       \\
HMPO23139    &   23:16:10.5 &    +59:55:28 & 111.256 & -0.770 &  4.8 &  HMPO & &  & Mambo       \\
HMPO23151    &   23:17:21.0 &    +59:28:49 & 111.236 & -1.238 &  5.7 &  HMPO & &  & Mambo       \\
HMPO23545    &   23:57:06.1 &    +65:24:48 & 117.315 & 3.136 &  0.8 &  HMPO & &  & Mambo       \\
HMC009.62    &   18:06:15.2 &    -20:31:37 & 9.621 & 0.193 &  5.7 &   HMC & &  & ATLASGAL    \\
\vspace{0.1cm}
HMC010.47    &   18:08:38.2 &    -19:51:50 & 10.472 & 0.027 &  5.8 &   HMC & &  & ATLASGAL    \\
HMC029.96    &   18:46:04.0 &    -02:39:21 & 29.956 & -0.017 &  7.4 &   HMC & &  & ATLASGAL    \\
HMC031.41    &   18:47:34.2 &    -01:12:45 & 31.412 & 0.308 &  7.9 &   HMC & &  & ATLASGAL    \\
HMC034.26    &   18:53:18.5 &    +01:14:58 & 34.257 & 0.154 &  4.0 &   HMC & &  & ATLASGAL    \\
HMC045.47    &   19:14:25.7 &    +11:09:26 & 45.466 & 0.045 &  6.0 &   HMC & &  & ATLASGAL    \\
\vspace{0.1cm}
HMC075.78    &   20:21:44.1 &    +37:26:40 & 75.783 & 0.343 &  4.1 &   HMC & &  & SCUBA   \\
NGC7538B     &   23:13:45.4 &    +61:28:11 & 111.542 & 0.777 &  2.65 (5.61)\tablefootmark{h} &   HMC & & & SCUBA \\
Orion-KL     &   05:35:14.4 &    -05:22:31 & 208.993 & -19.385 &  0.44 &   HMC & &  & SCUBA   \\
W3IRS5       &   02:25:40.7 &    +62:05:52 & 133.715 & 1.215 &  1.8 &   HMC & &  & SCUBA   \\
W3H$_2$O     &   02:27:04.6 &    +61:52:25 & 133.949 & 1.065 &  2.0 &   HMC & &  & SCUBA   \\
\vspace{0.1cm}
UCH005.89    &   18:00:30.4 &    -24:04:00 & 5.886 & -0.392 &  2.5 &   UCH{\sc ii} & & & ATLASGAL    \\
UCH010.10    &   18:05:13.1 &    -19:50:35 & 10.099 & 0.739 &  4.4 &   UCH{\sc ii} & & & ATLASGAL    \\
UCH010.30    &   18:08:55.8 &    -20:05:55 & 10.300 & -0.147 &  6.0 &   UCH{\sc ii} & & & ATLASGAL    \\
UCH012.21    &   18:12:39.7 &    -18:24:20 & 12.208 & -0.102 & 13.5 &   UCH{\sc ii} & & & ATLASGAL    \\
UCH013.87    &   18:14:35.8 &    -16:45:43 & 13.872 & 0.280 &  4.4 &   UCH{\sc ii} & & & ATLASGAL    \\
\vspace{0.1cm}
UCH030.54    &   18:46:59.3 &    -02:07:24 & 30.535 & 0.021 &  6.1 &   UCH{\sc ii} & & & ATLASGAL    \\
UCH035.20    &   19:01:46.4 &    +01:13:25 & 35.200 & -1.741 &  3.2 &   UCH{\sc ii} & & & SCUBA   \\
UCH045.12    &   19:13:27.8 &    +10:53:37 & 45.122 & 0.132 &  6.9 &   UCH{\sc ii} & & & ATLASGAL    \\
UCH045.45    &   19:14:21.3 &    +11:09:14 & 45.454 & 0.060 &  6.0 &   UCH{\sc ii} & & & ATLASGAL    \\
\hline                                   
\end{tabular}
\tablefoot{For the IRDCs we indicate whether or not they show embedded 24 or 70$\mu$ m point sources with y(es) or n(o) (or ``-'' if there are no data available). In the last column the sources of the dust continuum data are presented.}
\tablefoottext{a}{unbracketed values are preferred, bracketed values are alternative values}
\tablefoottext{b}{very weak emission compared with the background located at the same position as 24$\mu$m emission source}
\tablefoottext{c}{for $v_{\rm lsr}=(97.6) 111.3$~km\,s$^{-1}$}
\tablefoottext{d}{for $v_{\rm lsr}=(52.8) 100.2$~km\,s$^{-1}$}
\tablefoottext{e}{for $v_{\rm lsr}=94.3 (98.4)$~km\,s$^{-1}$}
\tablefoottext{f}{no embedded central point source found, a nearby extended source with emission inside the beam is detected}
\tablefoottext{g}{for the near (far) kinematic solution}
\tablefoottext{h}{parallactic (kinematic) distance}
\end{table*}

The source sample contains 59 high-mass star-forming regions, consisting of 19 IRDCs and 20 HMPOs as well as 11 HMCs and 9 UCH{\sc ii}s. 
The complete sample is listed in Table~\ref{tbl:sourcelist} with coordinates and distances. The telescope pointings were centered on the column density peaks derived from the dust continuum observations obtained either with Mambo (at 1.2\,mm) or from the ATLASGAL and SCUBA surveys at 870~$\mu$m and 850~$\mu$m, respectively (\citealt{schuller2009,difrancesco2008}).
From these maps we determined the corresponding H$_2$ column densities (see Section~\ref{sec:column_density}). The sources are mostly located within the galactic plane, with an average heliocentric distance of $\sim4\,$~kpc.

The sources were selected from different source lists. The lists of the IRDCs were first presented in \citet{carey2000} and \citet{sridharan2005} and are part of the {\it Herschel} guaranteed time key project EPOS \citep[Early Phase of Star Formation, ][]{ragan2012}. This sample consists of 6 IRDCs showing no internal point sources shortward of 70~$\mu$m and 13 IRDCs that have internal point sources at 24~$\mu$m and 70~$\mu$m (see source list in Table~\ref{tbl:sourcelist} for individual internal point source detections). The HMPOs were taken from the well-studied sample by \citet{sridharan2002}, and \citet{beuther2002a,beuther2002b}. HMC sources are selected from the line-rich sample of \citet{hatchell1998b} including a few additional well-known HMCs, W3IRS5 and W3(H$_2$O) and Orion-KL. For the UCH{\sc ii}s, we selected line-poor high-mass star-forming regions from \citet{hatchell1998b}, and additional sources from \citet{wc1989b}.

\section{Observations with the IRAM 30\,m}\label{sec:observations}

\subsection{Data collection}
The 59 sources were partly observed in the winter semester (20.1.2011-17.3.2011) and mainly in the summer semester (28.10.2011-30.10.2011, 23.11.2011) with the IRAM 30\,m telescope at Pico Veleta (Spain). We performed a spectral line survey in two setups at 3\,mm and 1\,mm using the EMIR receiver with the FFTS backends. The system temperatures of the observations taken in the two different semesters are comparable with $T_{\rm sys}\sim100$~K in the 3\,mm band and $T_{\rm sys}\sim250$~K in the 1\,mm band. The large 16~GHz bandwidth allowed us to efficiently study the chemical properties of this relatively large sample of high-mass star-forming regions over a broad range of simultaneously observed molecular transitions. Due to changes in the instruments and thus the coverage in frequency during distinct observation runs we only analyzed the frequency ranges at which all sources were observed, namely 86-94~GHz, 217-221~GHz, and 241-245~GHz.
The frequency ranges were chosen to cover transitions of important molecules containing nitrogen, oxygen, sulfur, carbon, and silicon. The observations were carried out in wobbler-switching mode with 1.25\,min on-source integration time with a spectral resolution of $\sim0.3$\,km\,s$^{-1}$ at 1\,mm and $\sim0.6$\,km\,s$^{-1}$ at 3\,mm. The beam sizes of the IRAM 30\,m telescope are $11\arcsec$ at 1\,mm and $29\arcsec$ at 3\,mm. Typical 1-~sigma rms values are on the order of $\sim0.1$\,K at 1\,mm and $\sim0.03$\,K at 3\,mm, respectively. To reduce all spectroscopic data, the standard GILDAS\footnote{http://www.iram.fr/IRAMFR/GILDAS} software package CLASS was used.

\subsection{Data analysis strategies} \label{sec:problems}
For the analysis we focused on a specific sample of 15 different molecular species for which we calculated the column densities (see Table \ref{tbl:moleculeslist} for a list of the molecules). To directly compare the derived values with the modeled values (see Section~\ref{sec:Model}) we relied on the most common molecules among the different sources at a distinct evolutionary stage (e.g., N$_2$H$^+$, HCN, C$_2$H, HNC, HCO$^+$). These molecules we complemented with more complex molecules that are typical of more evolved stages of high-mass star-forming regions, for
instance, the HMC phase (e.g., CH$_3$CN, OCS). The homogeneous list of molecules from various chemical ``families'' made our search for the best-fit model parameters ($T$, $n_{\rm H}$, age) more reliable.
The spectra of these molecules were fitted with a Gaussian profile and the corresponding column densities were obtained (see below). For that we excluded all lines that deviate extremely from Gaussian profiles, because this makes the derivation of column densities, based on the assumptions described in Section~\ref{sec:column_density}, unreliable. The reasons for such complex line shapes might be optical depth effects, self-absorption, flux accidentally measured at an $240\arcsec$ off-position due to the limited distance in the wobbler-switching mode, or peculiar gas kinematics.
Furthermore, the spectra of very abundant molecules might be affected by foreground and background emission. Infall or outflow motions also lead to a departure of a line profile from Gaussian shape toward double-peaked profiles.
In the source W3(H$_2$O) a masering methanol line caused a high intensity in one channel around the line peak of that line. Other problems in fitting the line profiles were broad line-wings that we were unable to fit with a single Gaussian, but only with a combination of a narrow main component and a broader underlying component. In these cases we fitted the narrow component, which most likely traces the bulk component of the molecular gas. Some of the lines with slightly asymmetric shapes still allowed Gaussian fitting, which we also included in the analysis, with the caution that the derived values are probably underestimated.

The hyperfine line of HCN (1-0) shows anomalies in the line ratios compared with the predicted theoretical values in a substantial number of sources, which was found by other groups as well \citep[see][]{cernicharo1984,loughnane2012}. In the optical thin LTE case the relative intensities of the three HCN(1-0) hyperfine components are F(0-1):F(2-1):F(1-1) = 1:5:3 \citep[e.g.,][]{wannier1974}.
In our spectra we see various deviations from these ratios with either F(0-1)/F(1-1) $\geqq 1,$ which cannot be solely explained by the optical thickness of the line. The probable reason for this behavior is scattering of radiation from the core in a moderate-density envelope \citep{cernicharo1984}. This is possibly enhanced by an inherently complex spatial distribution of HCN through the cloud or clumpiness of the cloud itself, which causes tricky absorption and re-emission of the radiation. In such a case we refrained from using this line in the data analysis.
In summary, the main problems in the analysis of the spectral lines were optical depth effects, a poor off-position, foreground and background emission, a complex velocity structure of the source, and anomalies in the HCN hyperfine structure.
The assumptions used to derive molecular abundances and column densities are presented in Section~\ref{sec:column_density}, and we discuss the uncertainties in Section~\ref{sec:uncertainties}.

\section{Results}\label{sec:results}

\subsection{Chemical characteristics of the four evolutionary stages of high-mass star-forming regions}
\begin{figure*}
\centering
\includegraphics[angle=0,width=0.5\textwidth]{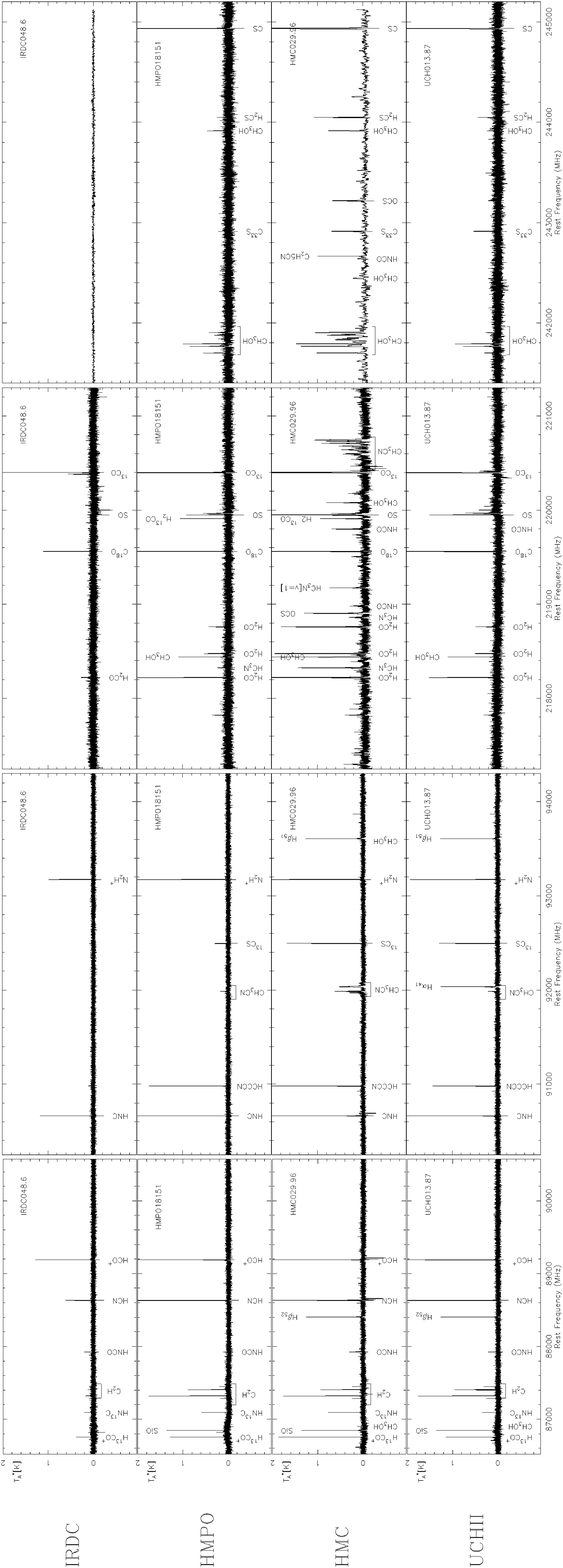}
\caption{From left to right, we show sample 16 GHz spectra of the sources IRDC048.6, HMPO18151, HMC029.96, and UCH013.87, representing the four selected evolutionary stages of high-mass star formation. The frequency resolution is $\sim0.2$\,MHz and typical rms values are$\sim0.1$\,K at 1\,mm and $\sim0.03$\,K at 3mm.
\label{fig:samplespectra}}
\end{figure*}

In Figure~\ref{fig:samplespectra} we show sample spectra for each of the four evolutionary stages. We now describe the general characteristics of each stage on the basis of these four examples.
In the object IRDC048.6 we only detected simple and common molecules such as isotopologues of CO or HCO$^+$ and tracers of cold and dense matter such as N$_2$H$^+$.
The spectra of HMPO18151 already shows a higher number of detected molecules and higher intensities of the observed lines. In this phase the temperature starts to increase, which allows radicals to become mobile on the dust grain surfaces, synthesizing complex (organic) ices and simpler species that may be able to evaporate to the gas phase. This also leads to an overall increase in the intensities of the detected lines (e.g., C$_2$H, H$_2$CO, HCN, HCO$^+$ and isotopologues), such that more species appear on the spectrum. We detected an organic species, methanol (CH$_3$OH), SiO -- a molecule that is a tracer of outflows and shocks, and we began to see sulfur-bearing molecules (e.g., CS, SO, H$_2$CS and isotopologues).
In the HMC stage the spectrum shows the largest number of lines, including transitions of long and complex molecules such as methyl formate CH$_3$OCHO. Some of the HMC spectra already show recombination lines (e.g. HMC029.96), which is typically only visible in the spectra of UCH{\sc ii} regions. In this case they are generated
by a nearby UCH{\sc ii} region \citep[see][]{beuther2007d} that is also covered by the single-dish beam.
During this late evolutionary phase many complex, photofragile molecules might be destroyed by the intense ionizing radiation of the central star(s). This enhances the amount of ionized medium and thus reduces the amount of the neutral medium compared with HMCs, leading to lower line fluxes and column densities. Correspondingly, the UCH{\sc ii} spectrum shows fewer molecular lines. Apart from additional recombination lines of hydrogen, the UCH{\sc ii} spectra looks qualitatively very similar to the HMPO spectrum.

We selected 15 different molecular species that trace the physical conditions and chemical evolution to analyze and compare them with a chemical model. A list of all the selected lines is given in Table~\ref{tbl:moleculeslist}. To exclude optically thick lines we used the rarer isotopologues, when available. Specifically, we did not analyze $^{13}$CO, HNC, HCO$^+$, or CS. The assumed isotopic ratios are given in Section~\ref{sec:column_density}. The detected lines in each source were fitted with a Gaussian profile to obtain the total line intensities. For sources in which we did not detect a specific transition we used the 3-~sigma rms value to estimate an upper limit of the corresponding column density.

\subsection{Calculation of the H$_2$ column densities}\label{sec:column_densityh2}
To derive abundances we first calculated the H$_2$ column densities for all observed regions. Many of our sources are covered by the galactic plane survey ATLASGAL \citep{schuller2009}, which observed dust emission at 870$\mu$m. Most of the HMPOs were observed at 1.2\,mm with the bolometer MAMBO at the IRAM 30m. The data for the remaining sources were taken from the SCUBA Legacy Catalog published by \citet{difrancesco2008} which are bolometer maps observed at 850$\mu$m with the JCMT. In Table~\ref{tbl:sourcelist} we list which continuum data were taken for each of the sources.

The coordinates for the spectral observations were centered on the highest intensity peak in the bolometer map. With the observed value of the peak intensity we derived H$_2$ column densities following Equation~\ref{equ:dusttoh2} taken from \citet{schuller2009}. The typical temperatures for the observed sources were assumed. Average IRDC temperatures are around 15~K based on several NH$_3$ IRDC surveys \citep[e.g., ][]{sridharan2005, pillai2006,chira2013}. Typical HMPO temperatures were estimated from SED fits and are $\sim 50$~K. HMCs are usually warmer and we chose $T=100$~K as the average temperature. The same temperature of $T=100$~K is assumed for the UCH{\sc ii} stage. Finally, the dust opacities were interpolated from \citet{ossenkopf1994}, assuming grains with thin ice mantles, gas densities of $n=10^5$~cm$^{-3}$, and a gas-to-dust mass ratio $R=100$. The actually used dust opacities are $\kappa_{850\mu \rm m}=1.48$, $\kappa_{870\mu \rm m}=1.42$, $\kappa_{1.2\rm mm}=0.97$. Using these 
assumptions and assuming that the emission is optically thin and at LTE, the H$_2$ column density is calculated as

\begin{equation}
N_{H_2}=\frac{F_{\nu} \cdot R}{B_{\nu} \cdot \Omega  \cdot \kappa_{\nu} \cdot \mu \cdot m_{H}}  
\label{equ:dusttoh2}
.\end{equation}

In this calculations we also assumed that the dust and the gas are collisionally coupled and thus have the same temperature.
The observations with the IRAM 30m at 1\,mm have a HPBW of $11\arcsec$ and at 3\,mm a HPBW of $29\arcsec$. The data obtained with MAMBO have a resolution of $11\arcsec$, the ATLASGAL maps $19.2\arcsec$ and the SCUBA data $22.9\arcsec$. For the molecules observed at 3\,mm we smoothed the maps with a Gaussian kernel to obtain the same resolution and derived the corresponding H$_2$ column density. For the molecules observed at 1\,mm we used the original dust continuum maps to calculate the H$_2$ column density. Thus we have two different H$_2$ column densities measured for each source on two different spatial scales.
While the H$_2$ column densities for the molecules at 3\,mm are beam matching, this is not the case for all sources for the molecules at 1\,mm. We slightly underpredict the H$_2$ column densities in the cases of the 1\,mm molecular line data, where we only have ATLASGAL or SCUBA data because of the larger beam.
To facilitate the modeling, we calculated for each angular resolution the corresponding physical radius, using estimates of the distances to the sources. Thus, we avoided effects from the heterogeneous nature in angular resolution of the continuum and spectroscopic data. For sources with a distance ambiguity we used the lowest value. Three of the sources show different velocity components in some transitions, resulting in different estimates of the kinematic distances. In these cases we chose the kinematic distance corresponding to the main velocity component that was detected in all transitions. 

Another uncertainty in the comparison of the continuum data and the molecular line data comes from the different techniques of measuring. The continuum data are observed with bolometer arrays and suffer for some degree from spatial filtering of the large-scale structures \citep[see][]{motte2001,schuller2009}. In contrast, the spectroscopic data are observed with heterodyne receivers and have no filtering problems. This is a general uncertainty in all comparisons between bolometer- and heterodyne-receiver-based measurements and leads to an underestimation of the true flux of the bolometer-based measurements. Hence, in our case the derived abundances are slightly overestimated. This effect is hard to quantify, but is not expected to play an important role in our case. It is reasonable to assume that the large-scale structure, which is filtered out, consists of diffuse gas with low density. A possible way to quantify this effect of missing large-scale structures in a first approximation is to take the 
canonical CO abundance of $10^{-4}$ and compare it with our derived values, which are about a factor 2-4 lower. Thus, the estimated H$_2$ column densities are even higher than the canonical values and the spatial filtering seems to have only a minor impact.

\subsection{Calculation of the molecular column densities} \label{sec:column_density}
To calculate column densities for the different molecules in Table~\ref{tbl:moleculeslist} we made several simplifying assumptions. The uncertainties introduced by these assumptions are discussed in Section~\ref{sec:uncertainties}. 1) Local thermodynamic equilibrium (LTE), since the densities in the observed high-mass star-forming regions are on the order of $10^5\,{\rm cm}^{-3}$ and higher, which exceeds the critical densities for many of the observed transitions. 2) A uniform gas kinetic temperature for all molecules observed in the same evolutionary stage. In the first iteration of the comparison with the model we assumed $T=15$~K for IRDCs, $T=50$~K for HMPOs, $T=100$~K for HMCs and $T=100$~K for UCH{\sc ii}s, respectively (iteration 0). In a second step we changed the assumed temperatures based on the outcome of the model (iteration 1). Due to the large beam sizes used to observe the high-mass star-forming regions, we took lower average temperature values for the later stages than the values observed 
with high-density tracers such as methyl cyanide. 3) The dust and gas temperatures are equal. 4) Line emission was assumed to be optically thin. For the species for which a rarer isotopologue line is also detected, we used this line to derive the column density (in particular, HN$^{13}$C, H$^{13}$CO$^+$, $^{13}$CS, C$^{33}$S, and C$^{18}$O). For N$_2$H$^+$ and HCN we fitted the hyperfine line structure and constrained the optical depth. 5) Derived column densities were beam averaged. 6) The medium is spatially homogeneous. 7) The Rayleigh-Jeans approximation of the Planck function was used. 8) No isotopic fractionation.
Because we only provide single-dish measurements with limited spatial resolution at single pointings, only mean values smoothed over the beam were derived.

\begin{table*}
\setlength{\tabcolsep}{5pt}
\small
\caption{List of analyzed molecules with transitions, frequencies, energies of the upper level, critical densities, and effective densities calculated for 10~K and 100~K. \tablefootmark{a}
}             
\label{tbl:moleculeslist}      
\centering                          
\begin{tabular}{lccccccp{1.5cm}}        
\hline\hline                 
Molecule & Transition & Frequency & E$_u$/k & n$_{\rm crit}$ for $T=10$~K & n$_{\rm crit}$ for $T=100$~K & n$_{\rm eff}$ for $T=10$~K & n$_{\rm eff}$ for $T=100$~K \\
 &  & $[{\rm GHz}]$ & $[\rm K]$ & $[10^5{\rm cm}^{-3}]$ & $[10^5{\rm cm}^{-3}]$ & $[10^3{\rm cm}^{-3}]$ & $[10^3{\rm cm}^{-3}]$ 
\\ 
\hline                        
H$^{13}$CO$^+$&   1-0     &    86.7543   &     4.2      &  1.5  & 2.1 & 3 & 0.9\\
SiO           &   2-1     &    86.8470   &     6.3      &  2.7  & 3.0 & 20 & 4 \\
HN$^{13}$C    &   1-0     &    87.0909   &     4.2      &    & ~10\tablefootmark{b} & - & - \\
C$_2$H        &   $1_{3/2,2}$-$0_{1/2,1}$ &    87.3169   & 4.2 &    &   ~1\tablefootmark{c} & - & - \\
HNCO          &   4(0,4)-3(0,3) &    87.9252   &   10.6 &  9.8\tablefootmark{d}  &   5.3\tablefootmark{e} & 30 & 3 \\
HCN           &   1-0     &    88.6316   &     4.3      &  10.1  &   26.6 & 29 & 5.1 \\
CH$_3$OCHO    &   7(2,5)-6(2,4) E &     90.1456   & 19.7 &    &   -   & - & - \\
CH$_3$CN      &   5-4 (K=2)   &    91.9800   &    41.8  &  4.2\tablefootmark{d}  &   2.7 & - & 20\\
$^{13}$CS     &   2-1     &    92.4943   &     6.7      &  3.3\tablefootmark{f}  & 3.8\tablefootmark{f} & - & - \\
N$_2$H$^+$    &   1-0     &    93.1737   &     4.5      &  1.4  &   2.0 & 4 & 2\\
H$_2$CO(para)\tablefootmark{g} &  3(0,3)-2(0,2) &   218.2221   &    21.0       &  23.5  &   50.3 & 100 & 15\\
CH$_3$OH\tablefootmark{h}     &   4(2,2)-3(1,2)   &     218.4401 & 45.5 & 133.9 &  781.1  & - & - \\
H$_2$CO(para)\tablefootmark{i} &  3(2,2)-2(2,1) &   218.4756   &    68.1       & 30.2   &   56.1 & - & 50 \\
OCS\tablefootmark{j}          &   18-17   &   218.9034   &    99.8      & 4.2  &  4.0 & - & - \\
C$^{18}$O     &   2-1     &   219.5604   &    15.8      & 0.08  &    0.10 & - & - \\
SO            &   6-5     &   219.9488   &    35.0      &  22.6\tablefootmark{k}  &   37.4 & - & 50\\
C$^{33}$S     &   5-4     &   242.9136   &    35.0      &  51.4\tablefootmark{f}  &    54.0\tablefootmark{f} & - & - \\
OCS\tablefootmark{l}          &   20-19   &   243.2180   &   122.6      & 5.9  &    5.6 & - & - \\
CH$_3$OH\tablefootmark{m}     &   5(1,4)-4(1,3) &   243.9158  & 49.7 & 8.0  &  15.5   & - & - \\
\hline                                   
\end{tabular}
\tablefoot{}
\tablefoottext{a}{Values are taken from the LAMDA database \citep{schoeier2005}, unless otherwise noted. Other species, visible in Figure~\ref{fig:samplespectra}, were not analyzed due to presumed high optical depth, as explained in the text. The effective densities were calculated with RADEX \citep{vandertak2007} assuming $\log \frac{N}{\Delta \nu}=13.5 \rm {cm}^{-2} (\rm {km s}^{-1})^{-1}$ \citep{evans1999}. For some molecules only the value for $T=100$~K is given, because at $T=10$~K the transition is too weak.}
\tablefoottext{b}{\citet{sakai2012}}
\tablefoottext{c}{\citet{sakai2010}}
\tablefoottext{d}{For $T=20$~K}
\tablefoottext{e}{For $T=80$~K}
\tablefoottext{f}{Value for the main isotopologue C$^{34}$S}
\tablefoottext{g}{Hereafter referred to as H$_2$CO-K0}
\tablefoottext{h}{Hereafter referred to as CH$_3$OH-4}
\tablefoottext{i}{Hereafter referred to as H$_2$CO-K2}
\tablefoottext{j}{Hereafter referred to as OCS-18}
\tablefoottext{k}{For $T=60$~K}
\tablefoottext{l}{Hereafter referred to as OCS-20}
\tablefoottext{m}{Hereafter referred to as CH$_3$OH-5}
\end{table*}

We calculated the molecular column densities following the equations in \citet{tielens2005}:

\begin{equation}
N_{u}=\frac{1.94 \cdot 10^3}{A_{ul}}\cdot \nu^{2}_{ul} \cdot \int T_{mb} \delta v
,\end{equation}

where the line frequency $\nu_{ul}$ is in GHz, the integrated intensity is in K\,km\,s$^{-1}$, and the Einstein coefficient $A_{ul}$ is in s$^{-1}$. Then the total column density can be calculated,

\begin{equation}
N_{tot} = N_{u} \cdot \frac{Q}{g_{u} exp[-E_{u}/kT_{ex}]}
,\end{equation}

where $Q$ is the partition function, $g_u$ the statistical weight of the upper level, $E_u$ the upper state energy, and $k$ the Boltzmann constant. $T_{ex}$ is the excitation temperature and assumed to be equal to $T_{kin}$.
Finally, we converted the derived column densities for the rarer isotopologues given in Table~\ref{tbl:moleculeslist} into the column densities of their main isotopologues. We hereby assumed for all sources the same relative isotopic ratios representative of the Sun and local ISM: $^{12}$C/$^{13}$C=89, $^{16}$O/$^{18}$O=499, and $^{32}$S/$^{33}$S=127 \citep{lodders2003}.
The molecular column densities were then divided by the H$_2$ column densities and the abundances were derived. The derived abundances for the molecules with transitions at 1\,mm are upper limits because of the differences in resolution between ATLASGAL/SCUBA and the 1\,mm molecular data. This does not affect the modeling, because we directly converted the beam sizes into physical sizes. The resulting median abundances including all detections and upper limits for each subsequent evolutionary phase are given in Tables~\ref{tbl:obsmedianabunirdc} -- \ref{tbl:obsmedianabunuch}. In addition to the molecules listed in Table~\ref{tbl:moleculeslist}, in some regions we detected also HC$_3$N, H$_2$CS, CH$_3$CHO, and other species. However, analyzing these is not a main focus of the current paper. The more detailed investigation of these species will be presented in future publications.

\subsection{Uncertainties in the observed values} \label{sec:uncertainties}
Our estimates of the column densities are based on simplifying basic assumptions, such as optically thin line limit, uniform excitation temperature for all observed molecules, and all sources being at a particular evolutionary stage, uncertainties in the derived H$_2$ column densities, which all need to be taken into account in the data analysis and further modeling.

We assumed a fixed dust opacity for all sources. In addition, we assumed a single excitation temperature for all sources and molecules in one evolutionary sample that is equal to the assumed kinetic temperature in this evolutionary stage. In this way, we did not account for source peculiarities or a possible misclassification of the stage or an overlap of several stages in a single source. Furthermore, the kinetic temperature of single molecules might deviate from the real excitation temperature in case of subthermally excited lines. This might especially be the case for the HMCs and UCHs, which have the highest kinetic temperatures. Comparing the calculated abundances derived with the high excitation temperature $T_{\rm ex}=100$~K with those derived with the lower excitation temperature of $T_{\rm ex}=50$~K, we found that the abundances decrease by  factors of $3-10$ in the case with the low $T_{\rm ex}$. HNCO, CH$_3$CN, CH$_3$OH, and CH$_3$OCHO are the most affected molecules.

We made the same exercise for the IRDCs and compared the abundances calculated with T$_{\rm ex}=15$~K with those calculated with $10$~K. The difference in the abundances reaches at most a factor of 3, with the exception of OCS, which shows a large difference of a factor of 15. We attributed this deviation to a non-LTE effect because a high J rotational line of OCS was employed, which might be subthermally excited even at typical densities of the high-mass star-forming regions. A more detailed discussion of the effect of uncertainties on the excitation temperature is given in Section~\ref{sec:comparison_obs_model}.
Furthermore, assuming optically thin lines implies that the calculated column densities may also be just lower limits.

Moreover, uncertainties in the conversion of the dust into gas densities constitute a factor of about 1.5 compared with \citet{draine2011}. The isotopic ratios vary between different environments and the uncertainties in the adopted isotopic ratios are a factor of $2-4$ \citep[e.g.,][]{wilson1994}. \citet{glassgold1985} found that the $^{12}$C/$^{13}$C ratio in PDRs might deviate even by more than a factor of 10 from the solar value. More recently, \citet{roellig2013} suggested strong fractionation effects in C-chemistry.

In combination, these systematic errors result in an overall uncertainty of about one order of magnitude in the derived molecular abundances. This level of uncertainty is typical for astrochemical studies. It is present in the assumptions made to analyze observational data as well as in the intrinsic uncertainties of chemical models \citep[e.g.,][]{wakelam2010,vasyunina2011,vasyunina2012,albertsson2013}. Improvements of this situation are an area of active study.

In this work, these uncertainties were taken into account during the data fitting, as explained in Section~\ref{sec:fitting}. To avoid suffering strongly from possible misclassifications and peculiarities of single sources, we treated the sample in a statistical way and were mainly interested in the median characteristics of a subsample. That might lead to larger spreads within one subsample, but makes the analysis and comparison with the model more robust and reliable.

\subsection{Chemical evolution}

\begin{table*}
\caption{Observed median abundances and standard deviation in $a(x)=a \cdot 10^{x}$ for the IRDCs for $T=15$~K and $T=20.9$~K. x(e) is the ionization fraction.}             
\label{tbl:obsmedianabunirdc}      
\centering                          
\begin{tabular}{lccccccc}        
\hline\hline                 
Molecule & Abundance & Standard deviation & Abundance & Standard deviation & det & lim & none \\
 & $T=15$~K & & $T=20.9$~K\tablefootmark{a} & & & & 
\\    
\hline                        
HCN            &  1.6(-09) &  1.1(-09) & 3.2(-09) & 2.1(-09) &  7  &   0  &  12 \\
HNCO           &  8.3(-11) &  7.3(-11) & 1.8(-10) & 1.5(-10) & 17  &   2  &   0 \\
HN$^{13}$C     &  5.8(-11) &  4.3(-11) & 1.2(-10) & 8.6(-11) & 19  &   0  &   0 \\
C$_2$H         &  1.2(-08) &  6.3(-09) & 2.3(-08) & 1.2(-08) & 19  &   0  &   0 \\
SiO            &  8.4(-11) &  1.3(-10) & 1.6(-10) & 2.5(-10) & 12  &   7  &   0 \\
H$^{13}$CO$^+$ &  4.5(-11) &  2.6(-11) & 9.1(-11) & 5.2(-11) & 19  &   0  &   0 \\
N$_2$H$^+$     &  6.0(-10) &  3.3(-10) & 9.6(-10) & 5.4(-10) & 19  &   0  &   0 \\
$^{13}$CS      &  $\leq$6.1(-11) &  4.5(-11) & $\leq$1.3(-10) & 8.7(-11) &  8  &  11  &   0 \\
SO             &  $\leq$1.9(-10) &  1.8(-10) & $\leq$2.4(-10) & 2.3(-10) &  8  &  11  &   0 \\%
C$^{18}$O      &  5.6(-08) &  2.1(-08) & 9.0(-08) & 3.4(-08) & 19  &   0  &   0 \\%
H$_2$CO        &  6.6(-10) &  1.4(-09) & 5.3(-10) & 9.9(-10) & 24  &  13  &   1 \\%
C$^{33}$S      &  $\leq$1.0(-10) &  1.4(-10) & $\leq$1.1(-10) & 1.7(-10) &  0  &  19  &   0 \\
OCS            &  $\leq$3.7(-08) &  7.3(-08) & $\leq$1.1(-08) & 1.6(-08) &  2  &  36  &   0 \\%
CH$_3$OH       &  $\leq$9.9(-10) &  1.3(-09) & $\leq$1.2(-09) & $\leq$1.7(-09) & 16  &  22  &   0 \\
CH$_3$OCHO     &  $\leq$2.1(-10) &  5.3(-10) & $\leq$3.9(-10) & 3.1(-10) &  0  &  19  &   0 \\%
CH$_3$CN       &  $\leq$1.5(-10) &  9.9(-11) & $\leq$1.8(-10) & 1.1(-10) &  2  &  17  &   0 \\%
x(e)           &  4.6(-09) &  2.6(-09) & 9.0(-09) & 5.0(-09) & 19  &   0  &   0 \\%
\hline                                   
\end{tabular}
\tablefoot{The last three columns show the number of sources with a detection (det), non-detection (lim), and detection but no derived column density (none) due to reasons described in Section~\ref{sec:problems}.}
\tablefoottext{a}{Mean value of $T$ for the best-fit model.}
\end{table*}


\begin{table*}
\caption{Observed median abundances and standard deviation in $a(x)=a \cdot 10^{x}$ for the HMPOs for $T=50$~K and $T=29.5$~K. x(e) is the ionization fraction.}             
\label{tbl:obsmedianabunhmpo}      
\centering                          
\begin{tabular}{lccccccc}        
\hline\hline                 
Molecule & Abundance & Standard deviation & Abundance & Standard deviation & det & lim & none\\
 & $T=50$~K & & $T=29.5$~K\tablefootmark{a} &  & & &
\\    
\hline                        
HCN            &   1.1(-08) &  7.1(-09)  & 3.8(-09) & 2.4(-09) & 14  &   0  &   6 \\
HNCO           &   1.8(-10) &  3.4(-10)  & 7.7(-11) & 1.5(-10) & 17  &   3  &   0 \\%
HN$^{13}$C     &   3.7(-10) &  1.9(-10)  & 1.3(-10) & 6.6(-11) & 18  &   2  &   0 \\
C$_2$H         &   1.4(-07) &  3.8(-08)  & 5.0(-08) & 1.4(-08) & 20  &   0  &   0 \\
SiO            &   4.0(-10) &  4.0(-10)  & 1.4(-10) & 1.4(-10) & 16  &   4  &   0 \\
H$^{13}$CO$^+$ &   4.5(-10) &  1.4(-10)  & 1.5(-10) & 4.8(-11) & 20  &   0  &   0 \\
N$_2$H$^+$     &   2.4(-09) &  1.4(-09)  & 1.1(-09) & 5.7(-10) & 20  &   0  &   0 \\%
$^{13}$CS      &   9.0(-10) &  7.5(-10)  & 3.2(-10) & 2.6(-10) & 17  &   3  &   0 \\%
SO             &   1.3(-09) &  9.4(-10)  & 6.4(-10) & 4.6(-10) & 20  &   0  &   0 \\
C$^{18}$O      &   2.7(-07) &  1.2(-07)  & 1.1(-07) & 4.9(-08) & 20  &   0  &   0 \\%
H$_2$CO        &   7.1(-10) &  5.6(-10)  & 4.7(-10) & 6.0(-10) & 38  &   1  &   1 \\%
C$^{33}$S      &   $\leq$1.3(-10) &  1.2(-10)  & $\leq$6.8(-11) & 6.5(-11) &  8  &  12  &   0 \\
OCS            &   $\leq$1.3(-09) &  1.6(-09)  & $\leq$2.0(-09) & 2.7(-09) & 17  &  23  &   0 \\
CH$_3$OH       &   9.1(-10) &  1.6(-09)  & 1.0(-09) & 1.5(-09) & 32  &   8  &   0 \\
CH$_3$OCHO     &   $\leq$2.5(-10) &  2.7(-10)  & $\leq$2.1(-10) & 2.3(-10) &  0  &  20  &   0 \\
CH$_3$CN       &   1.3(-10) &  1.6(-10)  & 1.0(-10) & 1.2(-10) & 13  &   7  &   0 \\%
x(e)           &   4.3(-08) &  1.3(-08)  & 1.5(-08) & 4.6(-09) & 20 &    0 &    0 \\%
\hline                                   
\end{tabular}
\tablefoot{The last three columns show the number of sources with a detection (det), non-detection (lim), and detection but no derived column density (none) due to reasons described in Section~\ref{sec:problems}.}
\tablefoottext{a}{Mean value of $T$ for the best-fit model.}
\end{table*}

\begin{table*}
\caption{Observed median abundances and standard deviation in $a(x)=a \cdot 10^{x}$ for the HMCs for $T=100$~K and $T=40.2$~K. x(e) is the ionization fraction.}             
\label{tbl:obsmedianabunhmc}      
\centering                          
\begin{tabular}{lccccccc}        
\hline\hline                 
Molecule & Abundance & Standard deviation & Abundance & Standard deviation & det & lim & none\\
 & $T=100$~K & & $T=40.2$~K\tablefootmark{a} &  &  &  & 
\\    
\hline                        
HCN            &  4.9(-08) &         0  & 7.4(-09) &        0 &  1  &   0  &  10 \\
HNCO           &  1.6(-09) &  1.3(-09)  & 1.5(-10) & 1.3(-10) & 10  &   1  &   0 \\
HN$^{13}$C     &  4.9(-10) &  4.9(-10)  & 7.5(-11) & 7.4(-11) & 11  &   0  &   0 \\
C$_2$H         &  3.1(-07) &  1.6(-07)  & 4.6(-08) & 2.4(-08) & 11  &   0  &   0 \\
SiO            &  1.3(-09) &  2.5(-09)  & 2.1(-10) & 3.9(-10) & 11  &   0  &   0 \\
H$^{13}$CO$^+$ &  8.2(-10) &  6.7(-10)  & 1.3(-10) & 1.0(-10) & 11  &   0  &   0 \\
N$_2$H$^+$     &  7.8(-10) &  2.5(-09)  & 2.7(-10) & 4.9(-10) & 11  &   0  &   0 \\
$^{13}$CS      &  5.6(-09) &  3.0(-09)  & 8.8(-10) & 4.7(-10) & 11  &   0  &   0 \\
SO             &  1.3(-08) &  2.4(-08)  & 2.8(-09) & 5.0(-09) & 11  &   0  &   0 \\
C$^{18}$O      &  1.7(-06) &  6.1(-07)  & 3.0(-07) & 1.1(-07) & 11  &   0  &   0 \\
H$_2$CO        &  9.3(-09) &  4.9(-09)  & 1.9(-09) & 1.4(-09) & 20  &   0  &   2 \\
C$^{33}$S      &  1.8(-09) &  1.9(-09)  & 4.3(-10) & 4.5(-10) & 11  &   0  &   0 \\
OCS            &  1.3(-08) &  1.1(-08)  & 9.6(-09) & 9.1(-09) & 22  &   0  &   0 \\
CH$_3$OH       &  2.6(-08) &  3.7(-08)  & 3.6(-09) & 4.3(-09) & 22  &   0  &   0 \\
CH$_3$OCHO     &  $\leq$2.5(-09) &  2.7(-09)  & $\leq$2.4(-10) & 2.5(-10) &  5  &   6  &   0 \\
CH$_3$CN       &  2.9(-09) &  1.9(-09)  & 4.4(-10) & 2.9(-10) & 10  &   1  &   0 \\
x(e)           &  7.7(-08) &  6.1(-08)  & 1.2(-08) & 9.5(-09) & 11  &   0  &   0 \\
\hline                                   
\end{tabular}
\tablefoot{The last three columns show the number of sources with a detection (det), non-detection (lim), and detection but no derived column density (none) due to reasons described in Section~\ref{sec:problems}.}
\tablefoottext{a}{Mean value of $T$ for the best-fit model.}
\end{table*}

\begin{table*}
\caption{Observed median abundances and standard deviation in $a(x)=a \cdot 10^{x}$ for the UCH{\sc ii}s for $T=100$~K and $T=36.0$~K. x(e) is the ionization fraction.}             
\label{tbl:obsmedianabunuch}      
\centering                          
\begin{tabular}{lccccccc}        
\hline\hline                 
Molecule & Abundance & Standard deviation & Abundance & Standard deviation & det & lim & none\\
 & $T=100$~K & & $T=36.0$~K\tablefootmark{a} &  & &  & 
\\    
\hline                        
HCN            &  4.6(-08) &  1.4(-08)  & 5.6(-09) & 1.6(-09) &  5  &   0  &   4 \\
HNCO           &  1.1(-09) &  1.3(-09)  & 8.6(-11) & 9.9(-11) &  6  &   3  &   0 \\
HN$^{13}$C     &  7.6(-10) &  1.0(-09)  & 9.2(-11) & 1.3(-10) &  9  &   0  &   0 \\
C$_2$H         &  3.9(-07) &  2.3(-07)  & 4.6(-08) & 2.7(-08) &  8  &   1  &   0 \\
SiO            &  3.7(-10) &  1.2(-09)  & 4.6(-11) & 1.6(-10) &  5  &   4  &   0 \\
H$^{13}$CO$^+$ &  1.0(-09) &  4.5(-10)  & 1.3(-10) & 5.4(-11) &  9  &   0  &   0 \\
N$_2$H$^+$     &  4.1(-09) &  4.1(-09)  & 7.4(-10) & 9.2(-10) &  9  &   0  &   0 \\
$^{13}$CS      &  6.3(-09) &  5.3(-09)  & 7.9(-10) & 6.7(-10) &  9  &   0  &   0 \\
SO             &  5.5(-09) &  2.2(-09)  & 1.0(-09) & 4.0(-10) &  5  &   1  &   3 \\
C$^{18}$O      &  1.3(-06) &  9.9(-07)  & 2.0(-07) & 1.5(-07) &  8  &   1  &   0 \\
H$_2$CO        &  4.9(-09) &  3.7(-09)  & 7.2(-10) & 8.8(-10) & 16  &   2  &   0 \\
C$^{33}$S      &  1.2(-09) &  6.3(-10)  & 2.5(-10) & 1.3(-10) &  7  &   2  &   0 \\
OCS            &  $\leq$1.6(-09) &  3.5(-09)  & $\leq$1.3(-09) & 3.2(-09) &  6  &  12  &   0 \\
CH$_3$OH       &  8.7(-09) &  2.4(-08)  & 9.7(-10) & 2.5(-09) & 10  &   8  &   0 \\
CH$_3$OCHO     &  $\leq$1.3(-09) &  3.0(-09)  & $\leq$9.7(-11) & 2.3(-10) &  0  &   9  &   0 \\
CH$_3$CN       &  1.3(-09) &  1.2(-09)  & 1.6(-10) & 1.6(-10) &  7  &   2  &   0 \\
x(e)           &  9.4(-08) &  4.3(-08)  & 1.2(-08) & 5.6(-09) &  9  &   0  &   0 \\
\hline                                   
\end{tabular}
\tablefoot{The last three columns show the number of sources with a detection (det), non-detection (lim), and detection but no derived column density (none) due to reasons described in Section~\ref{sec:problems}.}
\tablefoottext{a}{Mean value of $T$ for the best-fit model.}
\end{table*}

\begin{figure*}
\centering
\includegraphics[angle=0,width=0.86\textwidth]{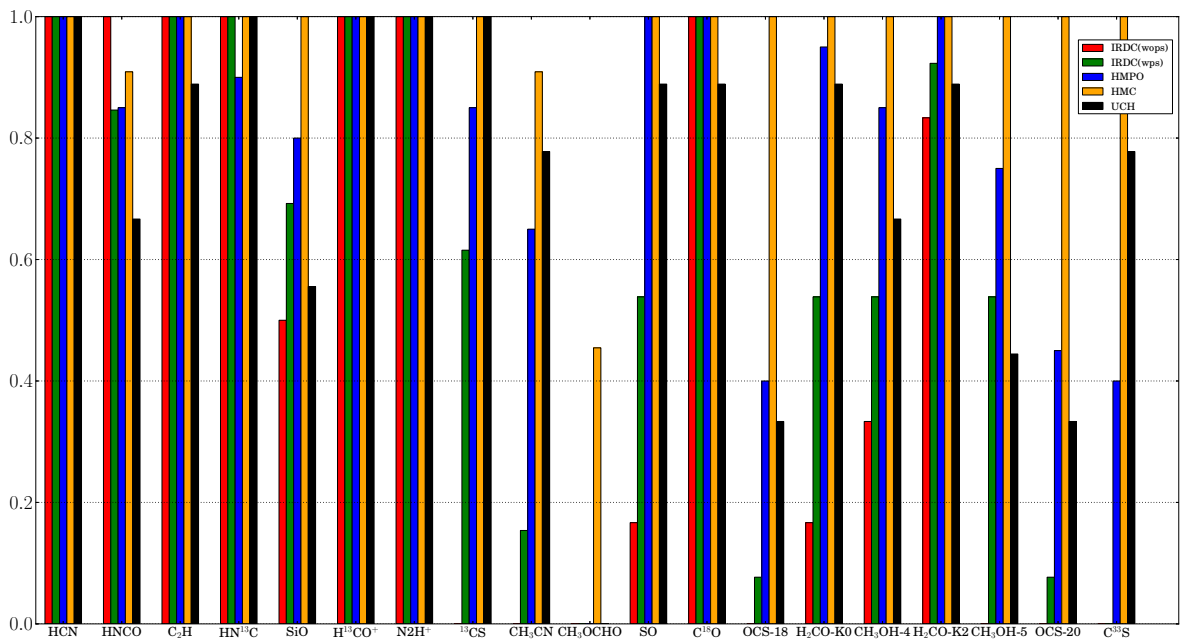}
\caption{Relative detection fractions for each analyzed molecular transition ordered from left to right: IRDCs without associated point sources at wavelengths below 70 $\mu$m, IRDCs with associated point sources, HMPOs, HMCs, and UCH{\sc ii}s. The color notation for the different source types is explained in the upper right corner.}
\label{fig:detectionfractions}
\end{figure*}

\begin{figure*}
\centering
\includegraphics[angle=0,width=0.68\textwidth]{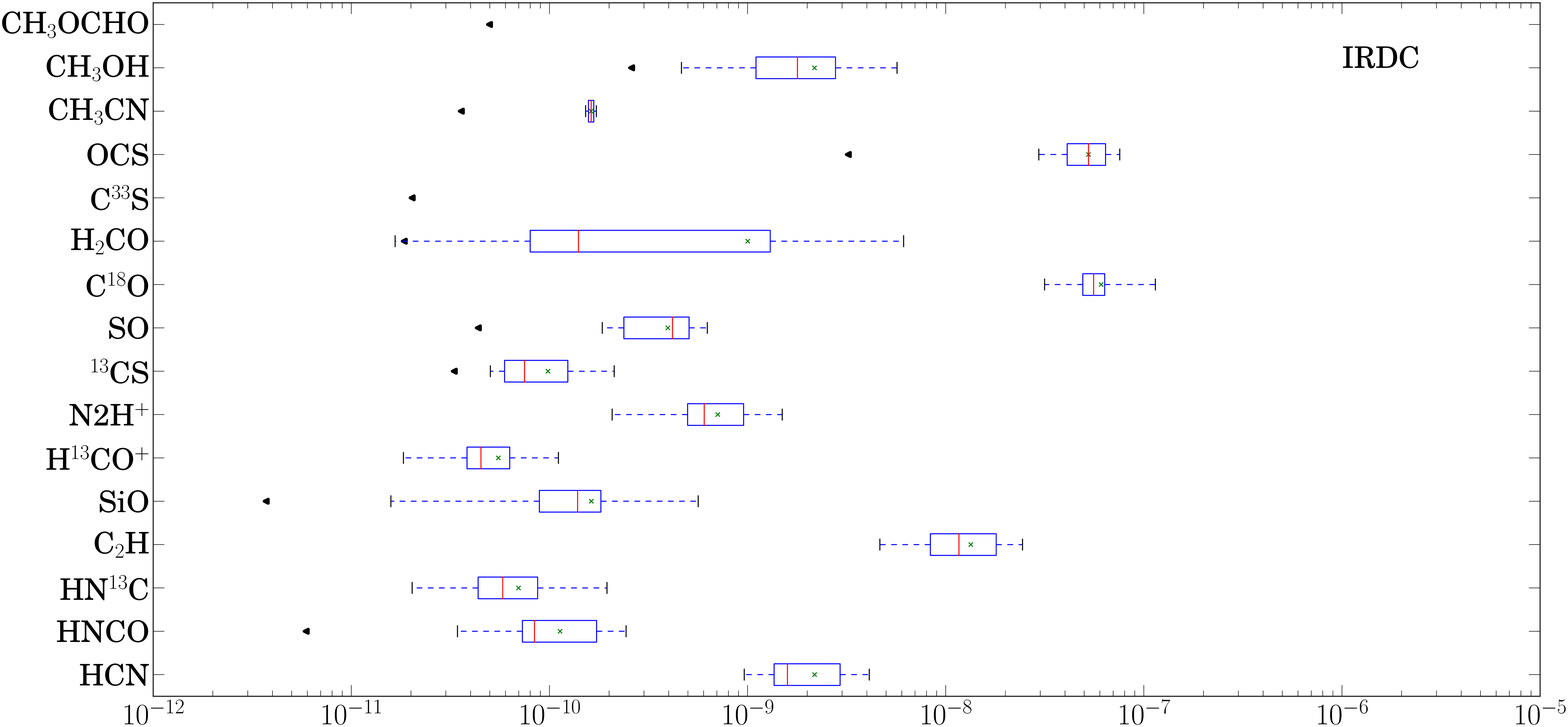}\\
\includegraphics[angle=0,width=0.68\textwidth]{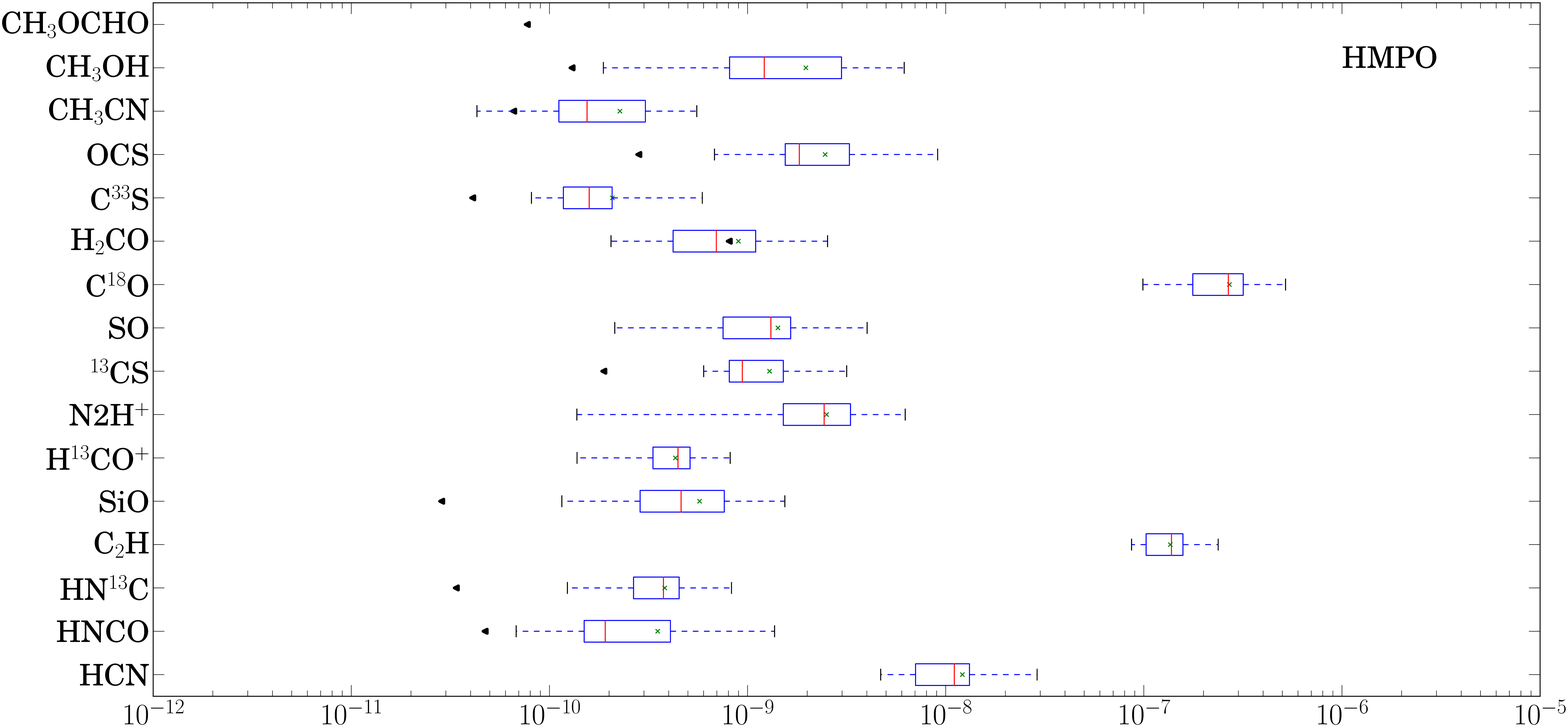}\\
\includegraphics[angle=0,width=0.68\textwidth]{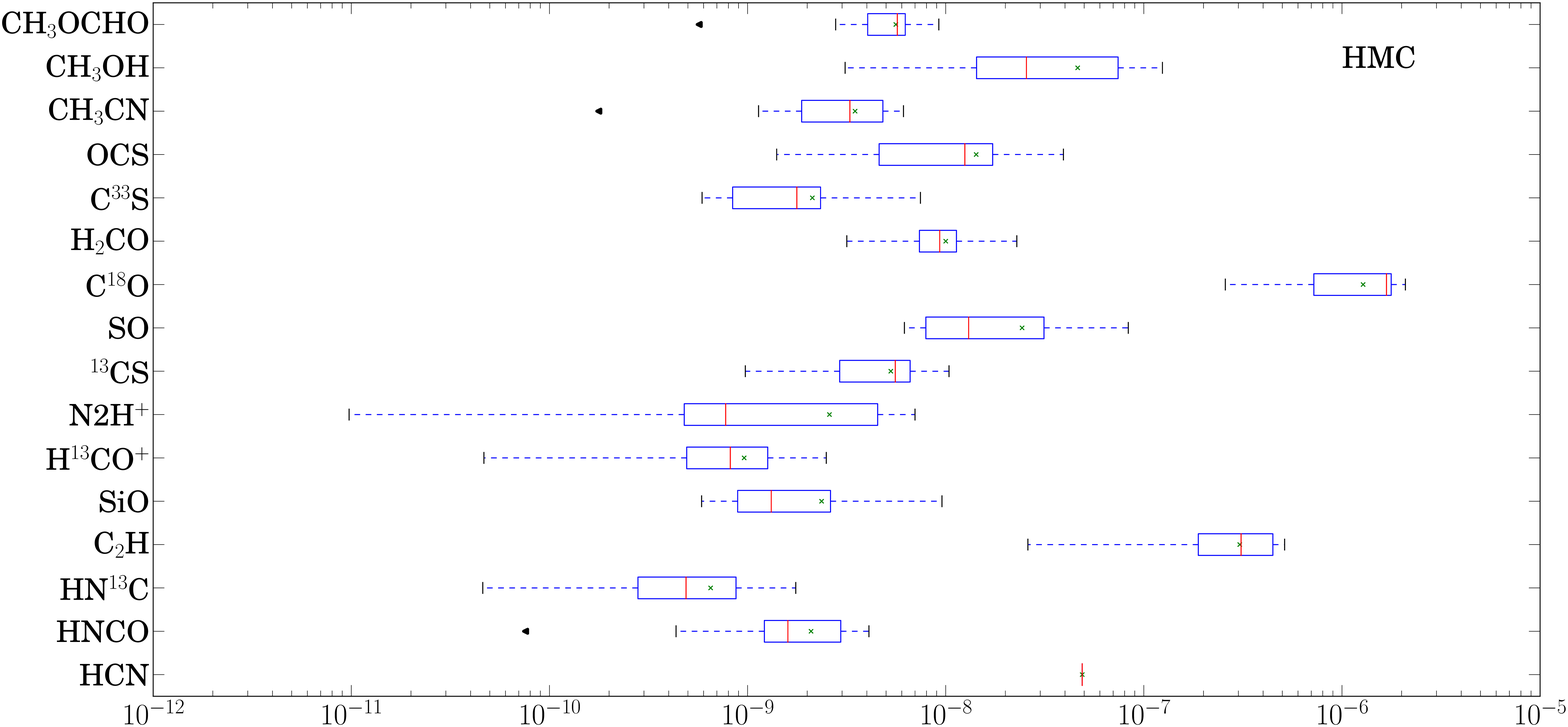}\\
\includegraphics[angle=0,width=0.68\textwidth]{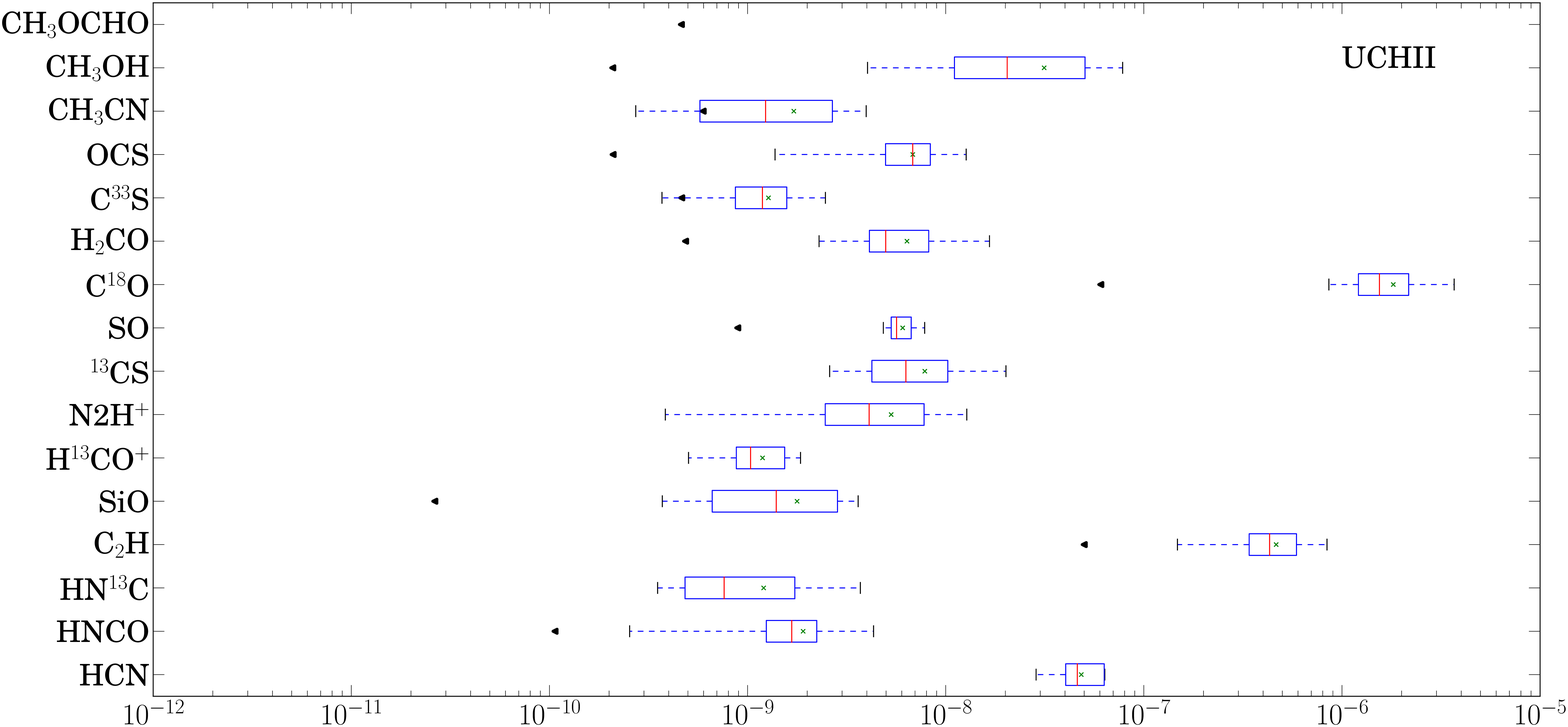}
\caption{From top to bottom abundances (with respect to H$_2$) of the analyzed molecules in the IRDC, HMPO, HMC and UCH{\sc ii} sample. The red line shows the median, the green cross is the mean, the bar indicates the inner 25\%-75\% range around the median and the whiskers mark the total range of all detections. The black arrows indicate the lowest upper limit of all calculated upper limits for this particular molecule and stage.}
\label{fig:detection}
\end{figure*}

In Sections~\ref{sec:column_densityh2} and \ref{sec:column_density} we calculated the column densities of H$_2$  and other molecular species. From these values we derived averaged abundances for each source. In the next step we combined all these abundances from all sources at a specific evolutionary stage and derived a characteristic chemical ''portrait`` for each phase. The results are given in Tables~\ref{tbl:obsmedianabunirdc} -- \ref{tbl:obsmedianabunuch} together with the detection fractions. The detection fractions are plotted in Figure~\ref{fig:detectionfractions} and the abundances are plotted in Figure~\ref{fig:detection} for the IRDCs, HMPOs, HMCs and UCH{\sc ii} regions. The individual column densities for each stage are presented in the online appendix in Figures~\ref{fig:coldens_irdc} -- \ref{fig:coldens_uch}.

All abundances in all four evolutionary stages have values ranging between $10^{-11}$ and $10^{-6}$. The spread of column densities within a particular stage is about one order of magnitude or lower, except for H$_2$CO in the IRDCs and N$_2$H$^+$ in the HMCs with spreads of almost three orders of magnitude.
Within the two subsamples of IRDCs (with and without associated point sources at wavelengths below 70~$\mu$m) the difference in median abundances does not exceed a factor of 2. But in terms of detected molecules we found a stronger difference. The molecules $^{13}$CS, CH$_3$CN, and OCS are only detected in the more evolved IRDCs. The detection fractions of SiO, SO, and CH$_3$OH are at least 20\% lower toward IRDCs without a point source than in sources with an embedded point source. This is a sign that the latter group is already at a slightly more advanced chemical evolutionary stage.

In general, the detected abundances over all evolutionary stages tend to increase with rising temperature during the evolution. To identify the chemical traces of the evolution we compared the abundances within the  25\%-75\% range of the median value between the different evolutionary phases. The abundances in the IRDC stage are around $10^{-10}$ and $10^{-9}$, with the
OCS, C$_2$H, and the CO isotopologue having magnitutedes higher
by one to two orders of magnitudes. Compared with the HMPOs, all abundances in the IRDCs are lower by about a half to one order of magnitude. Exceptions are the organic species CH$_3$OH, CH$_3$CN, and H$_2$CO, which have abundances similar to those in the HMPOs. Moreover, OCS is remarkably much more abundant, but only detected in IRDC028.2. The overall detection rate in the IRDCs is lower than that of the HMPOs.
Next, in the HMPOs the abundances have values mainly in the range of $10^{-10}$ to $10^{-8}$. They are even higher in the HMCs. Only N$_2$H$^+$ has similar abundances across these stages. The abundances of H$^{13}$CO$^+$ and HN$^{13}$C, while still higher in the HMC phase, are closer to the respective HMPO values than the values of other molecules.
The abundances of most of the molecules in the HMCs have values between $10^{-9}$ and $10^{-7}$. This is the only phase where we detected the most complex molecule analyzed in our sample, methyl formate CH$_3$OCHO. However, methyl formate was detected in IRDC028.2 by \citet{vasyunina2013}, which means that this non-detection might be a combination of lower sensitivity and a weaker transition.

From the HMCs to the UCH{\sc ii} stage, abundances of CH$_3$OH, CH$_3$CN, OCS, C$^{33}$S, H$_2$CO, SO, and SiO become lower. The remaining molecules have similar or slightly higher abundances in the UCH{\sc ii} stage. However, the overall detection rate of the different molecules for the HMCs is much higher than for the UCH{\sc ii}s. Complex or heavy molecules such as  CH$_3$OCHO, CH$_3$OH, SiO, HNCO, and OCS are not detected in $\ga 30\%$ of the UCH{\sc ii}s and are almost always found in the HMCs.

\subsection{Ionization degree} \label{sec:ionizationdegree}
In our spectral setup we observed two molecular ions, H$^{13}$CO$^+$ and N$_2$H$^+$. Together with the observed H$_2$ densities this enables us to calculate lower limits of the electron fraction following \citet{caselli2002a}. This method is based on calculations from \citet{caselli1998} using a simple chemical model and is only valid for homogeneous clouds. Since the deuterated counterparts are not covered by our setup we were only able to derive the electron fraction based on these two molecular species. The median electron fractions are given in Tables~\ref{tbl:obsmedianabunirdc} -- \ref{tbl:obsmedianabunuch}. The median value increases with evolutionary phase from $5 \cdot 10^{-9}$ in the IRDCs to $\sim10^{-8}-10^{-7}$ in the UCH{\sc ii} regions (depending on the assumed excitation temperature). This behavior agrees with the scenario of an increasing ionization radiation during the evolution of the central source that finally reaches its maximum in the UCH{\sc ii} phase. The derived values are also 
consistent with a lower limit 
from Caselli et al. of $~10^{-9}$ and the work by \citet{miettinen2011}, who found lower limits of $\sim10^{-8}-10^{-7}$ in massive clumps associated with IRDCs.

\section{Discussion}\label{sec:discussion}
In the previous section we described the observational results about the chemical composition of the different evolutionary phases. In the next step we model the chemistry over the full evolutionary range. A detailed background of the presumed evolutionary sequence is outlined in the introduction in Section~\ref{sec:introduction}. Furthermore, combining the information from observations and model, we give a comprehensive description of this evolutionary sequence in Section~\ref{sec:chemevolsequence}.

\subsection{Model}
\label{sec:Model}
In this section we describe our iterative chemical fitting model `MUSCLE' (`MUlti Stage CLoud codE'), which we used to
constrain mean physical properties and chemical ages during the evolution from the earliest IRDC phase till the 
late UCH{\sc ii} phase of the high-mass star formation.

\subsubsection{Physical model}
\label{sec:Phys_model}
We modeled each evolutionary stage in a 1D approximation. Each environment is spherically symmetric, homogeneous, and has a fixed outer radius of $r_\mathrm{out}=0.5$~pc. This radius is close to a typical value for dense parts of high-mass star-forming regions and also represents the largest $29\arcsec$ IRAM beam size used in our single-dish observations as derived for the mean distance of the sample. The radial density and temperature profiles are modeled by modified power laws,
\begin{equation}
\label{eq:rho_r}
\begin{array}{ll}

 \rho(r) = \rho_{\rm {in}}(r/r_{\rm {in}})^{-p},  & r\ge r_{\rm {in}};\\
 \rho(r) = \rho_{\rm {in}},              &r<r_{\rm {in}}
 
\end{array}
\end{equation}

and

\begin{equation}
\label{eq:temp_r}
\begin{array}{ll}

 T(r) = T_{\rm {in}}(r/r_{\rm {in}})^{-q},  & r\ge r_{\rm {in}};\\
 T(r) = T_{\rm {in}},              &r<r_{\rm {in}},

\end{array}
\end{equation}
respectively. Here, $r_{\rm {in}}$ is the inner radius below which temperature and density are assumed to be constants and $p$ and $q$ are the corresponding radial profiles of density and temperature. The parameters $r_{\rm {in}}$, $p$, $\rho_{\rm {in}}$ and $T_{\rm {in}}$ are the fitted physical quantities. $r_{\rm {in}}$ was varied between $10^{-4}-10^{-2}$ of $r_{\rm {out}}$. The gas and dust were assumed to be in thermal equilibrium, therefore we did not calculate dust and gas temperatures
separately. Eq.~\ref{eq:rho_r} was numerically integrated from $r=0$ till $r=r_{\rm out}$ to derive the total modeled H$_2$ column density of a cloud.
For that, a grid of $30-40$ radial cells was found to provide sufficient accuracy, and hence was adopted in large-scale chemical modeling.

We assumed that at each stage a cloud is furthermore embedded in large-scale, low-density matter that shields it from the interstellar FUV radiation by visual extinction of 10~mag.
For UCH{\sc ii} regions we assumed that the size of a forming Str\"{o}mgren zone is still small, $\la 0.1$~pc, such that it can be neglected for calculations of bulk beam-averaged quantities such as molecular column densities.

\subsubsection{Chemical model}
\label{sec:Chem_model}
The adopted time-dependent gas-grain chemical model based on the `ALCHEMIC' code is fully described in \citet{semenov2010}.
A brief summary of the updates is provided below. 
The chemical rate file is based on the 2007 realization of the OSU network\footnote{See:
\url{http://www.physics.ohio-state.edu/~eric/research.html}}. The network is supplied with
a set of $\sim 1000$ reactions with high-temperature barriers from \citet{Harada_ea10} and \citet{Harada_ea12}.
The recent updates as of September 2012 to the reaction rates are implemented (e.g., from KIDA
database\footnote{\url{http://kida.obs.u-bordeaux1.fr}}), see \citet{albertsson2013}.
We considered cosmic-ray particles (CRP) and CRP-induced FUV radiation as the only external ionizing sources.
The  CRP ionization rate $\zeta_{\rm CR}=5 \cdot 10^{-17}$~s$^{-1}$ was used.
The UV dissociation and ionization photorates from \citet{vDea_06}\footnote{\url{http://www.strw.leidenuniv.nl/~ewine/photo/}}
were adopted, assuming the case corresponding to the spectral shape of the interstellar FUV radiation field.

In addition to pure gas-phase chemical processes, the chemical network includes gas-grain interactions.
We assumed that molecules  stick to grain surfaces at low temperatures with 100\% probability.
We did not allow H$_2$ to stick to grains
because the binding energy of H$_2$ to pure H$_2$ mantle is low, $\sim 100$~K \citep{Lee_72}.
The ices are released back to the gas phase by thermal, CRP-, and CRP-induced UV-photodesorption. The grain
re-charging was modeled by dissociative recombination and radiative neutralization of ions on grains, electron sticking to
grains. Chemisorption of surface molecules was not considered. We used the UV photodesorption yield for ices of $10^{-3}$
\citep[e.g.,][]{Oeberg_ea09b,Oeberg_ea09a}.

The synthesis of complex  molecules was included using a set of surface reactions (together with desorption energies) and
photodissociation reactions of ices from \citet{Garrod_Herbst06} \citep[see also][]{Semenov_Wiebe11}.
We assumed that each $0.1\,\mu$m spherical olivine grain provides $\approx 1.88 \cdot 10^6$ surface sites for accreting gaseous species.
The surface recombination  solely through the Langmuir-Hinshelwood formation mechanism is considered.
Upon a surface recombination, the reaction products are assumed to remain on the grain as the grain lattice would absorb all energy
released during the recombination.
Following interpretations of experimental results on the formation of molecular hydrogen on dust grains \citep{Katz_ea99}, we employed the standard rate equation approach to the surface chemistry without H and H$_2$ tunneling  through the potential walls of the surface sites.
Overall, our chemical network consists of 656 species made of 13 elements, and 7907 reactions. No chemical parameter was
varied during the iterative fitting of the observational data.

Our ALCHEMIC Fortran code \citep{semenov2010}
is based on the double-precision variable-coefficient ordinary differential equation solver
with the preconditioned Krylov (DVODPK) method GMRES for the solution of linear
systems\footnote{\url{http://www.netlib.org/ode/vodpk.f}}. The approximate Jacobi matrix is generated automatically
from the supplied chemical network. For astrochemical models
dominated by hydrogen reactions the Jacobi matrix is sparse, with $\la
1\%$ of non-zero elements. The corresponding linearized system of algebraic equations is solved using a high-performance sparse
unsymmetric MA48 solver from the Harwell Mathematical Software Library\footnote{\url{http://www.hsl.rl.ac.uk/}}.

\subsubsection{Initial abundances}
The initial abundances very close to the low-metals set of \citet{lee1998} were used to model the chemistry at the IRDC evolutionary stage (see Table~\ref{tab:inabun}). The only difference was the initially atomic abundance of Si ($3\cdot10^{-9}$ with respect to H) and S ($8\cdot10^{-7}$ with respect to H), which we had to use to achieve a better fit to the observational data. Other initial abundance sets with different amounts of oxygen and sulfur, which have been proposed in studies of chemistry in molecular clouds and protoplanetary disks, did not produce as good a fit to our IRDC data \citep[see e.g.,][]{Wakelam_Herbst08,Graedel_ea82,Jenkins_09,Hincelin_ea11,Dutrey_ea11}. For modeling the chemistry of the evolutionary stages after the IRDC stage, the final chemical abundances from the best-fit model of the previous evolutionary stage were used as initial abundances (i.e., to model  HMPOs we used the molecular abundances of the best-fit IRDC model at the best-fit chemical age). This 
allowed us to reproduce an approximately steady warming of the matter during the evolution of the high-mass star-forming clouds.

\begin{table}
\caption{Initial atomic and molecular abundances.}             
\label{tab:inabun}      
\centering                          
\begin{tabular}{ll}        
\hline\hline
Species & Relative abundance 
\\    
\hline                        
H$_2$&   $0.499$     \\
H    &   $2.00 (-3)$  \\
He   &   $9.75 (-2)$  \\
C    &   $7.86 (-5)$  \\
N    &   $2.47 (-5)$  \\
O    &   $1.80 (-4)$  \\
S    &   $8.00 (-7)$  \\
Si   &   $3.00 (-9)$  \\
Na   &   $2.25 (-9)$  \\
Mg   &   $1.09 (-8)$  \\
Fe   &   $2.74 (-9)$  \\
P    &   $2.16 (-10)$ \\
Cl   &   $1.00 (-9)$  \\
\hline                                   
\end{tabular}
\end{table}

\subsubsection{Iterative fitting of the data}\label{sec:fitting}
Using the physical and chemical models described above, we iteratively fitted the mean observed molecular column densities for each individual evolutionary stage by varying the inner radius $r_{\rm {in}}$, the density at the inner radius $\rho_{\rm {in}}$, the temperature at the inner radius $T_{\rm {in}}$ and the density slope $p$. First, for each 1D cloud model with a grid of 30 cells we derived appropriate ranges of $r_{\rm {in}}$, the $\rho_{\rm {in}}$, and the density profile $\rho$ by calculating beam-averaged H$_2$ column densities such that they are within a factor of 3 to the observationally derived values. To limit the computational time and storage space required for our large-scale modeling, we restricted the density profile parameter $p$ to lie between the values of 1.5 and 2.0, as seen in a variety of observations by \citet{beuther2002a,mueller2002} and \citet{hatchell2003}. The inner radius was varied between $5\cdot10^{-5}-5\cdot10^{-2}$~pc.
Another parameter whose value was iteratively varied was the temperature at the inner radius, $T_{\rm {in}}$ (between $15-27$~K for IRDCs, $50-250$~K for HMPOs, $150-500$~K for HMCs and UCH{\sc ii} regions).  All other physical parameters were fixed during iterations, including the temperature profile $q$, which we set prior to the modeling. In the first stage we assumed an isothermal sphere with $q=0$ 
and in the later stages a temperature profile with fixed $q=0.4$ as a standard value. Simulations of massive star formation by \citet{vandertak2000c} found that the temperature profiles have steeper indices in the inner regions, but flatten outside the inner 2000 AU and run asymptotically into a distribution with $q=0.4$. Typically, by varying the parameters $r_{\rm {in}}$, $\rho_{\rm {in}}$, $T_{\rm {in}}$ and $p$ for each stage, one model realization consists of 500-3\,000 cloud models.

Using this large set of cloud models and our chemical model described above, we calculated corresponding time-dependent chemical structures over 1~Myr (with 99 logarithmically taken time steps between $10^4$ and $10^6$~years). After that, we calculated the corresponding beam-averaged column densities and the goodness of the fit for each cloud model, observed molecule,
and each time step by using a confidence criterion from Equation~3 in \citet{Garrod_ea07}. As the standard deviation $\sigma$ for all observed molecules we assumed an order of magnitude variation between the modeled and the observed column densities. For molecules for which only the upper limits of the observed column densities
were available in more than 50\% of the sources  (due to non-detections), we set the goodness-of-fit to be 0 when the modeled column density was lower than the upper limit multiplied by 10. Then, the confidence criteria for all molecules at a given time step and for a given cloud model were summed and divided 
by the number of observed molecules, and a total goodness-of-fit was obtained. Finally, a code searched for the lowest value that
represented the best-fit physical model and the best-fit chemical age. Because we modeled averaged data, the chemical age is also an average for each of the evolutionary stages. These ages can be considered as a typical mean time for an object with a certain chemical composition to evolve to the stage with the subsequent characteristic chemical composition. A typical execution time of the MUSCLE model for one evolutionary stage is between 2 and 10~hours on a server with 24 Intel Xeon 3.2~GHz CPUs.

\subsection{Radial distributions of molecules in the model}

\begin{figure*}
\includegraphics[width=0.43\textwidth]{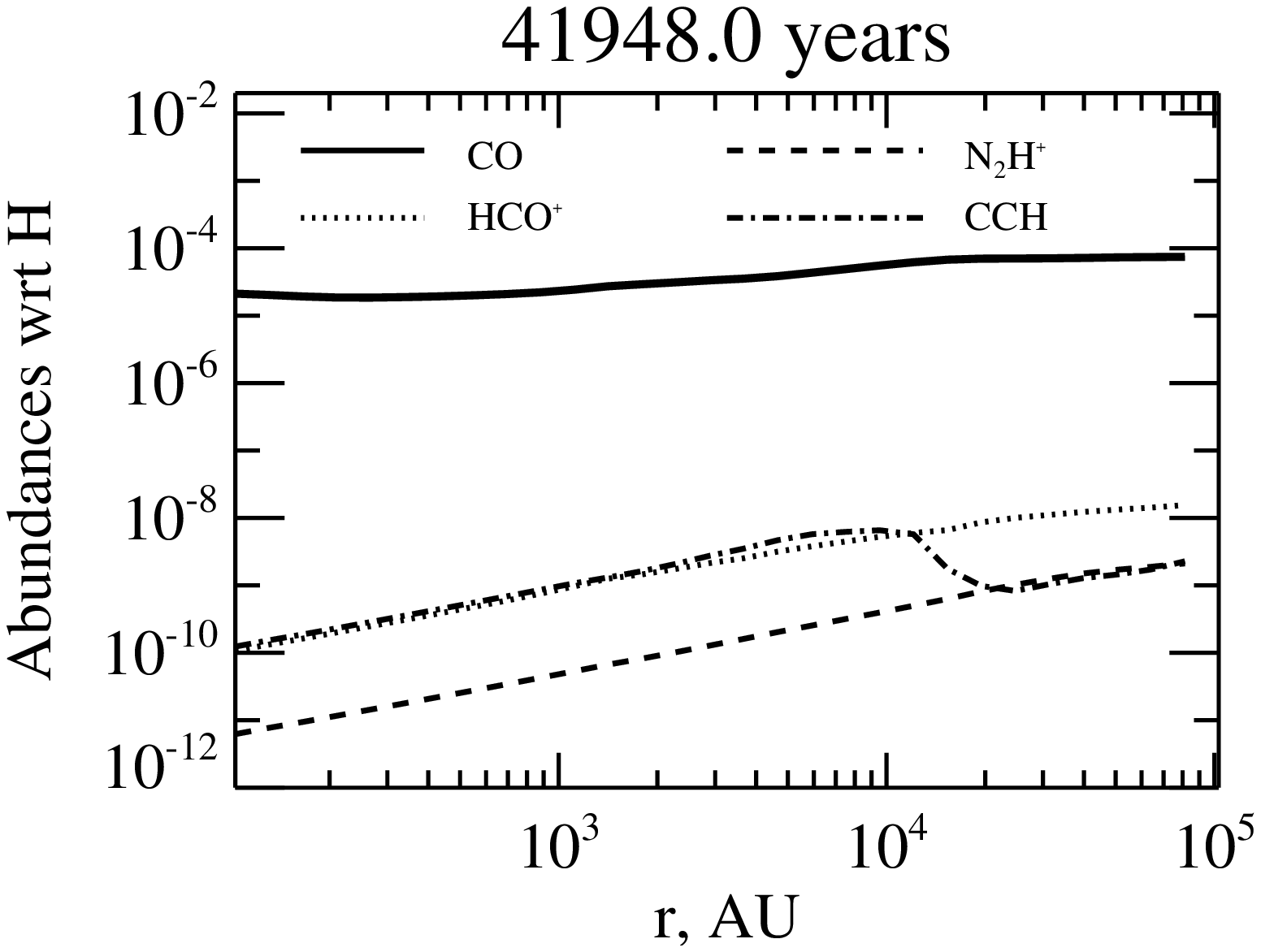}
\includegraphics[width=0.43\textwidth]{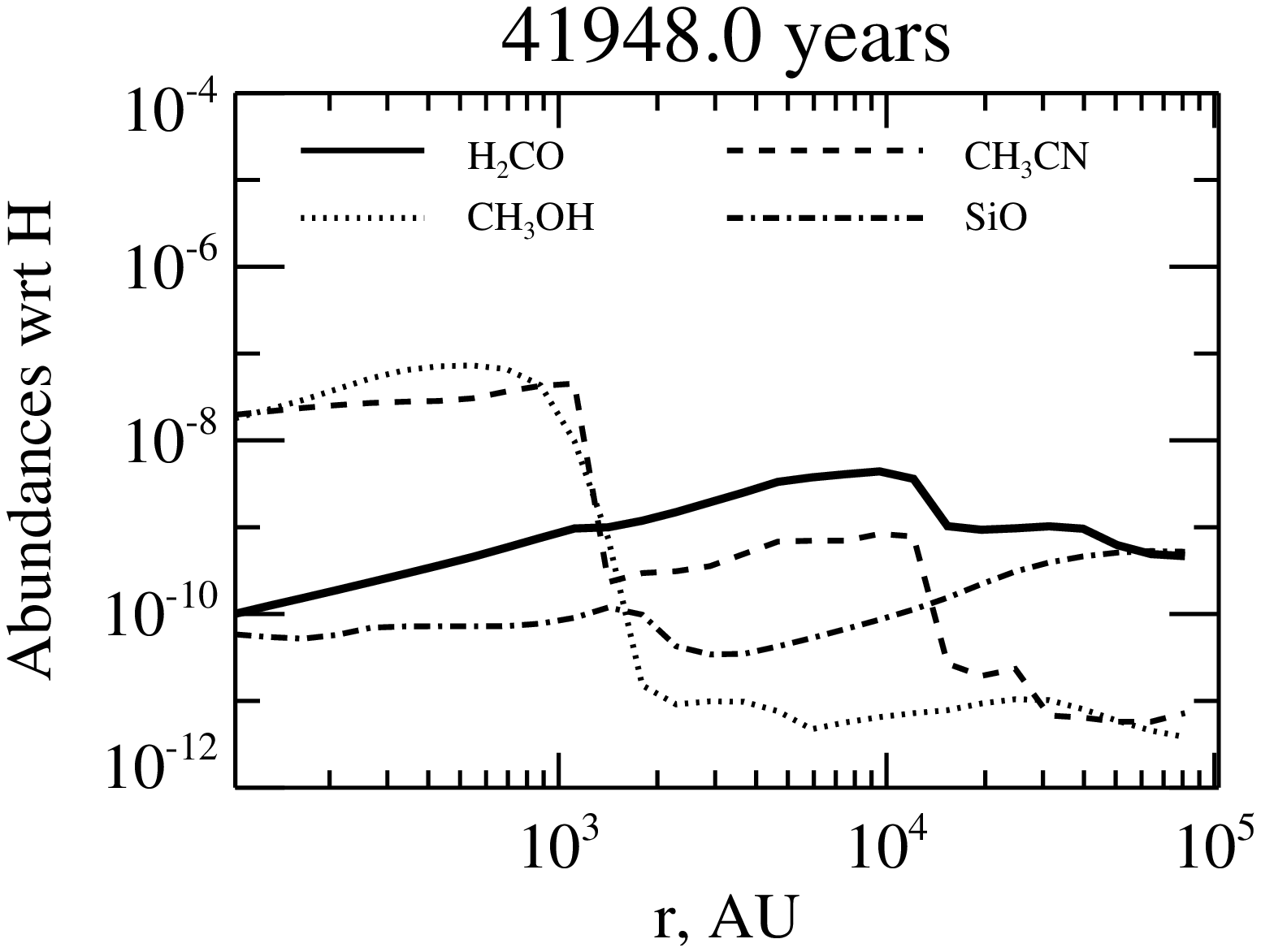}
\caption{Snapshot of the radial distributions of CO, HCO$^+$, N$_2$H$^+$, C$_2$H, H$_2$CO, CH$_3$OH, CH$_3$CN, and SiO at the time of the best-fit model of the HMC phase. The time is given relative to the beginning of the HMC phase.}
\label{fig:radial_distr}
\end{figure*}

In Figure~\ref{fig:radial_distr} we show the radial distribution of several molecules for the best-fit model of the HMC stage. Simpler molecules like N$_2$H$^+$ and HCO$^+$  , which are produced in the gas-phase, have a flatter distribution or even increase with radius, whereas more complex molecules like CH$_3$OH and CH$_3$CN are predominantly abundant in the warm inner region, where ices are able to thermally desorb from dust grain surfaces. This leads to the conclusion that to properly account for these effects it is also necessary to assume different beam-filling factors for different molecules. Interferometric observations with high spatial resolutions are needed to address this problem in detail.

\subsection{Comparison of observations and the best-fit models} \label{sec:comparison_obs_model}
\begin{figure*}
\includegraphics[width=0.32\textwidth]{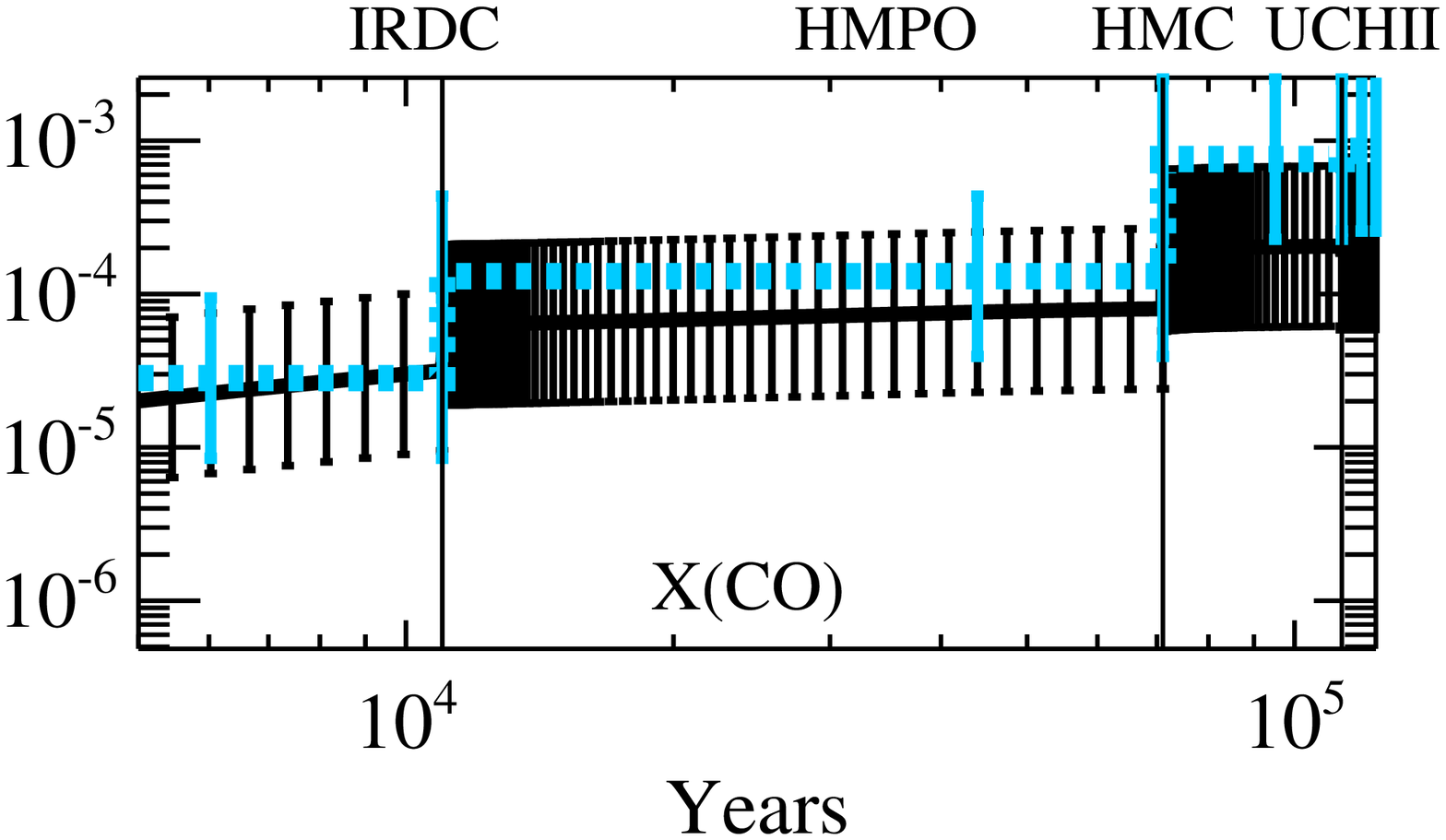}
\includegraphics[width=0.32\textwidth]{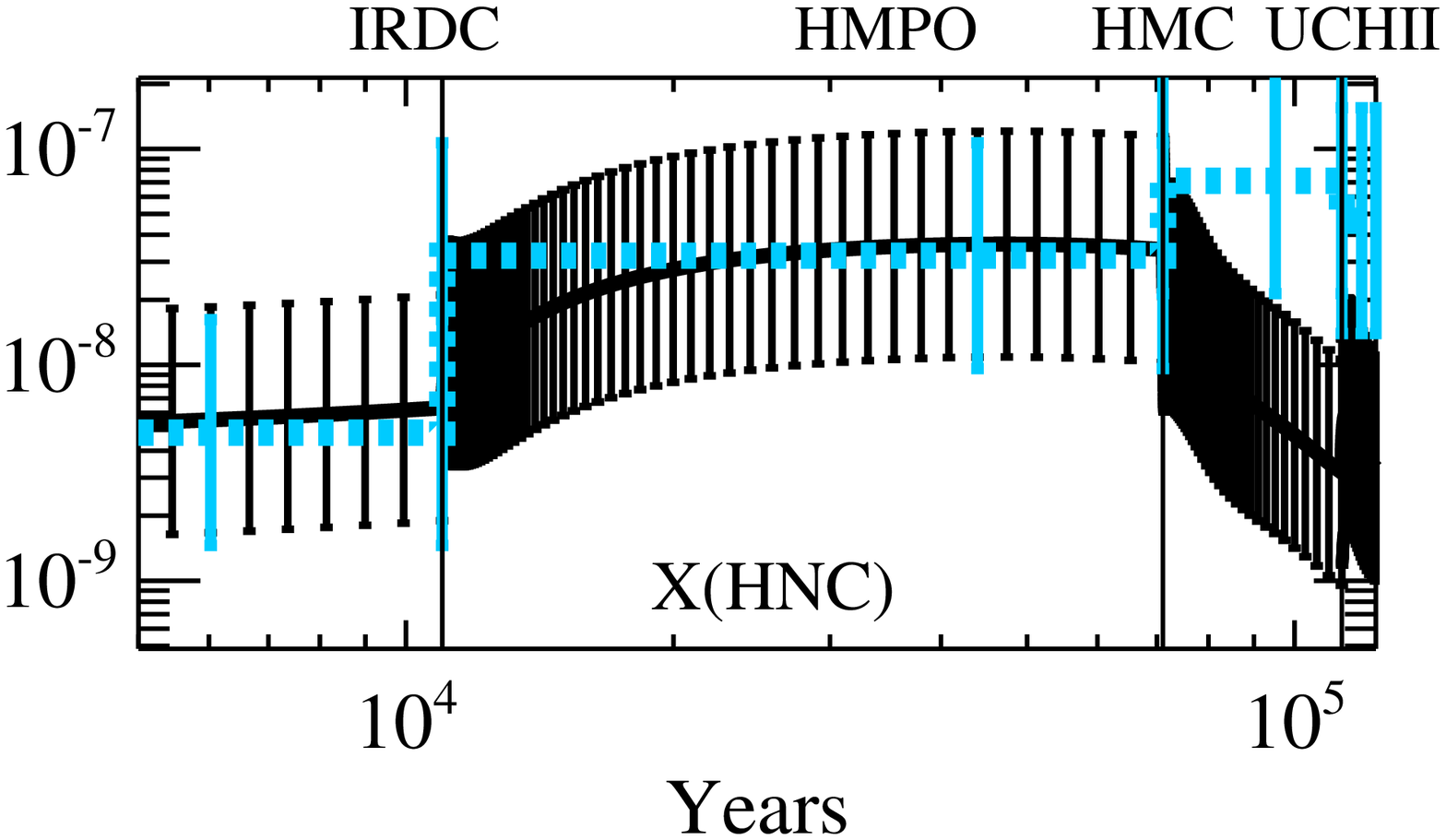}
\includegraphics[width=0.32\textwidth]{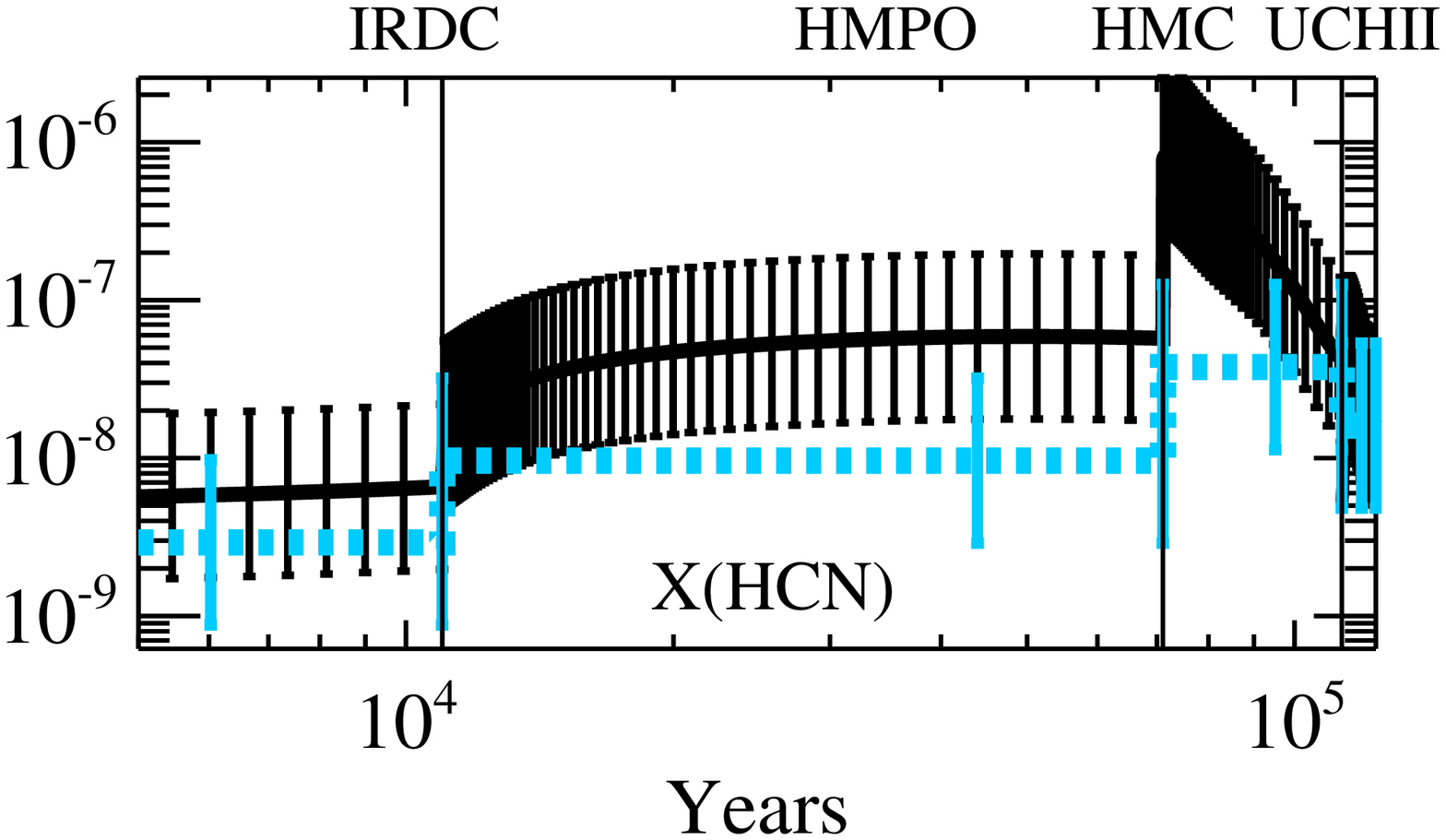}\\
\includegraphics[width=0.32\textwidth]{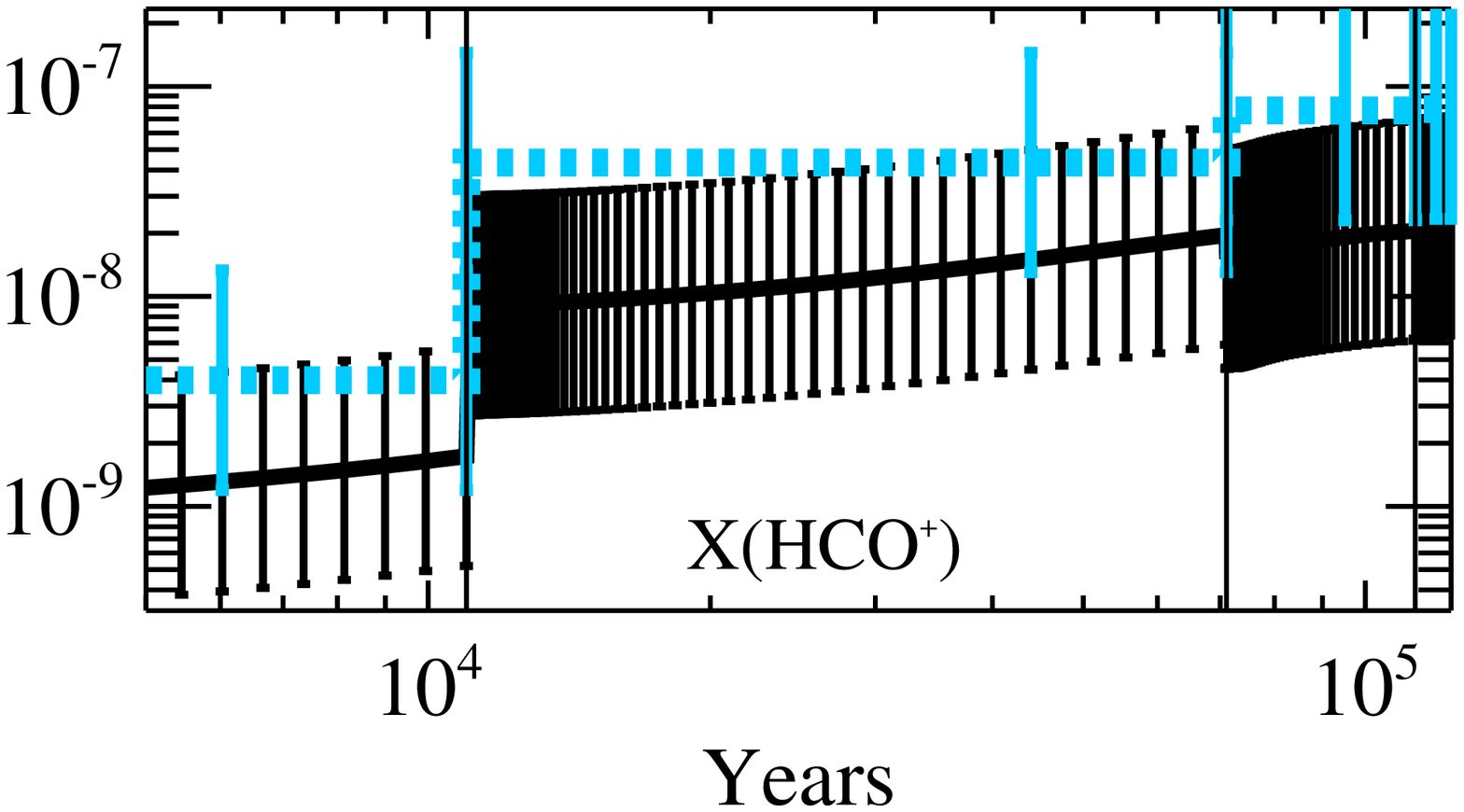}
\includegraphics[width=0.32\textwidth]{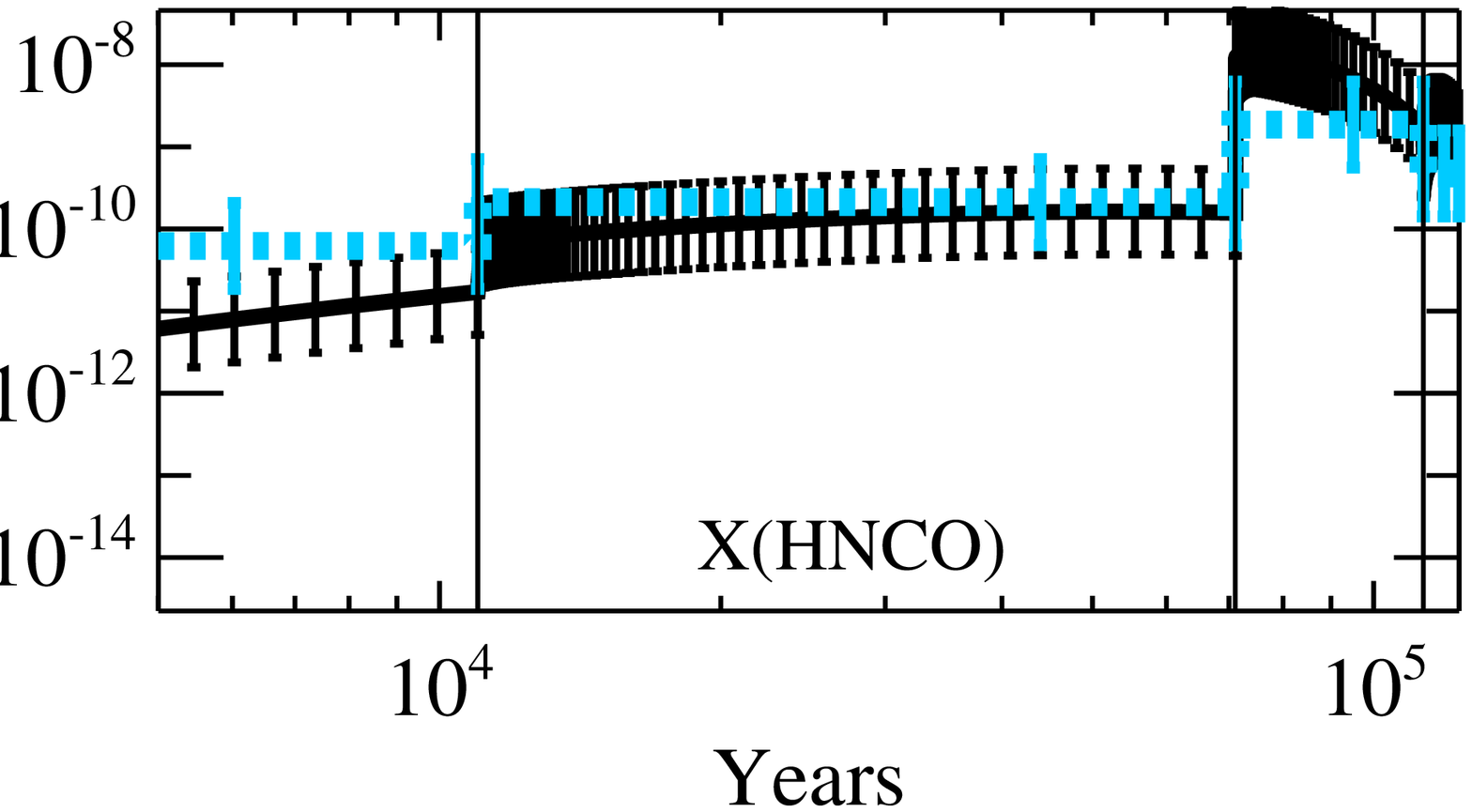}
\includegraphics[width=0.32\textwidth]{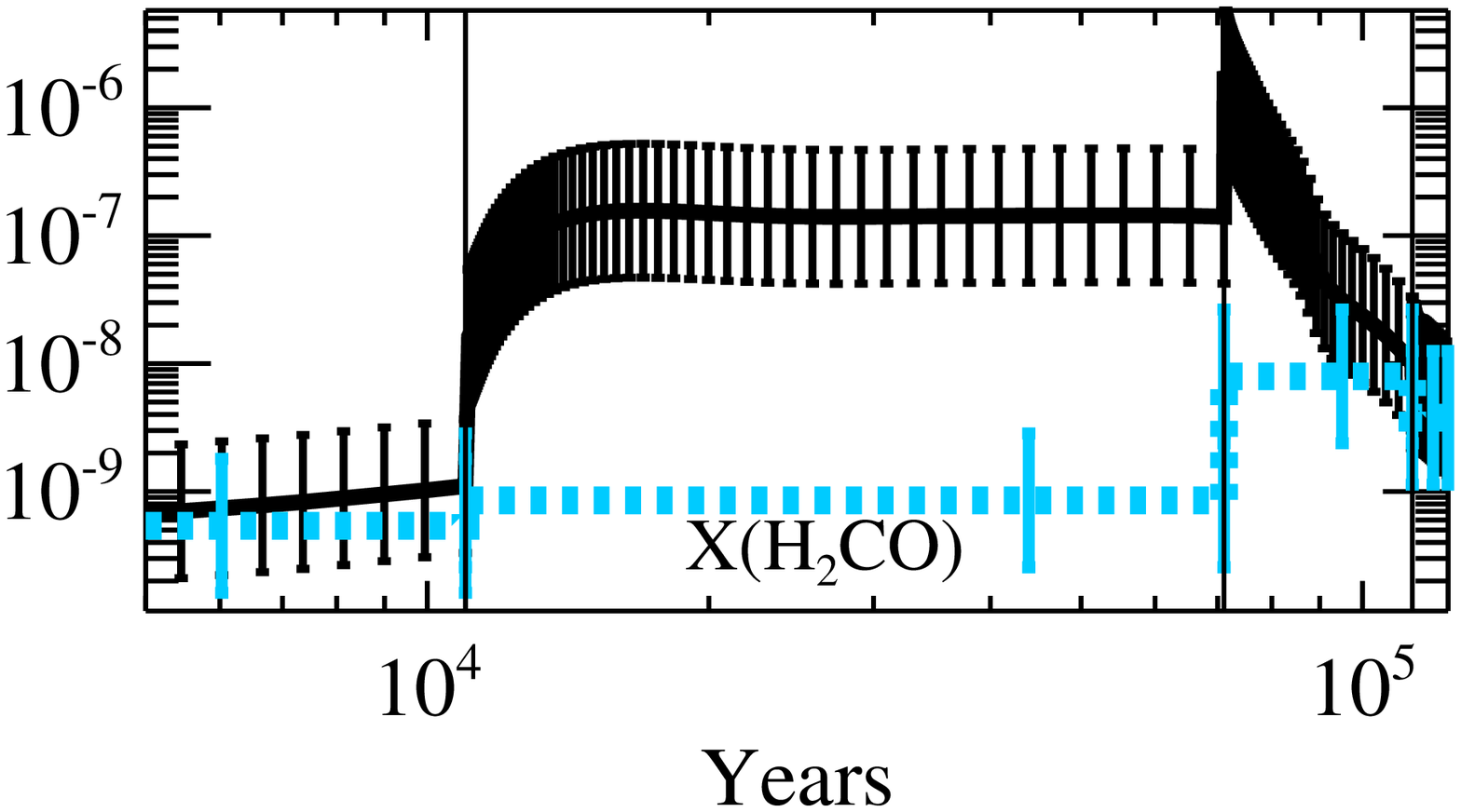}\\
\includegraphics[width=0.32\textwidth]{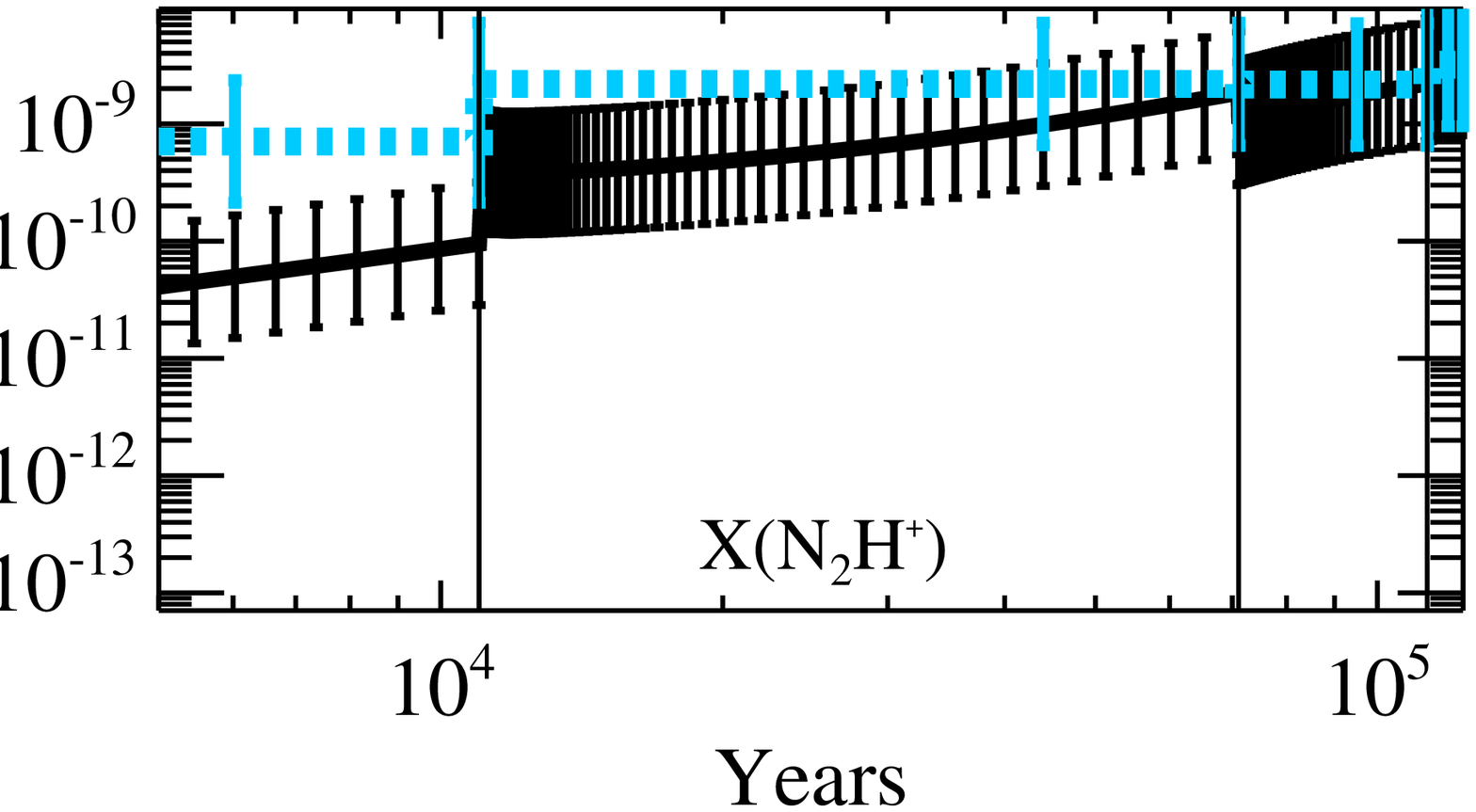}
\includegraphics[width=0.32\textwidth]{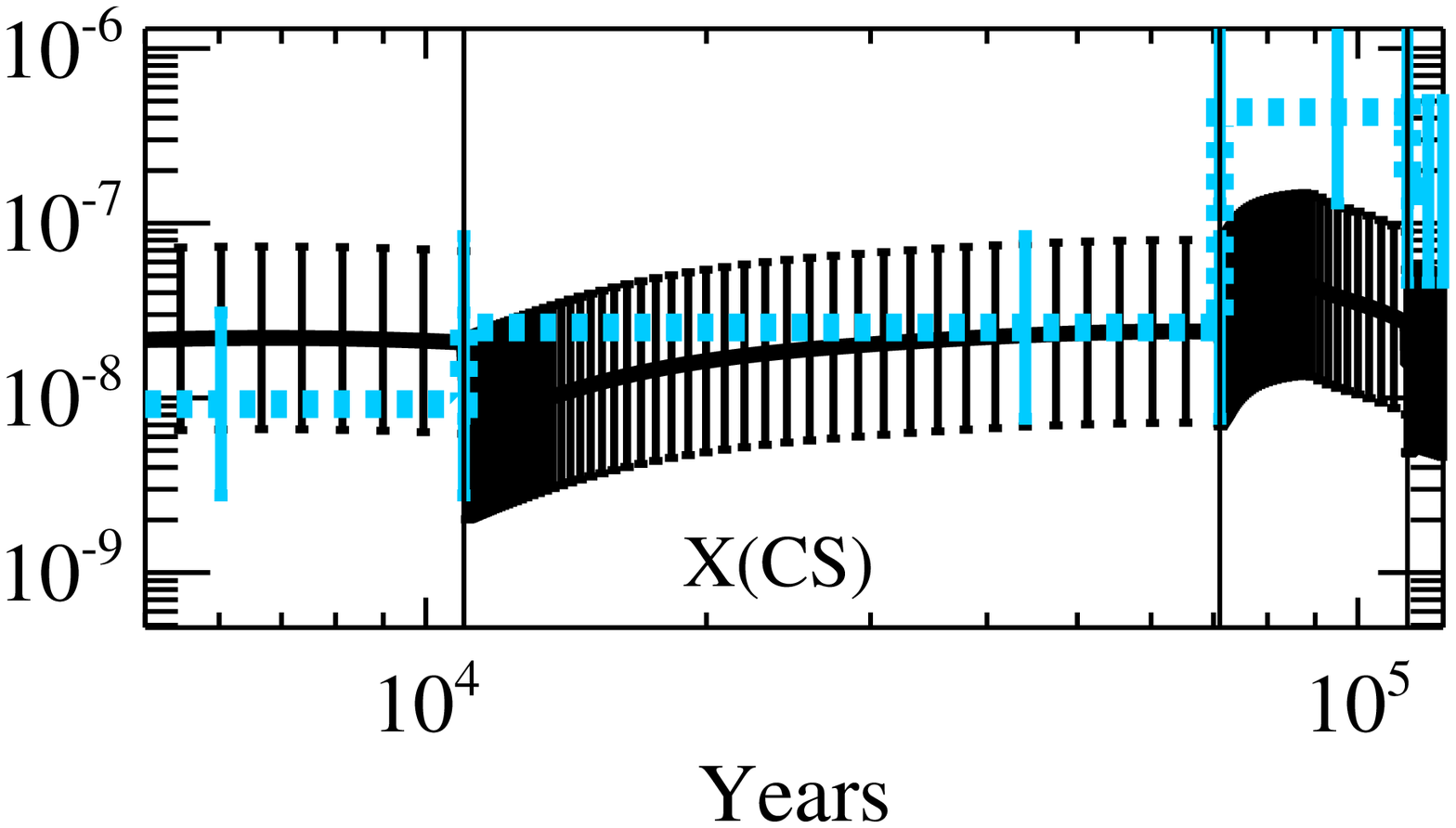}
\includegraphics[width=0.32\textwidth]{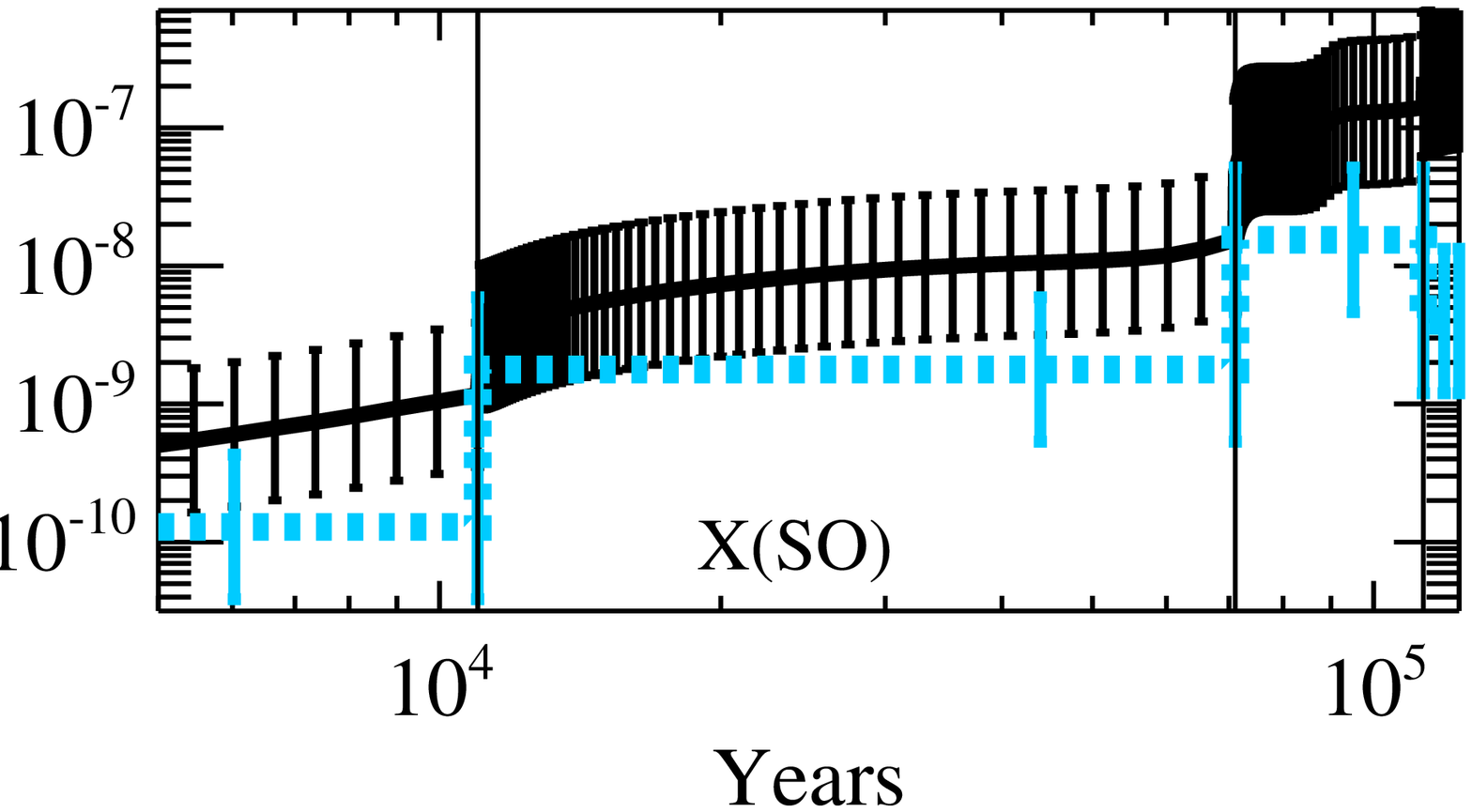}\\
\includegraphics[width=0.32\textwidth]{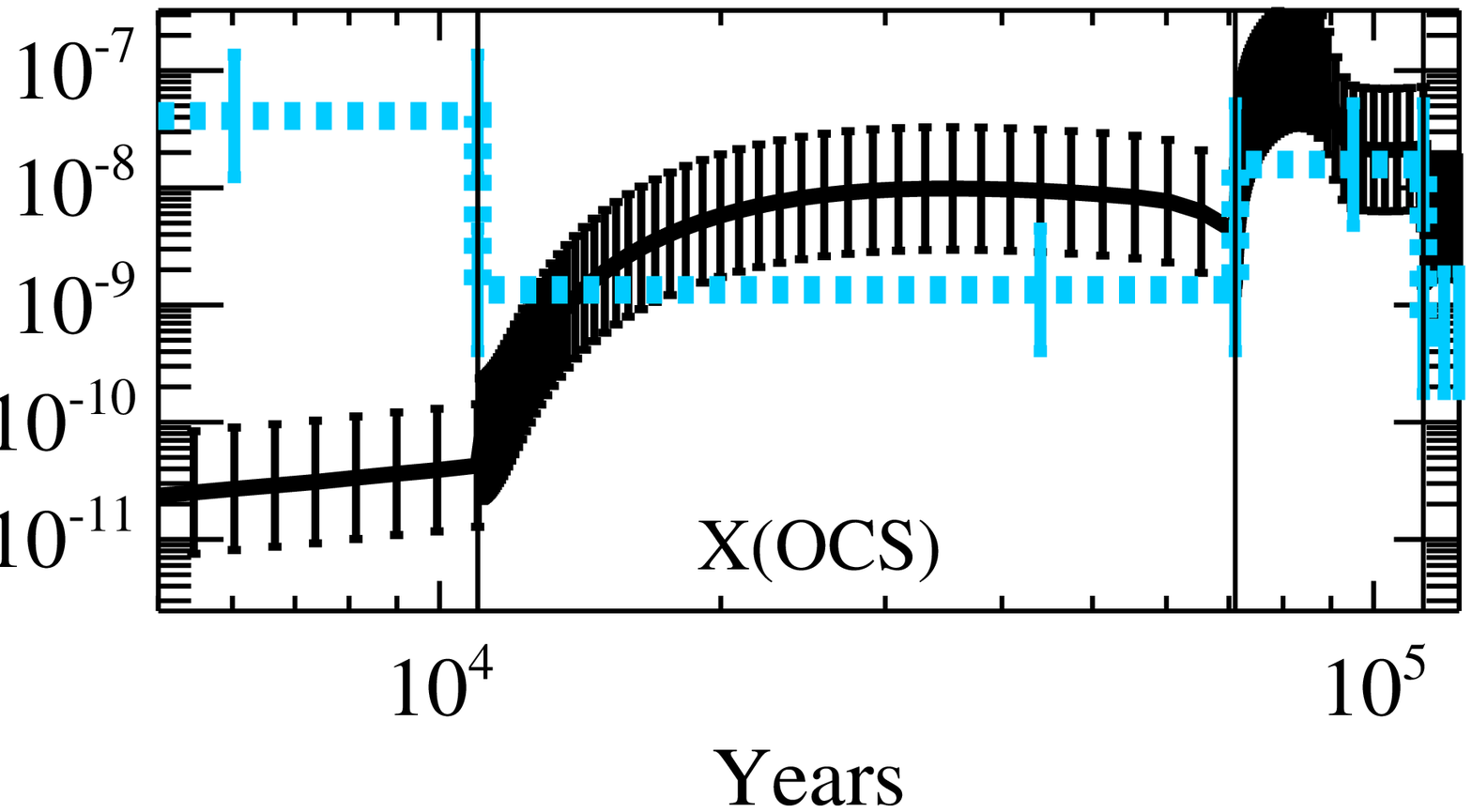}
\includegraphics[width=0.32\textwidth]{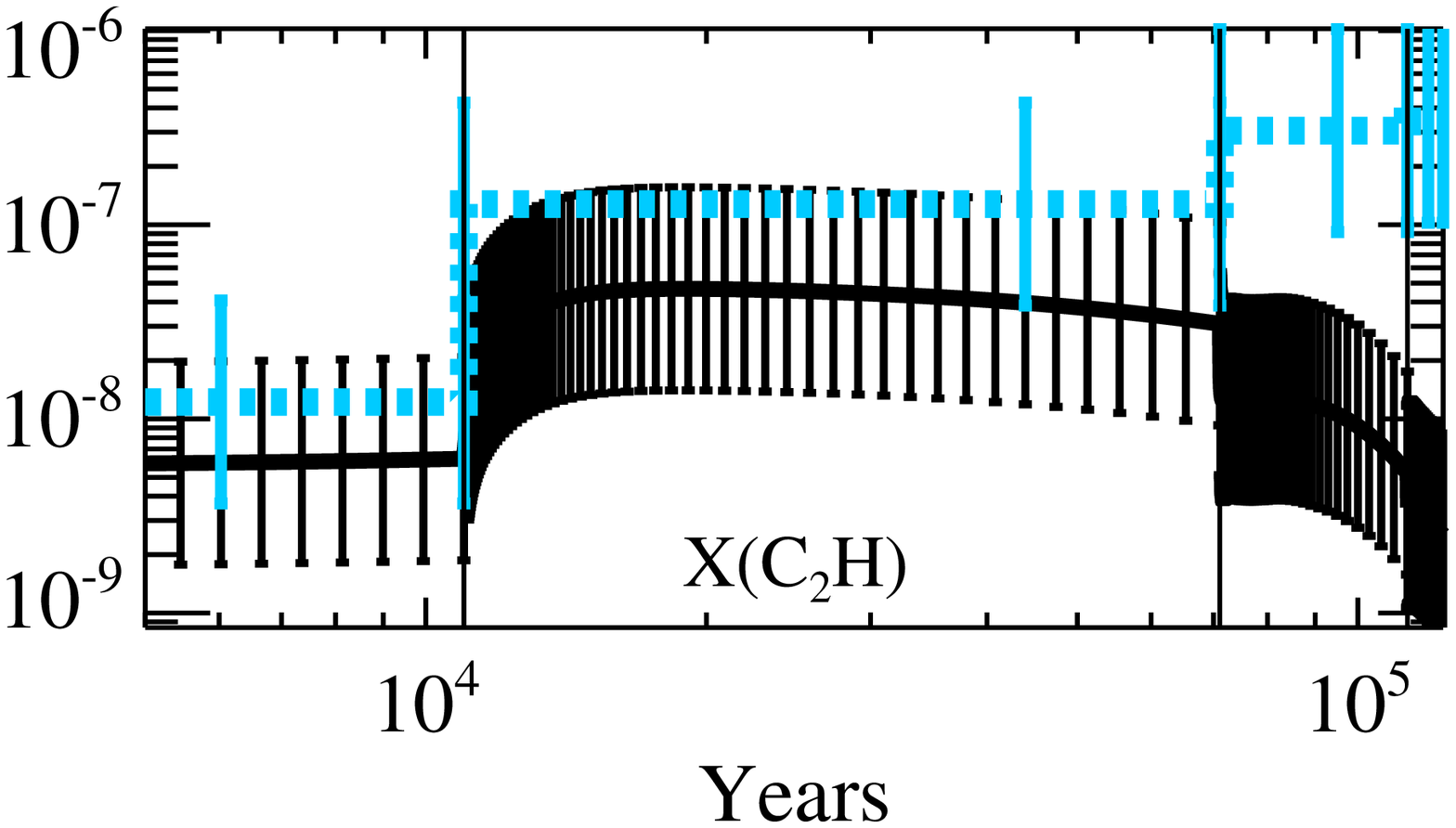}
\includegraphics[width=0.32\textwidth]{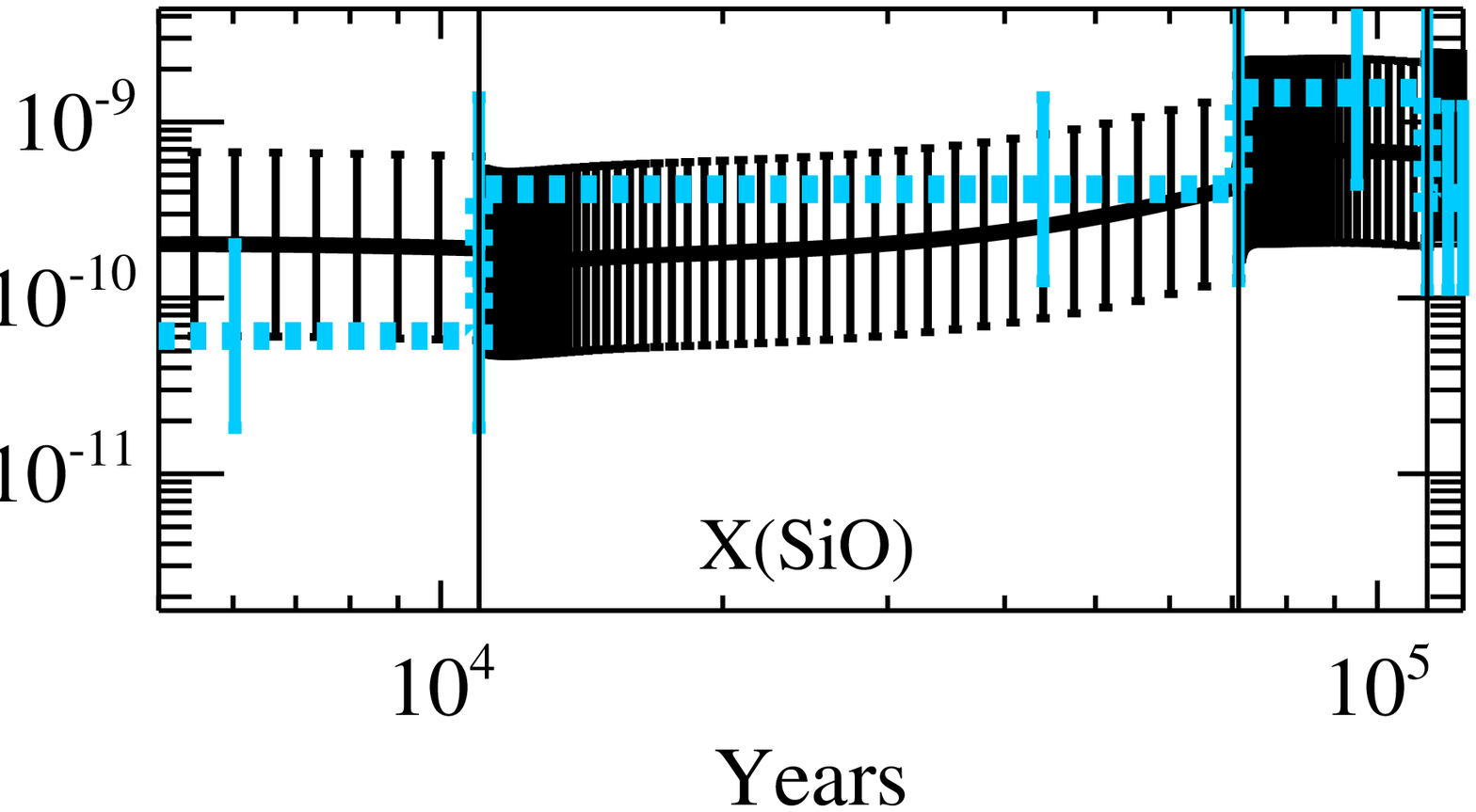}\\
\includegraphics[width=0.32\textwidth]{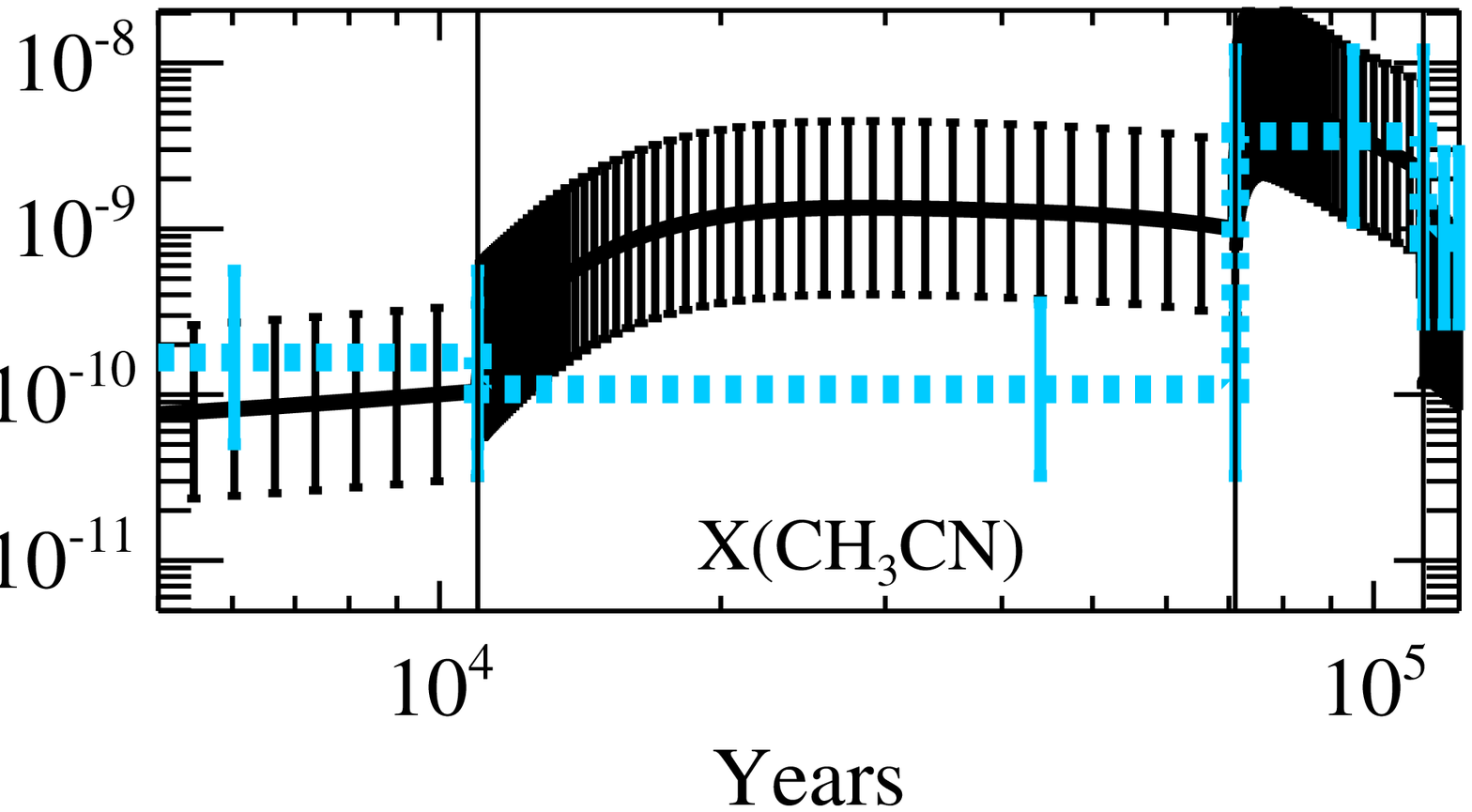}
\includegraphics[width=0.32\textwidth]{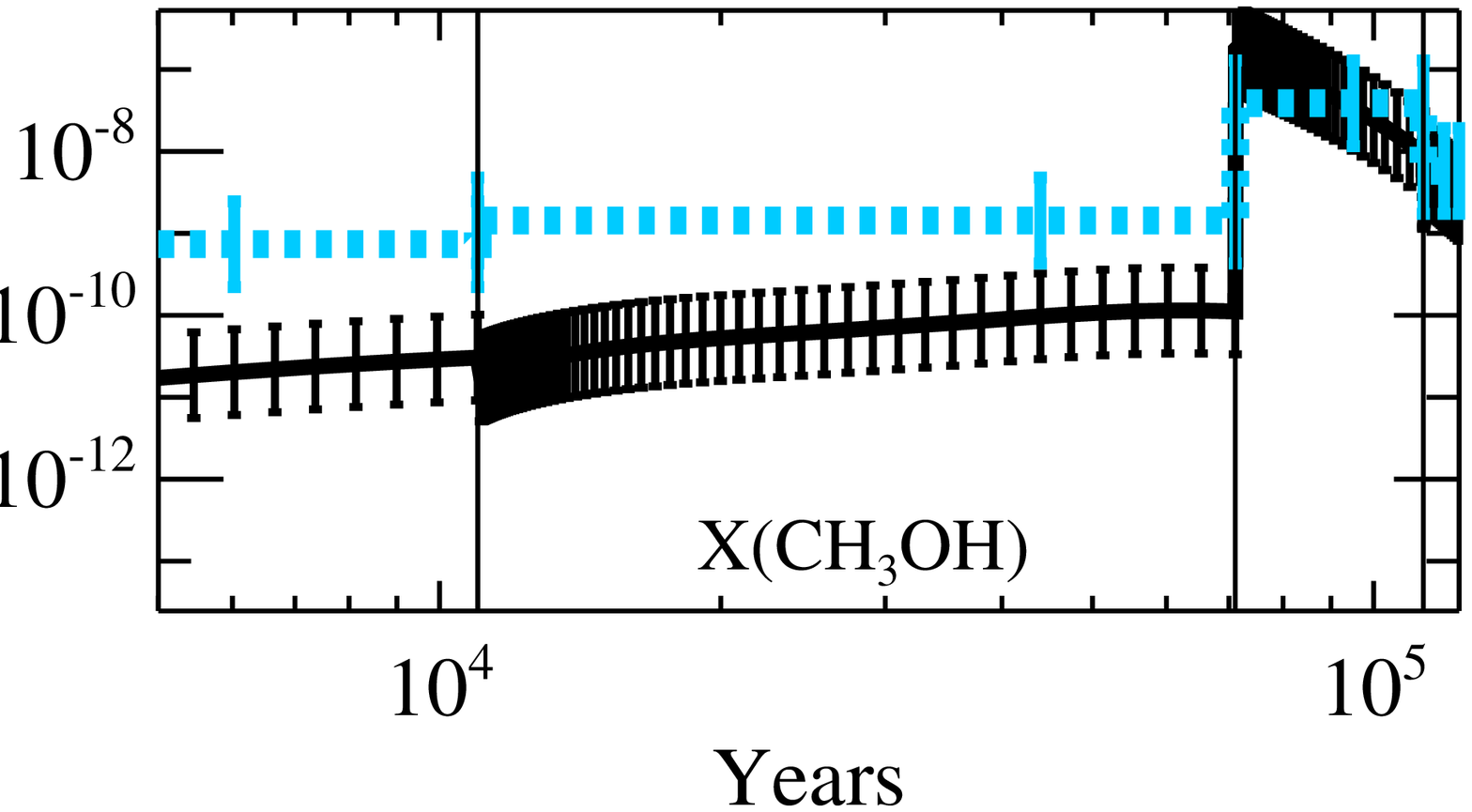}
\caption{Relative abundances to H$_2$  plotted for all four stages. The modeled values are shown as a black solid line, the observed values are depicted by a blue dashed line. The error bars are indicated by the vertical marks.}
\label{fig:coldens_comp}
\end{figure*}

The comparison of the data with the best-fit model described in Section~\ref{sec:fitting} for our samples of IRDCs, HMPOs, HMCs, and UCH{\sc ii}s, shows good overall agreement between the observed and modeled column densities. The best-fit parameters are shown in Tables~\ref{tab:IRDC_fit} -- \ref{tab:UCHII_fit} and their corresponding column densities in Tables~\ref{tbl:bestfit_IRDC} -- \ref{tbl:bestfit_UCH}. Furthermore, we list the corresponding radial temperature and density profiles as well as the chemical
age of the individual stages derived with the time-dependent chemical code. Figure~\ref{fig:coldens_comp} shows the observed and modeled abundances as a function of evolved time. In this plot, the modeled abundances between the different stages seem to have discontinuities. This is because the best-fit models of various stages do not have similar physical structures. The chemistry reacts quickly to these changes in the physical structure at the beginning of a new stage. The abundance 
at each subsequent stage is plotted 100 years after the starting time of the new stage and thus the abundance occurs as a discontinuity.

\begin{table}
\tiny
\caption{Parameters of the best-fit IRDC model.}             
\label{tab:IRDC_fit}      
\begin{tabular}{llll}        
\hline\hline                 
Parameter & Symbol & Value
\\    
\hline                        
Inner radius & $r_{\rm {in}}$ & $102$~AU\\
Outer radius & $r_{\rm {out}}$ & $0.5$~pc\tablefootmark{\rm \bf a}\\
Density at the inner radius & $\rho_{\rm {in}}$ & $1.2\cdot10^{9}$~cm$^{-3}$\\
Average density with a beam of $26\,000$~AU& $\bar{\rho}$ & $1.0\cdot10^{6}$~cm$^{-3}$ \\
Average density with a beam of $54\,000$~AU& $\bar{\rho}$ & $3.0\cdot10^{5}$~cm$^{-3}$ \\
Density profile & $p$ & $1.8$\\
Temperature at the inner radius & $T_{\rm {in}}$ & $20.9$~K \\
Average temperature & $\bar{T}$ & $20.9$~K \\
Temperature profile & $q$ & $0$\\
\hline                                   
\end{tabular}
\tablefoot{}
\tablefoottext{a}{This value is limited by the largest $29\arcsec$ IRAM beam size used in
our observations.}
\end{table}

\begin{table}
\tiny
\caption{Parameters of the best-fit HMPO model.}             
\label{tab:HMPO_fit}      
\begin{tabular}{llll}        
\hline\hline                 
Parameter & Symbol & Value
\\    
\hline                        
Inner radius & $r_{\rm {in}}$ & $1\,130$~AU\\
Outer radius & $r_{\rm {out}}$ & $0.5$~pc\tablefootmark{\rm \bf a}\\
Density at the inner radius & $\rho_{\rm {in}}$ & $5.0\cdot10^{6}$~cm$^{-3}$\\
Average density with a beam of $21\,725$~AU& $\bar{\rho}$ & $4.1\cdot10^{5}$~cm$^{-3}$ \\
Average density with a beam of $57\,275$~AU& $\bar{\rho}$ & $9.8\cdot10^{4}$~cm$^{-3}$ \\
Density profile & $p$ & $1.5$\\
Temperature at the inner radius & $T_{\rm {in}}$ & $77.3$~K \\
Average temperature & $\bar{T}$ & $29.5$~K \\
Temperature profile & $q$ & $0.4$\\
\hline                                   
\end{tabular}
\tablefoot{}
\tablefoottext{a}{This value is limited by the largest $29\arcsec$ IRAM beam size used in
our observations.}
\end{table}

\begin{table}
\tiny
\caption{Parameters of the best-fit HMC model.}             
\label{tab:HMC_fit}      
\begin{tabular}{llll}        
\hline\hline                 
Parameter & Symbol & Value
\\    
\hline                        
Inner radius & $r_{\rm {in}}$ & $102$~AU\\
Outer radius & $r_{\rm {out}}$ & $0.5$~pc\tablefootmark{\rm \bf a}\\
Density at the inner radius & $\rho_{\rm {in}}$ & $2.7\cdot10^{9}$~cm$^{-3}$\\
Average density with a beam of $45\,400$~AU& $\bar{\rho}$ & $7.6\cdot10^{5}$~cm$^{-3}$ \\
Average density with a beam of $63\,100$~AU& $\bar{\rho}$ & $5.0\cdot10^{5}$~cm$^{-3}$ \\
Density profile & $p$ & $1.9$\\
Temperature at the inner radius & $T_{\rm {in}}$ & $268.3$~K \\
Average temperature & $\bar{T}$ & $40.2$~K \\
Temperature profile & $q$ & $0.4$\\
\hline                                   
\end{tabular}
\tablefoot{}
\tablefoottext{a}{This value is limited by the largest $29\arcsec$ IRAM beam size used in
our observations.}
\end{table}

\begin{table}
\tiny
\caption{Parameters of the best-fit UCH{\sc ii} model.}             
\label{tab:UCHII_fit}      
\begin{tabular}{llll}        
\hline\hline                 
Parameter & Symbol & Value
\\    
\hline                        
Inner radius & $r_{\rm {in}}$ & $102$~AU\\
Outer radius & $r_{\rm {out}}$ & $0.5$~pc\tablefootmark{\rm \bf a}\\
Density at the inner radius & $\rho_{\rm {in}}$ & $2.4\cdot10^{8}$~cm$^{-3}$\\
Average density with a beam of $57\,800$~AU& $\bar{\rho}$ & $1.8\cdot10^{5}$~cm$^{-3}$ \\
Average density with a beam of $85\,400$~AU& $\bar{\rho}$ & $8.7\cdot10^{4}$~cm$^{-3}$ \\
Density profile & $p$ & $1.5$\\
Temperature at the inner radius & $T_{\rm {in}}$ & $293.1$~K \\
Average temperature & $\bar{T}$ & $36.0$~K \\
Temperature profile & $q$ & $0.4$\\
\hline                                   
\end{tabular}
\tablefoot{}
\tablefoottext{a}{This value is limited by the largest $29\arcsec$ IRAM beam size used in
our observations.}
\end{table}

\begin{table}
\caption{Median column densities in $a(x)=a \cdot 10^{x}$ for observations and best-fit IRDC model.When the molecule was detected in fewer than $50\%$ of the sources, we marked it as an upper limit.}             
\label{tbl:bestfit_IRDC}      
\centering                          
\begin{tabular}{lrr}        
\hline\hline                 
Molecule & Observed col. den. & Modeled col. den.\\
 & $[{\rm cm}^{-2}]$ & $[{\rm cm}^{-2}]$
\\    
\hline                        
CO & 1.9(18) & 2.1(18)  \\
HNC & 1.5(14)& 2.0(14)  \\
HCN & 9.1(13) & 2.1(14)  \\
HCO$^+$ & 1.2(14) & 5.4(13)  \\
HNCO & 1.9(12) & 5.3(11)  \\
H$_2$CO & 3.5(13) & 7.2(13)  \\
N$_2$H$^+$ & 2.2(13) & 3.0(12)  \\
CS & $\leq$6.0(14) & 1.4(15)  \\
SO & $\leq$8.4(12) & 7.6(13)  \\
OCS & $\leq$2.7(15) & 2.8(12)  \\
C$_2$H & 3.8(14) & 1.9(14)  \\
SiO & 1.9(12) & 6.0(12)  \\
CH$_3$CN & $\leq$5.2(12) & 3.3(12) \\
CH$_3$OH & $\leq$4.8(13) & 2.0(12) \\
\hline
Agreement & & 14/14 = 100\%   \\ 
\hline                                   
\end{tabular}
\end{table}

\begin{table}
\caption{Median column densities in $a(x)=a \cdot 10^{x}$ for observations and best-fit HMPO model. Modeled best-fit values in italics disagree with the observed values within one order of magnitude. When the molecule was detected in fewer than $50\%$ of the sources, we marked it as an upper limit.}             
\label{tbl:bestfit_HMPO}      
\centering                          
\begin{tabular}{lrr}        
\hline\hline                 
Molecule  & Observed col. den. & Modeled col. den.\\
 & $[{\rm cm}^{-2}]$ & $[{\rm cm}^{-2}]$
\\    
\hline                        
CO & 6.7(18) & 4.1(18)  \\
HNC & 7.0(14) & 7.5(14)  \\
HCN & 2.2(14) & 1.3(15)  \\
HCO$^+$ & 9.6(14) & 4.3(14)  \\
HNCO & 4.7(12) & 3.5(12)  \\
H$_2$CO & 4.4(13) & \it{7.2(15)}  \\
N$_2$H$^+$ & 4.8(13) & 4.1(13)  \\
CS & 1.3(15) & 1.2(15)  \\
SO & 9.1(13) & 7.9(14)  \\
OCS & $\leq$6.9(13) & 2.2(14)  \\
C$_2$H & 2.8(15) & 6.8(14)  \\
SiO & 9.1(12) & 9.4(12)  \\
CH$_3$CN & 2.4(12) & 2.2(13)  \\
CH$_3$OH & 7.3(13) & \it{5.7(12)}  \\
\hline
Agreement & & 12/14 = 86\%  \\  
\hline                                   
\end{tabular}
\end{table}

\begin{table}
\caption{Median column densities in $a(x)=a \cdot 10^{x}$ for observations and best-fit HMC model. Modeled best-fit values in italics disagree with the observed values within one order of magnitude. When the molecule was detected in fewer than $50\%$ of the sources, we marked it as an upper limit.}             
\label{tbl:bestfit_HMC}      
\centering                          
\begin{tabular}{lrr}        
\hline\hline                 
Molecule & Observed col. den. & Modeled col. den.\\
 & $[{\rm cm}^{-2}]$ & $[{\rm cm}^{-2}]$
\\    
\hline                        
CO & 4.2(19) & 1.1(19)  \\
HNC & 2.4(15) & \it{1.1(14)}  \\
HCN & 1.3(15) & 1.4(15)  \\
HCO$^+$ & 2.6(15) & 6.9(14)  \\
HNCO & 6.3(13) & 5.9(13)  \\
H$_2$CO & 4.4(14) & 5.6(14)  \\
N$_2$H$^+$ & 7.3(13) & 7.4(13)  \\
CS & 2.4(16) & \it{1.5(15)}  \\
SO & 8.6(14) & 7.8(15)  \\
OCS & 8.8(14) & 1.2(15)  \\
C$_2$H & 1.0(16) & \it{1.8(14)}  \\
SiO & 5.0(13) & 2.2(13)  \\
CH$_3$CN & 1.2(14) & 7.6(13)  \\
CH$_3$OH & 2.2(15) & 5.1(14)  \\
\hline
Agreement & & 11/14 = 79\%   \\ 
\hline                                   
\end{tabular}
\end{table}

\begin{table}
\caption{Median column densities in $a(x)=a \cdot 10^{x}$ for observations and best-fit UCH{\sc ii} model. Modeled best-fit values in italics disagree with the observed values within one order of magnitude. When the molecule was detected in fewer than $50\%$ of the sources, we marked it as an upper limit.}             
\label{tbl:bestfit_UCH}      
\centering                          
\begin{tabular}{lrr}        
\hline\hline                 
Molecule  & Observed col. den. & Modeled col. den.\\
 & $[{\rm cm}^{-2}]$ & $[{\rm cm}^{-2}]$
\\    
\hline                        
CO & 2.8(19) & 6.4(18)  \\
HNC & 1.1(15) & \it{9.0(13)}  \\
HCN & 3.7(14) & 4.0(14)  \\
HCO$^+$ & 1.8(15) & 4.8(14)  \\
HNCO & 1.1(13) & 1.8(13)  \\
H$_2$CO & 1.2(14) & 2.4(14)  \\
N$_2$H$^+$ & 7.3(13) & 5.4(13)  \\
CS & 5.0(15) & 5.3(14)  \\
SO & 1.3(14) & \it{6.7(15)} \\
OCS & $\leq$1.9(13) & 1.6(14)  \\
C$_2$H & 7.8(15) & \it{8.0(13)}  \\
SiO & 8.5(12) & 1.7(13)  \\
CH$_3$CN & 2.0(13) & 8.7(12)  \\
CH$_3$OH & 1.9(14) & 1.2(14)  \\
\hline
Agreement & & 11/14 = 79\% \\  
\hline                                   
\end{tabular}
\end{table}

In the IRDC stage the best-fit model was able to successfully explain 14 out of 14 different species. The best-fit age is $\sim11\,000$~years (with an uncertainty of a factor of 2-3). The chemical age of a given stage is the time the model needed to evolve from the starting conditions to the best-fit model of this stage. For the later stages, the chemical composition of the best-fit model of the previous stage was set as the starting conditions for the next one.

In the second stage of an HMPO the best-fit model was able to successfully explain 12 out of 14 different species. The two species that are not explained are H$_2$CO and CH$_3$OH. The model strongly overproduces H$_2$CO by a factor of more than 100. Methanol is underproduced, but only slightly lower than the assumed criteria of one order of magnitude. In general, problems with fitting the data are due to environmental effects that are not taken into account in the model, such as shocks and UV-penetration. CH$_3$OH and SiO are known shock tracers and thus more likely to be not explained by the model. A not yet fully understood surface chemistry is another source of uncertainty in the model. H$_2$CO is partly produced in the gas phase and on the grains. The two different formation routes are strongly time dependent, and so far there is no consensus in the community about how to implement this dependency or the implementation of different desorption mechanisms that allow molecules to return to the 
gas phase \citep[see e.g.,][]{Garrod_ea07,vasyunin2013}. In this case the timescale seems to be too short to transform H$_2$CO into CH$_3$OH, which can partly explain the overproduction and underproduction, respectively. The best-fit age of the HMPO phase is $\sim60\,000$~years (with an uncertainty of a factor of 2-3).

In the third stage of an HMC the best-fit model was able to successfully explain 11 out of 14 different species. The three species that differ are HNC, CS, and C$_2$H. While HNC and CS are only slightly underproduced compared with the agreement criteria, C$_2$H is underproduced by almost more than a factor of 50. In the model the HMC is still surrounded by an external shell with a visual extinction of 10~mag with no external UV-field, which leads to lower C$_2$H abundances. In reality, the cloud might be highly fractal and fragmented such that UV-photons can penetrate deeper and produce more C$_2$H. The best-fit age of the HMC phase is $\sim42\,000$~years (with an uncertainty of a factor of 2-3).

In the fourth stage of an UCH{\sc ii} the best-fit model was
able to successfully explain 11 out of 14 different species. The three species that are not reproduced are HNC, SO and C$_2$H. A caveat for our models of UCH{\sc ii} regions is that it has an internal UV-source that is not properly taken into account in the model. Furthermore, since this stage depends on all best-fit models of the previous stages, it is the most doubtful one. The successful fitting of 80\% of the molecules is very remarkable
already. The best-fit age of the UCH{\sc ii} regions is $\sim11\,500$~years (with an uncertainty of a factor of 2-3).

This is the first time that the chemical evolution was modeled and reasonably well fitted over the entire span of the early evolution of high-mass star formation regions. The mean temperatures of the best-fit model temperature profile for the last three stages (see Tables~\ref{tab:IRDC_fit} -- \ref{tab:UCHII_fit}) are considerably lower, by a factor of 2-3, than the excitation temperatures used to calculate the observed column densities. This difference is much smaller in the IRDC stage.

In a second step we took the mean temperatures from each best-fit model as excitation temperatures to recalculate the observed column densities. The difference in abundances for the two different excitation temperatures for the IRDCs is at most a factor of 2 for all molecules, except for OCS, which differs by a factor of 3. The abundances in the HMPOs change at most by a factor of 3. The discrepancy between the temperatures originally used and the mean temperatures derived in the best-fit for the HMCs and UCH{\sc ii} regions is larger than for the IRDCs and HMPOs. Thus, the change in the abundances is higher by about a factor of 6 for the HMCs (only CH$_3$OH and CH$_3$OCHO differ by factors of 7 and 10, respectively) and 6-9 for the UCH{\sc ii}s (only HNCO and CH$_3$OCHO differ by a factor of 13). These values are shown in Tables~\ref{tbl:obsmedianabunirdc} -- \ref{tbl:obsmedianabunuch} as well.

In this iterative step we sought again to find best-fit solutions for the newly derived values with more consistent temperatures. To keep the model physically meaningful, the temperature profile of the last three stages was set as a power-law with a constant temperature of the IRDC best-fit value in the outer parts. This resulted in a stable solution for the IRDCs. But for the later stages the model tends to produce a small and dense inner hot part and a large isothermal outer part. A decrease in the  modeled best-fit temperature leads to higher observationally derived H$_2$ and lower observationally derived molecular column densities. Fitting the new higher H$_2$ and lower molecular column densities lowers the mean temperatures even more and leads to a non-converging feedback. However, temperature measurements of tracers of the inner dense part such as CH$_3$CN show a hot inner component with $T>50$~K.

The discrepancy in temperatures between model and observations is caused by uncertainties and simplifications in both. On the observational side we have single-pointing single-dish spectra for which we assumed a single excitation temperature. However, the observed sources are most likely inhomogeneous and have structures of differing environments. We partly took this into account in the model by assuming a 1D physical structure with a temperature and density profile. Another reason for this mismatch could be subthermal excitation. To avoid the effect of non-converging feedback and to account for all the caveats, we stopped the modeling process and report the results of the first iteration of the best-fit model.

Based on the probable explanations for the mismatch between the temperatures used for the observations and derived with the model, improvements on both sides might help to solve the discrepancy. Detailed gas and dust temperature maps and dust continuum radiative transfer calculations might solve the degeneracy in constraining the temperature. Interferometric observations with higher resolutions would help to decrease uncertainties in the beam filling fraction. Moreover, extending the observed species toward deuterated molecules might help to better constrain temperatures since deuterium chemistry is sensitive to the previous thermal history of the environment \citep[see][]{caselli2012}. In addition, the model might be improved by including radiative transfer, a gradual warm-up phase \citep[see][]{garrod2008,vasyunin2013b}, and in general more physics like the proper treatment of shocks, outflows and UV radiation transport. Furthermore, at least 2D models would be necessary to be able to explain observations 
with higher resolutions.

\subsection{Comparison of observed and best-fit molecular ratios}

In addition to the information from column densities of single molecules, the ratios between different molecules provide another interesting tool for investigating the chemical evolution. In Figure~\ref{fig:coldenseratios} the observed and best-fit ratios of several molecular pairs are shown. Due to the limited sensitivity in the observations the overall dynamic range of observed ratios is smaller than in the model. In general, the ratios of the observed median column densities show no obvious trends with evolution that cannot be explained within the uncertainties. Because of the poorly explained molecular column densities in the best-fit model for a few species, such as H$_2$CO or C$_2$H, and the agreement criteria between model and observations, some of the ratios deviate strongly, depending on the specific molecule and evolutionary phase. 

\subsubsection{HCN/HNC}
Previous observations showed that the ratio of HCN/HNC depends on the temperature of the medium, for example, observations of Orion-KL show a decreasing ratio from the warm inner core toward the cold outer edge from 80 to 5 \citep{goldsmith1986,schilke1992}. The best-fit model follows this trend and increases until it
reaches the highest value of $\sim50$ in the beginning of the HMC phase, whereas our observations do not show significant differences between the four stages with a ratio between $0.3-0.6$. The total value of the observed ratio $<1$ might be underpredicted due to the assumptions of the isotopic ratio in the conversion from HN$^{13}$C to HNC. Furthermore, due to self absorption, high optical depth, and other effects discussed in Section \ref{sec:problems}, we refrained from using the observed HCN lines in the analysis in many cases. In these cases, the sources more likely have higher column densities. That might bias the derived column densities toward lower optical depths and therefore lower column 
densities.

\subsubsection{HCO$^+$/N$_2$H$^+$}
The ratio of the two molecules N$_2$H$^+$ and HCO$^+$ is a candidate for a chemical clock. There is evidence that their abundances are anticorrelated in chemical models of low-mass star-forming regions \citep[e.g.,][]{joergensen2004b} and massive star formation \citep[][]{barnes2013}, since both molecules are sensitive to the amount of CO. According to the models, N$_2$H$^+$ is more abundant during the early phases when CO is frozen out and is destroyed after CO returns into the gas phase. This behavior of the anticorrelation between CO and N$_2$H$^+$ is seen by \citet{bergin2002} toward the low-mass star-forming region B68 and in the ratio of N$_2$H$^+$ and CO of this work. The abundance of HCO$^+$ is directly dependent on the CO abundance and the ionization degree and thus is anticorrelated with N$_2$H$^+$. Our observations reproduce this trend of an increasing ratio with evolutionary stage. However, in the model the ratio is almost constant within the given errors. A possible explanation are the 
uncertainties in the observed values, which are larger than the total change of that particular ratio. Nevertheless, observations and model agree within the uncertainties and show a total spread of the ratio between 10 and 100.

\subsubsection{CH$_3$OH}
In our model, the ratios of the more complex molecule methanol with simple molecules such as CO is almost constant at $~10^{-6}$ during the IRDC and HMPO phase and jumps at the beginning of the hot core phase to $~10^{-3}$. During the IRDC and HMPO phases CH$_3$OH is steadily produced on and in ice mantles of dust grains. One of our model restrictions is that the methanol ice can only be desorbed to the gas by cosmic-ray particles (CRP) and CRP-UV heating. In the HMC phase the entire solid reservoir of dust mantles is released into the gas phase, leading to a sudden increase of the gas-phase concentration of many complex molecules. Their abundance may decrease with time after that event if they are actively destroyed or converted to other species in the gas, because surface chemistry becomes insignificant at $\ga 100-150$~K. Afterwards it steadily declines until it reaches $~10^{-5}$ at the end of the evolution. This behavior is imprinted by the abundance variability of CH$_3$OH and is thus seen in most of 
its ratios. It agrees with the 
observational ratio for the last three stages, but underestimates the IRDC stage.

\subsubsection{S-bearing molecules}
The ratio of the S-bearing molecules CS and SO is poorly described by the model and represents a well-known problem with sulfur chemistry in modern astrochemical databases \citep{druard2012,herpin2009,loison2012,Dutrey_ea11}. While in the observations for the IRDC stage both molecules have only an upper limit, the later stages show an increasing trend of the ratio between 10 and 100. In contrast, the model predicts a strongly decreasing CS/SO ratio over about three orders of magnitude. The ratio of SO with the early type molecule N$_2$H$^+$ is also poorly described by the model. Only the HMPO and HMC phase agree slightly, with a decreasing ratio with time. Since the model for CS and SO alone does not agree very well for both, it is natural that the ratio  does not agree either.

\begin{figure*}
\includegraphics[width=0.32\textwidth]{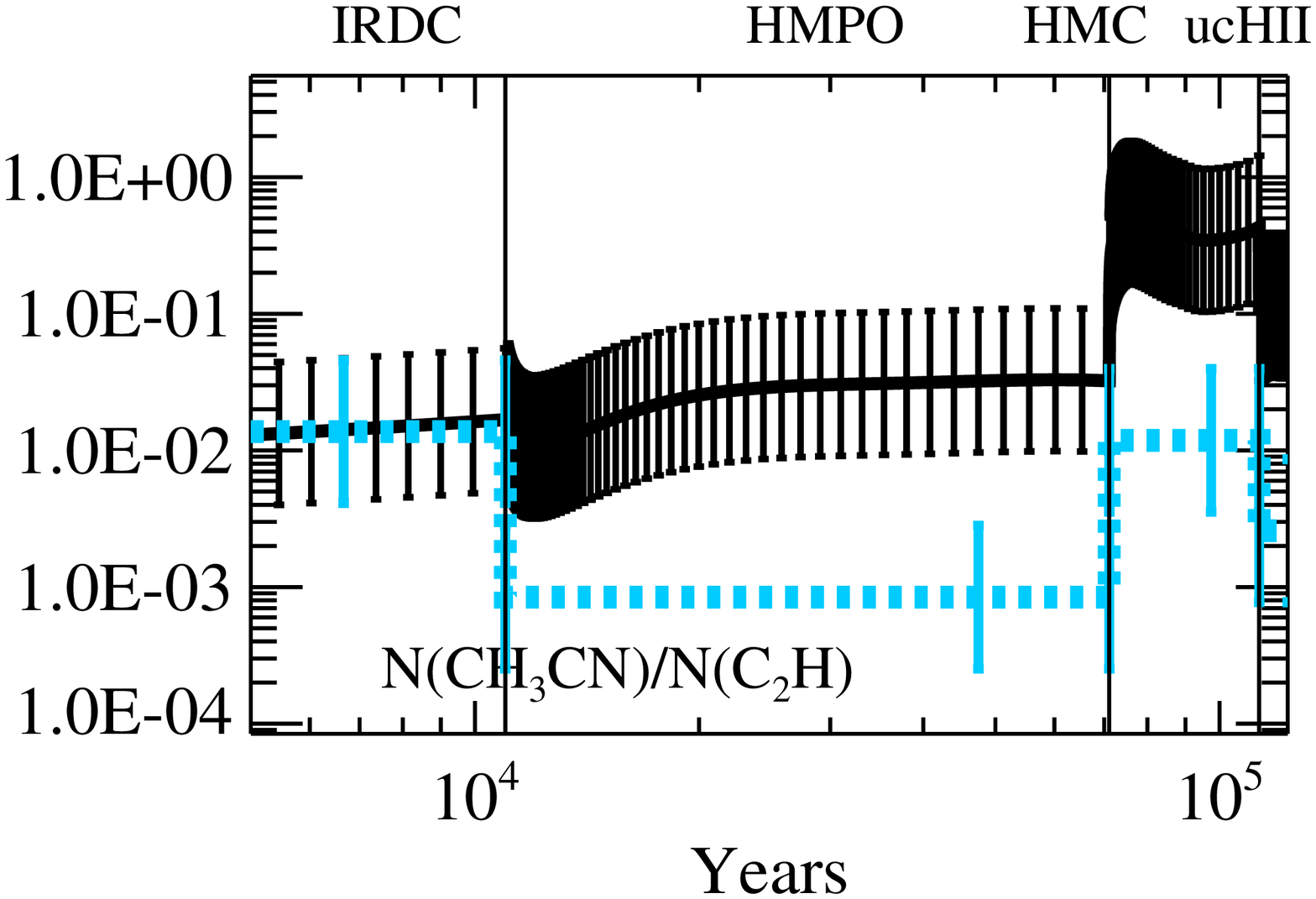}
\includegraphics[width=0.32\textwidth]{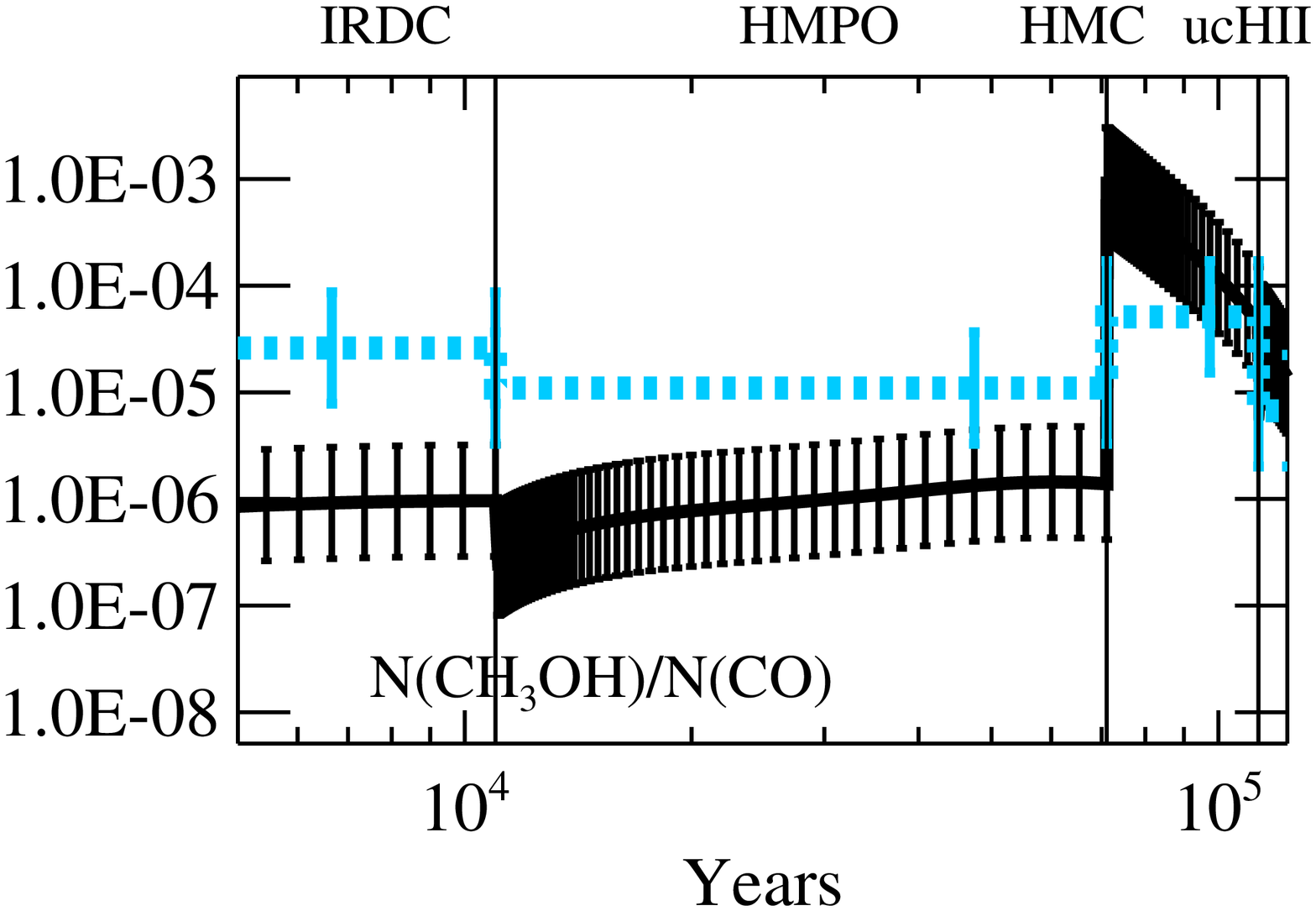}
\includegraphics[width=0.32\textwidth]{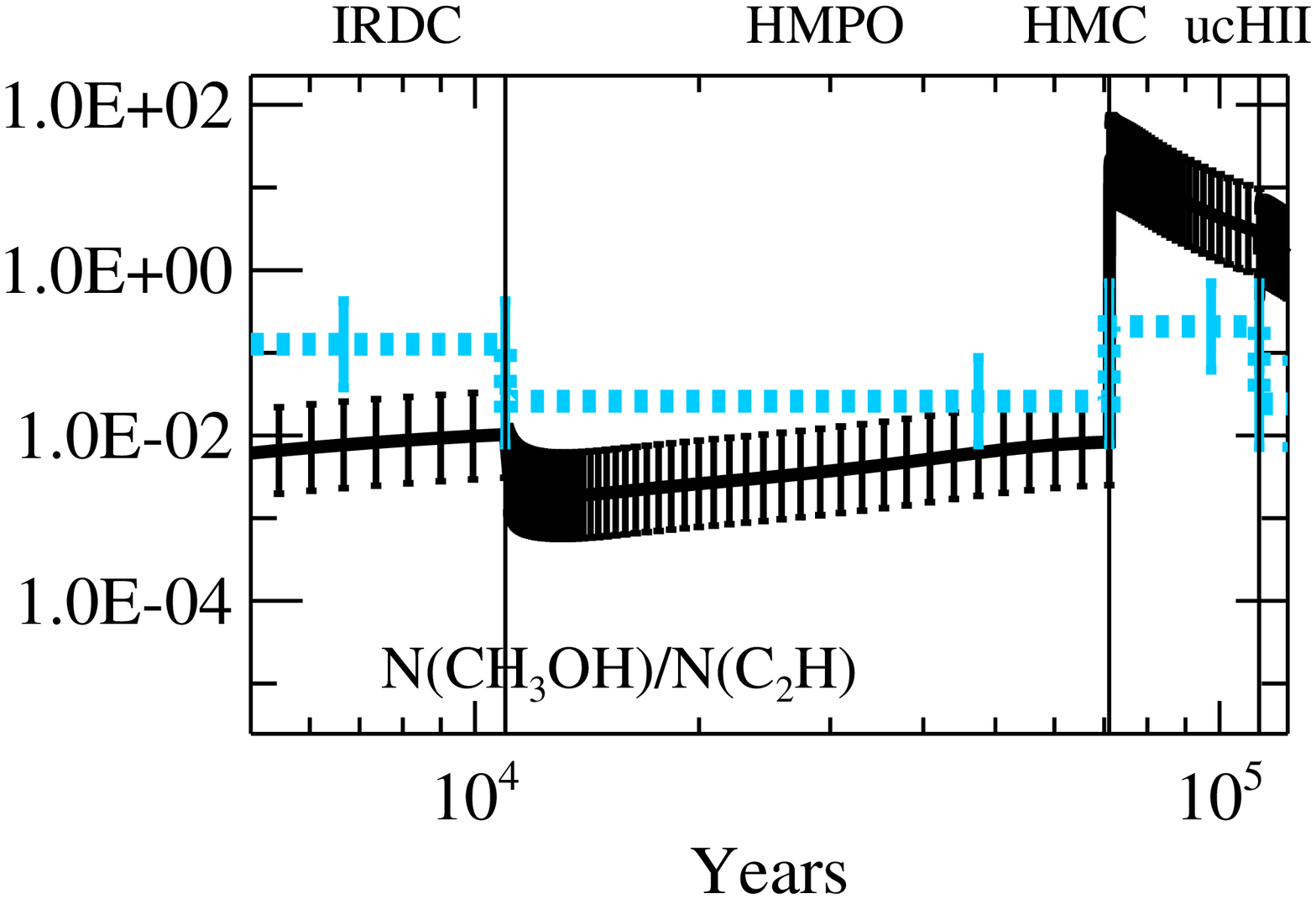}\\
\includegraphics[width=0.32\textwidth]{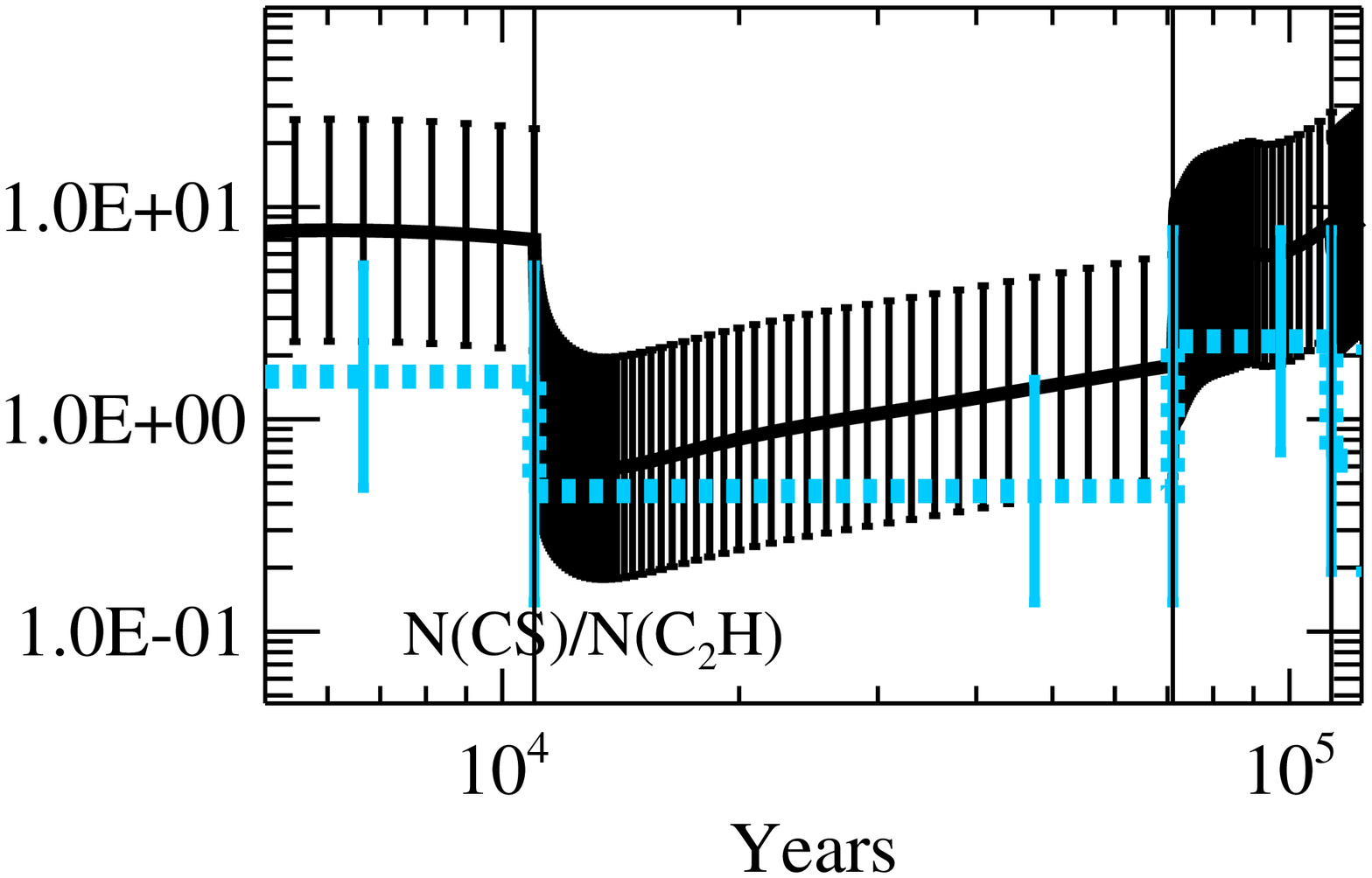}
\includegraphics[width=0.32\textwidth]{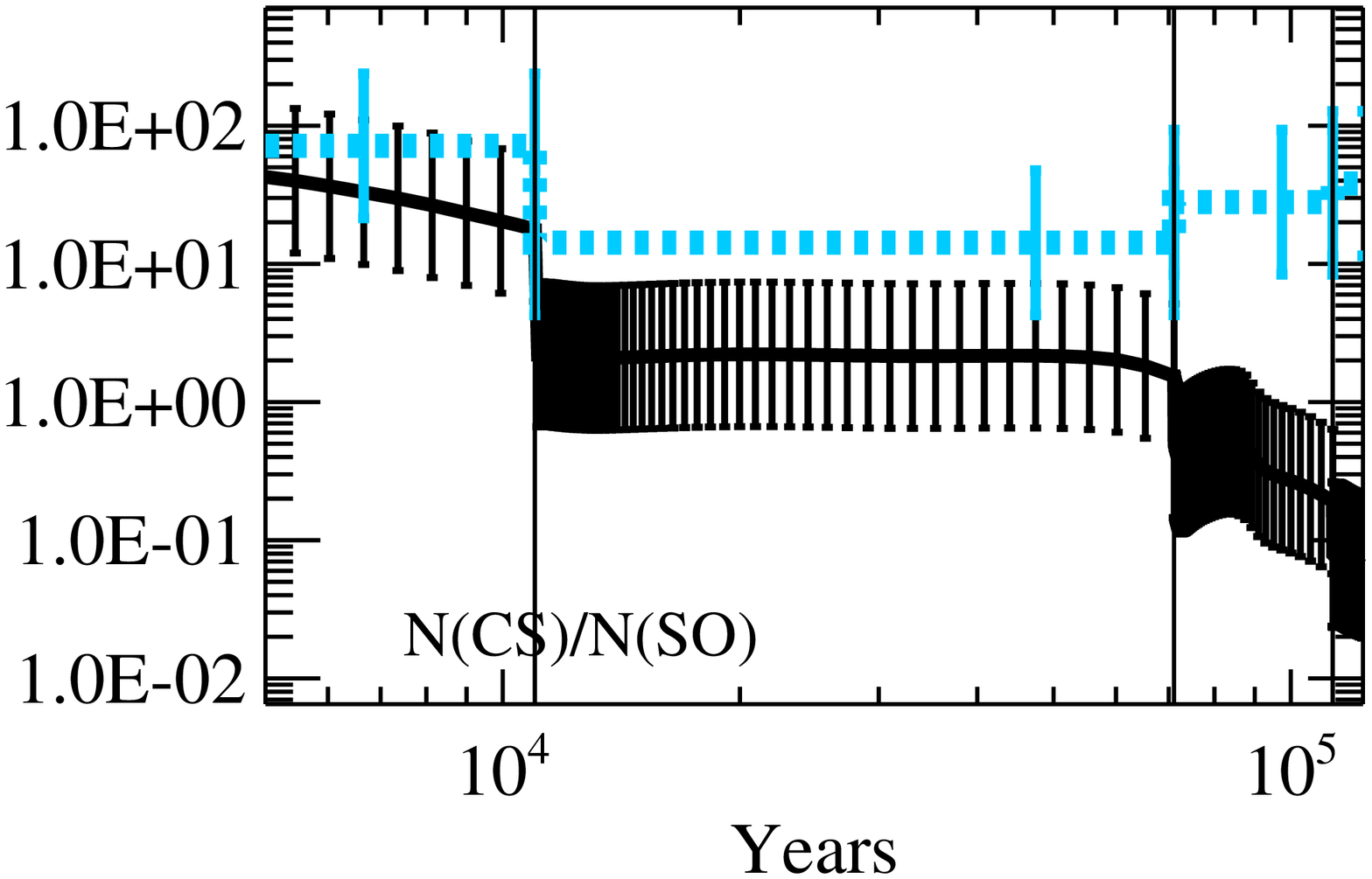}
\includegraphics[width=0.32\textwidth]{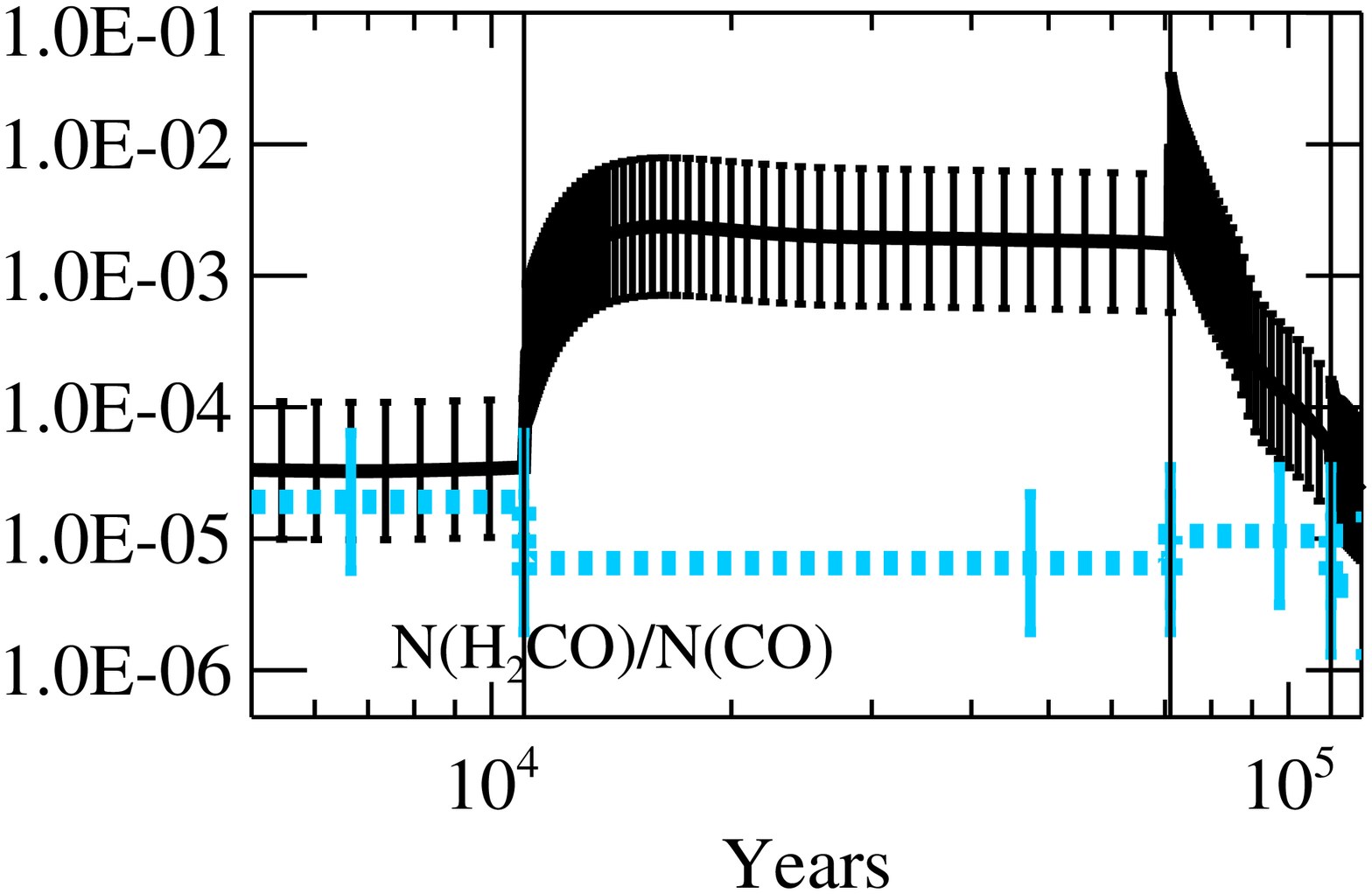}\\
\includegraphics[width=0.32\textwidth]{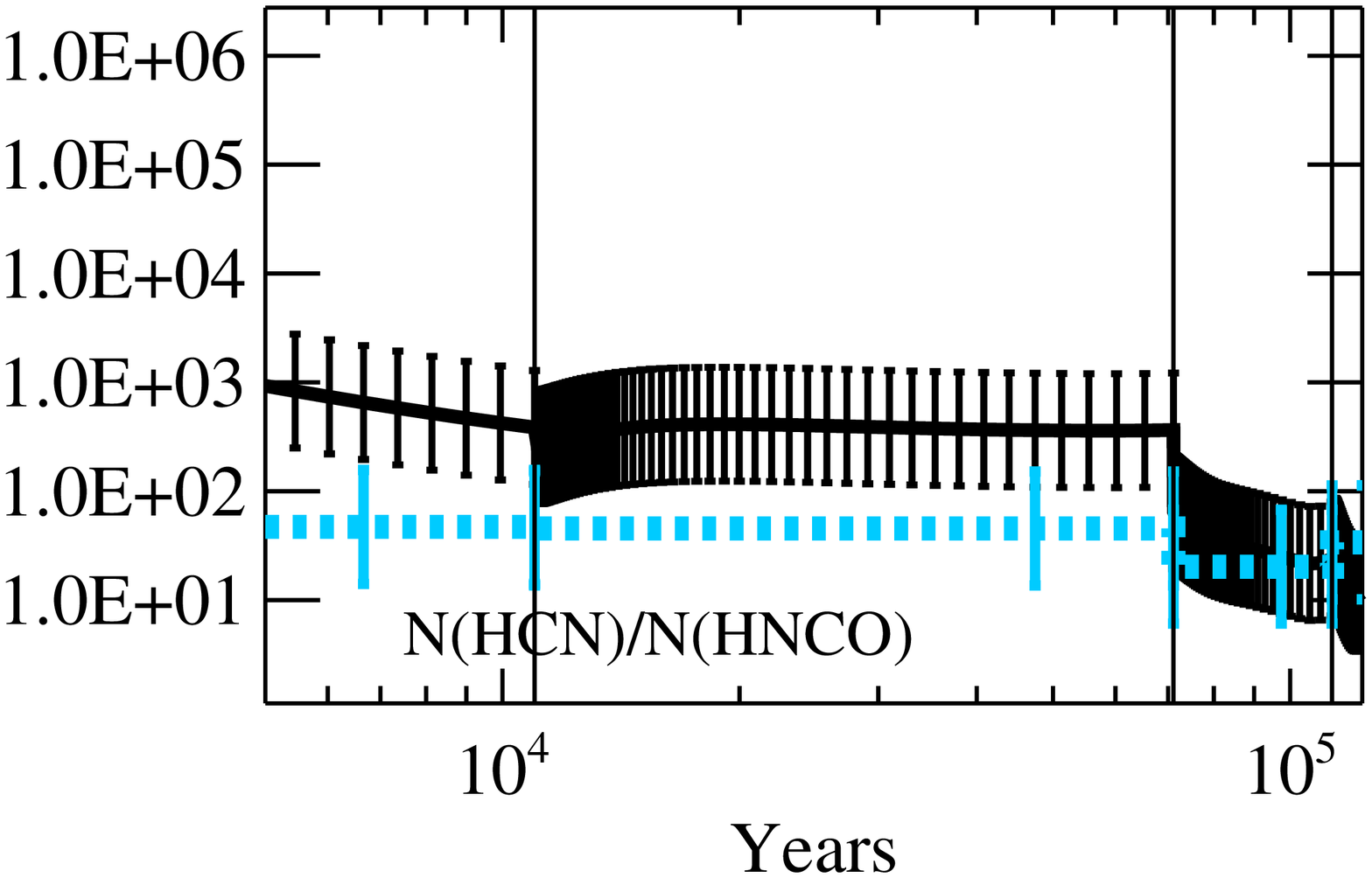}
\includegraphics[width=0.32\textwidth]{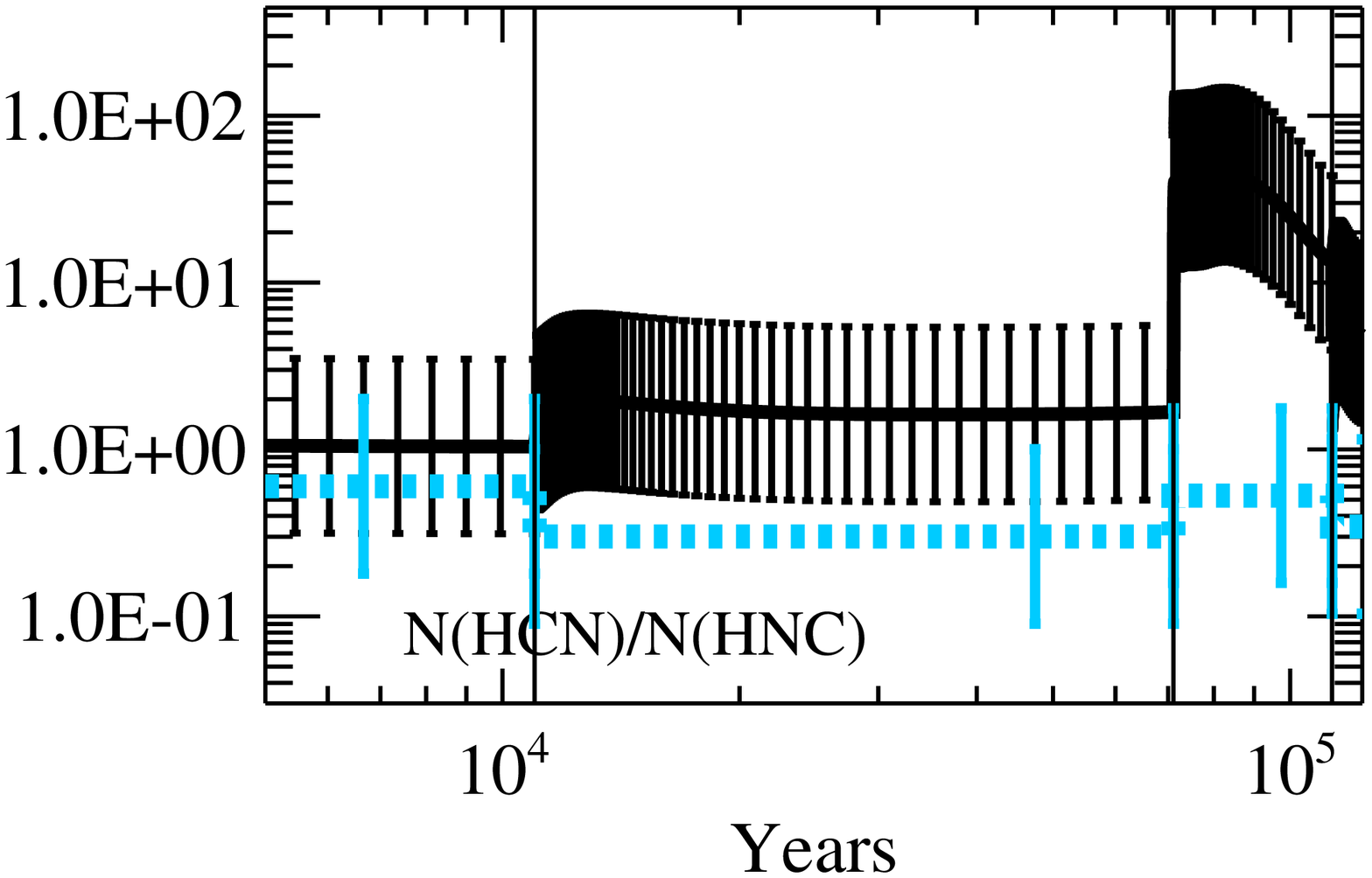}
\includegraphics[width=0.32\textwidth]{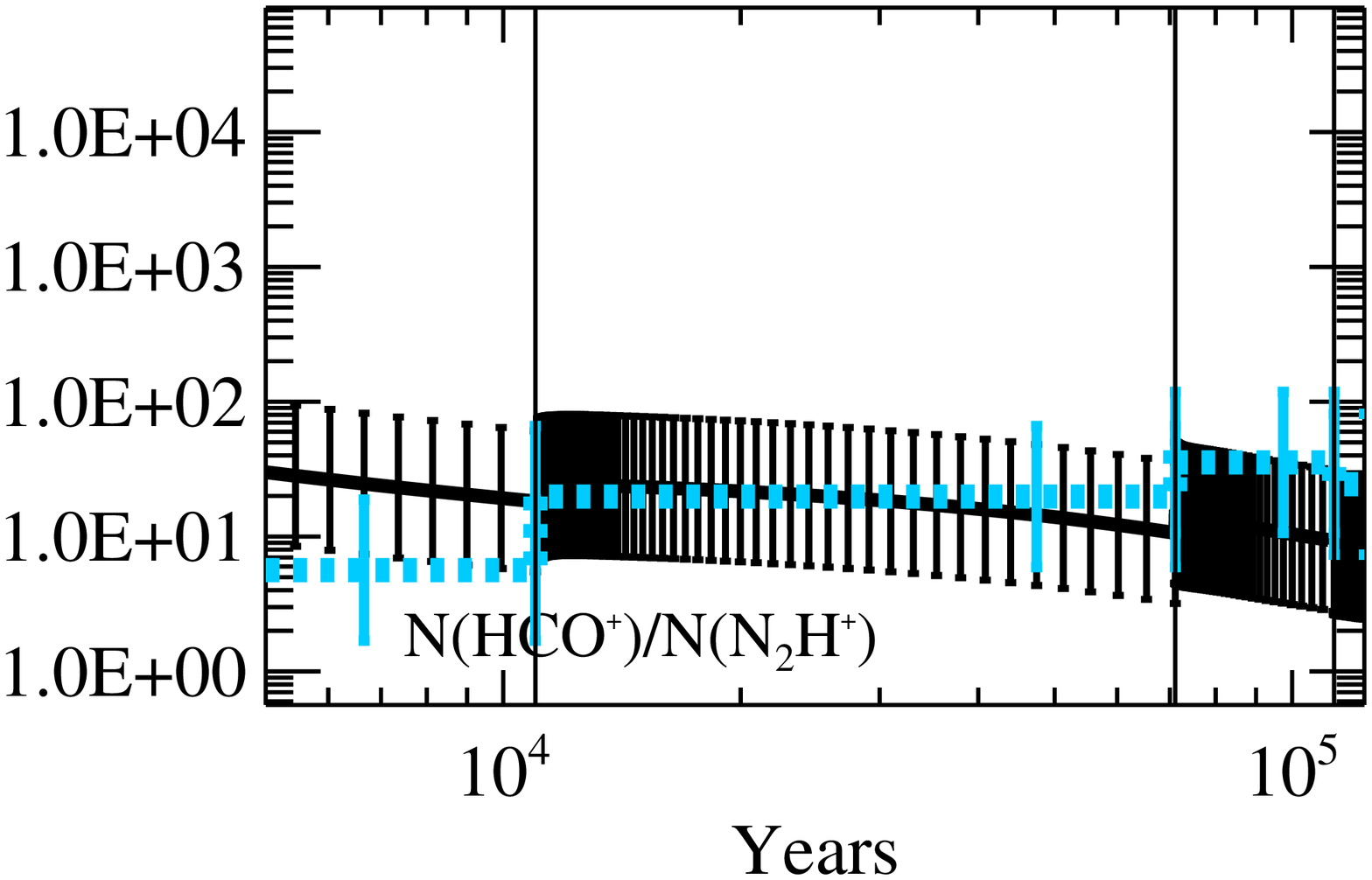}\\
\includegraphics[width=0.32\textwidth]{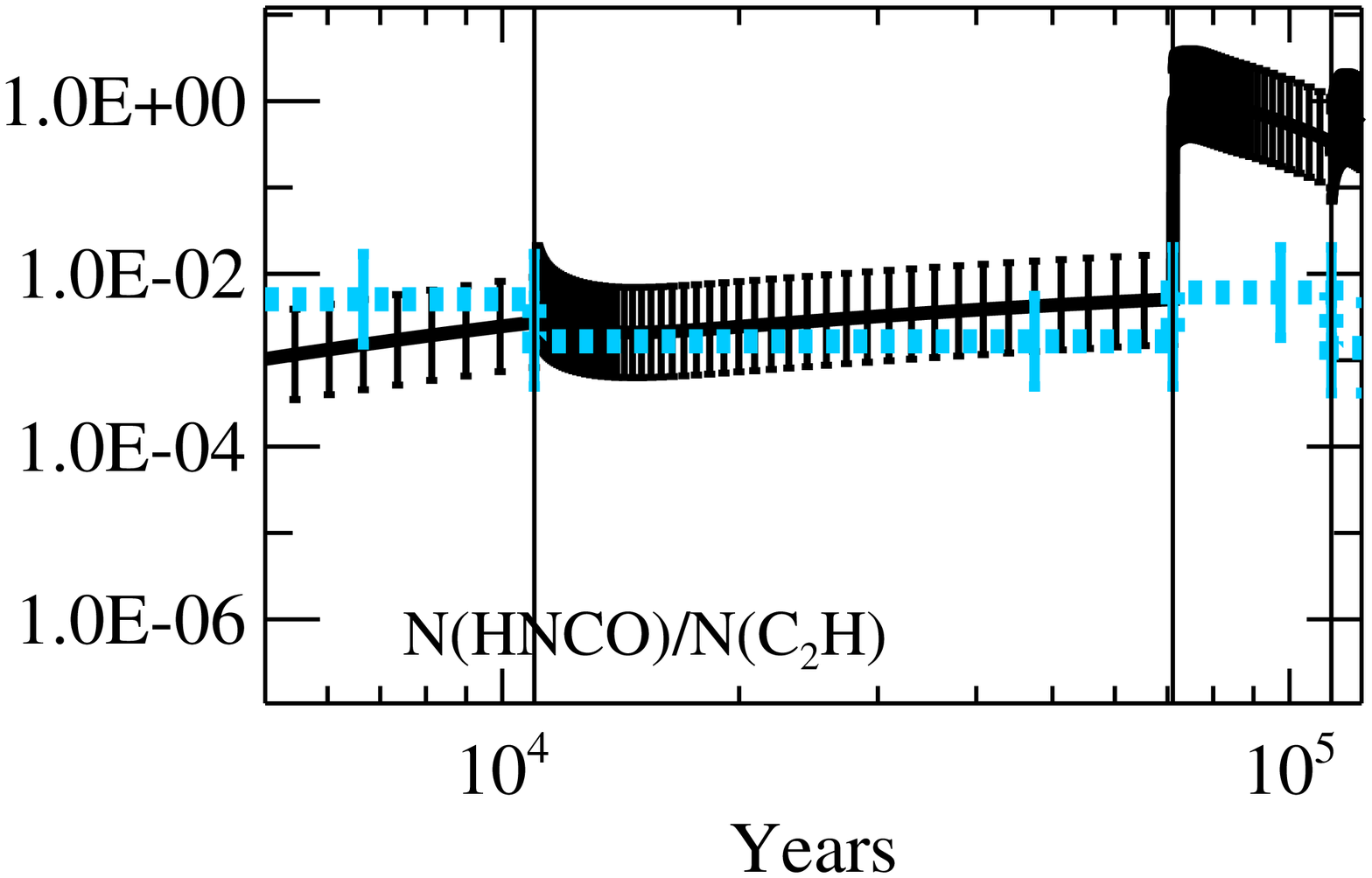}
\includegraphics[width=0.32\textwidth]{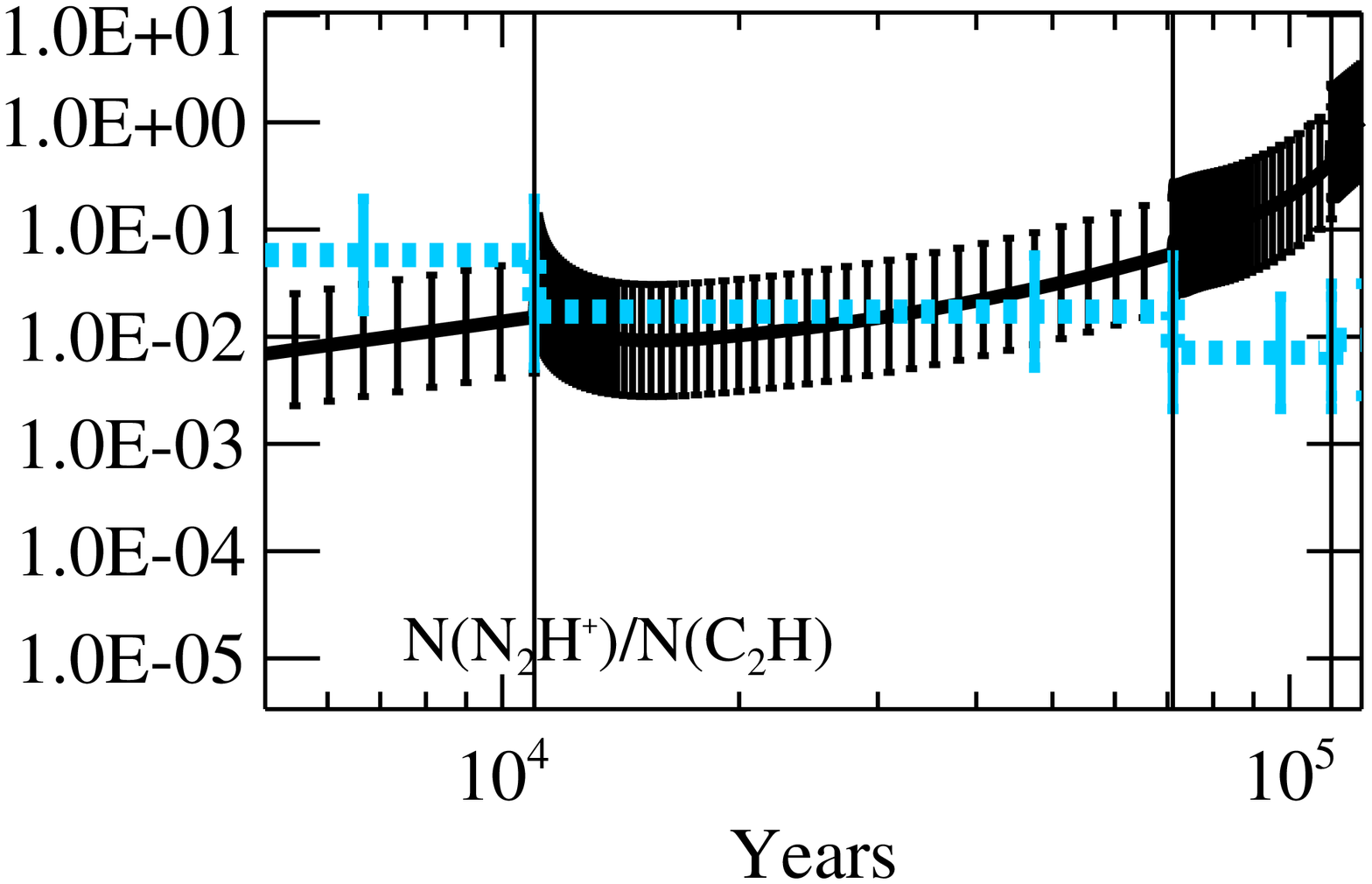}
\includegraphics[width=0.32\textwidth]{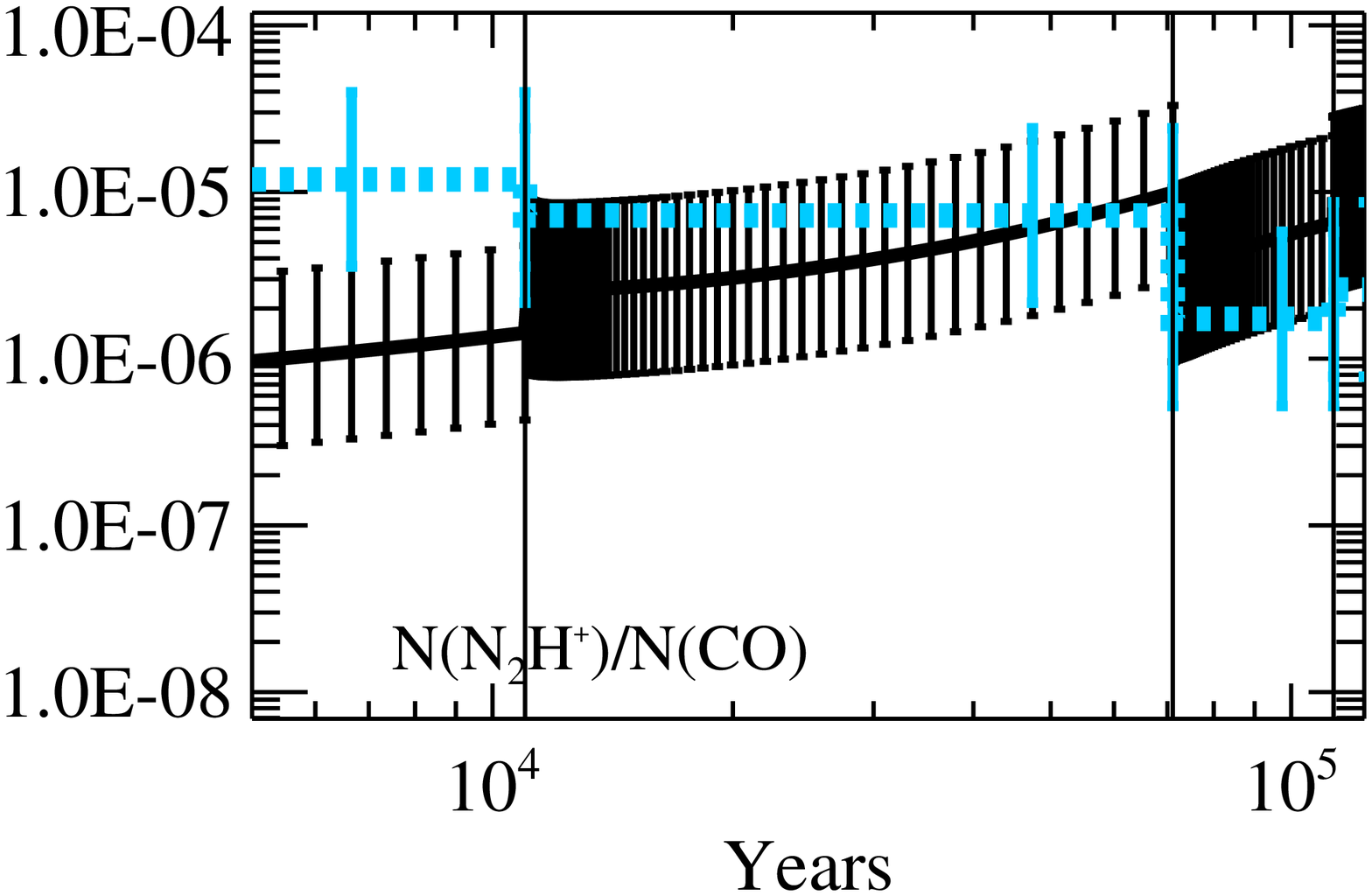}\\
\includegraphics[width=0.32\textwidth]{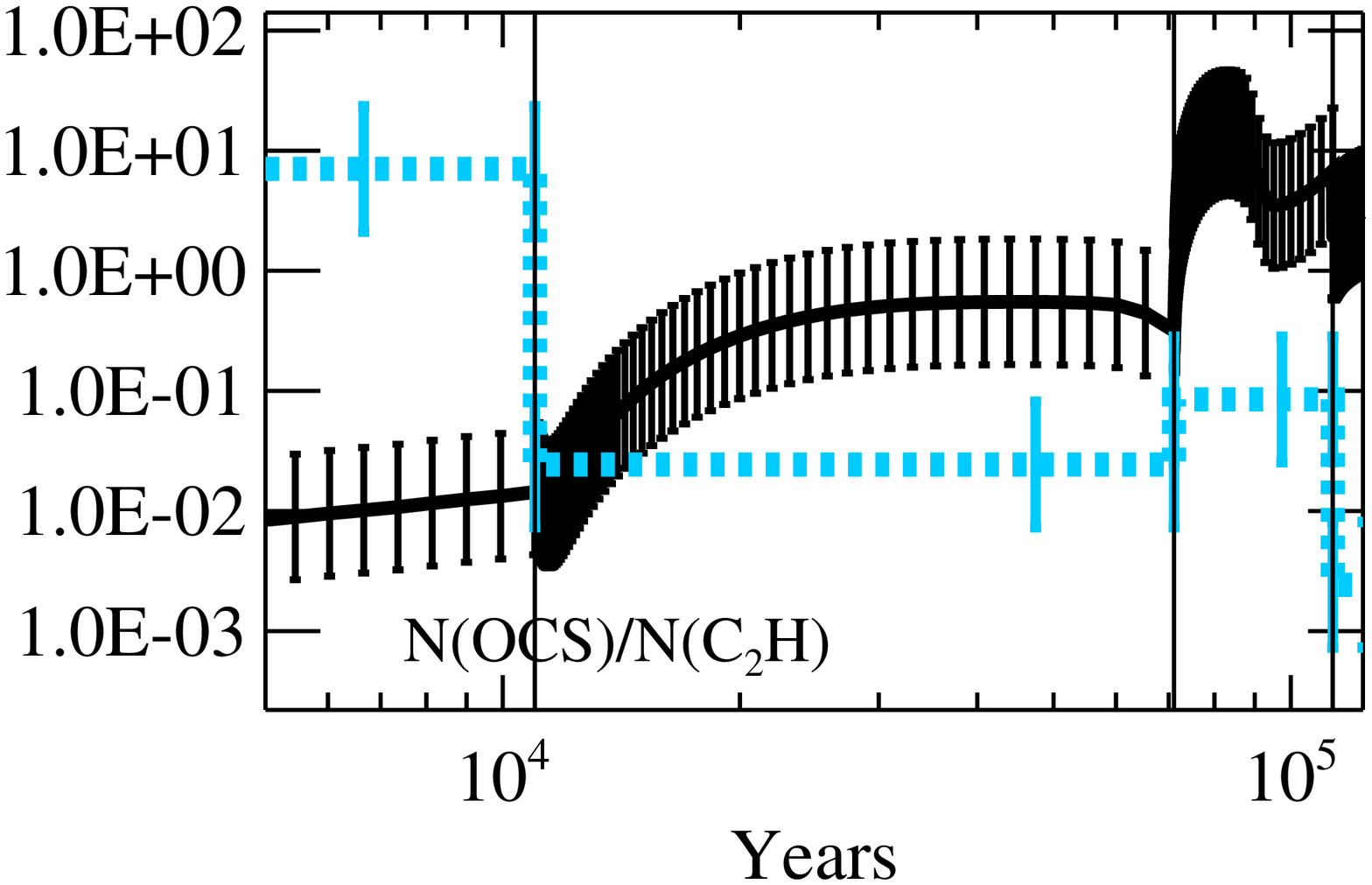}
\includegraphics[width=0.32\textwidth]{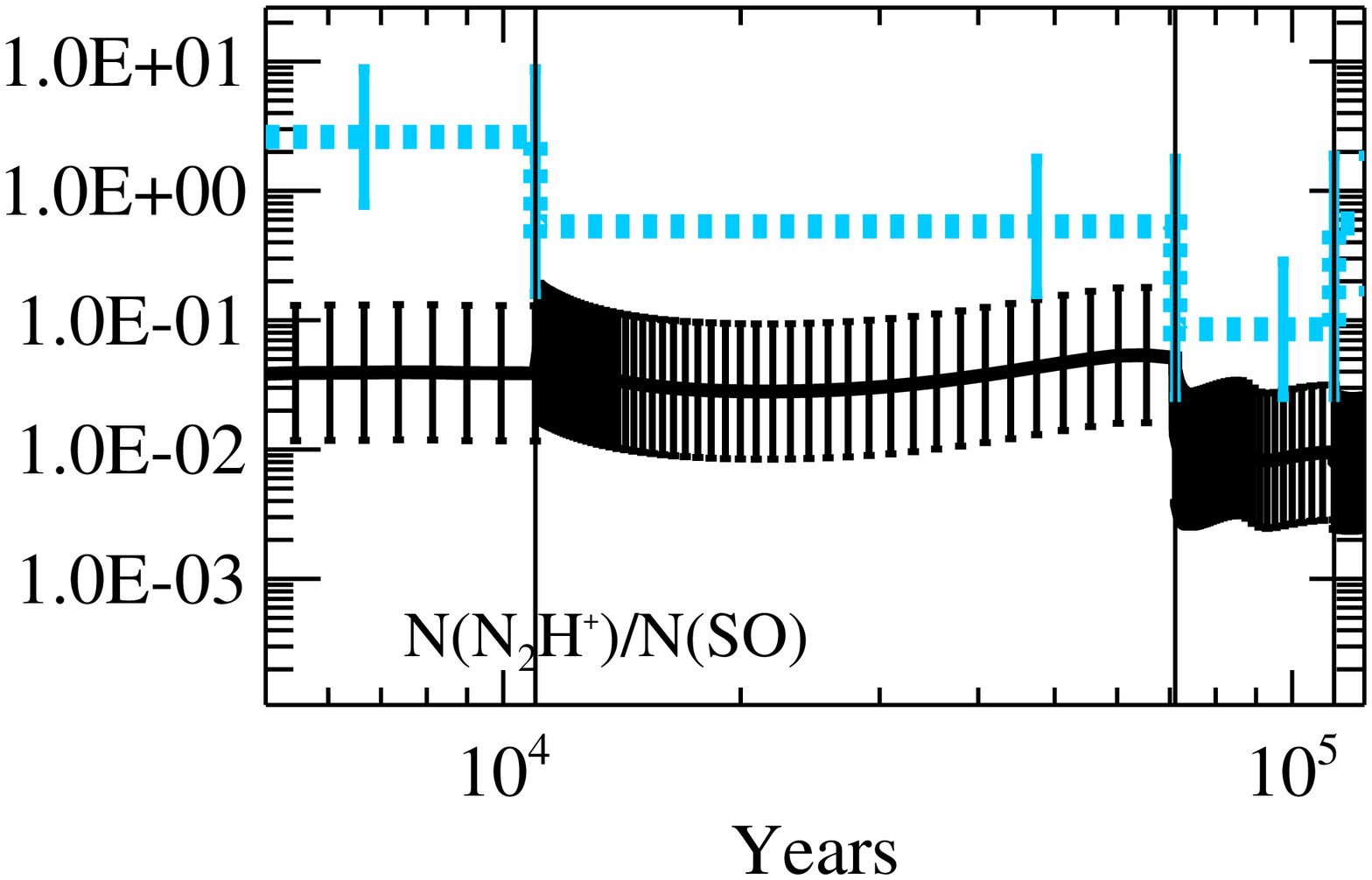}
\includegraphics[width=0.32\textwidth]{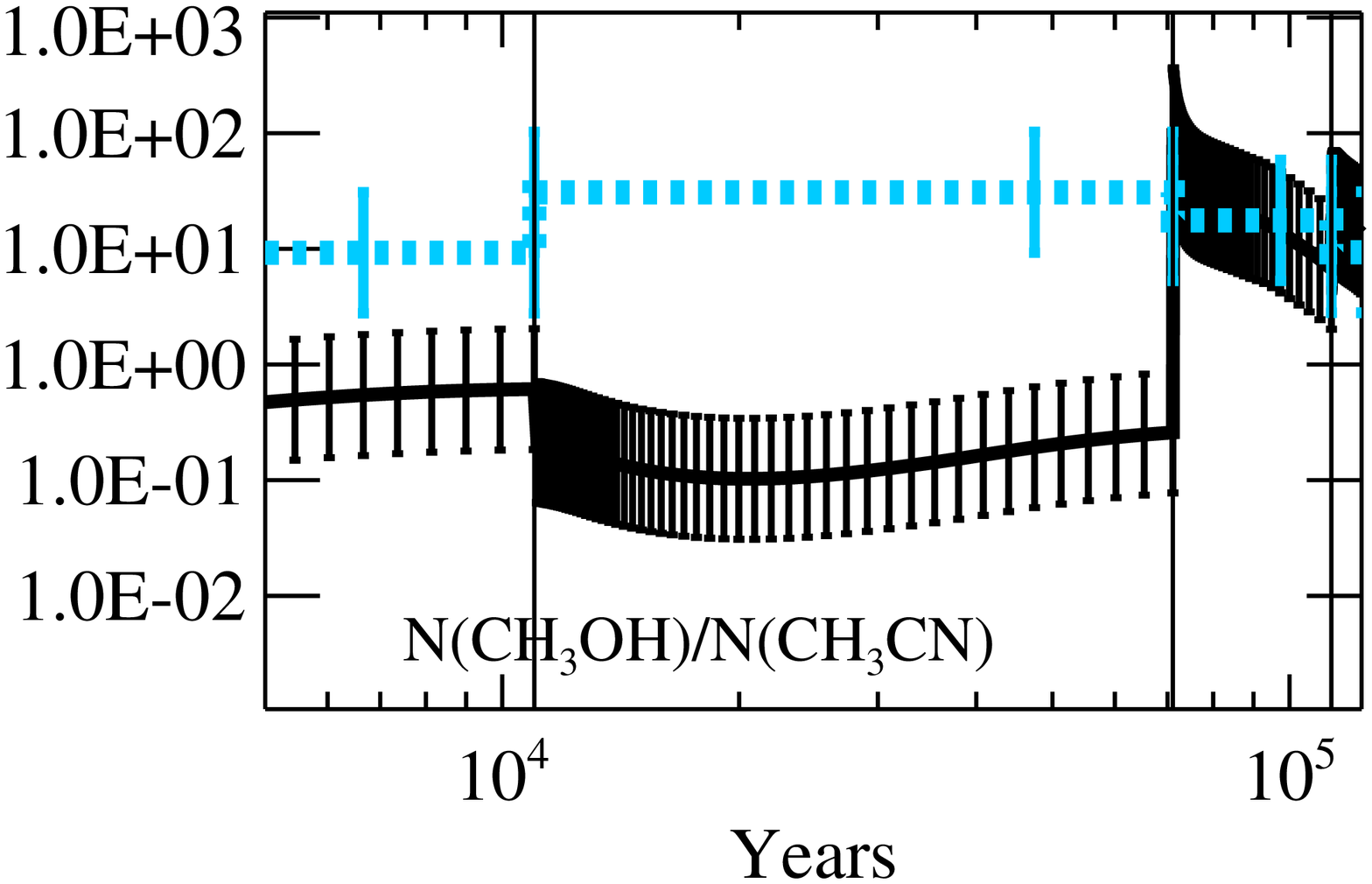}
\caption{Column density ratios for 15 different combinations of molecules plotted for all four stages. The modeled values are shown by the black solid line, the observed values are depicted by the blue dashed line. The error bars are indicated by the vertical marks.}
\label{fig:coldenseratios}
\end{figure*}

For completeness and to outline the strengths and caveats of our modeling approach, Figure~\ref{fig:coldenseratios} presents a few more ratios that we do not discuss in detail here.

\subsection{Chemical evolutionary sequence -- synthesis of observations and models}\label{sec:chemevolsequence}
With observational and modeled data at hand we can derive a general picture of the chemical evolution in high-mass star-forming regions. The starting condition at the year 0 in our model is a cold dense molecular cloud, which is still close to isothermal with the initial chemical composition given in Table~\ref{tab:inabun}. Any evolution prior to the IRDC stage to these initial conditions is not considered in this work. In our IRDC model the temperature remains constant, and the chemical evolution proceeds under these conditions for about 11\,000 years. At this time the best overall agreement between the observed and the calculated molecular column densities is reached. The modeled abundance of the best-fit model at the best-fit age are used as input abundances for the next HMPO stage of the evolution.
From that time the HMPO phase begins in our model. An internal source heats the surrounding medium, such that the temperature increases toward the inner region. This is implemented in the modeling by a temperature profile $q=0.4$. Increased temperature results in a higher mobility of the surface molecules and a more efficient desorption of ices into the gas phase. Consequently, the molecular abundances increase. Our best-fit model matches the observed values for the HMPOs after 60\,000~years (with a total age of 71\,000~years from time zero).
As temperatures and densities continue to rise due to formation of a central protostar(s), the object enters the HMC phase. The subsequent increase in temperature releases most of the molecules locked along with water ice in the grain mantles into the gas phase. This increases gas-phase abundances of already abundant molecules even further and observed transitions become stronger. The detection rate of molecular lines increases as well, including those from complex and heavy molecules. According to our best-fit model, the HMC phase lasts for about 42\,000~years, leading to a total age of 113\,000~years.
Finally, the last considered stage of high-mass star-forming regions begins. For the UCH{\sc ii} phase the temperature and density structures are re-adjusted again. The best-fit age derived for this last stage is 11\,500~years. Thus, we predict that the total evolution from the formation of an IRDC until the formation of an UCH{\sc ii} region takes about 124\,500~years (with an uncertainty of a factor of 2-3). It appears to be consistent with the estimates obtained from modeling of the formation of high-mass stars \citep[e.g.,][]{mckee2003}.

\subsection{Observed ionization degree compared with model data}\label{sec:ionratecomparison}

\begin{figure*}
\includegraphics[width=0.43\textwidth]{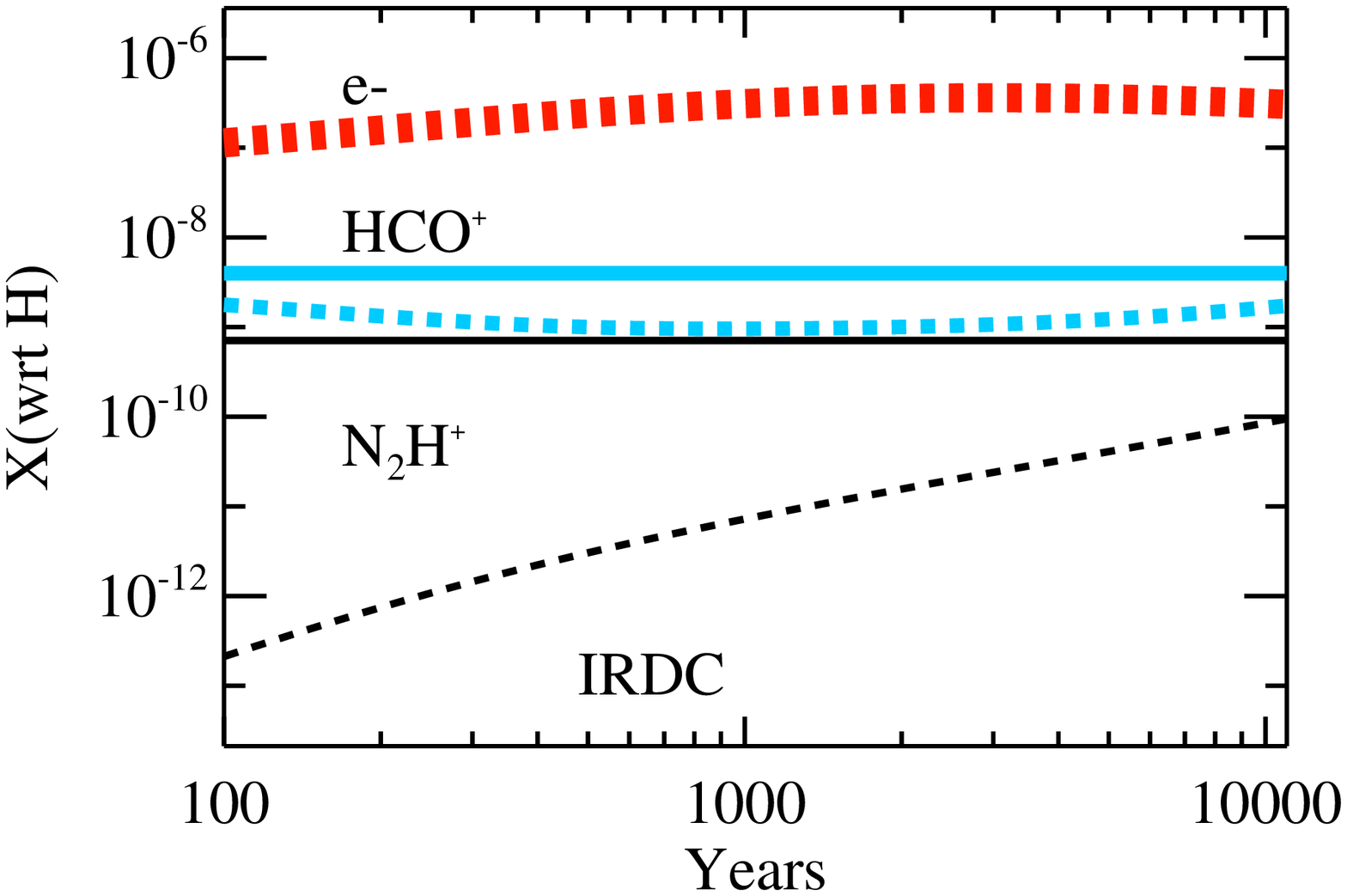}
\includegraphics[width=0.43\textwidth]{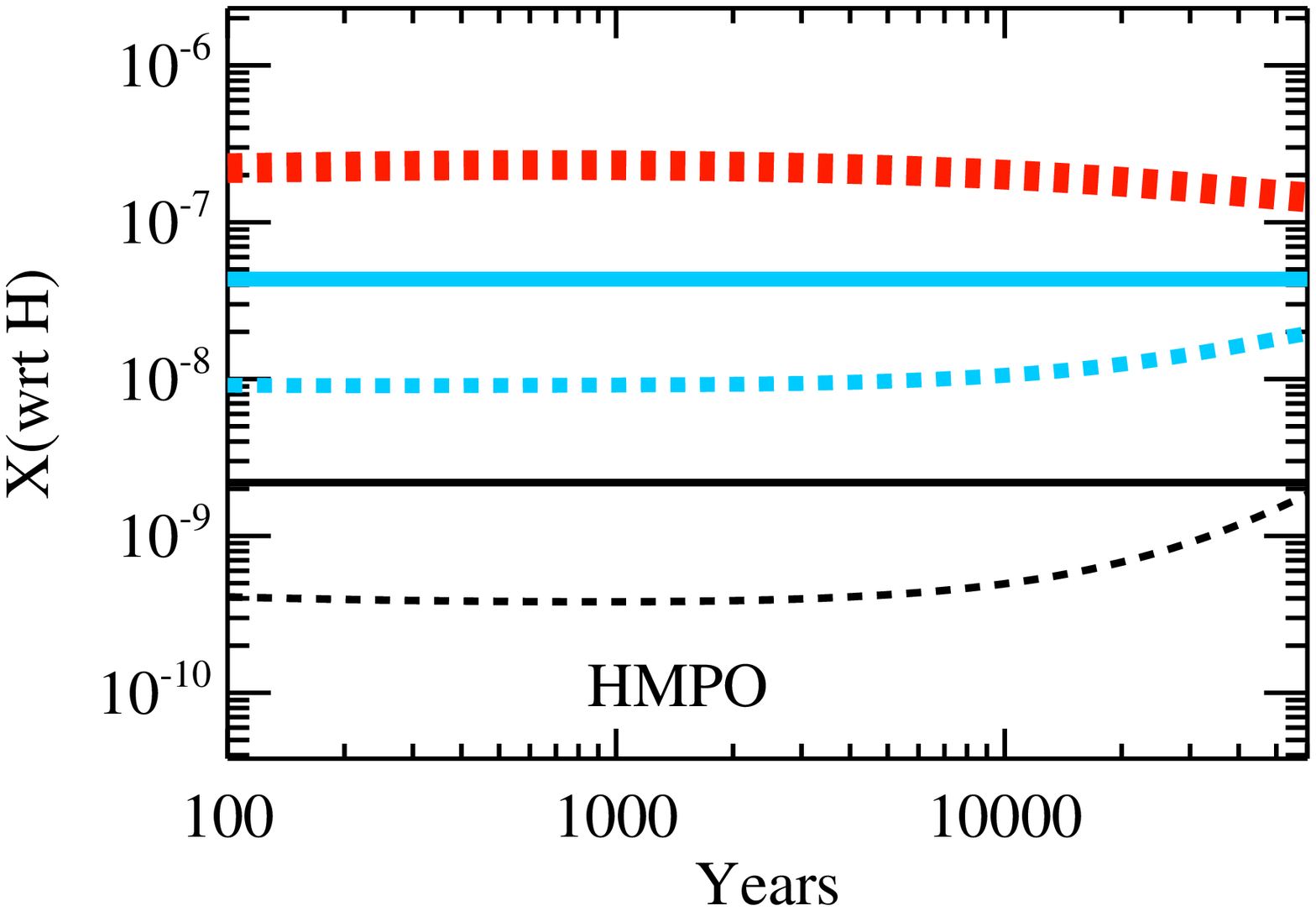}\\
\includegraphics[width=0.43\textwidth]{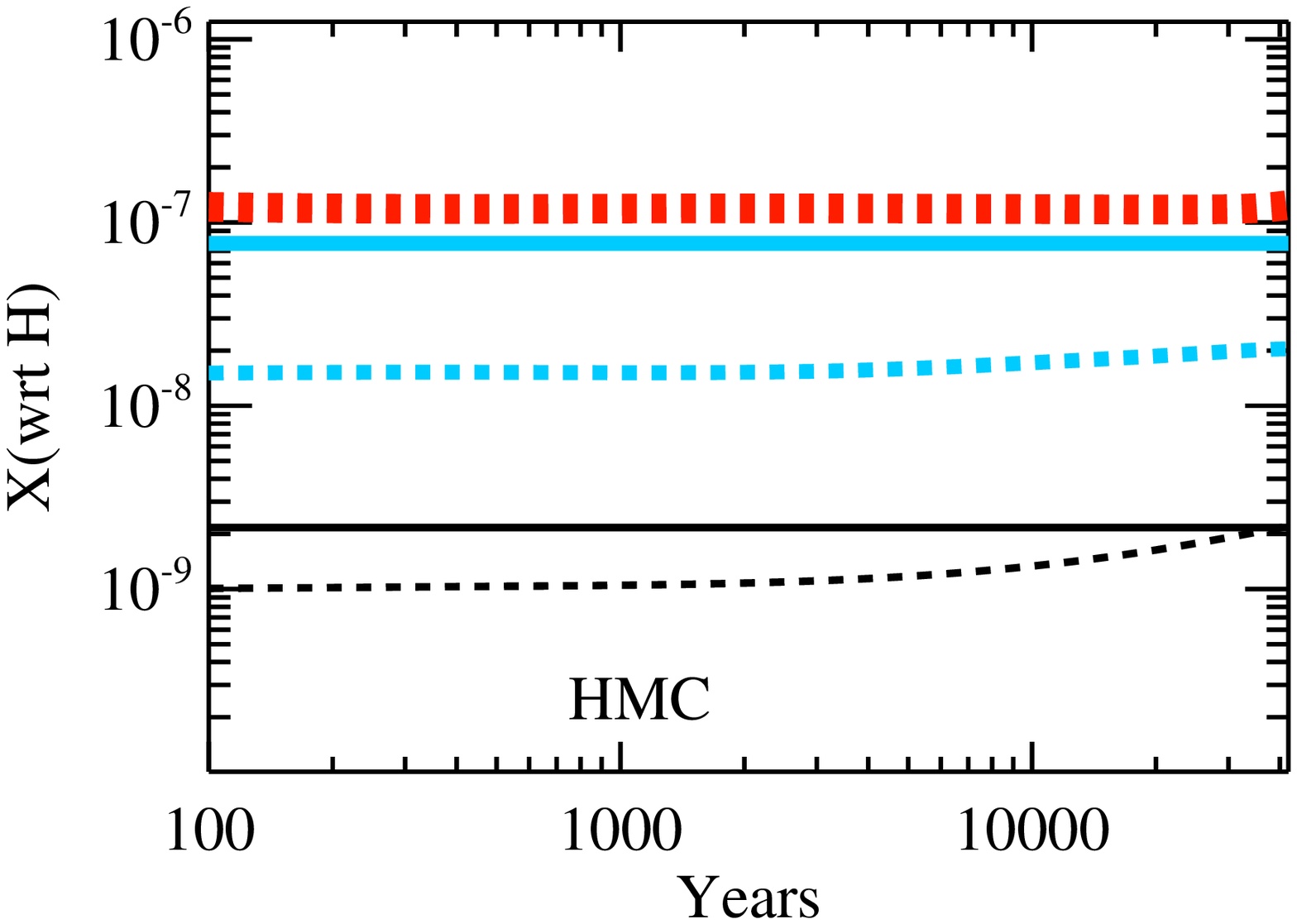}
\includegraphics[width=0.43\textwidth]{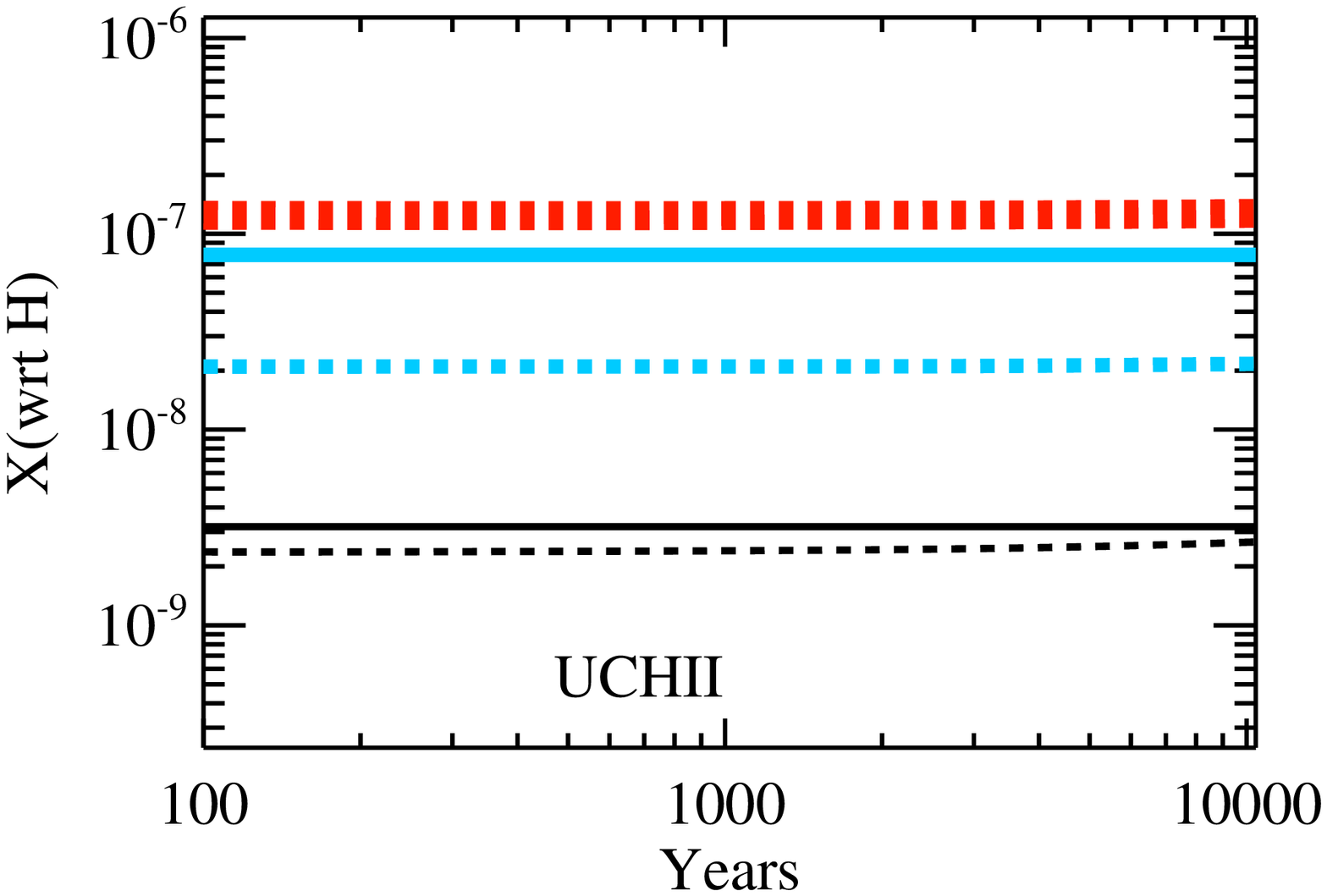}
\caption{Averaged relative abundances to H$_2$ for electrons, HCO$^+$, and N$_2$H$^+$  plotted for the all four stages. The observed
values are shown as solid lines, the modeled values are depicted as dashed lines. The electron abundances are plotted in red, N$_2$H$^+$ abundances in black, and HCO$^+$ abundances in blue. The error bars for both the model and the observational data are a factor of about 3 (not shown).}
\label{fig:iondegree}
\end{figure*}

In Section~\ref{sec:ionizationdegree} we calculated lower limits for the ionization degree in the four evolutionary stages of high-mass star formation. In addition, the best-fit model provides all necessary information to derive the exact modeled ionization degrees to compare. The modeled abundances of the main constituents, namely, electrons, N$_2$H$^+$, and HCO$^+$ and the observed abundances for N$_2$H$^+$ and HCO$^+$ , are shown in Figure~\ref{fig:iondegree}. While the observed lower limit increases with evolutionary phase, the ionization degree in the best-fit model remains almost constant at $\sim10^{-7}$ throughout the whole evolution. In comparison with the observed data it is off by about one order of magnitude in the first two stages and agrees for the last two stages.
This leads to the conclusion that a lower limit for the ionization degree derived from HCO$^+$ and N$_2$H$^+$ is a good first approximation. However, it also indicates that in particular in the early evolutionary phases, other ions contribute more strongly to the total ionization degree. These are deuterated ions in the dense cores, which we will observe in a follow-up study, but  the singly ionized carbon is also important during these stages. For example, \citet{miettinen2011} took deuterated ions into account as well and found slightly higher lower limits for x(e)~$\sim10^{-8}-10^{-7}$ , which are comparable with the results of our model. In addition to this, more sophisticated methods are needed to estimate the ionization degree with an uncertainty better than one order of magnitude. Here, any deviation from the assumption of a homogeneous medium used in our model (e.g., clumpiness on various scales) will tend to increase the modeled ionization fraction for the later evolutionary stages when some high-
energy photons 
might already be produced by forming protostar(s).


\subsection{Comparison with literature}


Our comprehensive study of the chemical evolution in the early phases of massive star formation can be set in context to previous studies focusing on distinct evolutionary phases.
The comparison of our work with the study of IRDCs from \citet{vasyunina2011}  qualitatively agrees. Assuming uncertainties of about one order of magnitude for each of the derived abundances, only 2 of 9 molecules show a stronger deviation, namely, HNCO and SiO with lower abundances of slightly more than one order of magnitude in our sample. We detected CH$_3$CN only in two sources in our IRDC sample, which was not detected by Vasyunina et al. either. In a more recent study, \citet{vasyunina2013} additionally derived abundances of H$_2$CO and CH$_3$OH for a sample of IRDCs. The range of abundances they found are overlapping with, but slightly higher than, our findings.
The observed sample of clumps in IRDCs from \citet{sanhueza2012} shows in general higher abundances for all observed molecules, but 5 out of 6 of them agree with our values. \citet{sakai2008} derived for a sample of IRDCs column densities of N$_2$H$^+$ with values that are similar to our results. For CH$_3$OH our findings fall into the lower end of their observed range of column densities, but are still consistent.
\citet{sakai2010} calculated the column densities for HMPO-like sources. Comparing their sample of MSX sources with our HMPO sample we find good agreement with their observed range of column densities and our median values. Only CH$_3$OH and CS isotopologues show a bigger deviation on the order of 10, which still agrees with the uncertainties of the observed values in our study.
Our derived median abundances of the detections for the HMCs and UCH{\sc ii}s are similar to each other within the uncertainties. However, we derived higher median column densities for the HMCs due to a smaller mean distance and higher H$_2$ column densities. \citet{hatchell1998b} inferred the H$_2$ column densities from  virial mass estimates of the observed gas, assuming a specific source size and extension. Therefore, their H$_2$ column densities are different for different molecules. Comparing our median column densities for the HMCs and UCH{\sc ii}s with their results, we find an overall good agreement. The only molecule that deviates by a factor of more than 10 is CH$_3$OH (for both phases) and CH$_3$OCHO (for the UCH{\sc ii} stage). \citet{reiter2011} derived molecular column densities of high-mass clumps coincident with water maser emission and mainly associated with H{\sc ii} regions for a larger set of molecules. These values we can compare with our more evolved sources. The range of values 
they found for HCO$^+$, C$_2$H, HNC and CS are slightly lower, by a factor up to 5, than the median values in our sample. The column densities of N$_2$H$^+$ and SO are similar to our findings.
The most prominent source in our sample is the well-studied Orion-KL hot core. A summary of abundances in this source is given in \citet{vandishoeck1998}. Our results also agree well with these values. The SiO abundances differ by a factor of 6, while all other eight molecules in comparison show differences by factors of 1-4.

\section{Conclusion}\label{sec:conclusion}
We observed the chemistry in 59 sources covering the early stages in massive star formation to study and characterize the physical and chemical evolution. This was the first consistent attempt to follow the evolution of physics and chemistry over several evolutionary steps in high-mass star formation. The underlying part of the evolutionary sequence assumed in this work started with an IRDC that evolved over the HMPO and HMC phase and ended with the formation of an UCH{\sc ii} region. Using this approach, we obtained a consistent sample of spectra for all of the sources. The spectra show an increasing number and intensity of the molecular transitions for more evolved stages. More complex and heavy molecules are formed with evolving age until the richness of molecular composition reaches its maximum in the HMC phase and declines for the UCH{\sc ii} stage, because these molecules are probably destroyed by the UV-radiation from the forming stars.

We calculated column densities for different molecular species that are in general increasing with evolutionary phase. Their ratios are less dependent on the temperature than the abundances. Thus we see a more obvious evolution in the abundances than in the ratios of these data. Furthermore, the derived lower limits of the ionization degree in the different stages increase along the evolutionary picture, as well.

In a second step we compared the characteristic median column densities of each evolutionary stage with a model. We calculated the best-fit parameters and ages, using a 1D physical model coupled to a gas-grain chemical model, in which the evolutionary sequence was approximately taken into account. The observed column densities and abundances agree reasonably well with the overall fit of our model. Furthermore, our results agree with previous observational studies. Based on the fit of the chemical model to 14 different molecular species of unresolved data, we were able to derive constrains on the underlying physical structure. The best-fit duration for the full evolution from an IRDC with atomic gas composition to the stage of an UCH{\sc ii} region is about 125\,000~years. This is consistent with estimates from theoretical models of the formation of high-mass stars. The results from this study can be considered as chemical templates for the different evolutionary stages during high-mass star formation.

\begin{acknowledgements}
The authors wish to thank the anonymous referee for carefully reading the manuscript; his or her comments and suggestions improved it significantly. T.G. is supported by the Sonderforschungsbereich SFB 881 ``The Milky Way System'' (subproject B3) of the German Research Foundation (DFG). T.G. is member of the IMPRS for Astronomy \& Cosmic Physics at the University of Heidelberg.
This research made use of NASA's Astrophysics Data System.
D.S. acknowledges support by the {\it Deutsche Forschungsgemeinschaft} through SPP~1385: ``The first ten million years of the solar system - a planetary materials approach'' (SE 1962/1-2).
T.V. wishes to thank the National Science Foundation (US) for its funding of the astrochemistry program at the University of Virginia.
In the reduction and analysis of the data we made use of the GILDAS software public available at http://www.iram.fr/IRAMFR/GILDAS .
\end{acknowledgements}

\bibliographystyle{aa}    
\bibliography{gerner_final.bbl}

\appendix

\section{Appendix material}

\clearpage
\onltab{
\begin{table*}
\tiny
\caption{Integrated intensity $\int T_{\rm mb}$ in K km\,s$^{-1}$. For N$_2$H$^+$ we give $T_{\rm {mb}}$ in K derived from the standard hyperfine fitting CLASS routine. Non-detections are indicated by $\leq$ and the 1$\sigma$~rms value is given. A blank entry means that the molecule is detected, but rejected from the analysis due to reasons discussed in Section~\ref{sec:problems}.}             
\label{tab:integrals_1}      
\centering                          
\begin{tabular}{llllllllllll}        
\hline\hline                 
Source & HCN & HNCO & C$_2$H & HN$^{13}$C & SiO & H$^{13}$CO$^+$ & N$_2$H$^+$ & $^{13}$CS & CH$_3$CN & CH$_3$OCHO & SO\\    
\hline                        
IRDC011.1 &      & 1.25 & 1.53 & 1.76 & 2.47 & 2.08 & 3.53 & 0.37 & 0.29 & $\leq$0.03 & $\leq$0.08\\ 
IRDC028.1 &      & 2.70 & 2.03 & 1.94 & 4.52 & 1.96 & 5.87 & 0.74 & $\leq$0.05 & $\leq$0.05 & 1.59\\ 
IRDC028.2 &      & 4.20 & 3.32 & 1.93 & 7.94 & 2.59 & 2.32 & 1.45 & 0.95 & $\leq$0.05 & 9.89 \\ 
IRDC048.6 & 3.47 & 0.34 & 0.24 & 0.38 & $\leq$0.03 & 0.79 & 1.99 & $\leq$0.03 & $\leq$0.03 & $\leq$0.03 & 0.26  \\ 
IRDC079.1 &      & $\leq$0.04 & 1.52 & 0.83 & $\leq$0.04 & 2.00 & 2.98 & 0.38 & $\leq$0.06 & $\leq$0.06 & 3.20  \\ 
IRDC079.3 & 5.73 & 0.44 & 1.96 & 1.25 & $\leq$0.04 & 1.87 & 2.26 & $\leq$0.06 & $\leq$0.06 & $\leq$0.06 & $\leq$0.19  \\ 
IRDC18151 & 47.03 & 0.85 & 5.28 & 1.38 & 1.97 & 2.88 & 7.62 & 0.40 & $\leq$0.03 & $\leq$0.03 & 3.34  \\ 
IRDC18182 &      & 0.49 & 1.37 & 0.82 & $\leq$0.03 & 0.91 & 2.09 & $\leq$0.03 & $\leq$0.03 & $\leq$0.03 & $\leq$0.07  \\ 
IRDC18223 & 17.08 & 0.86 & 2.33 & 3.05 & 2.56 & 2.82 & 5.27 & 0.37 & $\leq$0.03 & $\leq$0.03 & 1.15  \\ 
IRDC18306 &      & 0.38 & 0.76 & 0.76 & $\leq$0.03 & 0.68 & 1.00 & $\leq$0.03 & $\leq$0.03 & $\leq$0.03 & $\leq$0.07  \\ 
IRDC18308 & 13.70 & 1.85 & 1.42 & 0.83 & 2.09 & 0.91 & 2.25 & 0.25 & $\leq$0.03 & $\leq$0.03 & $\leq$0.07  \\ 
IRDC18310 &      & 1.69 & 0.82 & 1.52 & 2.52 & 1.14 & 5.40 & $\leq$0.03 & $\leq$0.03 & $\leq$0.03 & $\leq$0.07  \\ 
IRDC18337 &      & 0.82 & 1.00 & 0.68 & 3.11 & 0.91 & 2.45 & 0.26 & $\leq$0.02 & $\leq$0.02 & 1.27  \\ 
IRDC18385 &      & 0.35 & 1.06 & 0.57 & 0.84 & 0.83 & 2.27 & $\leq$0.05 & $\leq$0.05 & $\leq$0.05 & $\leq$0.16  \\ 
IRDC18437 & 7.13 & $\leq$0.04 & 1.04 & 0.36 & $\leq$0.04 & 0.47 & 2.41 & $\leq$0.06 & $\leq$0.06 & $\leq$0.06 & $\leq$0.16  \\ 
IRDC18454.1 &      & 0.54 & 1.22 & 0.62 & $\leq$0.04 & 0.70 & 1.88 & $\leq$0.06 & $\leq$0.06 & $\leq$0.06 & $\leq$0.17  \\ 
IRDC18454.3 &      & 2.13 & 2.75 & 1.89 & 2.43 & 2.74 & 3.37 & $\leq$0.05 & $\leq$0.05 & $\leq$0.05 & $\leq$0.16  \\ 
IRDC19175 &      & 0.33 & 0.56 & 0.30 & 0.54 & 0.51 & 1.02 & $\leq$0.05 & $\leq$0.05 & $\leq$0.05 & $\leq$0.12  \\ 
IRDC20081 & 6.90 & 0.35 & 1.32 & 0.31 & 0.22 & 1.10 & 2.64 & $\leq$0.06 & $\leq$0.06 & $\leq$0.06 & 2.42  \\ 
HMPO18089 &      & 1.44 & 5.57 & 3.35 & 3.02 & 4.99 & 6.52 & 5.13 & 1.80 & $\leq$0.03 & 18.35  \\ 
HMPO18102 &      & 4.23 & 2.93 & 2.83 & 4.25 & 3.20 & 4.26 & 0.87 & 0.82 & $\leq$0.03 & 2.92  \\ 
HMPO18151 & 49.32 & 0.44 & 5.13 & 1.62 & 1.91 & 3.80 & 7.92 & 0.81 & 0.07 & $\leq$0.03 & 3.96  \\ 
HMPO18182 & 34.55 & 1.32 & 4.53 & 2.08 & 2.60 & 4.06 & 7.41 & 2.37 & 0.66 & $\leq$0.03 & 8.67  \\ 
HMPO18247 & 10.49 & 0.36 & 1.88 & 1.02 & 0.32 & 1.42 & 4.66 & 0.39 & $\leq$0.03 & $\leq$0.03 & 0.77  \\ 
HMPO18264 & 76.18 & 1.04 & 6.93 & 2.04 & 6.32 & 5.17 & 13.61 & 1.35 & 0.70 & $\leq$0.03 & 17.25  \\ 
HMPO18310 &      & 0.89 & 3.25 & 1.49 & 1.23 & 2.24 & 4.69 & 0.82 & 0.16 & $\leq$0.03 & 0.81  \\ 
HMPO18488 &      & 1.10 & 3.94 & 1.29 & 3.49 & 2.65 & 6.12 & 1.21 & 0.27 & $\leq$0.06 & 8.43  \\ 
HMPO18517 & 49.32 & 1.00 & 6.90 & 1.42 & 2.05 & 5.11 & 7.80 & 1.14 & 0.36 & $\leq$0.05 & 10.68  \\ 
HMPO18566 &      & 2.09 & 3.26 & 2.09 & 3.85 & 3.34 & 5.71 & 1.77 & 0.46 & $\leq$0.05 & 7.36  \\ 
HMPO19217 &      & 1.51 & 3.08 & 1.31 & 2.93 & 3.46 & 4.26 & 1.39 & 0.16 & $\leq$0.05 & 12.40  \\ 
HMPO19410 & 39.12 & 0.57 & 6.01 & 1.82 & 0.83 & 3.21 & 15.50 & 1.80 & 0.13 & $\leq$0.05 & 6.03  \\ 
HMPO20126 & 53.19 & 0.69 & 6.48 & 2.75 & $\leq$0.05 & 4.03 & 10.28 & $\leq$0.08 & 0.35 & $\leq$0.08 & 10.79  \\ 
HMPO20216 & 16.24 & $\leq$0.05 & 2.42 & 0.60 & $\leq$0.05 & 1.33 & 2.50 & $\leq$0.07 & $\leq$0.07 & $\leq$0.07 & 0.61  \\ 
HMPO20293 & 29.35 & 0.43 & 5.29 & 1.75 & 3.83 & 2.65 & 14.77 & 0.90 & $\leq$0.06 & $\leq$0.06 & 4.86  \\ 
HMPO22134 & 13.64 & 0.12 & 2.81 & 0.30 & $\leq$0.03 & 1.14 & 2.55 & 0.32 & $\leq$0.04 & $\leq$0.03 & 2.79  \\ 
HMPO23033 & 35.88 & 0.53 & 5.84 & 1.05 & 1.03 & 3.47 & 7.37 & 0.71 & 0.18 & $\leq$0.03 & 3.09  \\ 
HMPO23139 & 30.56 & 0.59 & 4.17 & 0.46 & 1.43 & 1.57 & 4.60 & 0.89 & $\leq$0.03 & $\leq$0.03 & 14.17  \\ 
HMPO23151 & 14.85 & $\leq$0.03 & 2.36 & $\leq$0.03 & 0.40 & 0.50 & 0.14 & 0.43 & $\leq$0.03 & $\leq$0.03 & 8.40  \\ 
HMPO23545 & 6.68 & $\leq$0.03 & 1.38 & $\leq$0.03 & $\leq$0.03 & 0.82 & 0.39 & $\leq$0.03 & $\leq$0.03 & $\leq$0.03 & 1.16  \\ 
HMC009.62 &      & 5.02 & 11.63 & 4.01 & 13.39 & 9.05 & 7.25 & 7.69 & 3.59 & $\leq$0.03 & 34.33  \\ 
HMC010.47 &      & 8.92 & 13.14 & 6.07 & 9.11 & 10.40 & 6.67 & 10.83 & 10.61 & 1.45 & 45.61  \\ 
HMC029.96 &      & 1.89 & 7.19 & 3.14 & 4.28 & 6.43 & 6.01 & 7.07 & 2.68 & $\leq$0.03 & 39.19  \\ 
HMC031.41 &      & 5.76 & 6.66 & 2.60 & 5.85 & 2.99 &      & 8.39 & 6.05 & 1.66 & 41.68  \\ 
HMC034.26 &      & 4.38 & 11.20 & 6.92 & 9.74 & 14.78 & 7.32 & 12.72 & 8.25 & 1.28 & 84.72  \\ 
HMC045.47 &      & 2.28 & 5.31 & 2.68 & 4.81 & 6.21 & 7.63 & 1.70 & 0.58 & $\leq$0.03 & 12.56  \\ 
HMC075.78 &      & 1.33 & 6.44 & 1.45 & 4.63 & 4.55 & 4.58 & 2.73 & 0.65 & $\leq$0.06 & 38.86  \\ 
W3H2O & 69.66 & 1.05 & 7.08 & 0.91 & 3.99 & 4.09 & 3.07 & 3.76 & 1.68 & 0.51 & 66.59  \\ 
W3IRS5 &      & $\leq$0.02 & 7.58 & 0.41 & 1.21 & 1.73 & 0.22 & 0.71 & $\leq$0.03 & $\leq$0.03 & 69.46  \\ 
NGC7538B &      & 0.69 & 6.41 & 0.82 & 3.15 & 3.19 & 2.75 & 1.81 & 1.10 & $\leq$0.03 & 19.45  \\ 
Orion-KL &      & 24.47 & 7.96 & 1.96 & 410.83 & 3.22 & 0.51 & 9.43 & 59.54 & 5.57 & 2468.78  \\ 
UCH005.89 &      & 2.68 & 21.91 & 12.18 & 23.42 & 16.05 & 18.64 & 10.24 & 4.41 & $\leq$0.03 &       \\ 
UCH010.10 & 4.84 & $\leq$0.02 & $\leq$0.02 & 0.71 & $\leq$0.02 & 0.57 &      & 0.87 & $\leq$0.02 & $\leq$0.02 & $\leq$0.06  \\ 
UCH010.30 &      & 3.92 & 14.53 & 4.16 & 6.89 & 6.71 & 9.55 & 6.68 & 2.18 & $\leq$0.03 &       \\ 
UCH012.21 &      & 2.59 & 7.97 & 2.97 & 4.01 & 4.73 & 5.04 & 4.08 & 1.53 & $\leq$0.03 & 16.31  \\ 
UCH013.87 & 32.39 & 0.98 & 6.60 & 1.20 & $\leq$0.03 & 3.77 & 4.33 & 3.67 & 0.44 & $\leq$0.03 & 6.06  \\ 
UCH030.54 & 23.59 & 0.39 & 2.75 & 0.32 & $\leq$0.03 & 1.04 & 1.22 & 0.69 & $\leq$0.03 & $\leq$0.03 & 2.24  \\ 
UCH035.20 &      & 0.31 & 3.46 & 1.14 & $\leq$0.03 & 2.66 & 4.17 & 2.17 & 0.50 & $\leq$0.03 & 11.69  \\ 
UCH045.12 & 43.79 & $\leq$0.03 & 5.40 & 0.80 & 0.82 & 3.13 & 1.11 & 1.30 & 0.14 & $\leq$0.03 &       \\ 
UCH045.45 & 22.46 & $\leq$0.03 & 2.99 & 0.60 & 0.83 & 1.38 & 2.32 & 1.19 & 0.14 & $\leq$0.03 & 6.55  \\ 
\hline                                   
\end{tabular}
\end{table*}
}
\clearpage
\onltab{
\begin{table*}
\tiny
\caption{Continuation of Table~\ref{tab:integrals_1} with additional species. The integrated intensity $\int T_{\rm mb}$ in K km\,s$^{-1}$. Non-detections are indicated by $\leq$ and the 1$\sigma$~rms value is given. A blank entry means that the molecule is detected, but rejected from the analysis due to reasons discussed in Section~\ref{sec:problems}.}             
\label{tab:integrals_2}      
\centering                          
\begin{tabular}{lllllllllll}        
\hline\hline                 
Source & C$^{18}$O & OCS(1) & H$_2$CO(1) & CH$_3$OH(1) & H$_2$CO(2) & CH$_3$OH(2) & OCS(2) & C$^{33}$S & H$_{2,29}$ & H$_{2,\rm beam}$
\\    
\hline                        
IRDC011.1 & 7.98 & $\leq$0.08 & 0.69 & 1.08 & 3.29 & 0.66 & $\leq$0.10 & $\leq$0.10 & 2.70 & 1.70 \\ 
IRDC028.1 & 8.36 & $\leq$0.09 & 1.29 & 2.38 & 5.54 & 1.15 & $\leq$0.10 & $\leq$0.10 & 2.34 & 1.40  \\ 
IRDC028.2 & 16.84 & 2.38 & 7.19 & 15.38 &      & 13.38 & 1.65 & $\leq$0.09 & 7.86 & 5.21  \\ 
IRDC048.6 & 2.19 & $\leq$0.07 & $\leq$0.07 & $\leq$0.07 & 0.37 & $\leq$0.23 & $\leq$0.23 & $\leq$0.23 & 0.63 & 0.43 \\ 
IRDC079.1 & 14.07 & $\leq$0.21 & $\leq$0.21 & $\leq$0.21 & 3.57 & $\leq$0.17 & $\leq$0.17 & $\leq$0.17 & 3.45 & 2.73 \\ 
IRDC079.3 & 4.80 & $\leq$0.19 & $\leq$0.19 & $\leq$0.19 & 1.24 & $\leq$0.17 & $\leq$0.17 & $\leq$0.17 & 2.16 & 1.55 \\ 
IRDC18151 & 13.34 & $\leq$0.07 & 3.42 & 4.80 & 11.72 & 6.92 & $\leq$0.10 & $\leq$0.10 & 1.52 & 0.62 \\ 
IRDC18182 & 6.03 & $\leq$0.07 & $\leq$0.07 & $\leq$0.07 & 1.48 & $\leq$0.10 & $\leq$0.10 & $\leq$0.10 & 0.40 & 0.14 \\ 
IRDC18223 & 9.86 & $\leq$0.07 & 0.78 & 1.07 & 3.35 & 1.96 & $\leq$0.10 & $\leq$0.10 & 0.61 & 0.29 \\ 
IRDC18306 & 6.31 & $\leq$0.07 & $\leq$0.07 & $\leq$0.07 & 0.40 & $\leq$0.10 & $\leq$0.10 & $\leq$0.10 & 0.50 & 0.15 \\ 
IRDC18308 & 6.86 & $\leq$0.07 & 0.59 & 1.31 & 3.02 & 0.69 & $\leq$0.07 & $\leq$0.07 & 0.72 & 0.22 \\ 
IRDC18310 & 4.43 & $\leq$0.07 & 0.52 & 0.87 & 2.86 & $\leq$0.10 & $\leq$0.10 & $\leq$0.10 & 0.48 & 0.13 \\ 
IRDC18337 & 3.94 & $\leq$0.05 & 1.59 & 2.59 & 5.87 & 1.95 & $\leq$0.07 & $\leq$0.07 & 0.23 & 0.12 \\ 
IRDC18385 & 3.24 & $\leq$0.16 & $\leq$0.16 & $\leq$0.16 & 2.56 & $\leq$0.10 & $\leq$0.10 & $\leq$0.10 & 0.23 & 0.08 \\ 
IRDC18437 & 6.94 & $\leq$0.16 & $\leq$0.16 & $\leq$0.16 & $\leq$0.16 & $\leq$0.11 & $\leq$0.11 & $\leq$0.11 & 0.31 & 0.09 \\ 
IRDC18454.1 & 6.75 & $\leq$0.17 & $\leq$0.17 & $\leq$0.17 & $\leq$0.17 & $\leq$0.11 & $\leq$0.11 & $\leq$0.11 & 0.41 & 0.18 \\ 
IRDC18454.3 & 15.88 & $\leq$0.16 & $\leq$0.16 & 1.37 & 4.62 & $\leq$0.10 & $\leq$0.10 & $\leq$0.10 & 0.59 & 0.20 \\ 
IRDC19175 & 4.39 & $\leq$0.12 & $\leq$0.12 & $\leq$0.12 & 1.13 & $\leq$0.24 & $\leq$0.24 & $\leq$0.24 & 0.27 & 0.09 \\ 
IRDC20081 & 10.67 & $\leq$0.19 & $\leq$0.19 & $\leq$0.19 & 2.42 & $\leq$0.19 & $\leq$0.19 & $\leq$0.19 & 0.59 & 0.24 \\ 
HMPO18089 & 37.63 & 6.01 & 7.65 & 12.39 &      & 4.81 & 2.75 & 2.96 & 3.80 & 2.37 \\ 
HMPO18102 & 8.20 & 1.12 & 2.95 & 4.87 & 10.35 & 3.12 & $\leq$0.10 & $\leq$0.10 & 1.93 & 0.63 \\ 
HMPO18151 & 2     & $\leq$0.07 & 1.69 & 1.54 & 8.16 & 1.90 & $\leq$0.10 & 0.74 & 2.00 & 0.98 \\ 
HMPO18182 & 38.84 & 1.95 & 4.06 & 8.46 & 11.59 & 6.42 & 1.94 & 1.83 & 2.55 & 1.44 \\ 
HMPO18247 & 18.50 & $\leq$0.07 & 0.58 & 0.90 & 2.96 & 0.57 & 0.57 & $\leq$0.11 & 1.28 & 0.60 \\ 
HMPO18264 & 20.42 & 1.82 & 10.92 & 12.97 & 31.28 & 7.82 & 1.27 & 1.11 & 4.32 & 2.09 \\ 
HMPO18310 & 18.79 & $\leq$0.07 & 0.43 & 0.52 & 3.20 & $\leq$0.10 & $\leq$0.10 & $\leq$0.10 & 1.35 & 0.37 \\ 
HMPO18488 & 25.16 & $\leq$0.10 & 2.99 & 6.06 & 12.85 & 3.81 & 1.23 & 0.88 & 1.48 & 0.56 \\ 
HMPO18517 & 26.29 & 1.50 & 4.12 & 5.82 & 14.24 & 5.54 & 1.28 & 0.95 & 3.25 & 1.15 \\ 
HMPO18566 & 16.62 & 1.84 & 2.63 & 5.54 & 9.41 & 5.65 & 3.44 & 2.39 & 1.17 & 0.62 \\ 
HMPO19217 & 23.58 & $\leq$0.08 & 4.65 & 8.87 & 13.54 & 8.03 & 1.72 & $\leq$0.23 & 1.92 & 0.79 \\ 
HMPO19410 & 13.58 & 0.68 & 2.56 & 3.02 & 9.86 & 2.30 & 0.40 & 0.74 & 3.11 & 0.96 \\ 
HMPO20126 & 14.28 & $\leq$0.19 & 5.99 & 4.54 & 15.81 & 5.61 & $\leq$0.12 & $\leq$0.12 & 2.61 & 1.34 \\ 
HMPO20216 & 9.23 & $\leq$0.24 & $\leq$0.24 & $\leq$0.24 & 2.73 & $\leq$0.18 & $\leq$0.18 & $\leq$0.18 & 0.60 & 0.30 \\ 
HMPO20293 & 11.50 & $\leq$0.18 & 3.13 & 3.80 & 11.57 & 1.94 & $\leq$0.18 & $\leq$0.18 & 1.99 & 0.55 \\ 
HMPO22134 & 12.63 & $\leq$0.06 & 0.83 & $\leq$0.06 & 3.39 & $\leq$0.16 & $\leq$0.16 & $\leq$0.16 & 1.12 & 0.34 \\ 
HMPO23033 & 13.12 & $\leq$0.06 & 2.22 & 1.86 & 10.49 & 1.63 & $\leq$0.17 & $\leq$0.17 & 1.84 & 0.73 \\ 
HMPO23139 & 15.22 & 0.73 & 4.31 & 4.22 & 14.95 & 2.56 & $\leq$0.17 & $\leq$0.17 & 1.39 & 0.59 \\ 
HMPO23151 & 15.77 & $\leq$0.06 & 1.23 & 0.93 & 4.42 & $\leq$0.11 & $\leq$0.11 & $\leq$0.11 & 1.05 & 0.48 \\ 
HMPO23545 & 10.22 & $\leq$0.07 & 0.38 & $\leq$0.07 & 1.92 & $\leq$0.17 & $\leq$0.17 & $\leq$0.17 & 0.70 & 0.20 \\ 
HMC009.62 & 72.23 & 7.25 & 9.71 & 19.93 & 27.00 & 6.71 & 1.78 & 1.82 & 16.63 & 11.71 \\ 
HMC010.47 & 79.58 & 34.20 & 26.58 & 55.86 &      & 39.48 & 31.17 & 16.65 & 40.09 & 32.81 \\ 
HMC029.96 & 69.00 & 12.11 & 14.25 & 15.96 & 36.67 & 9.94 & 10.65 & 8.68 & 16.49 & 12.01 \\ 
HMC031.41 & 60.31 & 24.88 & 19.60 & 43.74 &      & 27.47 & 21.38 & 17.30 & 28.41 & 21.69 \\ 
HMC034.26 & 110.54 & 40.34 & 40.66 & 64.47 & 67.23 & 46.42 & 36.10 & 26.00 & 72.36 & 51.50 \\ 
HMC045.47 & 27.98 & 1.47 & 4.50 & 6.99 & 14.89 & 3.71 & 1.32 & 1.09 & 7.84 & 4.99 \\ 
HMC075.78 & 39.72 & 3.07 & 7.58 & 8.08 & 22.05 & 6.65 & 3.85 & 3.11 & 12.03 & 10.32 \\ 
W3H2O & 44.78 & 11.89 & 14.69 & 18.70 & 33.35 & 27.64 & 18.67 & 15.87 & 13.99 & 12.52 \\ 
W3IRS5 & 4     & 0.49 & 4.03 & 0.81 & 11.26 & 1.06 & 1.34 & 0.95 & 11.42 & 8.98 \\ 
NGC7538B & 29.60 & 1.76 & 7.64 & 4.91 & 21.93 & 5.82 & 2.07 & 2.85 & 23.60 & 19.59 \\ 
Orion-KL & 112.27 & 272.95 & 212.37 & 219.77 & 425.95 & 231.06 & 297.76 & 56.14 & 237.20 & 204.20 \\ 
UCH005.89 & 80.78 & 12.59 & 31.45 & 24.53 & 77.61 & 16.57 & 9.61 & 10.40 & 32.87 & 23.75 \\ 
UCH010.10 & $\leq$0.06 & $\leq$0.06 & $\leq$0.06 & $\leq$0.06 & $\leq$0.06 & $\leq$0.11 & $\leq$0.11 & $\leq$0.11 & 0.98 & 0.84 \\ 
UCH010.30 & 49.32 & 1.91 & 5.62 & 9.59 & 25.68 & 4.67 & 0.59 & 2.28 & 12.29 & 7.29 \\ 
UCH012.21 & 45.54 & 4.38 & 8.31 & 15.80 & 22.14 & 22.94 & 8.09 & 7.06 & 14.57 & 10.84 \\ 
UCH013.87 & 42.76 & $\leq$0.08 & 1.91 & 2.03 & 7.33 & 1.21 & $\leq$0.10 & 1.47 & 8.07 & 5.08 \\ 
UCH030.54 & 25.04 & $\leq$0.07 & 0.91 & $\leq$0.07 & 3.67 & $\leq$0.08 & $\leq$0.08 & $\leq$0.08 & 3.18 & 2.06 \\ 
UCH035.20 & 40.11 & $\leq$0.07 & 3.52 & 1.13 & 12.05 & $\leq$0.05 & $\leq$0.05 & 1.55 & 18.15 & 14.29 \\ 
UCH045.12 & 20.04 & $\leq$0.07 & 1.80 & $\leq$0.07 & 6.68 & $\leq$0.03 & $\leq$0.03 & 0.69 & 11.24 & 7.06 \\ 
UCH045.45 & 21.85 & $\leq$0.07 & 2.40 & 1.35 & 7.67 & $\leq$0.03 & $\leq$0.03 & 1.46 & 6.32 & 3.72 \\ 
\hline                                   
\end{tabular}
\end{table*}
}
\clearpage
\onltab{
\begin{table*}
\tiny
\caption{Iteration 0 column densities in $a(x)=a \cdot 10^{x}$~cm$^{-2}$ derived with the initially chosen typical temperatures (see Section~\ref{sec:column_density} and Tables~\ref{tbl:obsmedianabunirdc}--\ref{tbl:obsmedianabunuch}). Non-detections are indicated by $\leq$ and the upper limit derived with the 3$\sigma$~rms value. A blank entry means that the molecule is detected, but rejected from the analysis due to reasons discussed in Section~\ref{sec:problems}.}             
\label{tab:columndensities_iteration0_1}      
\centering                          
\begin{tabular}{llllllllllll}        
\hline\hline                 
Source & HCN & HNCO & C$_2$H & HN$^{13}$C & SiO & H$^{13}$CO$^+$ & N$_2$H$^+$ & $^{13}$CS & CH$_3$CN & CH$_3$OCHO & SO\\    
\hline                        
IRDC011.1 &         & 4.5(12) & 4.2(14) & 3.6(12) & 5.5(12) & 2.6(12) & 2.8(13) & 3.7(12) & 9.3(12) & $\leq$4.9(12) & $\leq$3.9(12)  \\ 
IRDC028.1 &         & 9.6(12) & 5.6(14) & 4.0(12) & 1.0(13) & 2.5(12) & 4.0(13) & 7.5(12) & $\leq$5.2(12) & $\leq$8.6(12) & 2.7(13)  \\ 
IRDC028.2 &         & 1.5(13) & 9.2(14) & 3.9(12) & 1.8(13) & 3.3(12) & 6.5(13) & 1.5(13) & 3.1(13) & $\leq$8.9(12) & 1.7(14)  \\ 
IRDC048.6 & 2.1(13) & 1.2(12) & 6.7(13) & 7.9(11) & $\leq$1.9(11) & 1.0(12) & 8.0(12) & $\leq$8.6(11) & $\leq$2.7(12) & $\leq$4.6(12) & 4.5(12)  \\ 
IRDC079.1 &         & $\leq$4.3(11) & 4.2(14) & 1.7(12) & $\leq$2.7(11) & 2.5(12) & 2.3(13) & 3.9(12) & $\leq$6.0(12) & $\leq$9.9(12) & 5.5(13)  \\ 
IRDC079.3 & 5.7(13) & 1.6(12) & 5.4(14) & 2.6(12) & $\leq$2.6(11) & 2.4(12) & 2.5(13) & $\leq$1.8(12) & $\leq$5.7(12) & $\leq$9.5(12) & $\leq$9.5(12)  \\ 
IRDC18151 & 1.3(14) & 3.0(12) & 1.5(15) & 2.8(12) & 4.4(12) & 3.6(12) & 4.9(13) & 4.1(12) & $\leq$2.9(12) & $\leq$4.8(12) & 5.7(13)  \\ 
IRDC18182 &         & 1.7(12) & 3.8(14) & 1.7(12) & $\leq$2.1(11) & 1.2(12) & 1.6(13) & $\leq$9.0(11) & $\leq$2.9(12) & $\leq$4.8(12) & $\leq$3.8(12)  \\ 
IRDC18223 & 9.5(13) & 3.1(12) & 6.4(14) & 6.3(12) & 5.7(12) & 3.6(12) & 3.7(13) & 3.8(12) & $\leq$2.9(12) & $\leq$4.7(12) & 2.0(13)  \\ 
IRDC18306 &         & 1.4(12) & 2.1(14) & 1.6(12) & $\leq$2.0(11) & 8.6(11) & 1.6(13) & $\leq$8.8(11) & $\leq$2.8(12) & $\leq$4.6(12) & $\leq$3.6(12)  \\ 
IRDC18308 & 1.6(14) & 6.6(12) & 3.9(14) & 1.7(12) & 4.7(12) & 1.2(12) & 2.2(13) & 2.6(12) & $\leq$2.6(12) & $\leq$4.3(12) & $\leq$3.5(12)  \\ 
IRDC18310 &         & 6.0(12) & 2.3(14) & 3.1(12) & 5.6(12) & 1.4(12) & 2.6(13) & $\leq$8.6(11) & $\leq$2.7(12) & $\leq$4.5(12) & $\leq$3.8(12)  \\ 
IRDC18337 &         & 2.9(12) & 2.8(14) & 1.4(12) & 7.0(12) & 1.1(12) & 1.8(13) & 2.6(12) & $\leq$2.1(12) & $\leq$3.4(12) & 2.2(13)  \\ 
IRDC18385 &         & 1.2(12) & 2.9(14) & 1.2(12) & 1.9(12) & 1.0(12) & 1.5(13) & $\leq$1.7(12) & $\leq$5.3(12) & $\leq$8.9(12) & $\leq$8.4(12)  \\ 
IRDC18437 & 1.6(13) & $\leq$4.0(11) & 2.9(14) & 7.3(11) & $\leq$2.5(11) & 5.9(11) & 8.9(12) & $\leq$1.7(12) & $\leq$5.4(12) & $\leq$8.9(12) & $\leq$8.2(12)  \\ 
IRDC18454.1 &         & 1.9(12) & 3.4(14) & 1.3(12) & $\leq$2.6(11) & 8.8(11) & 1.9(13) & $\leq$1.7(12) & $\leq$5.5(12) & $\leq$9.1(12) & $\leq$8.9(12)  \\ 
IRDC18454.3 &         & 7.6(12) & 7.6(14) & 3.9(12) & 5.5(12) & 3.5(12) & 3.2(13) & $\leq$1.7(12) & $\leq$5.3(12) & $\leq$8.8(12) & $\leq$8.3(12)  \\ 
IRDC19175 &         & 1.2(12) & 1.5(14) & 6.2(11) & 1.2(12) & 6.4(11) & 5.6(12) & $\leq$1.5(12) & $\leq$4.7(12) & $\leq$7.7(12) & $\leq$6.2(12)  \\ 
IRDC20081 & 9.1(13) & 1.2(12) & 3.7(14) & 6.4(11) & 5.0(11) & 1.4(12) & 6.5(12) & $\leq$1.8(12) & $\leq$5.7(12) & $\leq$9.5(12) & 4.1(13)  \\ 
HMPO18089 &         & 1.0(13) & 3.9(15) & 1.8(13) & 1.7(13) & 1.7(13) & 4.9(13) & 1.3(14) & 2.5(13) & $\leq$4.2(12) & 2.5(14)  \\ 
HMPO18102 &         & 3.1(13) & 2.0(15) & 1.5(13) & 2.3(13) & 1.1(13) & 6.5(13) & 2.1(13) & 1.1(13) & $\leq$4.2(12) & 4.0(13)  \\ 
HMPO18151 & 2.8(14) & 3.2(12) & 3.6(15) & 8.8(12) & 1.0(13) & 1.3(13) & 3.6(13) & 2.0(13) & 1.0(12) & $\leq$4.1(12) & 5.4(13)  \\ 
HMPO18182 & 2.0(14) & 9.6(12) & 3.1(15) & 1.1(13) & 1.4(13) & 1.4(13) & 1.0(14) & 5.8(13) & 9.1(12) & $\leq$4.1(12) & 1.2(14)  \\ 
HMPO18247 & 1.9(14) & 2.6(12) & 1.3(15) & 5.6(12) & 1.7(12) & 4.8(12) & 3.7(13) & 9.5(12) & $\leq$1.2(12) & $\leq$4.0(12) & 1.0(13)  \\ 
HMPO18264 & 1.4(15) & 7.6(12) & 4.8(15) & 1.1(13) & 3.5(13) & 1.7(13) & 1.8(14) & 3.3(13) & 9.6(12) & $\leq$3.9(12) & 2.3(14)  \\ 
HMPO18310 &         & 6.5(12) & 2.3(15) & 8.1(12) & 6.7(12) & 7.5(12) & 3.1(13) & 2.0(13) & 2.2(12) & $\leq$3.8(12) & 1.1(13)  \\ 
HMPO18488 &         & 8.0(12) & 2.7(15) & 7.0(12) & 1.9(13) & 8.9(12) & 4.1(13) & 3.0(13) & 3.7(12) & $\leq$7.8(12) & 1.1(14)  \\ 
HMPO18517 & 2.8(14) & 7.3(12) & 4.8(15) & 7.7(12) & 1.1(13) & 1.7(13) & 7.6(13) & 2.8(13) & 5.0(12) & $\leq$7.3(12) & 1.4(14)  \\ 
HMPO18566 &         & 1.5(13) & 2.3(15) & 1.1(13) & 2.1(13) & 1.1(13) & 8.5(13) & 4.3(13) & 6.3(12) & $\leq$7.3(12) & 1.0(14)  \\ 
HMPO19217 &         & 1.1(13) & 2.1(15) & 7.1(12) & 1.6(13) & 1.2(13) & 7.4(13) & 3.4(13) & 2.2(12) & $\leq$7.1(12) & 1.7(14)  \\ 
HMPO19410 & 2.3(14) & 4.1(12) & 4.2(15) & 9.9(12) & 4.6(12) & 1.1(13) & 1.0(14) & 4.4(13) & 1.8(12) & $\leq$6.5(12) & 8.2(13)  \\ 
HMPO20126 & 3.1(14) & 5.0(12) & 4.5(15) & 1.5(13) & $\leq$8.7(11) & 1.4(13) & 1.2(14) & $\leq$5.8(12) & 4.7(12) & $\leq$1.1(13) & 1.5(14)  \\ 
HMPO20216 & 9.3(13) & $\leq$1.1(12) & 1.7(15) & 3.3(12) & $\leq$7.9(11) & 4.5(12) & 1.0(13) & $\leq$5.2(12) & $\leq$2.9(12) & $\leq$9.7(12) & 8.3(12)  \\ 
HMPO20293 & 1.7(14) & 3.2(12) & 3.7(15) & 9.5(12) & 2.1(13) & 8.9(12) & 1.1(14) & 2.2(13) & $\leq$2.3(12) & $\leq$7.8(12) & 6.6(13)  \\ 
HMPO22134 & 1.8(14) & 8.9(11) & 1.9(15) & 1.6(12) & $\leq$4.4(11) & 3.8(12) & 1.7(13) & 7.9(12) & $\leq$1.5(12) & $\leq$3.6(12) & 3.8(13)  \\ 
HMPO23033 & 2.8(14) & 3.9(12) & 4.0(15) & 5.7(12) & 5.6(12) & 1.2(13) & 4.5(13) & 1.7(13) & 2.4(12) & $\leq$3.6(12) & 4.2(13)  \\ 
HMPO23139 & 4.7(14) & 4.3(12) & 2.9(15) & 2.5(12) & 7.8(12) & 5.3(12) & 4.7(13) & 2.2(13) & $\leq$1.1(12) & $\leq$3.6(12) & 1.9(14)  \\ 
HMPO23151 & 8.5(13) & $\leq$5.8(11) & 1.6(15) & $\leq$4.2(11) & 2.2(12) & 1.7(12) & 1.7(12) & 1.1(13) & $\leq$1.0(12) & $\leq$3.5(12) & 1.1(14)  \\ 
HMPO23545 & 3.8(13) & $\leq$6.1(11) & 9.6(14) & $\leq$4.5(11) & $\leq$4.6(11) & 2.8(12) & 2.3(12) & $\leq$2.0(12) & $\leq$1.1(12) & $\leq$3.7(12) & 1.6(13)  \\ 
HMC009.62 &         & 1.4(14) & 1.7(16) & 4.2(13) & 1.4(14) & 5.8(13) & 2.4(14) & 3.5(14) & 1.6(14) & $\leq$2.8(13) & 7.1(14)  \\ 
HMC010.47 &         & 2.5(14) & 1.9(16) & 6.3(13) & 9.4(13) & 6.6(13) & 3.6(14) & 4.9(14) & 4.8(14) & 4.6(14) & 9.4(14)  \\ 
HMC029.96 &         & 5.2(13) & 1.0(16) & 3.3(13) & 4.4(13) & 4.1(13) & 1.3(14) & 3.2(14) & 1.2(14) & $\leq$2.9(13) & 8.1(14)  \\ 
HMC031.41 &         & 1.6(14) & 9.6(15) & 2.7(13) & 6.0(13) & 1.9(13) & 4.4(13) & 3.8(14) & 2.7(14) & 5.3(14) & 8.6(14)  \\ 
HMC034.26 &         & 1.2(14) & 1.6(16) & 7.2(13) & 1.0(14) & 9.4(13) & 9.7(13) & 5.8(14) & 3.7(14) & 4.1(14) & 1.7(15)  \\ 
HMC045.47 &         & 6.3(13) & 7.6(15) & 2.8(13) & 4.9(13) & 4.0(13) & 7.3(13) & 7.7(13) & 2.6(13) & $\leq$2.6(13) & 2.6(14)  \\ 
HMC075.78 &         & 3.7(13) & 9.3(15) & 1.5(13) & 4.8(13) & 2.9(13) & 1.3(14) & 1.2(14) & 2.9(13) & $\leq$5.7(13) & 8.0(14)  \\ 
W3H$_2$O & 1.3(15) & 2.9(13) & 1.0(16) & 9.4(12) & 4.1(13) & 2.6(13) & 2.0(13) & 1.7(14) & 7.6(13) & 1.6(14) & 1.4(15)  \\ 
W3IRS5 &         & $\leq$1.6(12) & 1.1(16) & 4.2(12) & 1.2(13) & 1.1(13) & 5.0(12) & 3.3(13) & $\leq$3.8(12) & $\leq$2.7(13) & 1.4(15)  \\ 
NGC7538B &         & 1.9(13) & 9.2(15) & 8.5(12) & 3.2(13) & 2.0(13) & 1.3(13) & 8.2(13) & 5.0(13) & $\leq$2.5(13) & 4.0(14)  \\ 
Orion-KL &         & 6.7(14) & 1.1(16) & 2.0(13) & 4.2(15) & 2.1(13) & 4.3(12) & 4.3(14) & 2.7(15) & 1.8(15) & 5.1(16)  \\ 
UCH005.89 &         & 7.4(13) & 3.2(16) & 1.3(14) & 2.4(14) & 1.0(14) & 5.2(14) & 4.7(14) & 2.0(14) & $\leq$3.1(13) &          \\ 
UCH010.10 & 9.1(13) & $\leq$1.9(12) & $\leq$9.9(13) & 7.3(12) & $\leq$7.1(11) & 3.7(12) & 2.1(13) & 4.0(13) & $\leq$3.0(12) & $\leq$2.1(13) & $\leq$3.5(12)  \\ 
UCH010.30 &         & 1.1(14) & 2.1(16) & 4.3(13) & 7.1(13) & 4.3(13) & 3.2(14) & 3.0(14) & 9.9(13) & $\leq$2.9(13) &          \\ 
UCH012.21 &         & 7.1(13) & 1.1(16) & 3.1(13) & 4.1(13) & 3.0(13) & 7.3(13) & 1.9(14) & 6.9(13) & $\leq$2.8(13) & 3.4(14)  \\ 
UCH013.87 & 1.0(15) & 2.7(13) & 9.5(15) & 1.2(13) & $\leq$9.2(11) & 2.4(13) & 9.8(13) & 1.7(14) & 2.0(13) & $\leq$2.9(13) & 1.2(14)  \\ 
UCH030.54 & 2.6(14) & 1.1(13) & 4.0(15) & 3.3(12) & $\leq$9.2(11) & 6.7(12) & 5.7(12) & 3.2(13) & $\leq$3.8(12) & $\leq$2.7(13) & 4.6(13)  \\ 
UCH035.20 &         & 8.6(12) & 5.0(15) & 1.2(13) & $\leq$8.9(11) & 1.7(13) & 9.2(13) & 9.9(13) & 2.3(13) & $\leq$2.8(13) & 2.4(14)  \\ 
UCH045.12 & 1.4(15) & $\leq$2.4(12) & 7.8(15) & 8.3(12) & 8.5(12) & 2.0(13) & 8.8(12) & 5.9(13) & 6.2(12) & $\leq$2.9(13) &          \\ 
UCH045.45 & 3.7(14) & $\leq$2.4(12) & 4.3(15) & 6.2(12) & 8.5(12) & 8.8(12) & 5.3(13) & 5.4(13) & 6.1(12) & $\leq$2.6(13) & 1.3(14)  \\ 
\hline                                   
\end{tabular}
\end{table*}
}
\clearpage
\onltab{
\begin{table*}
\tiny
\caption{Continuation of Table~\ref{tab:columndensities_iteration0_1} with additional species. Iteration 0 column densities in $a(x)=a \cdot 10^{x}$~cm$^{-2}$ derived with the initially chosen typical temperatures (see Section~\ref{sec:column_density} and Tables~\ref{tbl:obsmedianabunirdc}--\ref{tbl:obsmedianabunuch}). Non-detections are indicated by $\leq$ and the upper limit derived with the 3$\sigma$~rms value. A blank entry means that the molecule is detected, but rejected from the analysis due to reasons discussed in Section~\ref{sec:problems}. The last column gives the lower limit for the ionization degree x(e).}             
\label{tab:columndensities_iteration0_2}      
\centering                          
\begin{tabular}{llllllllllll}        
\hline\hline                 
Source & C$^{18}$O & OCS(1) & H$_2$CO(1) & CH$_3$OH(1) & H$_2$CO(2) & CH$_3$OH(2) & OCS(2) & C$^{33}$S & H$_{2,29}$ & H$_{2,\rm beam}$ & x(e)
\\    
\hline                        
IRDC011.1 & 4.3(15) & $\leq$7.6(14) & 1.2(14) & 1.1(14) & 7.6(12) & 4.1(13) & $\leq$3.8(15) & $\leq$6.0(12) & 6.1(22) & 8.8(22) & 4.3(-09) \\ 
IRDC028.1 & 4.5(15) & $\leq$8.8(14) & 2.2(14) & 2.3(14) & 1.3(13) & 7.1(13) & $\leq$3.6(15) & $\leq$5.7(12) & 5.3(22) & 7.2(22) & 4.9(-09) \\ 
IRDC028.2 & 9.1(15) & 7.9(15) & 1.3(15) & 1.5(15) &         & 8.3(14) & 2.0(16) & $\leq$5.5(12) & 1.8(23) & 2.7(23) & 2.0(-09) \\ 
IRDC048.6 & 1.2(15) & $\leq$6.8(14) & $\leq$3.5(13) & $\leq$2.0(13) & 8.5(11) & $\leq$4.3(13) & $\leq$8.5(15) & $\leq$1.3(13) & 1.4(22) & 2.2(22) & 6.8(-09) \\ 
IRDC079.1 & 7.6(15) & $\leq$2.1(15) & $\leq$1.1(14) & $\leq$6.2(13) & 8.3(12) & $\leq$3.2(13) & $\leq$6.4(15) & $\leq$1.0(13) & 7.3(22) & 9.2(22) & 3.4(-09) \\ 
IRDC079.3 & 2.6(15) & $\leq$1.9(15) & $\leq$9.8(13) & $\leq$5.5(13) & 2.9(12) & $\leq$3.2(13) & $\leq$6.5(15) & $\leq$1.0(13) & 4.6(22) & 5.2(22) & 5.2(-09) \\ 
IRDC18151 & 7.2(15) & $\leq$7.4(14) & 6.0(14) & 4.7(14) & 2.7(13) & 4.3(14) & $\leq$3.7(15) & $\leq$5.9(12) & 8.1(22) & 2.3(23) & 4.6(-09) \\ 
IRDC18182 & 3.3(15) & $\leq$7.4(14) & $\leq$3.9(13) & $\leq$2.2(13) & 3.4(12) & $\leq$1.8(13) & $\leq$3.6(15) & $\leq$5.7(12) & 2.1(22) & 5.3(22) & 5.6(-09) \\ 
IRDC18223 & 5.3(15) & $\leq$7.3(14) & 1.4(14) & 1.1(14) & 7.8(12) & 1.2(14) & $\leq$3.7(15) & $\leq$5.9(12) & 3.2(22) & 1.1(23) & 1.1(-08) \\ 
IRDC18306 & 3.4(15) & $\leq$7.1(14) & $\leq$3.7(13) & $\leq$2.1(13) & 9.4(11) & $\leq$1.8(13) & $\leq$3.6(15) & $\leq$5.7(12) & 2.7(22) & 5.6(22) & 3.5(-09) \\ 
IRDC18308 & 3.7(15) & $\leq$6.8(14) & 1.0(14) & 1.3(14) & 7.0(12) & 4.3(13) & $\leq$2.7(15) & $\leq$4.3(12) & 3.8(22) & 8.0(22) & 3.3(-09) \\ 
IRDC18310 & 2.4(15) & $\leq$7.3(14) & 9.1(13) & 8.6(13) & 6.7(12) & $\leq$1.9(13) & $\leq$3.7(15) & $\leq$5.9(12) & 2.5(22) & 4.9(22) & 6.0(-09) \\ 
IRDC18337 & 2.1(15) & $\leq$5.4(14) & 2.8(14) & 2.6(14) & 1.4(13) & 1.2(14) & $\leq$2.7(15) & $\leq$4.2(12) & 1.2(22) & 4.5(22) & 9.7(-09) \\ 
IRDC18385 & 1.8(15) & $\leq$1.6(15) & $\leq$8.6(13) & $\leq$4.9(13) & 5.9(12) & $\leq$1.8(13) & $\leq$3.7(15) & $\leq$5.8(12) & 1.2(22) & 2.8(22) & 8.9(-09) \\ 
IRDC18437 & 3.7(15) & $\leq$1.6(15) & $\leq$8.4(13) & $\leq$4.7(13) & $\leq$1.1(12) & $\leq$2.0(13) & $\leq$4.0(15) & $\leq$6.3(12) & 1.6(22) & 3.4(22) & 3.8(-09) \\ 
IRDC18454.1 & 3.6(15) & $\leq$1.7(15) & $\leq$9.1(13) & $\leq$5.1(13) & $\leq$1.2(12) & $\leq$2.0(13) & $\leq$3.9(15) & $\leq$6.2(12) & 2.2(22) & 6.5(22) & 4.4(-09) \\ 
IRDC18454.3 & 8.6(15) & $\leq$1.6(15) & $\leq$8.4(13) & 1.4(14) & 1.1(13) & $\leq$1.9(13) & $\leq$3.9(15) & $\leq$6.1(12) & 3.1(22) & 7.5(22) & 1.1(-08) \\ 
IRDC19175 & 2.4(15) & $\leq$1.2(15) & $\leq$6.3(13) & $\leq$3.6(13) & 2.6(12) & $\leq$4.5(13) & $\leq$9.0(15) & $\leq$1.4(13) & 1.4(22) & 3.2(22) & 4.3(-09) \\ 
IRDC20081 & 5.8(15) & $\leq$1.9(15) & $\leq$9.7(13) & $\leq$5.5(13) & 5.6(12) & $\leq$3.6(13) & $\leq$7.1(15) & $\leq$1.1(13) & 3.1(22) & 9.0(22) & 4.2(-09) \\ 
HMPO18089 & 3.1(16) & 6.3(14) & 2.1(14) & 5.6(14) &         & 7.7(13) & 3.7(14) & 3.7(13) & 4.5(22) & 1.9(23) & 3.5(-08) \\ 
HMPO18102 & 6.8(15) & 1.2(14) & 8.3(13) & 2.2(14) & 3.5(13) & 5.0(13) & $\leq$4.1(13) & $\leq$3.9(12) & 2.3(22) & 5.1(22) & 4.5(-08) \\ 
HMPO18151 & 1.7(16) & $\leq$2.3(13) & 4.7(13) & 6.9(13) & 2.8(13) & 3.0(13) & $\leq$4.0(13) & 9.3(12) & 2.3(22) & 8.0(22) & 5.0(-08) \\ 
HMPO18182 & 3.2(16) & 2.0(14) & 1.1(14) & 3.8(14) & 3.9(13) & 1.0(14) & 2.6(14) & 2.3(13) & 3.0(22) & 1.2(23) & 4.4(-08) \\ 
HMPO18247 & 1.5(16) & $\leq$2.2(13) & 1.6(13) & 4.1(13) & 1.0(13) & 9.2(12) & 7.6(13) & $\leq$4.0(12) & 1.5(22) & 4.9(22) & 3.1(-08) \\ 
HMPO18264 & 1.7(16) & 1.9(14) & 3.1(14) & 5.8(14) & 1.1(14) & 1.3(14) & 1.7(14) & 1.4(13) & 5.1(22) & 1.7(23) & 3.4(-08) \\ 
HMPO18310 & 1.6(16) & $\leq$2.2(13) & 1.2(13) & 2.3(13) & 1.1(13) & $\leq$4.7(12) & $\leq$3.9(13) & $\leq$3.7(12) & 1.6(22) & 3.0(22) & 4.4(-08) \\ 
HMPO18488 & 2.1(16) & $\leq$3.1(13) & 8.4(13) & 2.7(14) & 4.4(13) & 6.1(13) & 1.6(14) & 1.1(13) & 1.7(22) & 4.5(22) & 4.8(-08) \\ 
HMPO18517 & 2.2(16) & 1.6(14) & 1.2(14) & 2.6(14) & 4.8(13) & 8.9(13) & 1.7(14) & 1.2(13) & 3.8(22) & 9.4(22) & 4.2(-08) \\ 
HMPO18566 & 1.4(16) & 1.9(14) & 7.4(13) & 2.5(14) & 3.2(13) & 9.1(13) & 4.6(14) & 3.0(13) & 1.4(22) & 5.1(22) & 7.9(-08) \\ 
HMPO19217 & 1.9(16) & $\leq$2.4(13) & 1.3(14) & 4.0(14) & 4.6(13) & 1.3(14) & 2.3(14) & $\leq$8.7(12) & 2.2(22) & 6.5(22) & 4.9(-08) \\ 
HMPO19410 & 1.1(16) & 7.1(13) & 7.2(13) & 1.4(14) & 3.3(13) & 3.7(13) & 5.3(13) & 9.2(12) & 3.7(22) & 7.8(22) & 2.9(-08) \\ 
HMPO20126 & 1.2(16) & $\leq$5.9(13) & 1.7(14) & 2.0(14) & 5.4(13) & 9.0(13) & $\leq$4.8(13) & $\leq$4.5(12) & 3.1(22) & 1.1(23) & 4.3(-08) \\ 
HMPO20216 & 7.6(15) & $\leq$7.4(13) & $\leq$2.0(13) & $\leq$3.2(13) & 9.2(12) & $\leq$8.5(12) & $\leq$7.1(13) & $\leq$6.6(12) & 7.1(21) & 2.5(22) & 5.8(-08) \\ 
HMPO20293 & 9.5(15) & $\leq$5.7(13) & 8.8(13) & 1.7(14) & 3.9(13) & 3.1(13) & $\leq$7.3(13) & $\leq$6.8(12) & 2.3(22) & 4.5(22) & 3.9(-08) \\ 
HMPO22134 & 1.0(16) & $\leq$2.0(13) & 2.3(13) & $\leq$8.6(12) & 1.1(13) & $\leq$7.8(12) & $\leq$6.5(13) & $\leq$6.1(12) & 1.3(22) & 2.7(22) & 2.7(-08) \\ 
HMPO23033 & 1.1(16) & $\leq$1.9(13) & 6.2(13) & 8.4(13) & 3.6(13) & 2.6(13) & $\leq$6.8(13) & $\leq$6.4(12) & 2.2(22) & 6.0(22) & 5.0(-08) \\ 
HMPO23139 & 1.3(16) & 7.6(13) & 1.2(14) & 1.9(14) & 5.1(13) & 4.1(13) & $\leq$6.9(13) & $\leq$6.4(12) & 1.6(22) & 4.8(22) & 3.2(-08) \\ 
HMPO23151 & 1.3(16) & $\leq$1.9(13) & 3.4(13) & 4.2(13) & 1.5(13) & $\leq$5.1(12) & $\leq$4.2(13) & $\leq$3.9(12) & 1.2(22) & 3.9(22) & 1.2(-08) \\ 
HMPO23545 & 8.4(15) & $\leq$2.1(13) & 1.1(13) & $\leq$8.8(12) & 6.5(12) & $\leq$8.2(12) & $\leq$6.9(13) & $\leq$6.4(12) & 8.2(21) & 1.6(22) & 3.0(-08) \\ 
HMC009.62 & 1.0(17) & 5.6(14) & 5.6(14) & 4.9(15) & 3.0(14) & 5.2(14) & 1.4(14) & 3.2(13) & 3.4(22) & 5.4(22) & 1.6(-07) \\ 
HMC010.47 & 1.1(17) & 2.6(15) & 1.5(15) & 1.4(16) &         & 3.1(15) & 2.4(15) & 2.9(14) & 8.1(22) & 1.5(23) & 7.7(-08) \\ 
HMC029.96 & 9.6(16) & 9.3(14) & 8.3(14) & 4.0(15) & 4.1(14) & 7.8(14) & 8.4(14) & 1.5(14) & 3.3(22) & 5.5(22) & 1.1(-07) \\ 
HMC031.41 & 8.4(16) & 1.9(15) & 1.1(15) & 1.1(16) &         & 2.1(15) & 1.7(15) & 3.0(14) & 5.8(22) & 1.0(23) & 3.0(-08) \\ 
HMC034.26 & 1.5(17) & 3.1(15) & 2.4(15) & 1.6(16) & 7.5(14) & 3.6(15) & 2.8(15) & 4.6(14) & 1.5(23) & 2.4(23) & 5.8(-08) \\ 
HMC045.47 & 3.9(16) & 1.1(14) & 2.6(14) & 1.7(15) & 1.7(14) & 2.9(14) & 1.0(14) & 1.9(13) & 1.6(22) & 2.3(22) & 2.3(-07) \\ 
HMC075.78 & 5.6(16) & 2.4(14) & 4.4(14) & 2.0(15) & 2.5(14) & 5.2(14) & 3.0(14) & 5.5(13) & 2.2(22) & 3.1(22) & 1.2(-07) \\ 
W3H$_2$O & 6.3(16) & 9.2(14) & 8.5(14) & 4.6(15) & 3.7(14) & 2.2(15) & 1.5(15) & 2.8(14) & 2.6(22) & 3.7(22) & 9.0(-08) \\ 
W3IRS5 & 5.6(16) & 3.8(13) & 2.3(14) & 2.0(14) & 1.3(14) & 8.3(13) & 1.1(14) & 1.7(13) & 2.1(22) & 2.7(22) & 4.7(-08) \\ 
NGC7538B & 4.1(16) & 1.4(14) & 4.4(14) & 1.2(15) & 2.5(14) & 4.5(14) & 1.6(14) & 5.0(13) & 4.4(22) & 5.8(22) & 4.2(-08) \\ 
Orion-KL & 1.6(17) & 2.1(16) & 1.2(16) & 5.4(16) & 4.8(15) & 1.8(16) & 2.3(16) & 9.8(14) & 4.4(23) & 6.1(23) & 4.2(-09) \\ 
UCH005.89 & 1.1(17) & 9.7(14) & 1.8(15) & 6.1(15) & 8.7(14) & 1.3(15) & 7.5(14) & 1.8(14) & 6.7(22) & 1.1(23) & 1.4(-07) \\ 
UCH010.10 & $\leq$2.3(14) & $\leq$1.3(13) & $\leq$9.7(12) & $\leq$4.2(13) & $\leq$1.9(12) & $\leq$2.6(13) & $\leq$2.6(13) & $\leq$5.8(12) & 2.0(21) & 3.9(21) & 1.8(-07) \\ 
UCH010.30 & 6.9(16) & 1.5(14) & 3.3(14) & 2.4(15) & 2.9(14) & 3.6(14) & 4.6(13) & 4.0(13) & 2.5(22) & 3.4(22) & 1.7(-07) \\ 
UCH012.21 & 6.4(16) & 3.4(14) & 4.8(14) & 3.9(15) & 2.5(14) & 1.8(15) & 6.4(14) & 1.2(14) & 3.0(22) & 5.0(22) & 9.4(-08) \\ 
UCH013.87 & 6.0(16) & $\leq$1.8(13) & 1.1(14) & 5.0(14) & 8.2(13) & 9.5(13) & $\leq$2.3(13) & 2.6(13) & 1.6(22) & 2.3(22) & 1.4(-07) \\ 
UCH030.54 & 3.5(16) & $\leq$1.7(13) & 5.3(13) & $\leq$5.4(13) & 4.1(13) & $\leq$2.0(13) & $\leq$2.0(13) & $\leq$4.4(12) & 6.4(21) & 9.5(21) & 9.3(-08) \\ 
UCH035.20 & 5.6(16) & $\leq$1.7(13) & 2.0(14) & 2.8(14) & 1.4(14) & $\leq$1.2(13) & $\leq$1.2(13) & 2.7(13) & 3.4(22) & 4.3(22) & 4.8(-08) \\ 
UCH045.12 & 2.8(16) & $\leq$1.6(13) & 1.0(14) & $\leq$5.1(13) & 7.5(13) & $\leq$6.8(12) & $\leq$6.8(12) & 1.2(13) & 2.3(22) & 3.3(22) & 7.8(-08) \\ 
UCH045.45 & 3.1(16) & $\leq$1.6(13) & 1.4(14) & 3.4(14) & 8.6(13) & $\leq$6.8(12) & $\leq$6.8(12) & 2.6(13) & 1.3(22) & 1.7(22) & 6.6(-08) \\ 
\hline                                   
\end{tabular}
\end{table*}
}
\clearpage
\onltab{
\begin{table*}
\tiny
\caption{Iteration 1 column densities in $a(x)=a \cdot 10^{x}$~cm$^{-2}$ derived with the mean temperatures from the best-fit models of iteration 0 (see Section~\ref{sec:column_density} and Tables~\ref{tbl:obsmedianabunirdc}--\ref{tbl:obsmedianabunuch}). Non-detections are indicated by $\leq$ and the upper limit derived with the 3$\sigma$~rms value. A blank entry means that the molecule is detected, A but rejected from the analysis due to reasons discussed in Section~\ref{sec:problems}.}             
\label{tab:columndensities_iteration1_1}      
\centering                          
\begin{tabular}{llllllllllll}        
\hline\hline                 
Source & HCN & HNCO & C$_2$H & HN$^{13}$C & SiO & H$^{13}$CO$^+$ & N$_2$H$^+$ & $^{13}$CS & CH$_3$CN & CH$_3$OCHO & SO
\\    
\hline                        
IRDC011.1 &         & 5.6(12) & 5.4(14) & 4.6(12) & 6.8(12) & 3.3(12) & 2.8(13) & 4.6(12) & 6.7(12) &  $\leq$5.4(12) & $\leq$3.0(12)  \\  
IRDC028.1 &         & 1.2(13) & 7.1(14) & 5.1(12) & 1.2(13) & 3.1(12) & 4.0(13) & 9.1(12) & $\leq$3.7(12) &  $\leq$9.6(12) & 2.1(13)  \\  
IRDC028.2 &         & 1.9(13) & 1.2(15) & 5.0(12) & 2.2(13) & 4.1(12) & 6.5(13) & 1.8(13) & 2.2(13) &  $\leq$9.9(12) & 1.3(14)  \\  
IRDC048.6 & 2.7(13) & 1.5(12) & 8.5(13) & 1.0(12) & $\leq$2.3(11) & 1.3(12) & 8.0(12) & $\leq$1.1(12) & $\leq$2.0(12) & $\leq$5.1(12) & 3.5(12)  \\  
IRDC079.1 &         & $\leq$5.4(11) & 5.4(14) & 2.2(12) & $\leq$3.3(11) & 3.2(12) & 2.3(13) & 4.8(12) & $\leq$4.3(12) & $\leq$1.1(13) & 4.3(13)  \\  
IRDC079.3 & 7.2(13) & 2.0(12) & 6.9(14) & 3.3(12) & $\leq$3.2(11) & 3.0(12) & 2.5(13) & $\leq$2.2(12) & $\leq$4.1(12) & $\leq$1.1(13) & $\leq$7.4(12)  \\  
IRDC18151 & 1.6(14) & 3.8(12) & 1.9(15) & 3.6(12) & 5.4(12) & 4.6(12) & 4.9(13) & 4.9(12) & $\leq$2.1(12) & $\leq$5.4(12) & 4.4(13)  \\  
IRDC18182 &         & 2.2(12) & 4.8(14) & 2.1(12) & $\leq$2.5(11) & 1.5(12) & 1.6(13) & $\leq$1.1(12) & $\leq$2.1(12) & $\leq$5.3(12) & $\leq$2.9(12)  \\  
IRDC18223 & 1.2(14) & 3.9(12) & 8.2(14) & 7.9(12) & 7.1(12) & 4.5(12) & 3.7(13) & 4.6(12) & $\leq$2.0(12) & $\leq$5.3(12) & 1.5(13)  \\  
IRDC18306 &         & 1.7(12) & 2.7(14) & 2.0(12) & $\leq$2.4(11) & 1.1(12) & 1.6(13) & $\leq$1.1(12) & $\leq$2.0(12) & $\leq$5.2(12) & $\leq$2.8(12)  \\  
IRDC18308 & 2.0(14) & 8.4(12) & 5.0(14) & 2.2(12) & 5.8(12) & 1.5(12) & 2.2(13) & 3.1(12) & $\leq$1.9(12) & $\leq$4.8(12) & $\leq$2.7(12)  \\  
IRDC18310 &         & 7.6(12) & 2.9(14) & 4.0(12) & 6.9(12) & 1.8(12) & 3.3(13) & $\leq$1.0(12) & $\leq$2.0(12) & $\leq$5.1(12) & $\leq$2.9(12)  \\  
IRDC18337 &         & 3.7(12) & 3.5(14) & 1.8(12) & 8.6(12) & 1.5(12) & 1.8(13) & 3.2(12) & $\leq$1.5(12) & $\leq$3.8(12) & 1.7(13)  \\  
IRDC18385 &         & 1.6(12) & 3.7(14) & 1.5(12) & 2.3(12) & 1.3(12) & 1.5(13) & $\leq$2.0(12) & $\leq$3.8(12) & $\leq$9.9(12) & $\leq$6.6(12)  \\  
IRDC18437 & 2.0(13) & $\leq$5.0(11) & 3.7(14) & 9.3(11) & $\leq$3.1(11) & 7.6(11) & 1.1(13) & $\leq$2.1(12) & $\leq$3.8(12) & $\leq$9.9(12) & $\leq$6.4(12)  \\ 
IRDC18454.1 &         & 2.4(12) & 4.3(14) & 1.6(12) & $\leq$3.1(11) & 1.1(12) & 1.9(13) & $\leq$2.1(12) & $\leq$3.9(12) & $\leq$1.0(13) & $\leq$6.9(12)  \\  
IRDC18454.3 &         & 9.6(12) & 9.7(14) & 4.9(12) & 6.7(12) & 4.4(12) & 3.2(13) & $\leq$2.0(12) & $\leq$3.8(12) & $\leq$9.8(12) & $\leq$6.4(12) \\  
IRDC19175 &         & 1.5(12) & 2.0(14) & 7.9(11) & 1.5(12) & 8.2(11) & 5.6(12) & $\leq$1.8(12) & $\leq$3.3(12) & $\leq$8.6(12) & $\leq$4.8(12)  \\  
IRDC20081 & 1.2(14) & 1.6(12) & 4.7(14) & 8.1(11) & 6.1(11) & 1.8(12) & 6.5(12) & $\leq$2.2(12) & $\leq$4.1(12) & $\leq$1.1(13) & 3.2(13)  \\  
HMPO18089 &         & 8.5(12) & 2.6(15) & 1.2(13) & 1.1(13) & 1.1(13) & 4.9(13) & 8.2(13) & 3.6(13) & $\leq$6.4(12) & 2.2(14)  \\  
HMPO18102 &         & 2.5(13) & 1.4(15) & 9.7(12) & 1.5(13) & 6.8(12) & 6.5(13) & 1.4(13) & 1.6(13) & $\leq$6.4(12) & 3.6(13)  \\  
HMPO18151 & 1.8(14) & 2.6(12) & 2.4(15) & 5.6(12) & 6.8(12) & 8.0(12) & 3.6(13) & 1.3(13) & 1.5(12) & $\leq$6.3(12) & 4.8(13)  \\  
HMPO18182 & 1.3(14) & 7.8(12) & 2.1(15) & 7.1(12) & 9.2(12) & 8.6(12) & 6.4(13) & 3.8(13) & 1.3(13) & $\leq$6.3(12) & 1.1(14)  \\  
HMPO18247 & 1.2(14) & 2.1(12) & 8.7(14) & 3.5(12) & 1.1(12) & 3.0(12) & 3.7(13) & 6.2(12) & $\leq$1.7(12) & $\leq$6.2(12) & 9.4(12)  \\  
HMPO18264 & 8.6(14) & 6.1(12) & 3.2(15) & 7.0(12) & 2.2(13) & 1.1(13) & 1.1(14) & 2.2(13) & 1.4(13) & $\leq$6.0(12) & 2.1(14)  \\  
HMPO18310 &         & 5.2(12) & 1.5(15) & 5.1(12) & 4.4(12) & 4.7(12) & 3.1(13) & 1.3(13) & 3.2(12) & $\leq$5.9(12) & 9.9(12)  \\  
HMPO18488 &         & 6.5(12) & 1.8(15) & 4.4(12) & 1.2(13) & 5.6(12) & 4.1(13) & 1.9(13) & 5.3(12) & $\leq$1.2(13) & 1.0(14)  \\  
HMPO18517 & 1.8(14) & 5.9(12) & 3.2(15) & 4.9(12) & 7.3(12) & 1.1(13) & 7.6(13) & 1.8(13) & 7.2(12) & $\leq$1.1(13) & 1.3(14)  \\  
HMPO18566 &         & 1.2(13) & 1.5(15) & 7.2(12) & 1.4(13) & 7.1(12) & 5.5(13) & 2.8(13) & 9.0(12) & $\leq$1.1(13) & 9.0(13)  \\  
HMPO19217 &         & 8.9(12) & 1.4(15) & 4.5(12) & 1.0(13) & 7.3(12) & 4.7(13) & 2.2(13) & 3.2(12) & $\leq$1.1(13) & 1.5(14)  \\  
HMPO19410 & 1.4(14) & 3.3(12) & 2.8(15) & 6.3(12) & 2.9(12) & 6.8(12) & 1.0(14) & 2.9(13) & 2.6(12) & $\leq$9.9(12) & 7.3(13)  \\  
HMPO20126 & 1.9(14) & 4.0(12) & 3.0(15) & 9.4(12) & $\leq$5.7(11) & 8.5(12) & 7.6(13) & $\leq$3.7(12) & 6.8(12) & $\leq$1.7(13) & 1.3(14)  \\  
HMPO20216 & 5.9(13) & $\leq$8.5(11) & 1.1(15) & 2.1(12) & $\leq$5.1(11) & 2.8(12) & 1.0(13) & $\leq$3.4(12) & $\leq$4.2(12) & $\leq$1.5(13) & 7.5(12)  \\  
HMPO20293 & 1.1(14) & 2.5(12) & 2.4(15) & 6.0(12) & 1.4(13) & 5.6(12) & 1.1(14) & 1.4(13) & $\leq$3.4(12) & $\leq$1.2(13) & 5.9(13)  \\  
HMPO22134 & 1.2(14) & 7.2(11) & 1.3(15) & 1.0(12) & $\leq$2.8(11) & 2.4(12) & 1.1(13) & 5.1(12) & $\leq$2.2(12) & $\leq$5.6(12) & 3.4(13)  \\  
HMPO23033 & 1.8(14) & 3.1(12) & 2.7(15) & 3.6(12) & 3.6(12) & 7.3(12) & 4.5(13) & 1.1(13) & 3.5(12) & $\leq$5.5(12) & 3.8(13)  \\  
HMPO23139 & 3.0(14) & 3.5(12) & 1.9(15) & 1.6(12) & 5.1(12) & 3.3(12) & 3.0(13) & 1.4(13) & $\leq$1.6(12) & $\leq$5.5(12) & 1.7(14)  \\  
HMPO23151 & 5.4(13) & $\leq$4.7(11) & 1.1(15) & $\leq$2.6(11) & 1.4(12) & 1.1(12) & 1.1(12) & 6.8(12) & $\leq$1.5(12) & $\leq$5.4(12) & 1.0(14)  \\  
HMPO23545 & 2.4(13) & $\leq$4.9(11) & 6.4(14) & $\leq$2.9(11) & $\leq$3.0(11) & 1.7(12) & 2.3(12) & $\leq$1.3(12) & $\leq$1.6(12) & $\leq$5.7(12) & 1.4(13)  \\  
HMC009.62 &         & 3.8(13) & 7.0(15) & 1.8(13) & 6.1(13) & 2.5(13) & 1.0(14) & 1.6(14) & 7.0(13) & $\leq$7.6(12) & 4.3(14)  \\  
HMC010.47 &         & 6.8(13) & 7.9(15) & 2.7(13) & 4.2(13) & 2.9(13) & 1.6(14) & 2.2(14) & 2.1(14) & 1.2(14) & 5.7(14)  \\  
HMC029.96 &         & 1.4(13) & 4.3(15) & 1.4(13) & 1.9(13) & 1.8(13) & 5.9(13) & 1.4(14) & 5.2(13) & $\leq$7.7(12) & 4.9(14)  \\  
HMC031.41 &         & 4.4(13) & 4.0(15) & 1.2(13) & 2.7(13) & 8.2(12) & 4.4(13) & 1.7(14) & 1.2(14) & 1.4(14) & 5.2(14)  \\  
HMC034.26 &         & 3.3(13) & 6.8(15) & 3.1(13) & 4.4(13) & 4.1(13) & 9.7(13) & 2.6(14) & 1.6(14) & 1.1(14) & 1.1(15)  \\  
HMC045.47 &         & 1.7(13) & 3.2(15) & 1.2(13) & 2.2(13) & 1.7(13) & 7.3(13) & 3.4(13) & 1.1(13) & $\leq$7.0(12) & 1.6(14)  \\  
HMC075.78 &         & 1.0(13) & 3.9(15) & 6.5(12) & 2.1(13) & 1.3(13) & 5.5(13) & 5.6(13) & 1.3(13) & $\leq$1.5(13) & 4.9(14)  \\  
W3H2O & 5.5(14) & 8.0(12) & 4.3(15) & 4.1(12) & 1.8(13) & 1.1(13) & 2.0(13) & 7.6(13) & 3.3(13) & 4.3(13) & 8.3(14)  \\  
W3IRS5 &         & $\leq$4.5(11) & 4.6(15) & 1.8(12) & 5.5(12) & 4.8(12) & 2.2(12) & 1.5(13) & $\leq$1.6(12) & $\leq$7.1(12) & 8.7(14)  \\  
NGC7538B &         & 5.3(12) & 3.9(15) & 3.7(12) & 1.4(13) & 8.8(12) & 1.3(13) & 3.7(13) & 2.1(13) & $\leq$6.7(12) & 2.4(14)  \\  
Orion-KL &         & 1.9(14) & 4.8(15) & 8.8(12) & 1.9(15) & 8.9(12) & 4.3(12) & 1.9(14) & 1.2(15) & 4.8(14) & 3.1(16)  \\  
UCH005.89 &         & 1.9(13) & 1.2(16) & 5.0(13) & 9.7(13) & 4.0(13) & 2.1(14) & 1.9(14) & 8.5(13) & $\leq$7.7(12) &          \\
UCH010.10 & 3.6(13) & $\leq$4.8(11) & $\leq$3.8(13) & 2.9(12) & $\leq$2.9(11) & 1.4(12) & 2.1(13) & 1.6(13) & $\leq$1.3(12) & $\leq$5.2(12) & $\leq$2.1(12)  \\
UCH010.30 &         & 2.7(13) & 8.0(15) & 1.7(13) & 2.9(13) & 1.7(13) & 1.3(14) & 1.2(14) & 4.2(13) & $\leq$7.1(12) &          \\
UCH012.21 &         & 1.8(13) & 4.4(15) & 1.2(13) & 1.7(13) & 1.2(13) & 7.3(13) & 7.6(13) & 3.0(13) & $\leq$7.1(12) & 2.0(14)  \\
UCH013.87 & 4.1(14) & 6.8(12) & 3.6(15) & 4.9(12) & $\leq$3.7(11) & 9.4(12) & 3.9(13) & 6.8(13) & 8.6(12) & $\leq$7.3(12) & 7.4(13)  \\
UCH030.54 & 1.0(14) & 2.7(12) & 1.5(15) & 1.3(12) & $\leq$3.7(11) & 2.6(12) & 5.8(12) & 1.3(13) & $\leq$1.6(12) & $\leq$6.7(12) & 2.7(13)  \\
UCH035.20 &         & 2.2(12) & 1.9(15) & 4.7(12) & $\leq$3.6(11) & 6.7(12) & 3.7(13) & 4.0(13) & 9.6(12) & $\leq$6.9(12) & 1.4(14)  \\
UCH045.12 & 5.6(14) & $\leq$6.1(11) & 3.0(15) & 3.2(12) & 3.4(12) & 7.8(12) & 8.7(12) & 2.4(13) & 2.6(12) & $\leq$7.1(12) &          \\
UCH045.45 & 1.4(14) & $\leq$6.0(11) & 1.6(15) & 2.4(12) & 3.4(12) & 3.5(12) & 2.1(13) & 2.2(13) & 2.6(12) & $\leq$6.6(12) & 8.0(13)  \\
\hline                                   
\end{tabular}
\end{table*}
}
\clearpage
\onltab{
\begin{table*}
\tiny
\caption{Continuation of Table~\ref{tab:columndensities_iteration1_1} with additional species. Iteration 1 column densities in $a(x)=a \cdot 10^{x}$~cm$^{-2}$ derived with the mean temperatures from the best-fit models of iteration 0 (see Section~\ref{sec:column_density} and Tables~\ref{tbl:obsmedianabunirdc}--\ref{tbl:obsmedianabunuch}). Non-detections are indicated by $\leq$ and the upper limit derived with the 3$\sigma$~rms value. A blank entry means that the molecule is detected, but rejected from the analysis due to reasons discussed in Section~\ref{sec:problems}. The last column gives the lower limit for the ionization degree x(e).}             
\label{tab:columndensities_iteration1_2}      
\centering                          
\begin{tabular}{lllllllllllllllllllllll}        
\hline\hline                 
Source & C$^{18}$O & OCS(1) & H$_2$CO(1) & CH$_3$OH(1) & H$_2$CO(2) & CH$_3$OH(2) & OCS(2) & C$^{33}$S & H$_{2,29}$ & H$_{2,\rm beam}$ & x(e)
\\    
\hline                        
IRDC011.1 & 4.4(15) & $\leq$1.6(14) & 5.5(13) & 9.1(13) & 8.5(12) & 2.7(13) & $\leq$5.2(14) & $\leq$4.3(12) & 3.7(22) & 5.3(22) & 8.9(-09) \\  
IRDC028.1 & 4.6(15) & $\leq$1.9(14) & 1.0(14) & 2.0(14) & 1.4(13) & 4.8(13) & $\leq$5.0(14) & $\leq$4.1(12) & 3.2(22) & 4.3(22) & 1.0(-08) \\  
IRDC028.2 & 9.3(15) & 1.7(15) & 5.8(14) & 1.3(15) &         & 5.6(14) & 2.8(15) & $\leq$3.9(12) & 1.1(23) & 1.6(23) & 4.1(-09) \\  
IRDC048.6 & 1.2(15) & $\leq$1.4(14) & $\leq$1.6(13) & $\leq$1.7(13) & 9.5(11) & $\leq$2.9(13) & $\leq$1.2(15) & $\leq$9.6(12) & 8.6(21) & 1.3(22) & 1.4(-08) \\  
IRDC079.1 & 7.7(15) & $\leq$4.4(14) & $\leq$5.0(13) & $\leq$5.3(13) & 9.3(12) & $\leq$2.2(13) & $\leq$8.9(14) & $\leq$7.3(12) & 4.3(22) & 5.5(22) & 7.1(-09) \\  
IRDC079.3 & 2.6(15) & $\leq$4.0(14) & $\leq$4.5(13) & $\leq$4.7(13) & 3.2(12) & $\leq$2.2(13) & $\leq$8.9(14) & $\leq$7.3(12) & 2.7(22) & 3.1(22) & 1.1(-08) \\  
IRDC18151 & 7.3(15) & $\leq$1.6(14) & 2.7(14) & 4.0(14) & 3.0(13) & 2.9(14) & $\leq$5.1(14) & $\leq$4.2(12) & 5.1(22) & 1.4(23) & 9.0(-09) \\  
IRDC18182 & 3.3(15) & $\leq$1.6(14) & $\leq$1.8(13) & $\leq$1.9(13) & 3.8(12) & $\leq$1.2(13) & $\leq$5.0(14) & $\leq$4.1(12) & 1.3(22) & 3.4(22) & 1.1(-08) \\  
IRDC18223 & 5.4(15) & $\leq$1.6(14) & 6.2(13) & 9.0(13) & 8.7(12) & 8.2(13) & $\leq$5.1(14) & $\leq$4.2(12) & 2.0(22) & 6.7(22) & 2.2(-08) \\  
IRDC18306 & 3.5(15) & $\leq$1.5(14) & $\leq$1.7(13) & $\leq$1.8(13) & 1.0(12) & $\leq$1.2(13) & $\leq$5.0(14) & $\leq$4.1(12) & 1.7(22) & 3.6(22) & 6.8(-09) \\  
IRDC18308 & 3.8(15) & $\leq$1.5(14) & 4.7(13) & 1.1(14) & 7.8(12) & 2.9(13) & $\leq$3.7(14) & $\leq$3.0(12) & 2.4(22) & 5.1(22) & 6.3(-09) \\  
IRDC18310 & 2.4(15) & $\leq$1.6(14) & 4.2(13) & 7.3(13) & 7.4(12) & $\leq$1.3(13) & $\leq$5.2(14) & $\leq$4.2(12) & 1.6(22) & 3.1(22) & 1.2(-08) \\  
IRDC18337 & 2.2(15) & $\leq$1.2(14) & 1.3(14) & 2.2(14) & 1.5(13) & 8.1(13) & $\leq$3.7(14) & $\leq$3.0(12) & 7.8(21) & 2.9(22) & 1.9(-08) \\  
IRDC18385 & 1.8(15) & $\leq$3.5(14) & $\leq$4.0(13) & $\leq$4.1(13) & 6.6(12) & $\leq$1.2(13) & $\leq$5.1(14) & $\leq$4.2(12) & 7.7(21) & 1.8(22) & 1.7(-08) \\  
IRDC18437 & 3.8(15) & $\leq$3.4(14) & $\leq$3.8(13) & $\leq$4.0(13) & $\leq$1.2(12) & $\leq$1.3(13) & $\leq$5.5(14) & $\leq$4.5(12) & 1.0(22) & 2.1(22) & 7.6(-09) \\ 
IRDC18454.1 & 3.7(15) & $\leq$3.7(14) & $\leq$4.2(13) & $\leq$4.3(13) & $\leq$1.3(12) & $\leq$1.3(13) & $\leq$5.4(14) & $\leq$4.4(12) & 1.4(22) & 4.1(22) & 8.6(-09) \\  
IRDC18454.3 & 8.7(15) & $\leq$3.4(14) & $\leq$3.9(13) & 1.1(14) & 1.2(13) & $\leq$1.3(13) & $\leq$5.4(14) & $\leq$4.4(12) & 2.0(22) & 4.8(22) & 2.1(-08) \\  
IRDC19175 & 2.4(15) & $\leq$2.6(14) & $\leq$2.9(13) & $\leq$3.0(13) & 2.9(12) & $\leq$3.1(13) & $\leq$1.2(15) & $\leq$1.0(13) & 9.1(21) & 2.0(22) & 8.6(-09) \\  
IRDC20081 & 5.9(15) & $\leq$3.9(14) & $\leq$4.5(13) & $\leq$4.7(13) & 6.3(12) & $\leq$2.4(13) & $\leq$9.8(14) & $\leq$8.0(12) & 2.0(22) & 5.7(22) & 8.2(-09) \\  
HMPO18089 & 2.3(16) & 1.5(15) & 3.7(14) & 9.9(14) &         & 1.6(14) & 1.2(15) & 3.6(13) & 8.2(22) & 3.6(23) & 1.2(-08) \\  
HMPO18102 & 5.0(15) & 2.8(14) & 1.4(14) & 3.9(14) & 3.2(13) & 1.0(14) & $\leq$1.3(14) & $\leq$3.7(12) & 4.2(22) & 9.5(22) & 1.6(-08) \\  
HMPO18151 & 1.2(16) & $\leq$5.4(13) & 8.2(13) & 1.2(14) & 2.5(13) & 6.4(13) & $\leq$1.3(14) & 8.9(12) & 4.3(22) & 1.5(23) & 1.7(-08) \\  
HMPO18182 & 2.4(16) & 4.8(14) & 2.0(14) & 6.7(14) & 3.5(13) & 2.2(14) & 8.4(14) & 2.2(13) & 5.5(22) & 2.2(23) & 1.5(-08) \\  
HMPO18247 & 1.1(16) & $\leq$5.3(13) & 2.8(13) & 7.2(13) & 9.0(12) & 1.9(13) & 2.5(14) & $\leq$3.9(12) & 2.8(22) & 9.1(22) & 1.1(-08) \\  
HMPO18264 & 1.3(16) & 4.5(14) & 5.3(14) & 1.0(15) & 9.6(13) & 2.6(14) & 5.5(14) & 1.3(13) & 9.4(22) & 3.2(23) & 1.2(-08) \\  
HMPO18310 & 1.2(16) & $\leq$5.2(13) & 2.1(13) & 4.1(13) & 9.8(12) & $\leq$9.8(12) & $\leq$1.3(14) & $\leq$3.5(12) & 2.9(22) & 5.5(22) & 1.5(-08) \\  
HMPO18488 & 1.5(16) & $\leq$7.3(13) & 1.5(14) & 4.8(14) & 3.9(13) & 1.3(14) & 5.4(14) & 1.1(13) & 3.2(22) & 8.4(22) & 1.7(-08) \\  
HMPO18517 & 1.6(16) & 3.7(14) & 2.0(14) & 4.6(14) & 4.4(13) & 1.9(14) & 5.5(14) & 1.1(13) & 7.0(22) & 1.7(23) & 1.5(-08) \\  
HMPO18566 & 1.0(16) & 4.6(14) & 1.3(14) & 4.4(14) & 2.9(13) & 1.9(14) & 1.5(15) & 2.9(13) & 2.5(22) & 9.4(22) & 2.7(-08) \\  
HMPO19217 & 1.4(16) & $\leq$5.8(13) & 2.3(14) & 7.1(14) & 4.1(13) & 2.7(14) & 7.5(14) & $\leq$8.4(12) & 4.2(22) & 1.2(23) & 1.7(-08) \\  
HMPO19410 & 8.3(15) & 1.7(14) & 1.3(14) & 2.4(14) & 3.0(13) & 7.7(13) & 1.7(14) & 8.9(12) & 6.8(22) & 1.4(23) & 1.0(-08) \\  
HMPO20126 & 8.8(15) & $\leq$1.4(14) & 2.9(14) & 3.6(14) & 4.8(13) & 1.9(14) & $\leq$1.6(14) & $\leq$4.3(12) & 5.7(22) & 2.0(23) & 1.5(-08) \\  
HMPO20216 & 5.7(15) & $\leq$1.8(14) & $\leq$3.5(13) & $\leq$5.6(13) & 8.3(12) & $\leq$1.8(13) & $\leq$2.3(14) & $\leq$6.4(12) & 1.3(22) & 4.6(22) & 2.0(-08) \\  
HMPO20293 & 7.1(15) & $\leq$1.4(14) & 1.5(14) & 3.0(14) & 3.5(13) & 6.5(13) & $\leq$2.4(14) & $\leq$6.6(12) & 4.3(22) & 8.3(22) & 1.4(-08) \\  
HMPO22134 & 7.8(15) & $\leq$4.7(13) & 4.0(13) & $\leq$1.5(13) & 1.0(13) & $\leq$1.6(13) & $\leq$2.1(14) & $\leq$5.8(12) & 2.4(22) & 5.1(22) & 9.3(-09) \\  
HMPO23033 & 8.1(15) & $\leq$4.5(13) & 1.1(14) & 1.5(14) & 3.2(13) & 5.5(13) & $\leq$2.2(14) & $\leq$6.2(12) & 4.0(22) & 1.1(23) & 1.7(-08) \\  
HMPO23139 & 9.4(15) & 1.8(14) & 2.1(14) & 3.4(14) & 4.6(13) & 8.6(13) & $\leq$2.2(14) & $\leq$6.2(12) & 3.0(22) & 8.8(22) & 1.1(-08) \\  
HMPO23151 & 9.7(15) & $\leq$4.6(13) & 6.0(13) & 7.4(13) & 1.4(13) & $\leq$1.1(13) & $\leq$1.4(14) & $\leq$3.8(12) & 2.3(22) & 7.2(22) & 4.2(-09) \\  
HMPO23545 & 6.3(15) & $\leq$4.9(13) & 1.8(13) & $\leq$1.6(13) & 5.9(12) & $\leq$1.7(13) & $\leq$2.2(14) & $\leq$6.2(12) & 1.5(22) & 3.1(22) & 1.0(-08) \\  
HMC009.62 & 5.2(16) & 9.9(14) & 3.7(14) & 1.7(15) & 9.9(13) & 2.1(14) & 3.5(14) & 2.2(13) & 9.5(22) & 1.5(23) & 2.4(-08) \\  
HMC010.47 & 5.7(16) & 4.7(15) & 1.0(15) & 4.7(15) &         & 1.2(15) & 6.1(15) & 2.0(14) & 2.3(23) & 4.3(23) & 1.2(-08) \\  
HMC029.96 & 5.0(16) & 1.7(15) & 5.5(14) & 1.3(15) & 1.3(14) & 3.1(14) & 2.1(15) & 1.0(14) & 9.4(22) & 1.6(23) & 1.7(-08) \\  
HMC031.41 & 4.3(16) & 3.4(15) & 7.5(14) & 3.7(15) &         & 8.7(14) & 4.2(15) & 2.1(14) & 1.6(23) & 2.8(23) & 4.8(-09) \\  
HMC034.26 & 8.0(16) & 5.5(15) & 1.6(15) & 5.4(15) & 2.5(14) & 1.5(15) & 7.1(15) & 3.1(14) & 4.1(23) & 6.7(23) & 9.0(-09) \\  
HMC045.47 & 2.0(16) & 2.0(14) & 1.7(14) & 5.8(14) & 5.5(13) & 1.2(14) & 2.6(14) & 1.3(13) & 4.5(22) & 6.5(22) & 3.6(-08) \\  
HMC075.78 & 2.9(16) & 4.2(14) & 2.9(14) & 6.7(14) & 8.1(13) & 2.1(14) & 7.5(14) & 3.7(13) & 6.3(22) & 8.7(22) & 1.8(-08) \\  
W3H2O & 3.2(16) & 1.6(15) & 5.6(14) & 1.6(15) & 1.2(14) & 8.7(14) & 3.7(15) & 1.9(14) & 7.4(22) & 1.1(23) & 1.4(-08) \\  
W3IRS5 & 2.9(16) & 6.7(13) & 1.5(14) & 6.8(13) & 4.1(13) & 3.4(13) & 2.6(14) & 1.1(13) & 6.0(22) & 7.6(22) & 7.1(-09) \\  
NGC7538B & 2.1(16) & 2.4(14) & 2.9(14) & 4.1(14) & 8.1(13) & 1.8(14) & 4.1(14) & 3.4(13) & 1.2(23) & 1.7(23) & 6.4(-09) \\  
Orion-KL & 8.1(16) & 3.7(16) & 8.2(15) & 1.8(16) & 1.6(15) & 7.3(15) & 5.8(16) & 6.7(14) & 1.3(24) & 1.7(24) & 6.4(-10) \\  
UCH005.89 & 5.5(16) & 2.1(15) & 1.3(15) & 2.0(15) & 2.7(14) & 5.3(14) & 2.4(15) & 1.2(14) & 2.2(23) & 3.6(23) & 1.8(-08) \\
UCH010.10 & $\leq$1.1(14) & $\leq$2.8(13) & $\leq$6.9(12) & $\leq$1.4(13) & $\leq$5.8(11) & $\leq$1.1(13) & $\leq$8.3(13) & $\leq$3.9(12) & 6.4(21) & 1.3(22) & 2.3(-08) \\
UCH010.30 & 3.3(16) & 3.1(14) & 2.3(14) & 7.8(14) & 8.8(13) & 1.5(14) & 1.5(14) & 2.7(13) & 8.1(22) & 1.1(23) & 2.0(-08) \\
UCH012.21 & 3.1(16) & 7.2(14) & 3.4(14) & 1.3(15) & 7.6(13) & 7.3(14) & 2.0(15) & 8.4(13) & 9.6(22) & 1.6(23) & 1.2(-08) \\
UCH013.87 & 2.9(16) & $\leq$3.8(13) & 7.8(13) & 1.6(14) & 2.5(13) & 3.9(13) & $\leq$7.3(13) & 1.7(13) & 5.3(22) & 7.6(22) & 1.7(-08) \\
UCH030.54 & 1.7(16) & $\leq$3.6(13) & 3.7(13) & $\leq$1.8(13) & 1.3(13) & $\leq$8.0(12) & $\leq$6.3(13) & $\leq$3.0(12) & 2.1(22) & 3.1(22) & 1.1(-08) \\
UCH035.20 & 2.7(16) & $\leq$3.6(13) & 1.4(14) & 9.2(13) & 4.1(13) & $\leq$5.0(12) & $\leq$3.9(13) & 1.8(13) & 1.1(23) & 1.4(23) & 5.7(-09) \\
UCH045.12 & 1.4(16) & $\leq$3.4(13) & 7.4(13) & $\leq$1.7(13) & 2.3(13) & $\leq$2.8(12) & $\leq$2.2(13) & 8.1(12) & 7.4(22) & 1.1(23) & 9.6(-09) \\
UCH045.45 & 1.5(16) & $\leq$3.4(13) & 9.9(13) & 1.1(14) & 2.6(13) & $\leq$2.8(12) & $\leq$2.2(13) & 1.7(13) & 4.2(22) & 5.6(22) & 7.9(-09) \\
\hline                                   
\end{tabular}
\end{table*}
}

\clearpage

\clearpage
\onlfig{
\begin{figure*}
\includegraphics[width=0.32\textwidth]{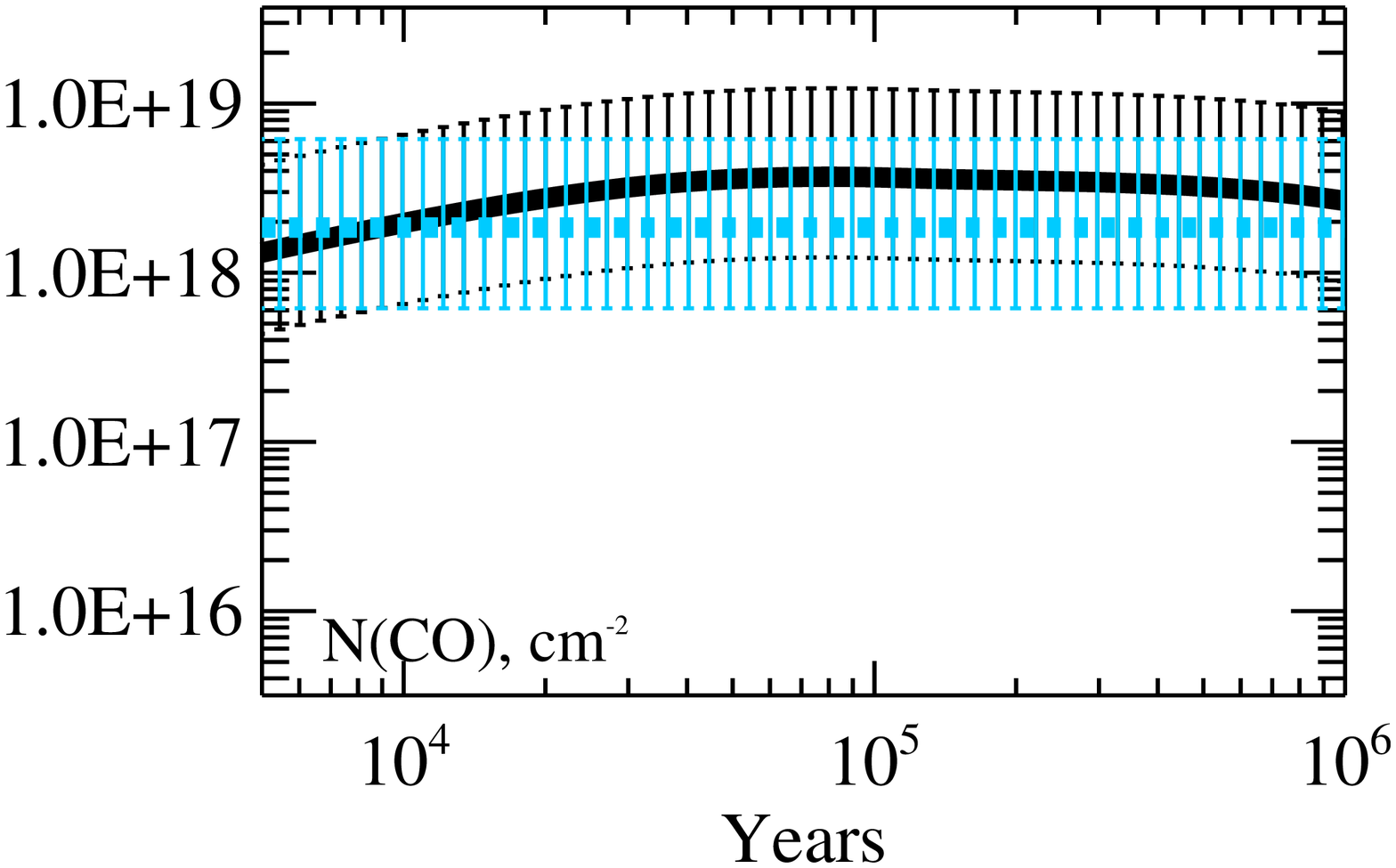}
\includegraphics[width=0.32\textwidth]{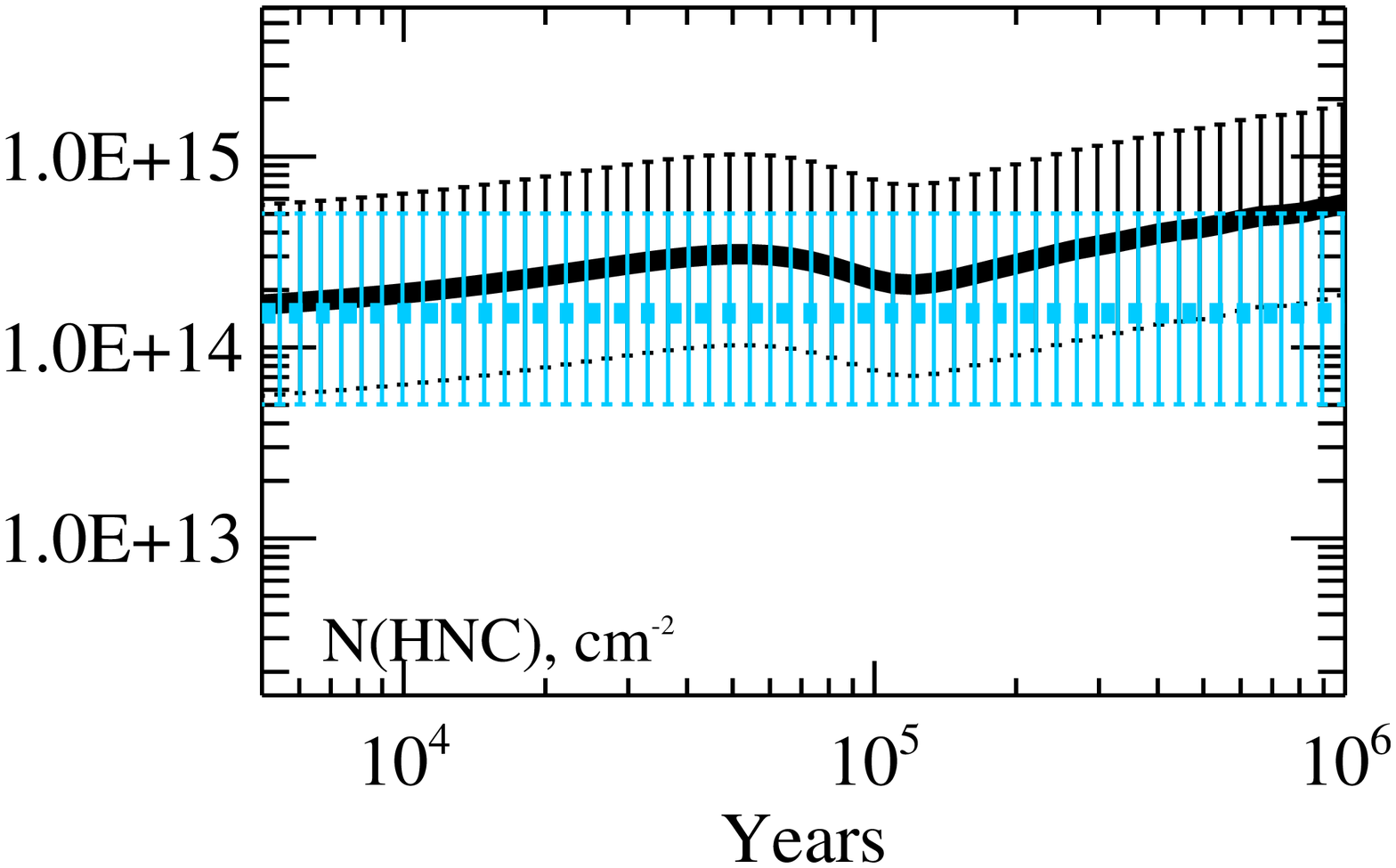}
\includegraphics[width=0.32\textwidth]{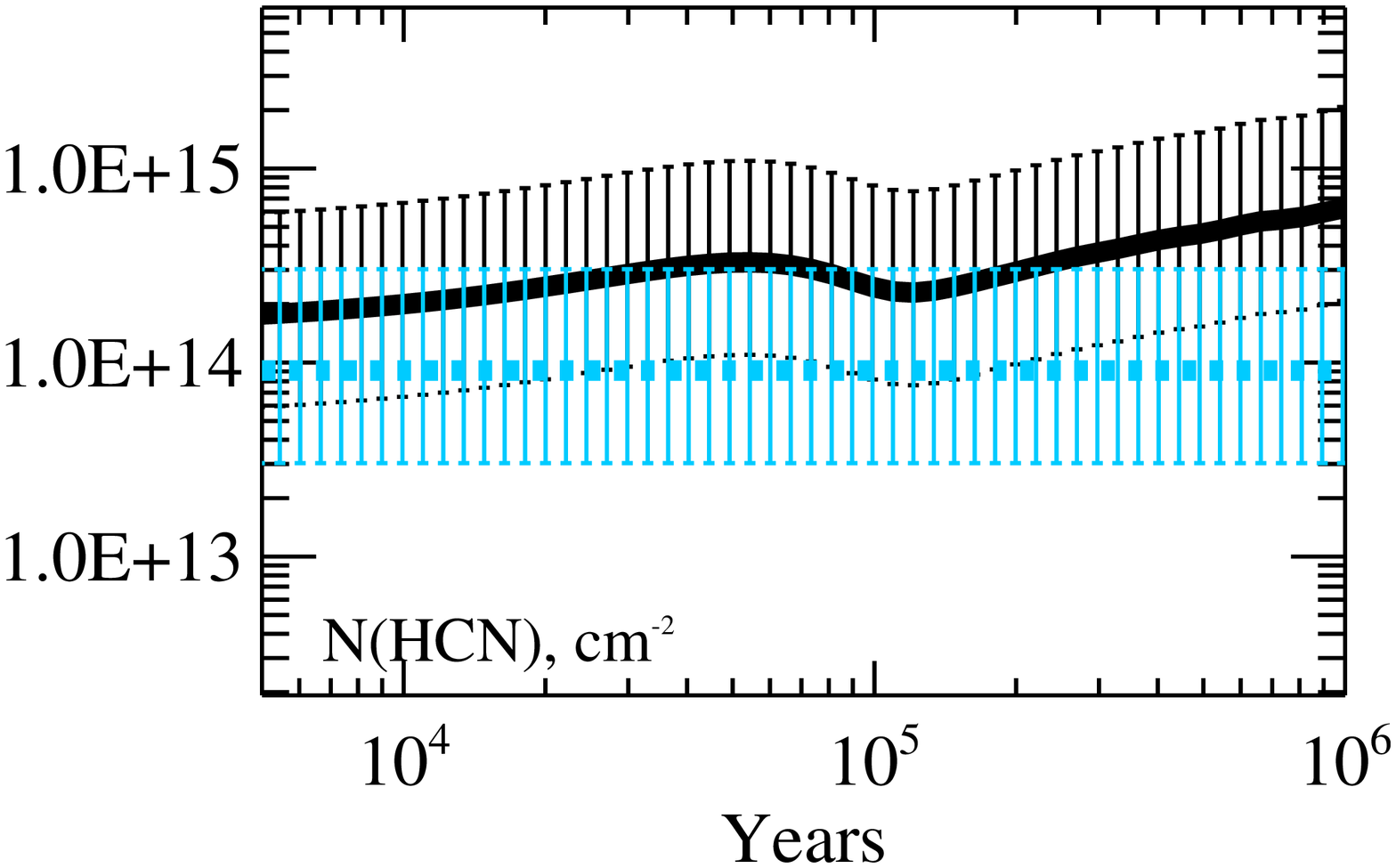}\\
\includegraphics[width=0.32\textwidth]{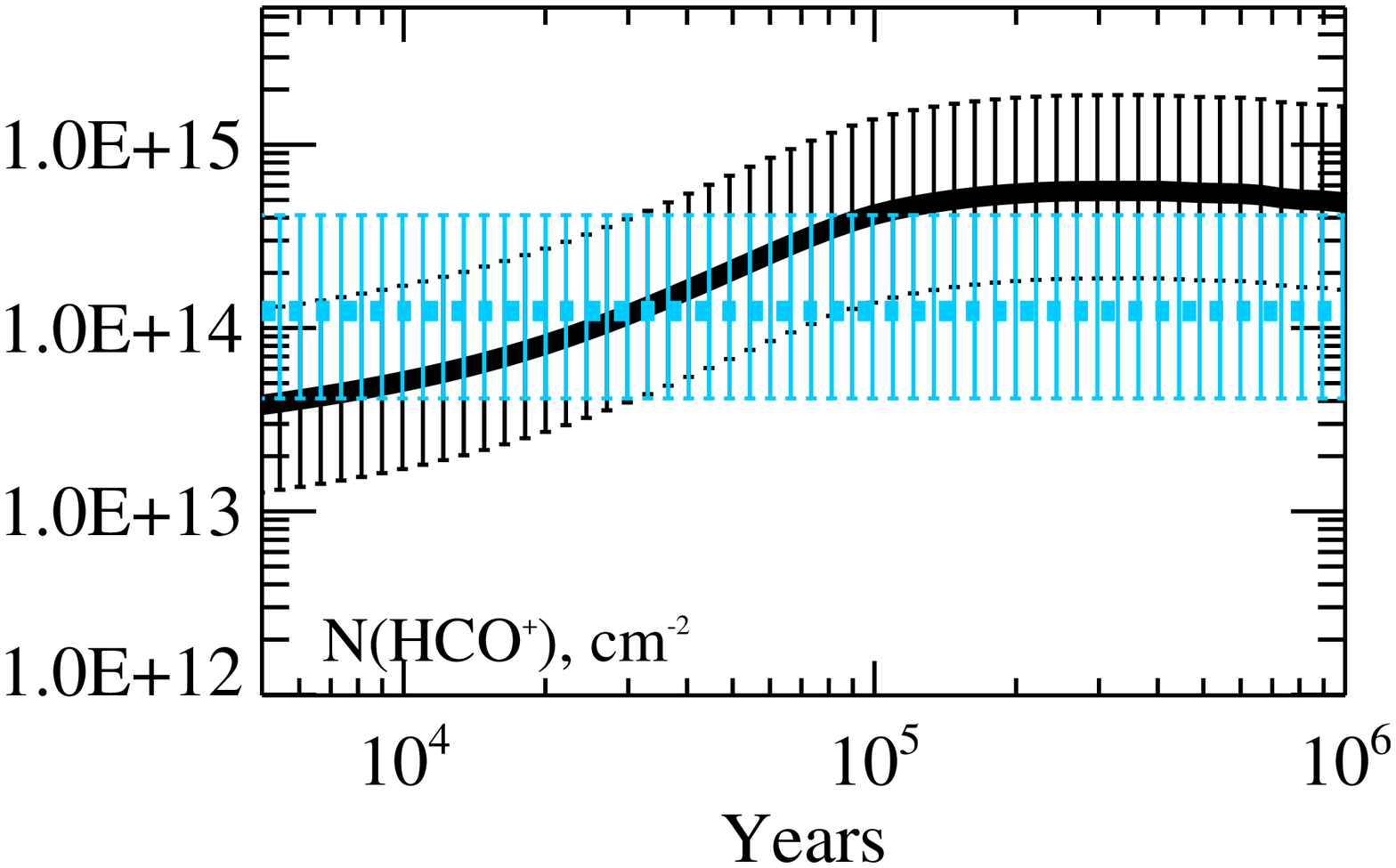}
\includegraphics[width=0.32\textwidth]{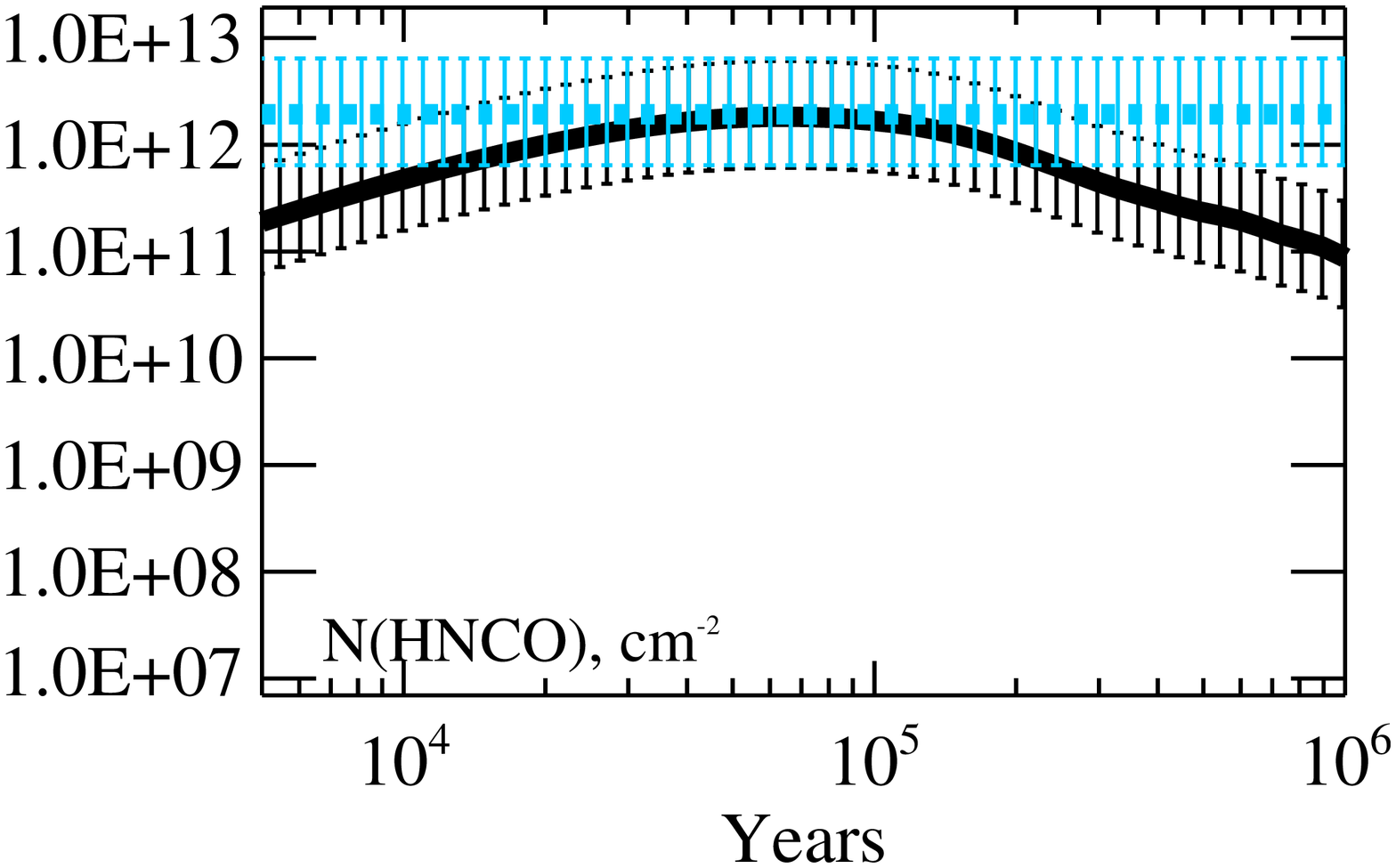}
\includegraphics[width=0.32\textwidth]{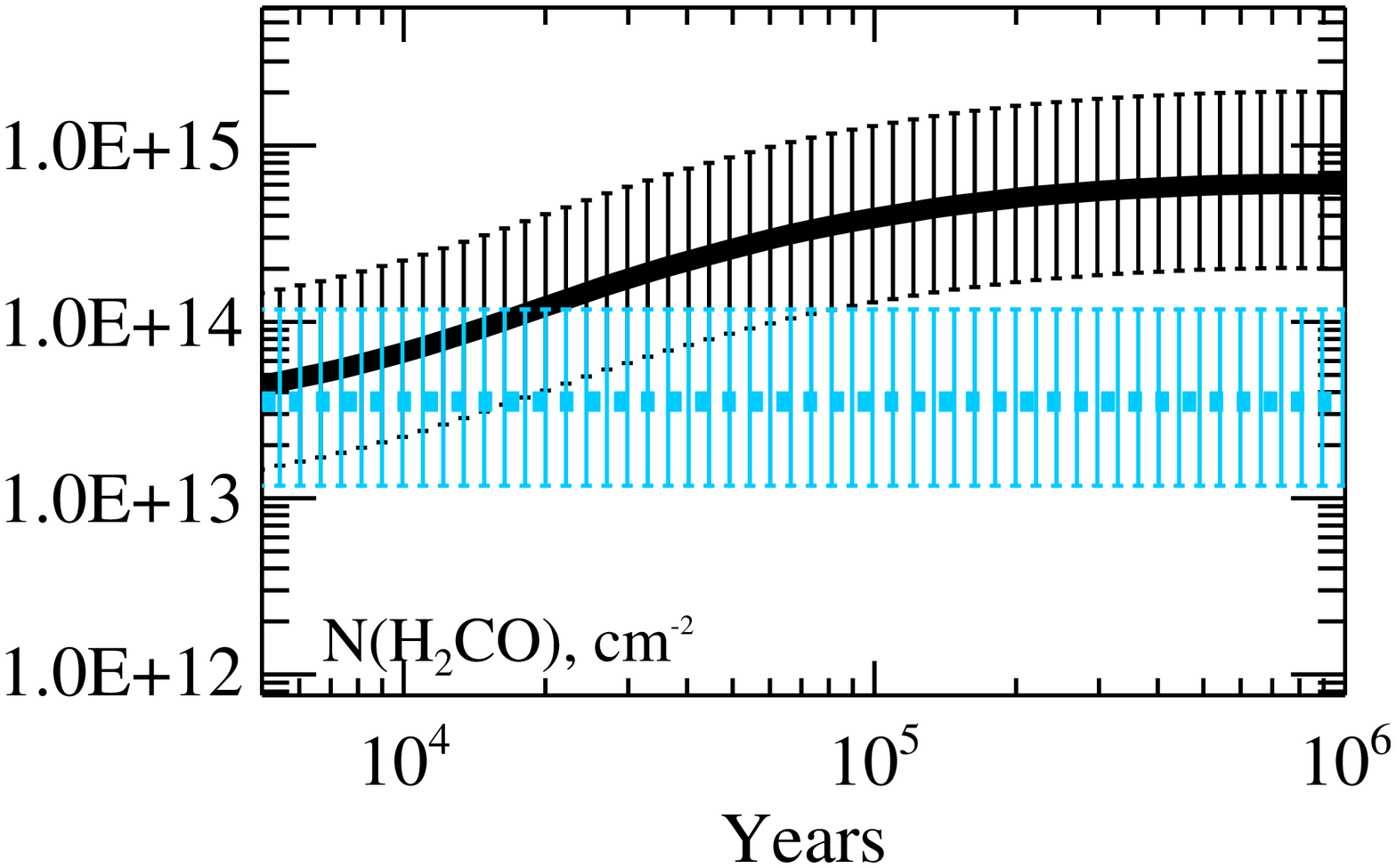}\\
\includegraphics[width=0.32\textwidth]{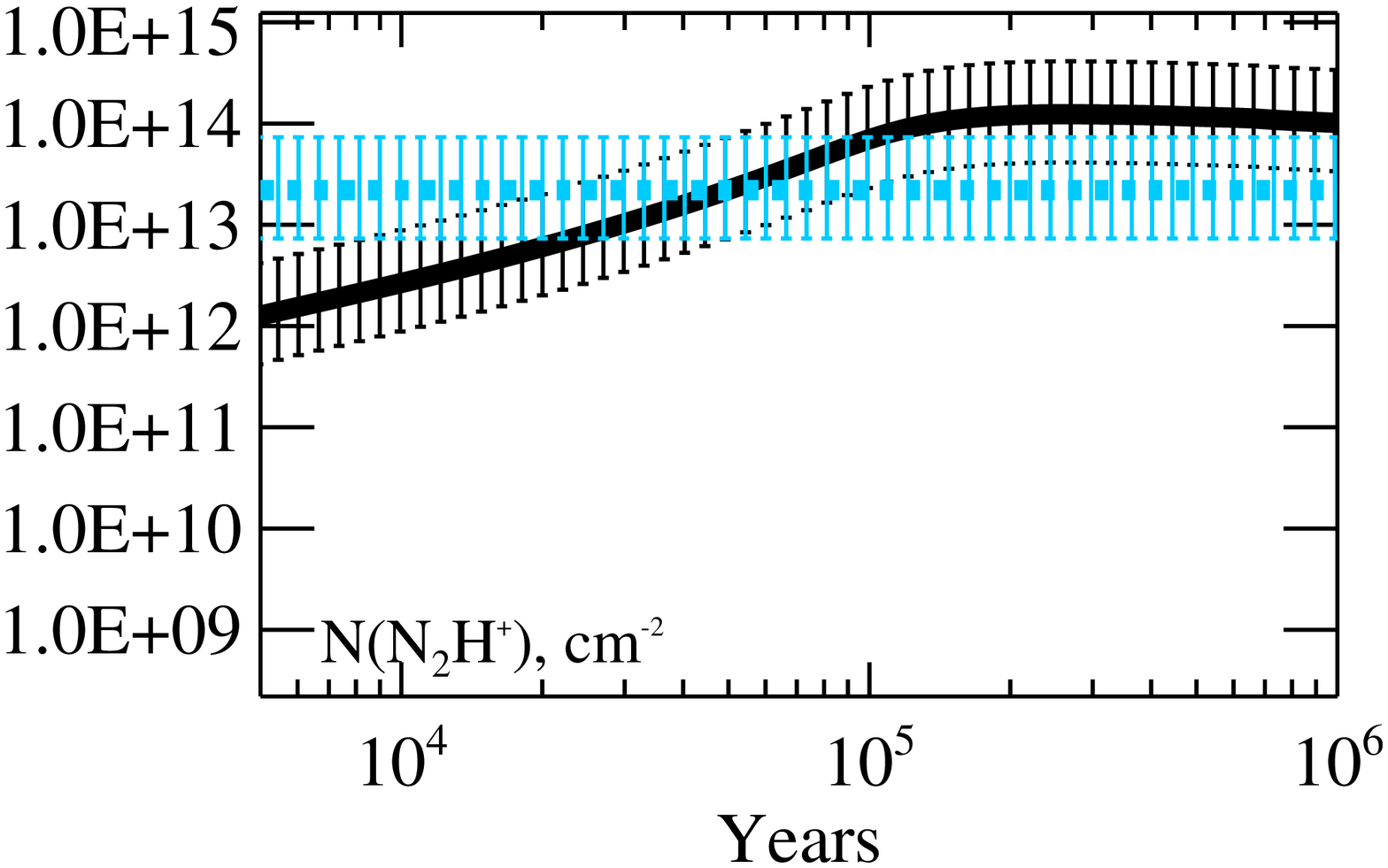}
\includegraphics[width=0.32\textwidth]{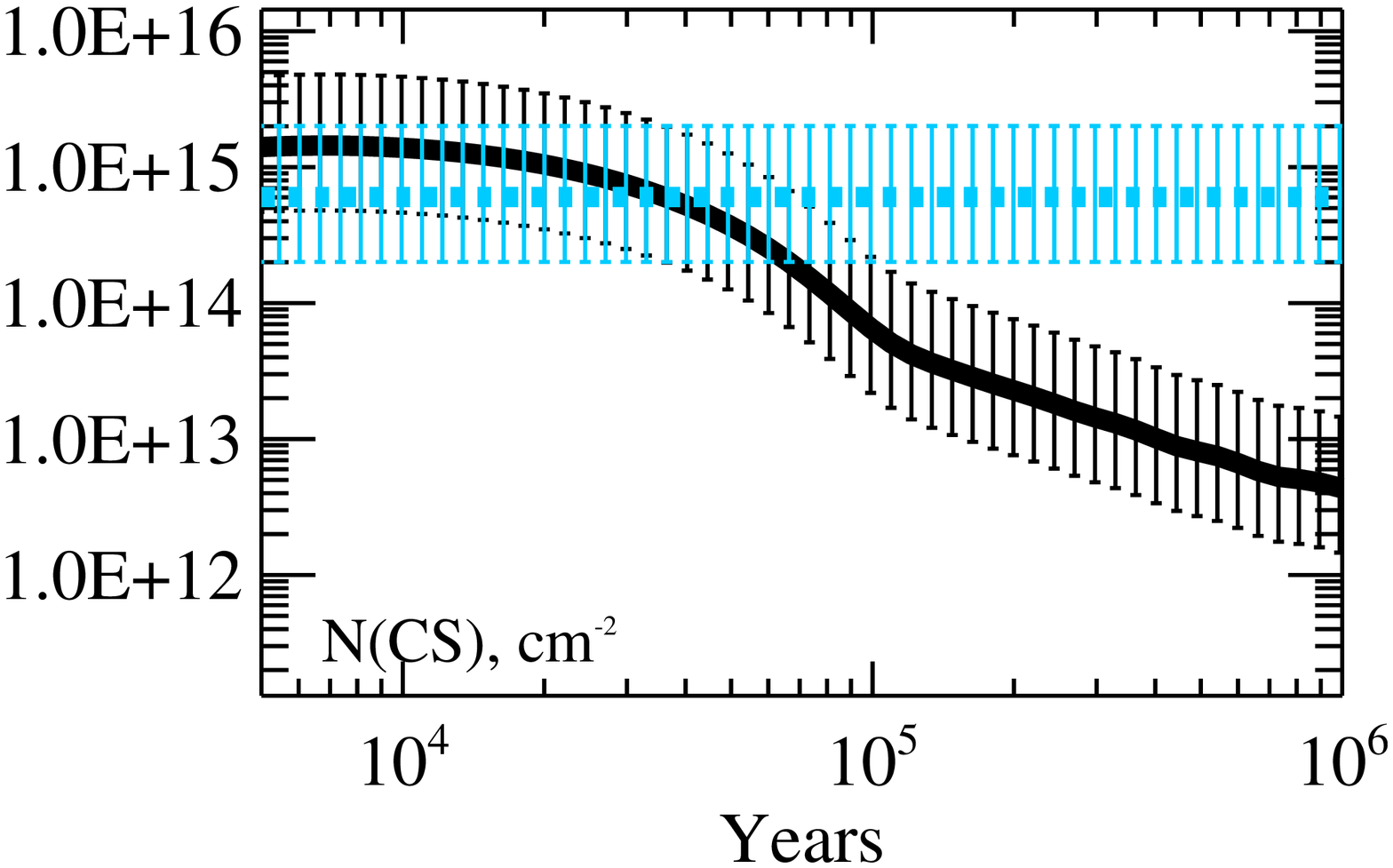}
\includegraphics[width=0.32\textwidth]{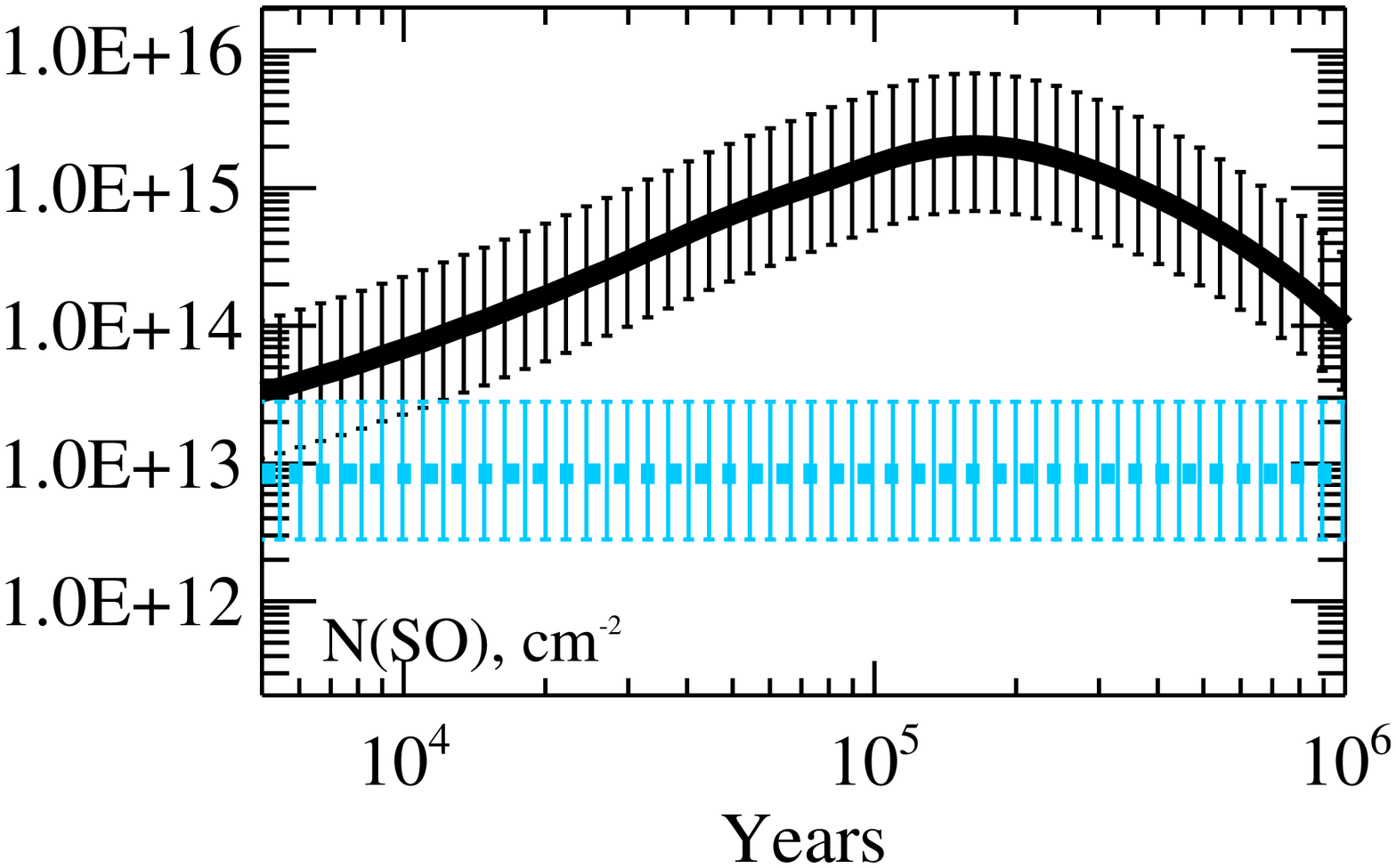}\\
\includegraphics[width=0.32\textwidth]{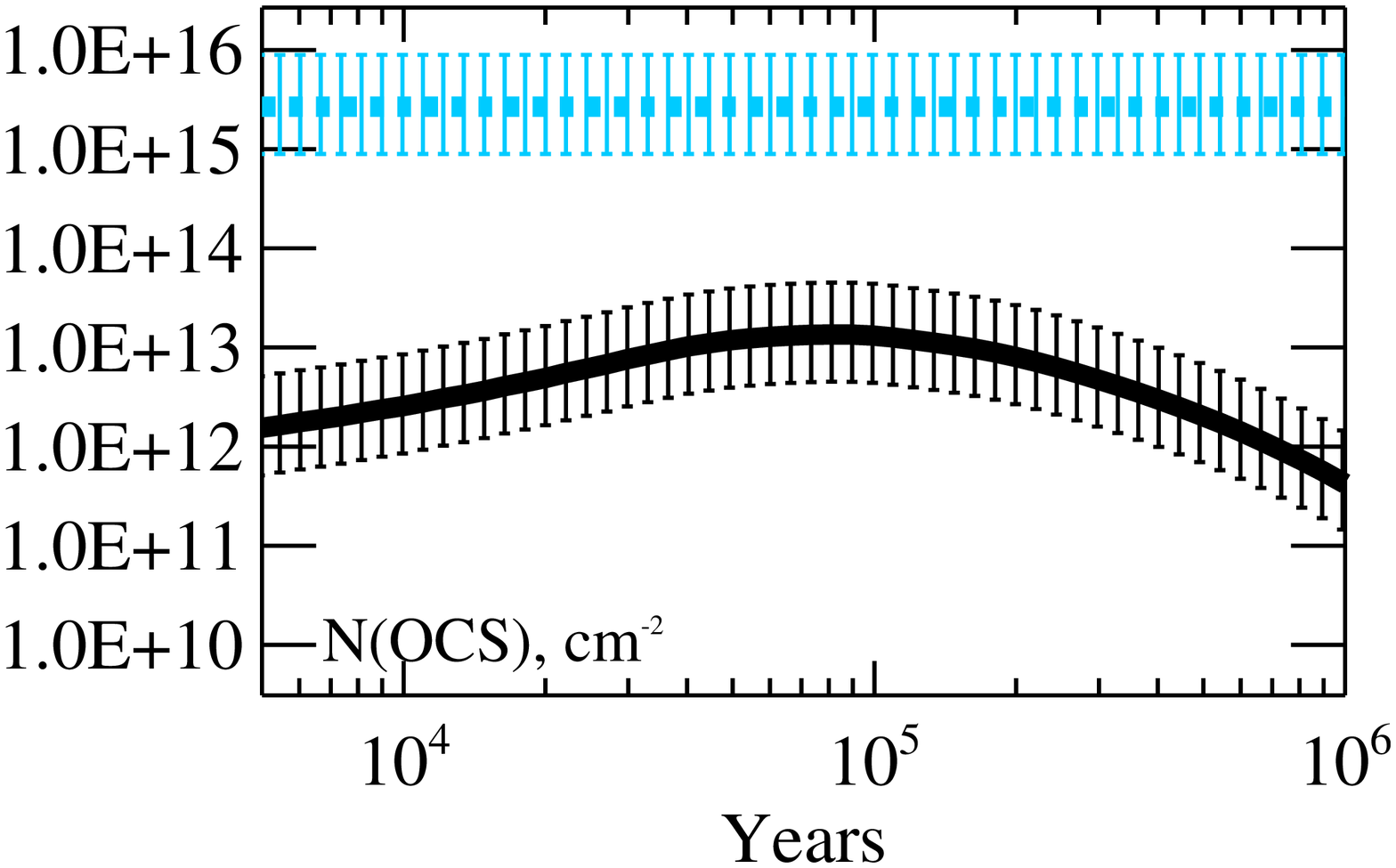}
\includegraphics[width=0.32\textwidth]{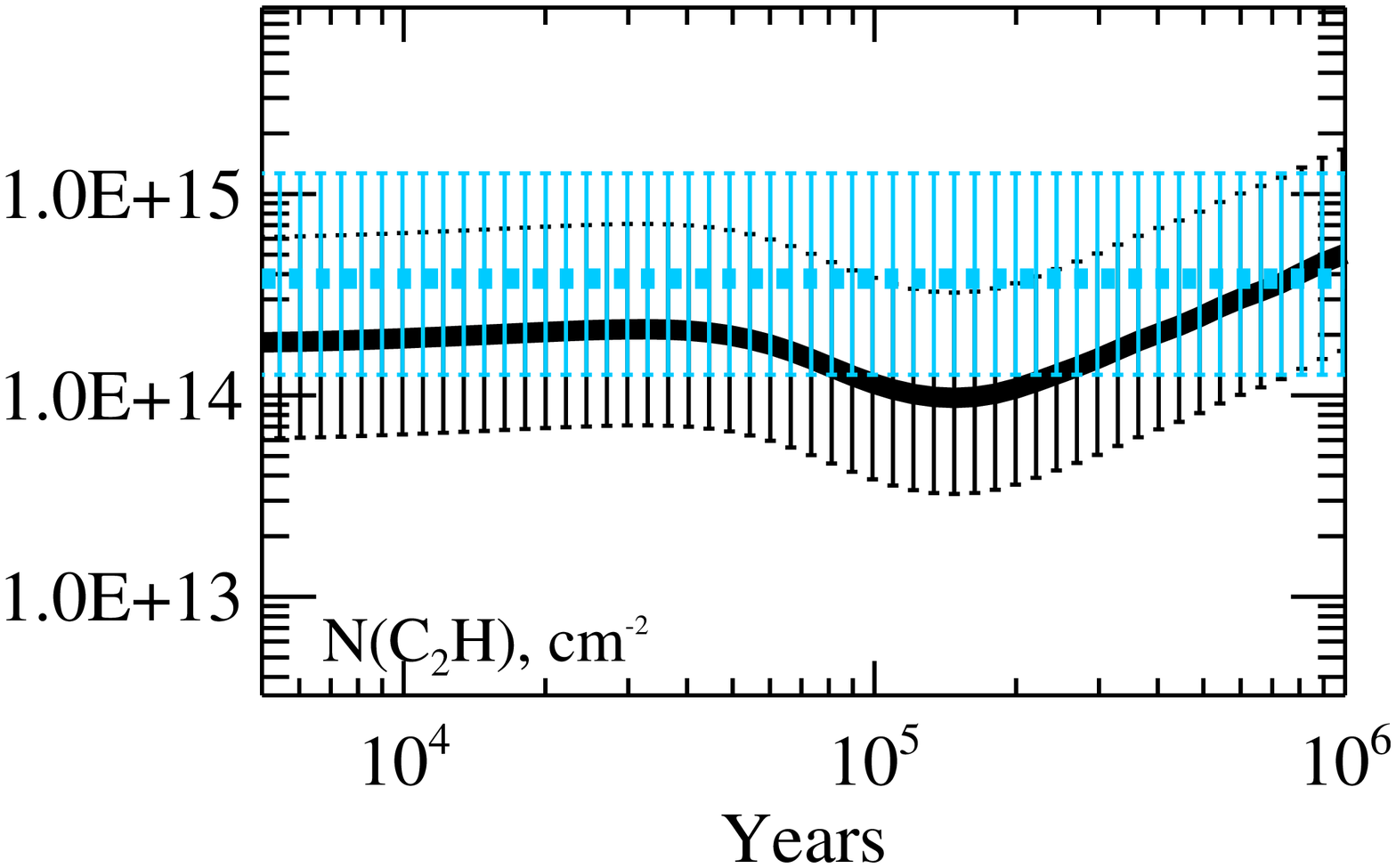}
\includegraphics[width=0.32\textwidth]{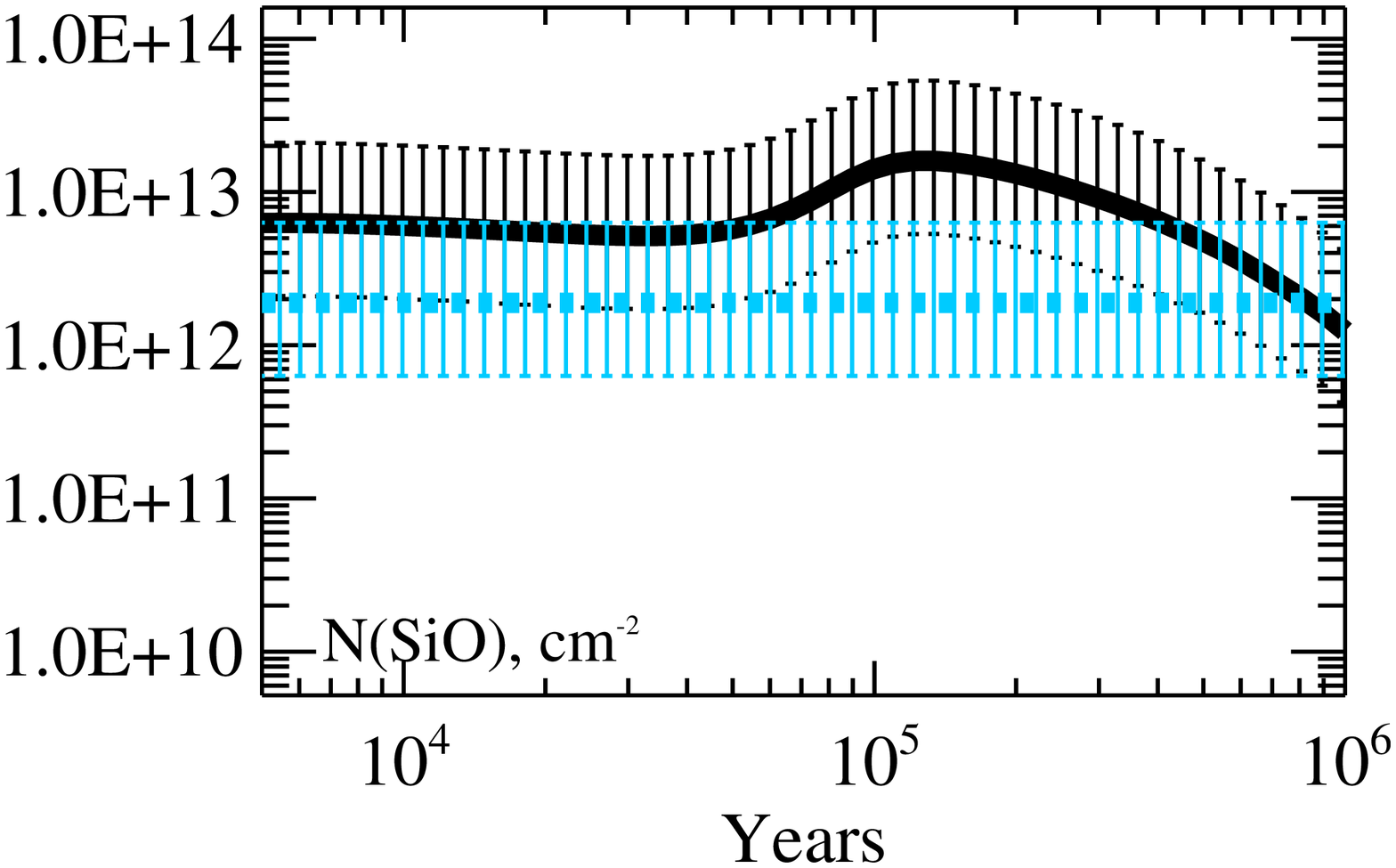}\\
\includegraphics[width=0.32\textwidth]{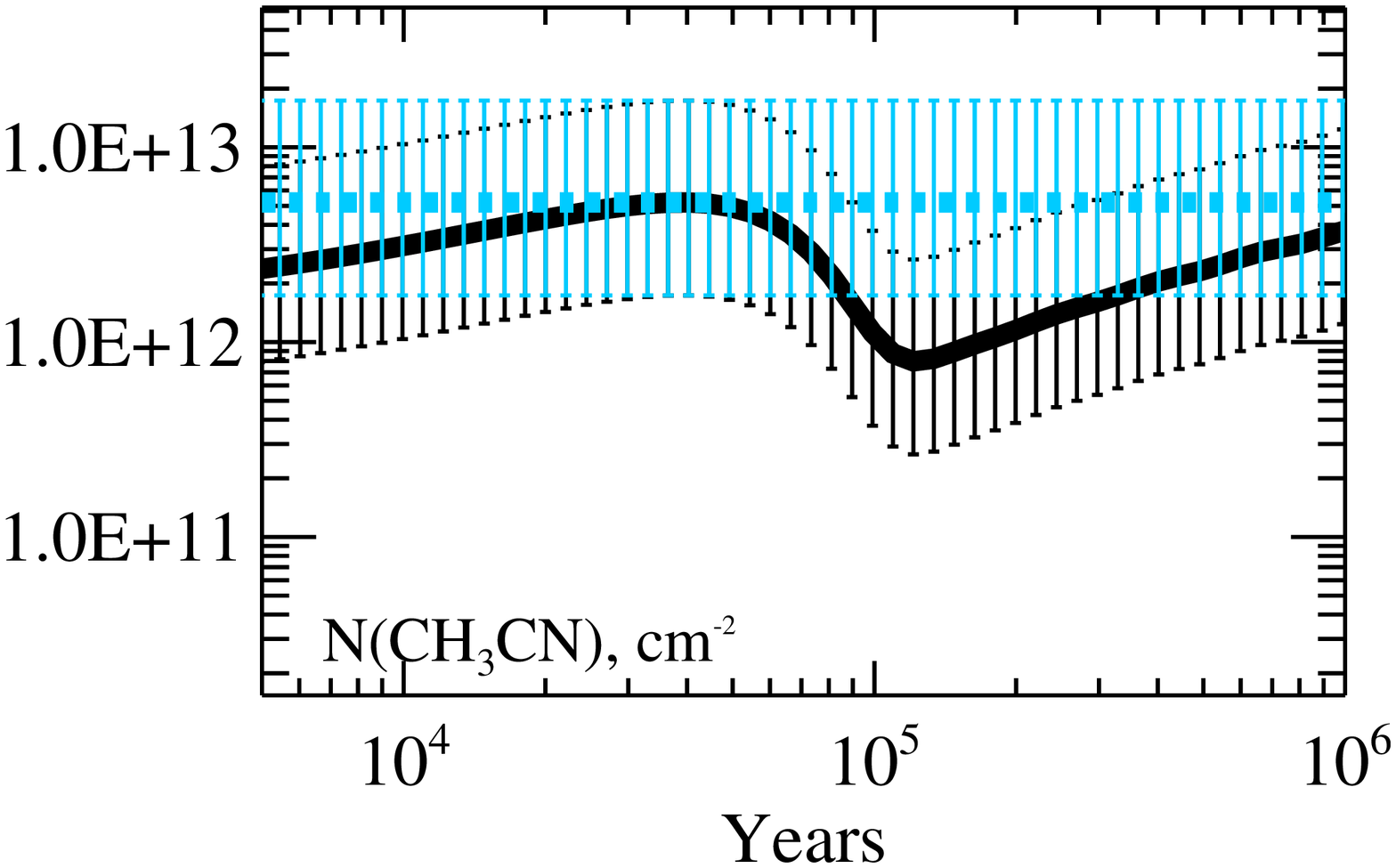}
\includegraphics[width=0.32\textwidth]{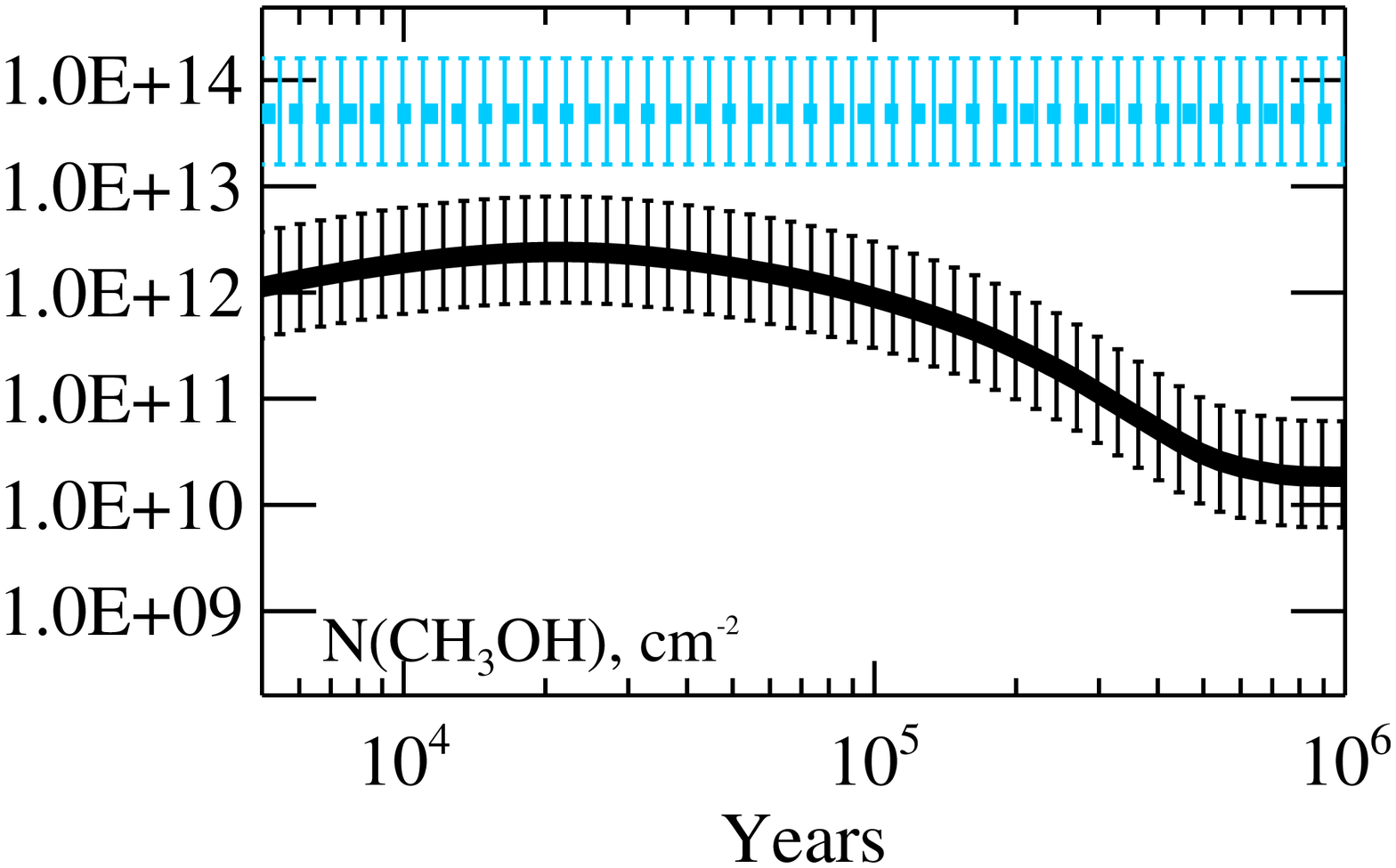}
\caption{Modeled and observed median column densities in cm$^{-2}$ in the IRDC stage. The modeled values are shown in black and the observed values in blue. The error bars are given by the vertical marks. The molecule name is labeled in the plot.}
\label{fig:coldens_irdc}
\end{figure*}
}
\clearpage
\onlfig{
\begin{figure*}
\includegraphics[width=0.32\textwidth]{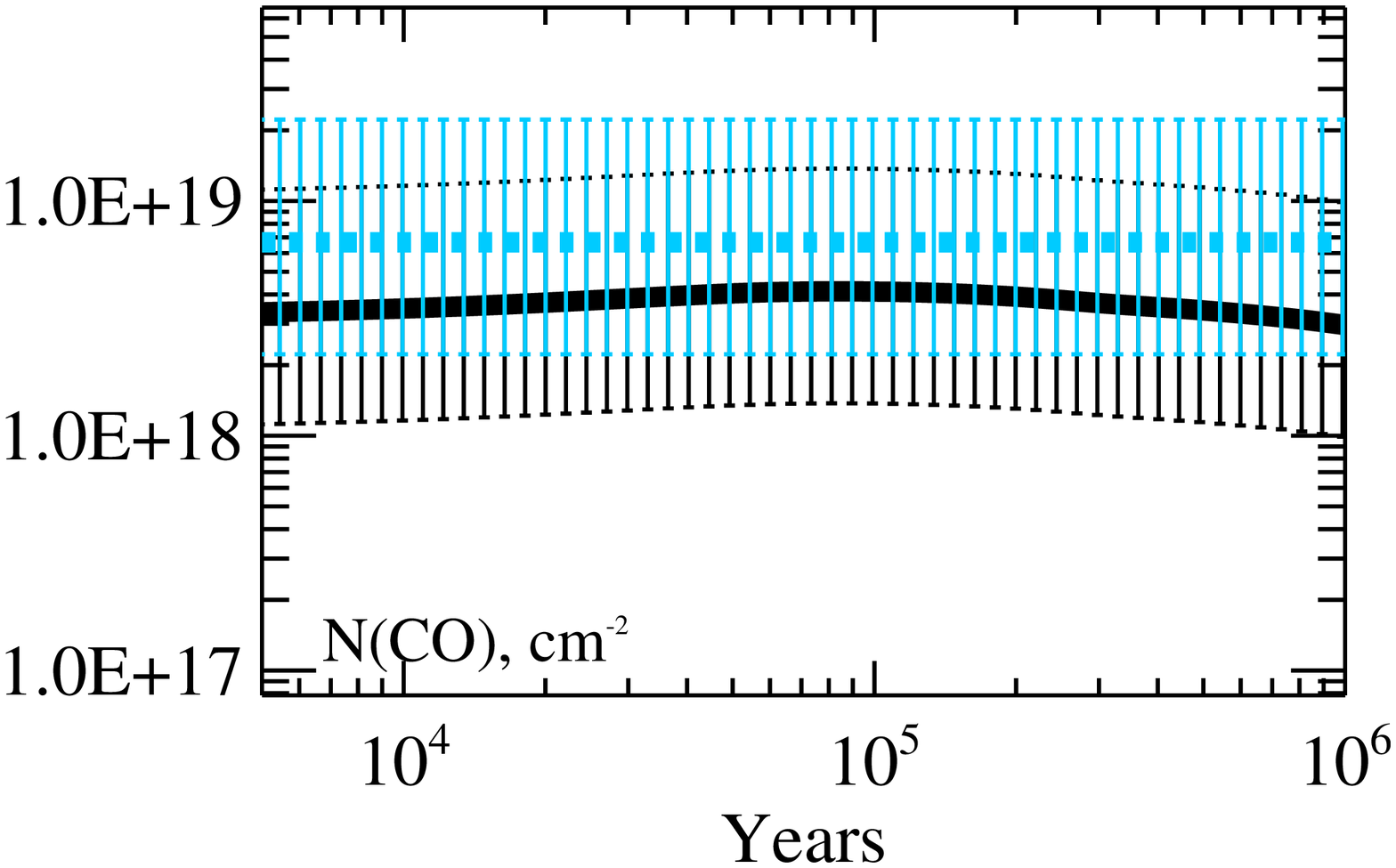}
\includegraphics[width=0.32\textwidth]{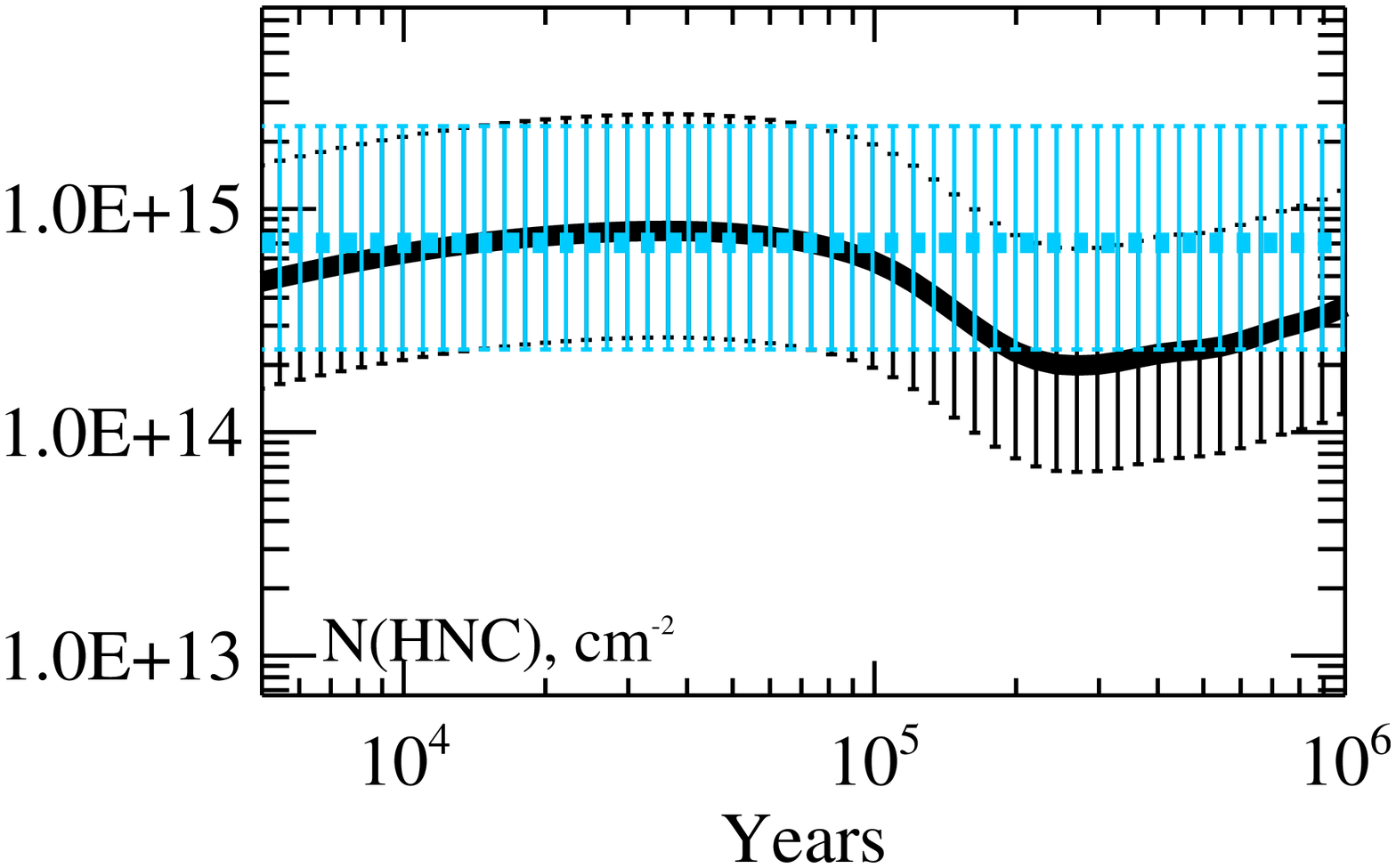}
\includegraphics[width=0.32\textwidth]{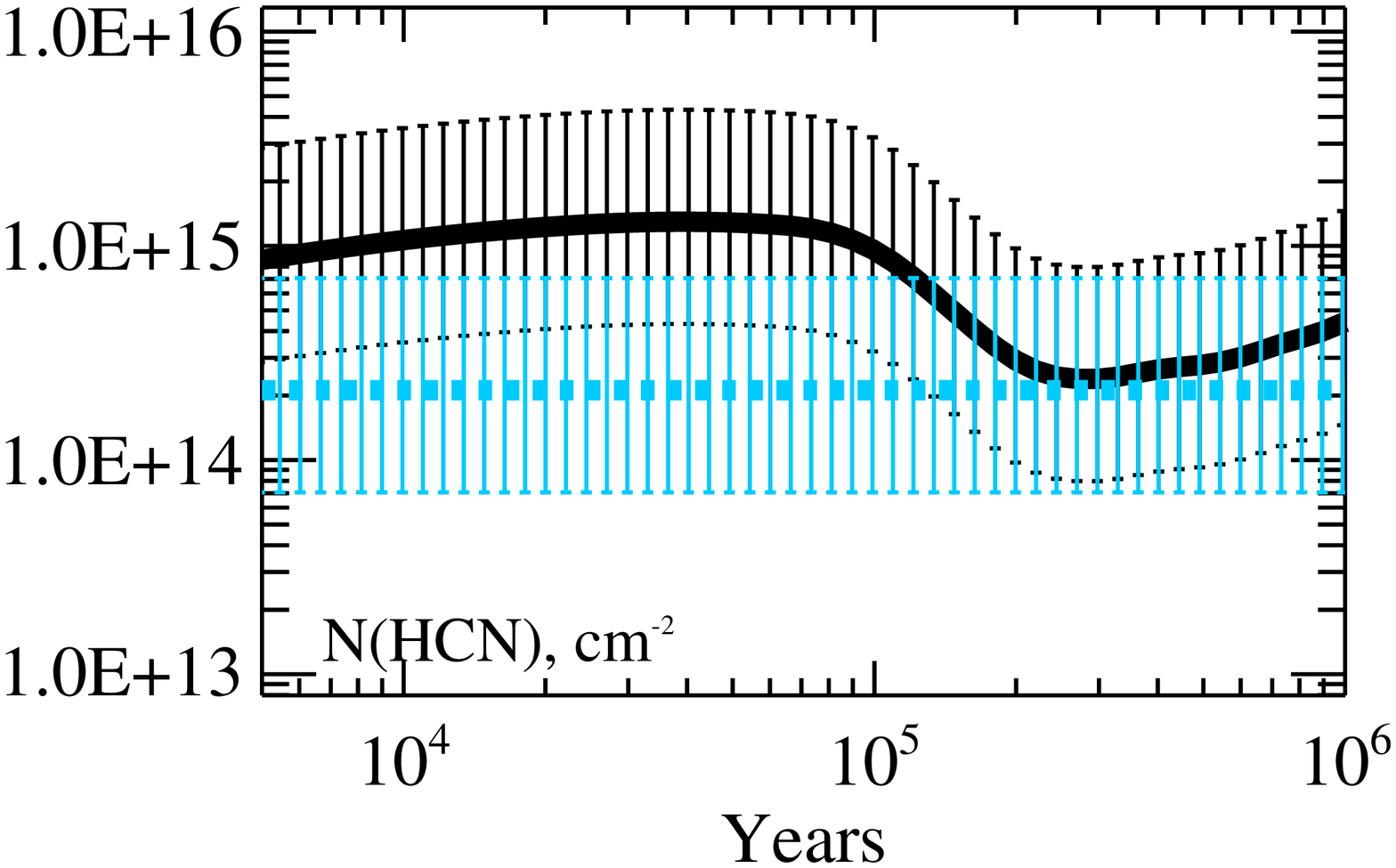}\\
\includegraphics[width=0.32\textwidth]{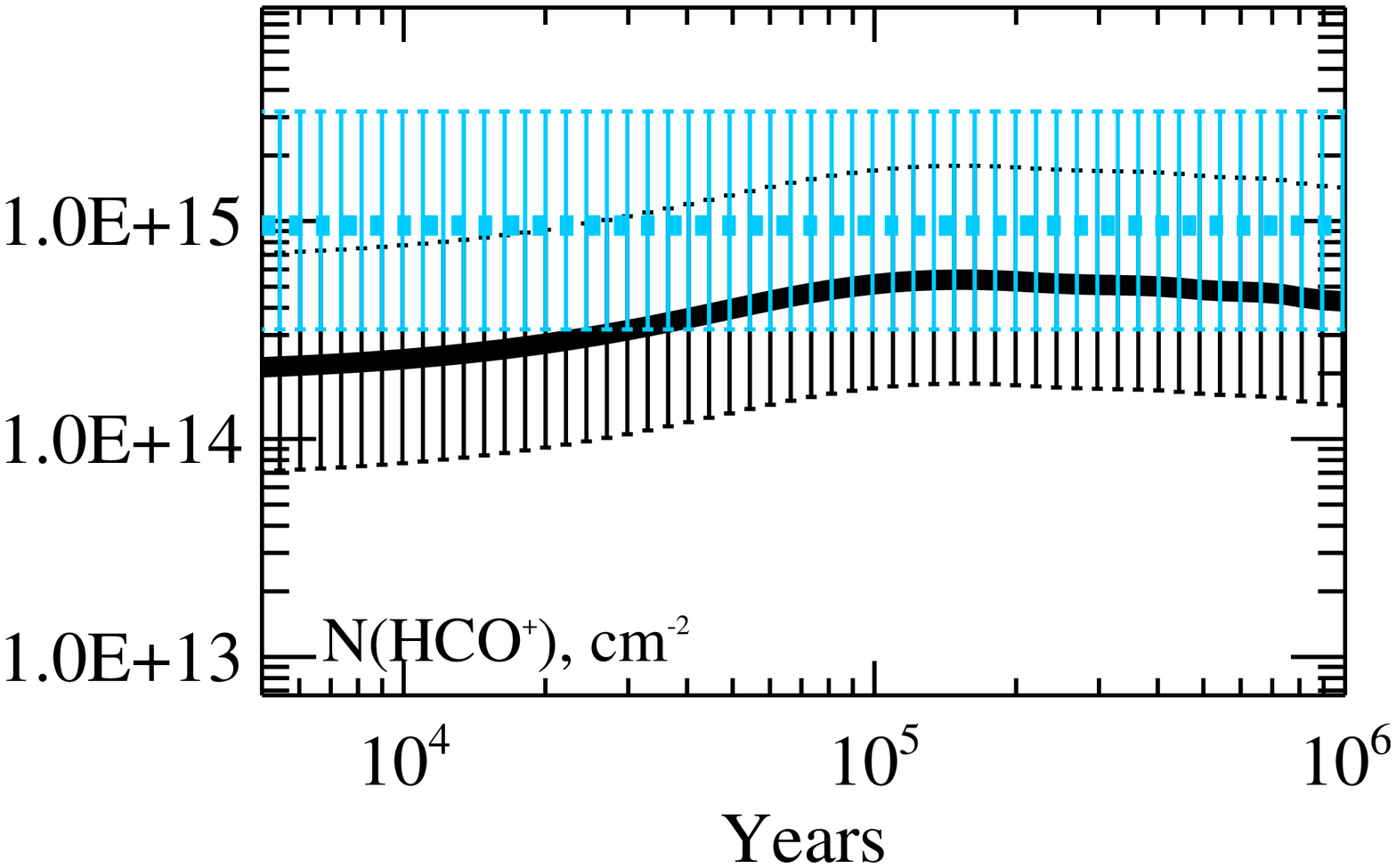}
\includegraphics[width=0.32\textwidth]{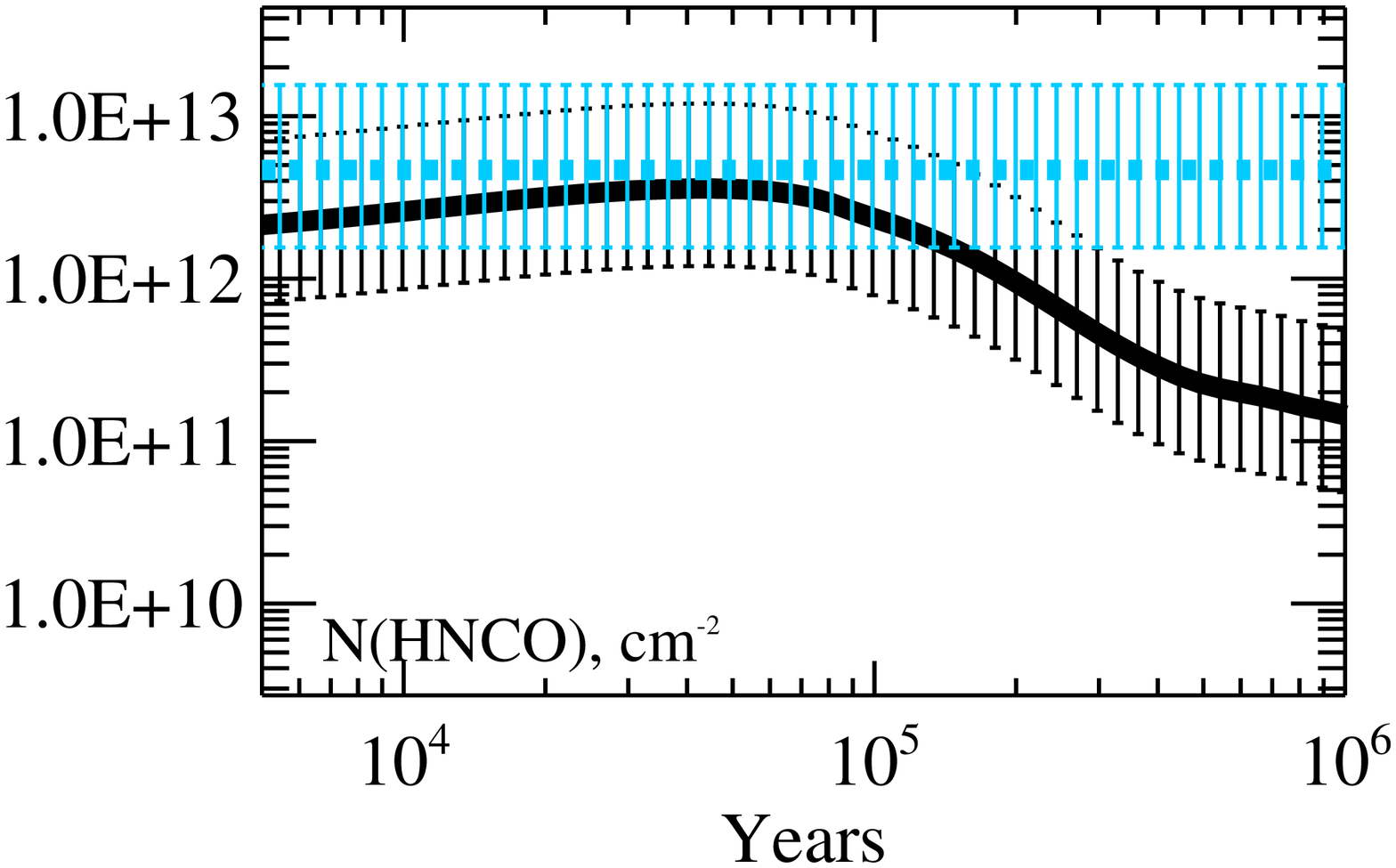}
\includegraphics[width=0.32\textwidth]{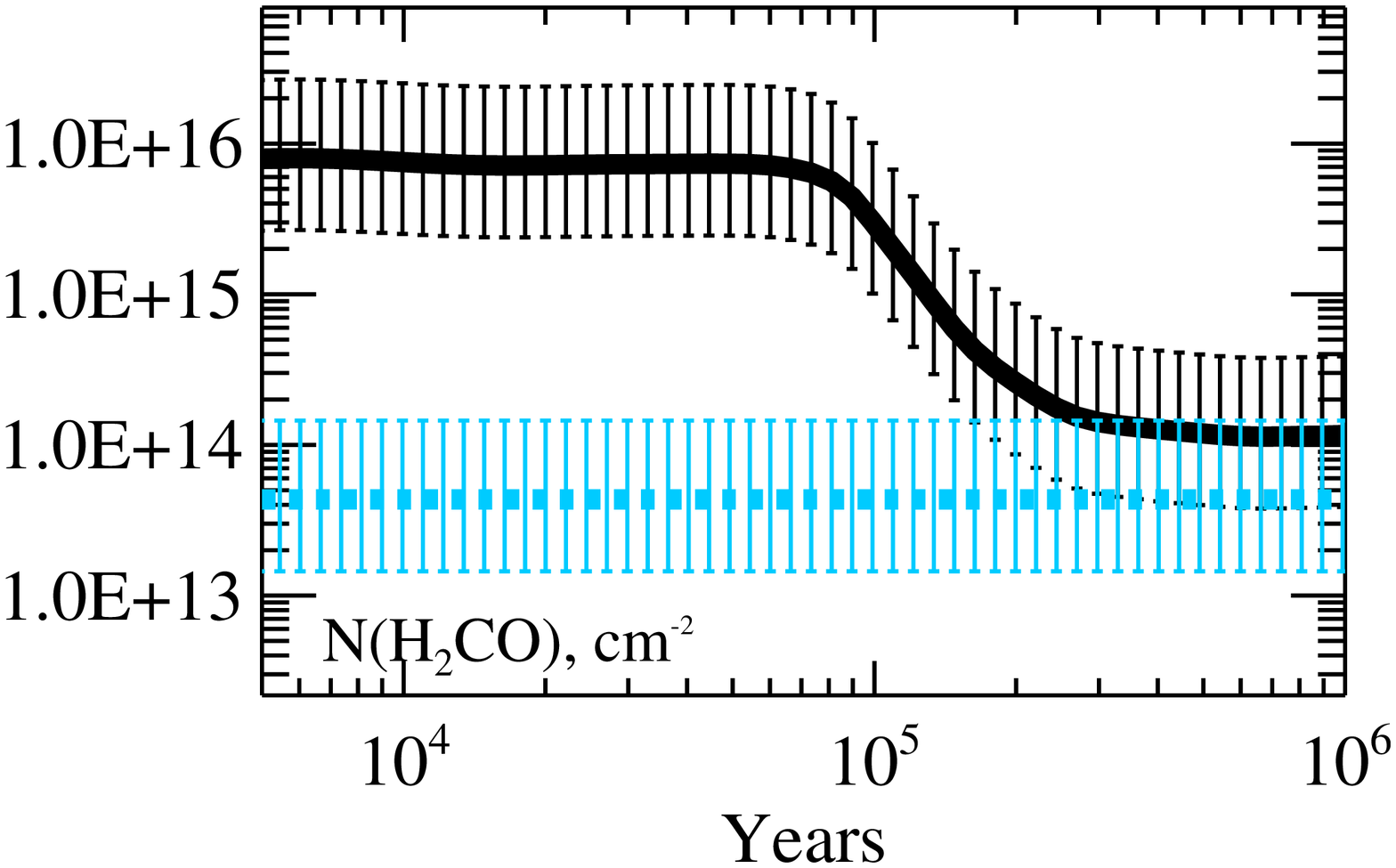}\\
\includegraphics[width=0.32\textwidth]{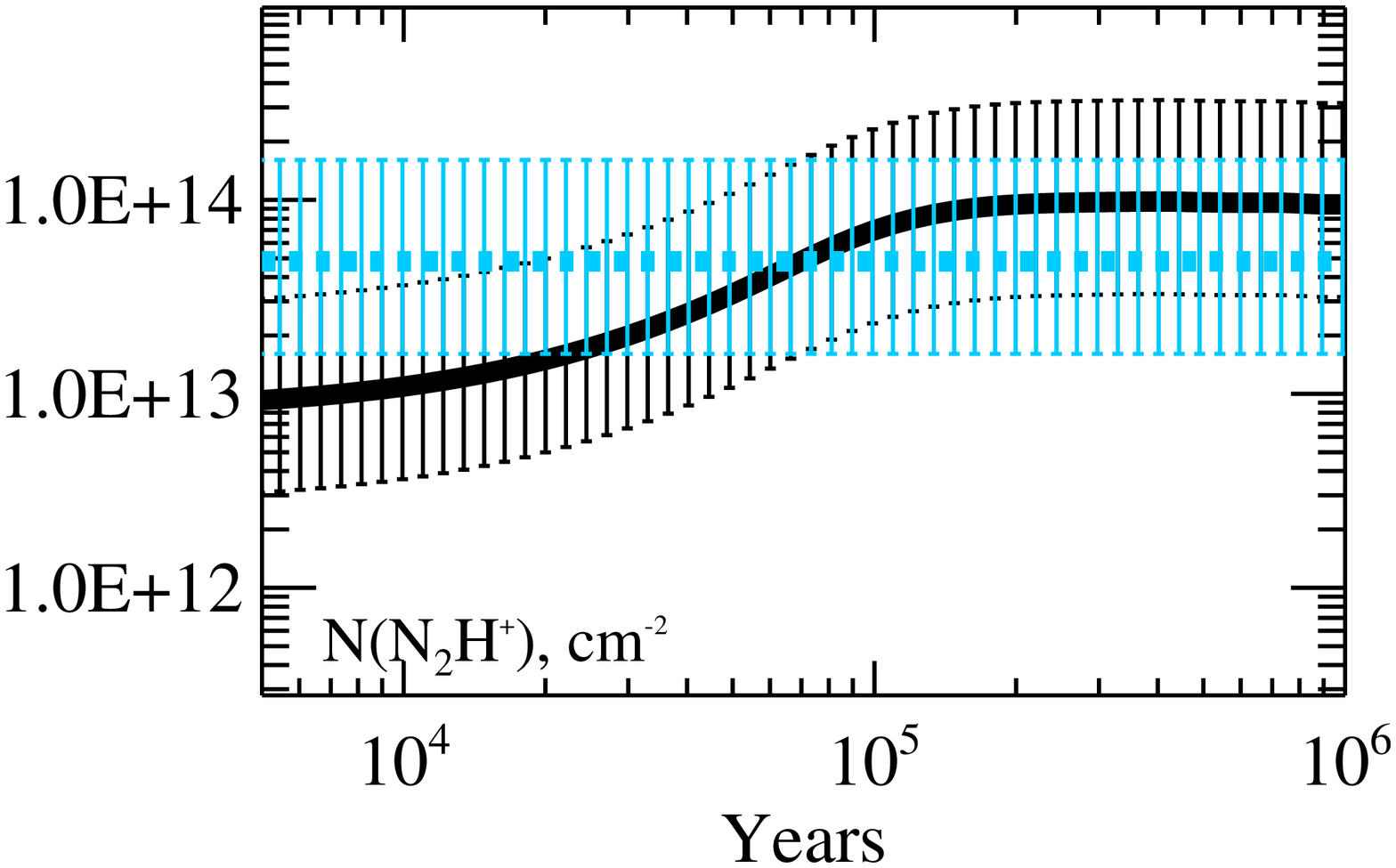}
\includegraphics[width=0.32\textwidth]{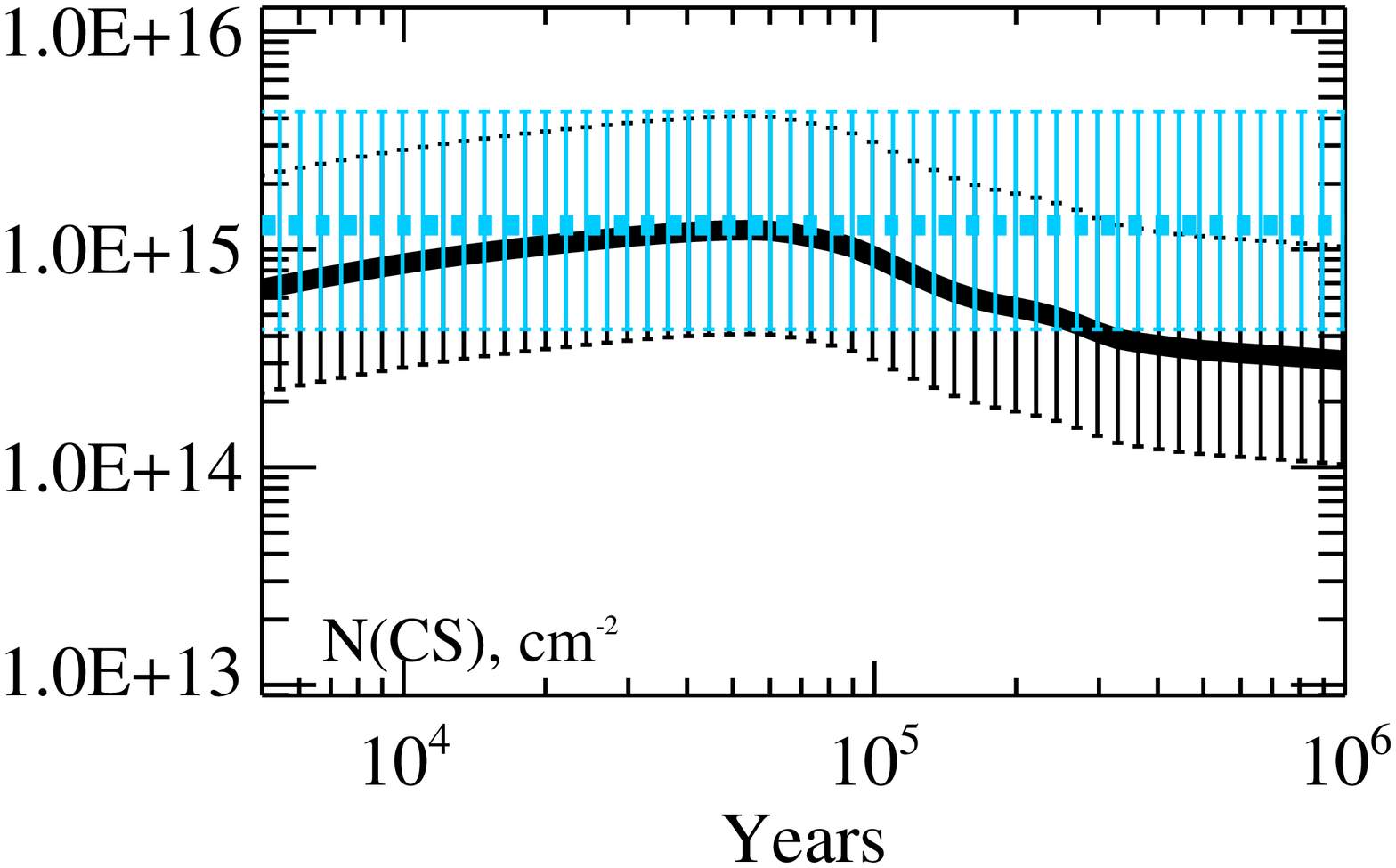}
\includegraphics[width=0.32\textwidth]{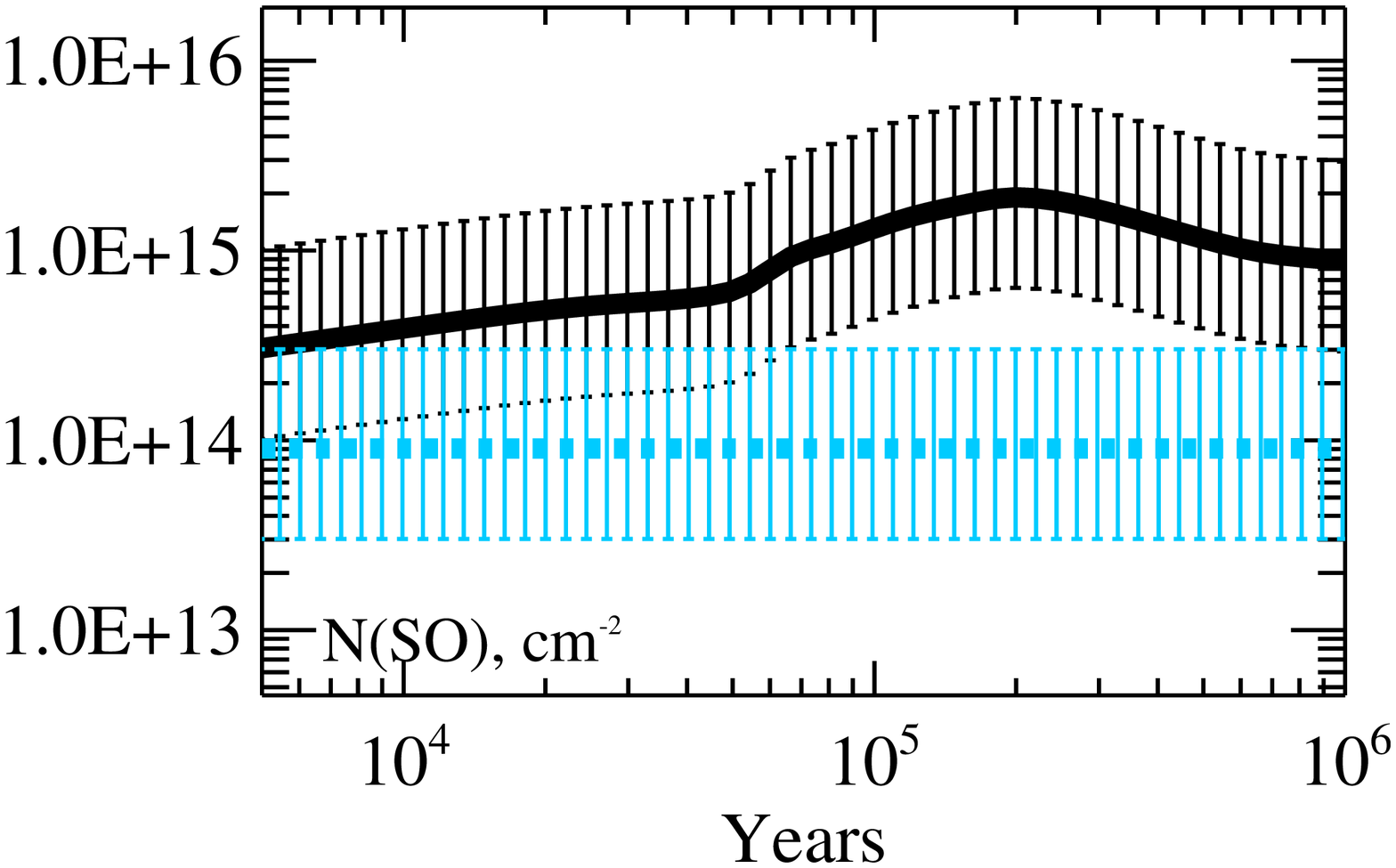}\\
\includegraphics[width=0.32\textwidth]{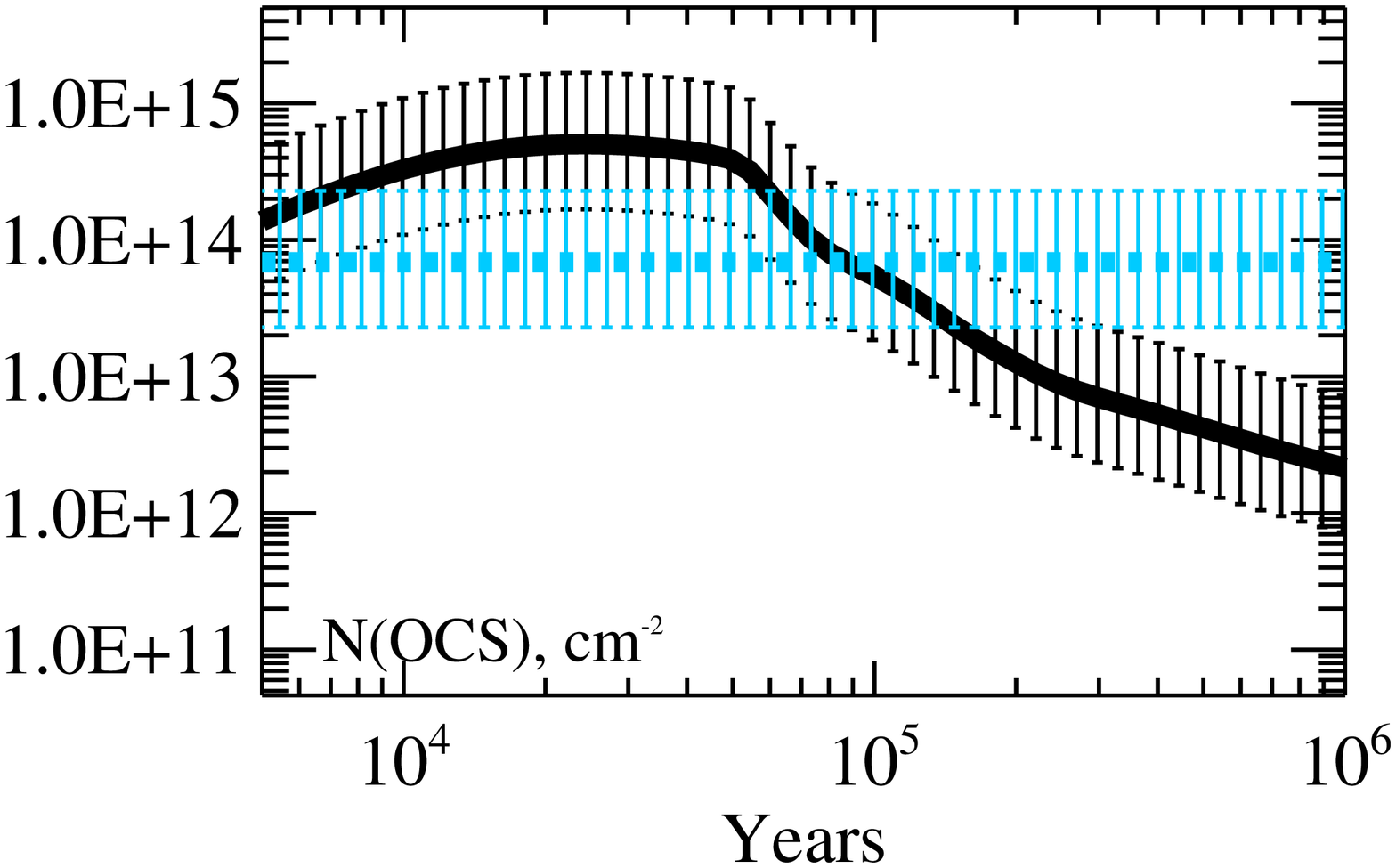}
\includegraphics[width=0.32\textwidth]{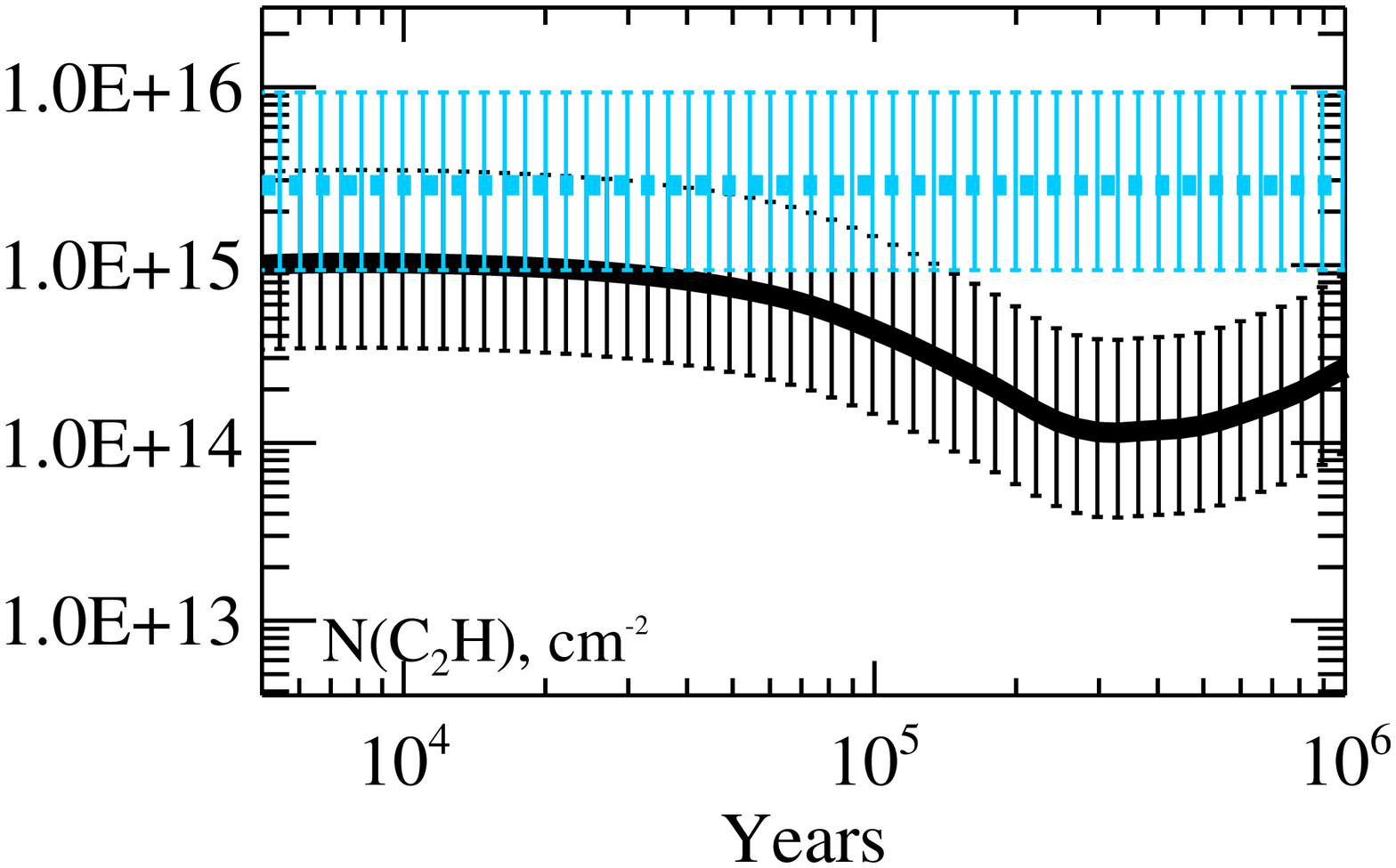}
\includegraphics[width=0.32\textwidth]{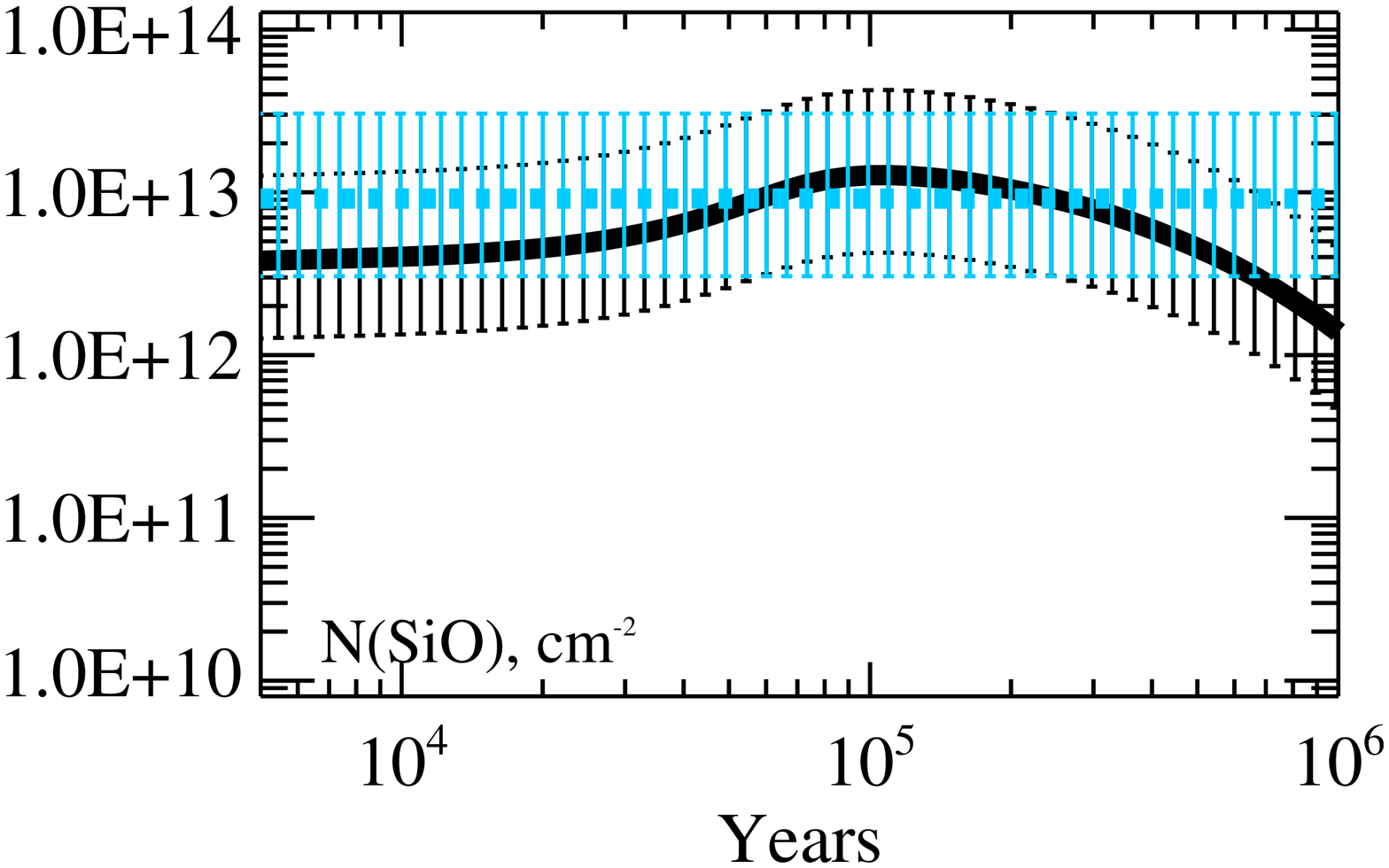}\\
\includegraphics[width=0.32\textwidth]{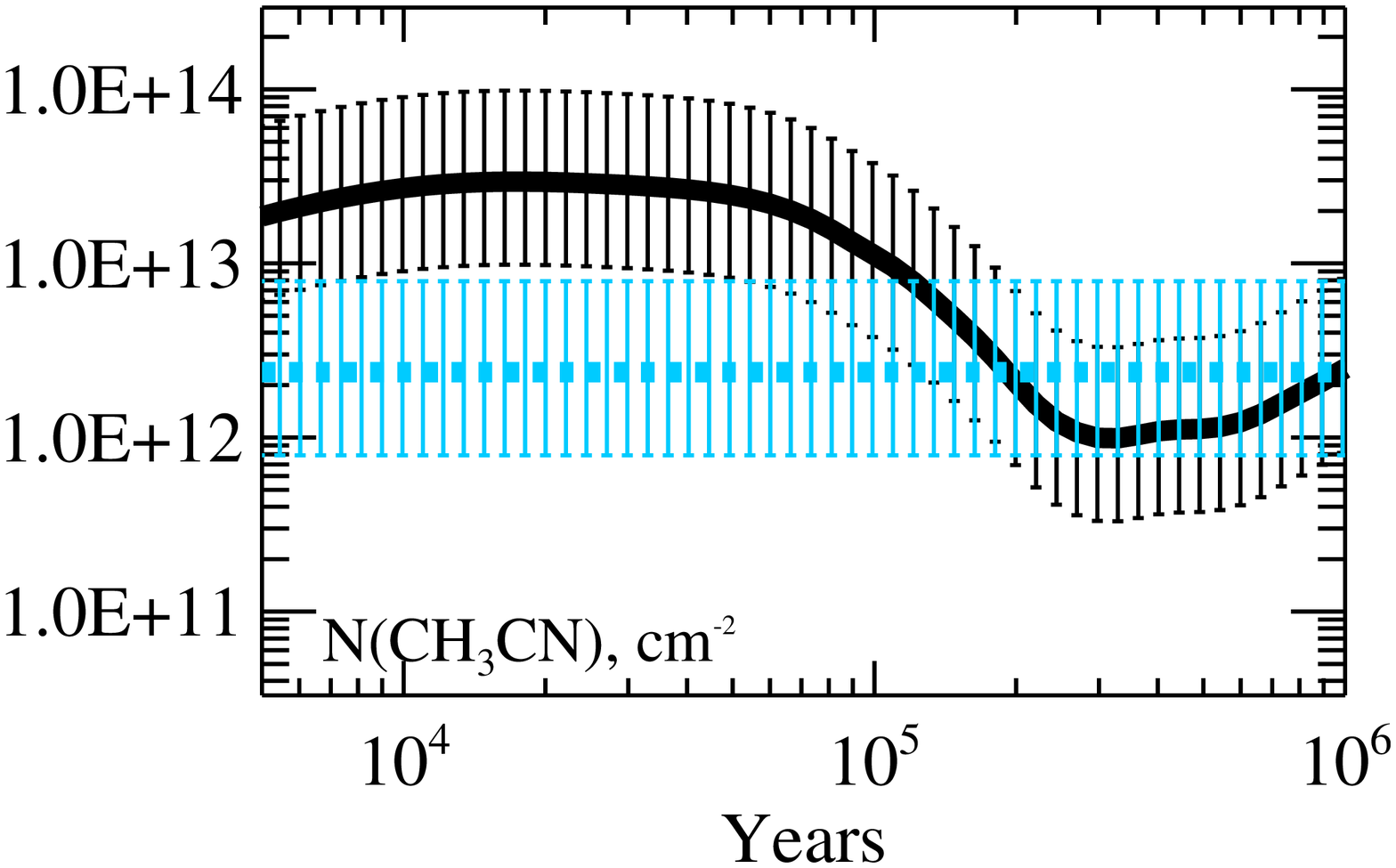}
\includegraphics[width=0.32\textwidth]{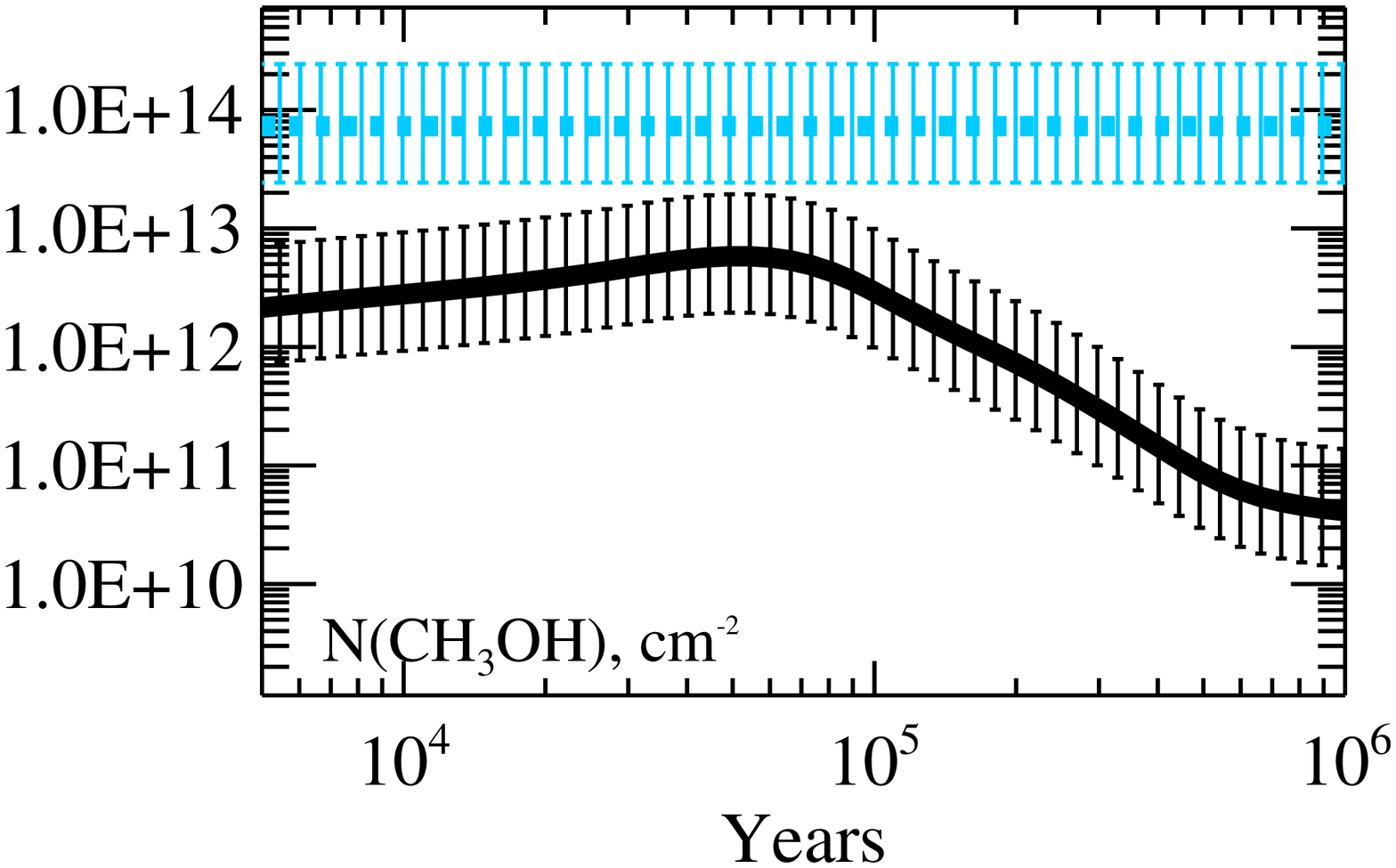}
\caption{Modeled and observed median column densities in cm$^{-2}$ in the HMPO stage. The modeled values are shown in black and the observed values in blue. The error bars are given by the vertical marks. The molecule name is labeled in the plot.}
\label{fig:coldens_hmpo}
\end{figure*}
}
\clearpage
\onlfig{
\begin{figure*}
\includegraphics[width=0.32\textwidth]{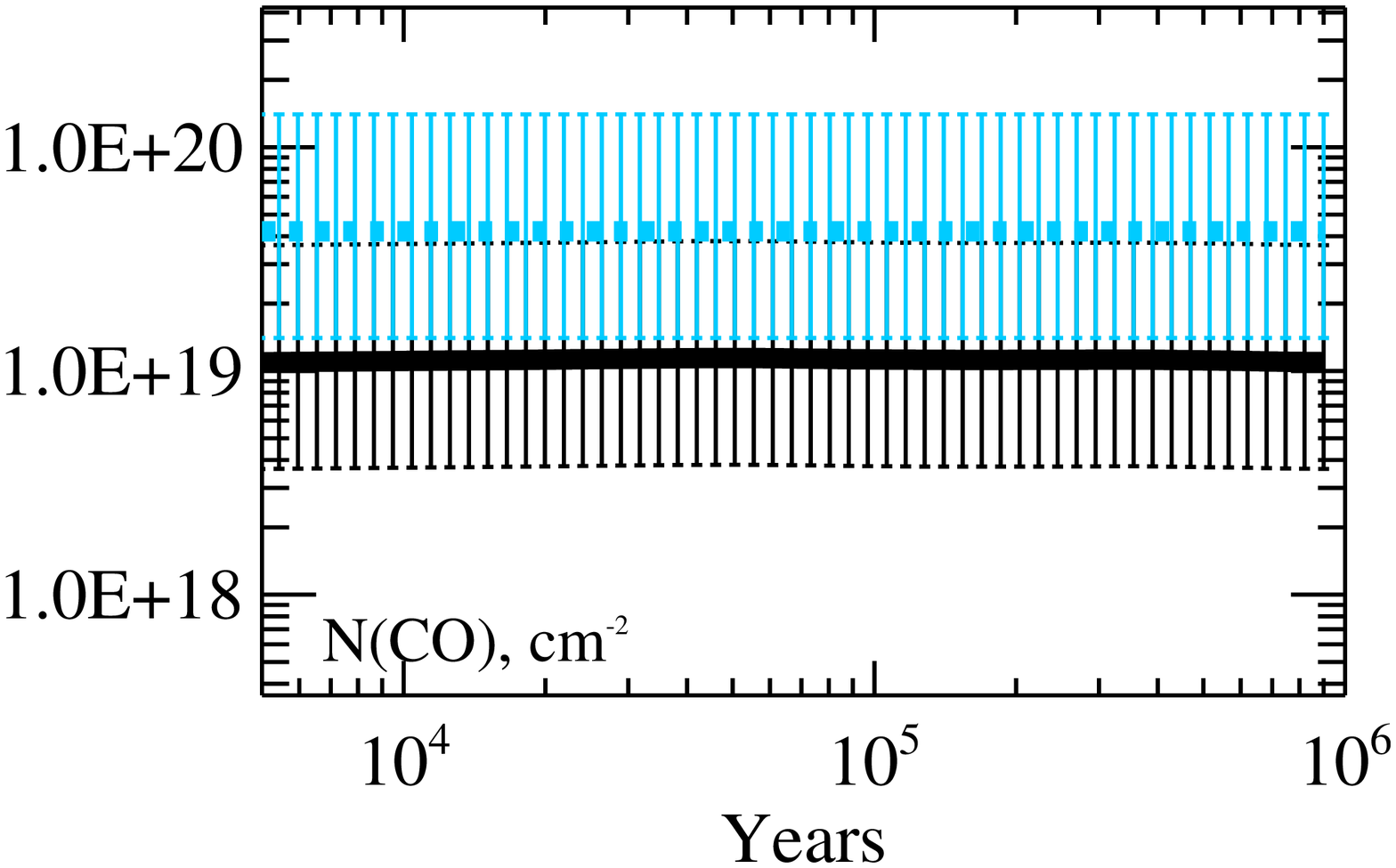}
\includegraphics[width=0.32\textwidth]{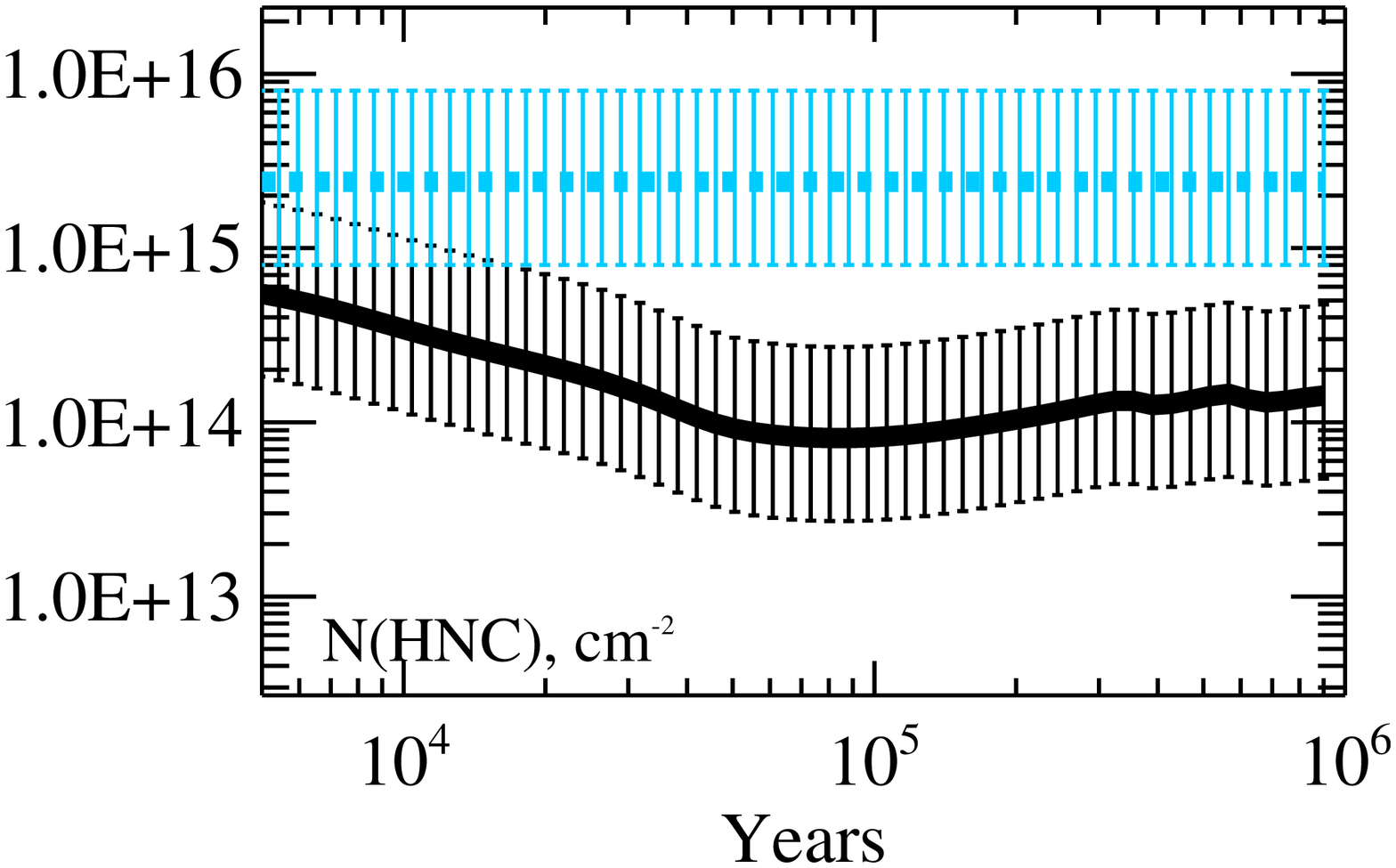}
\includegraphics[width=0.32\textwidth]{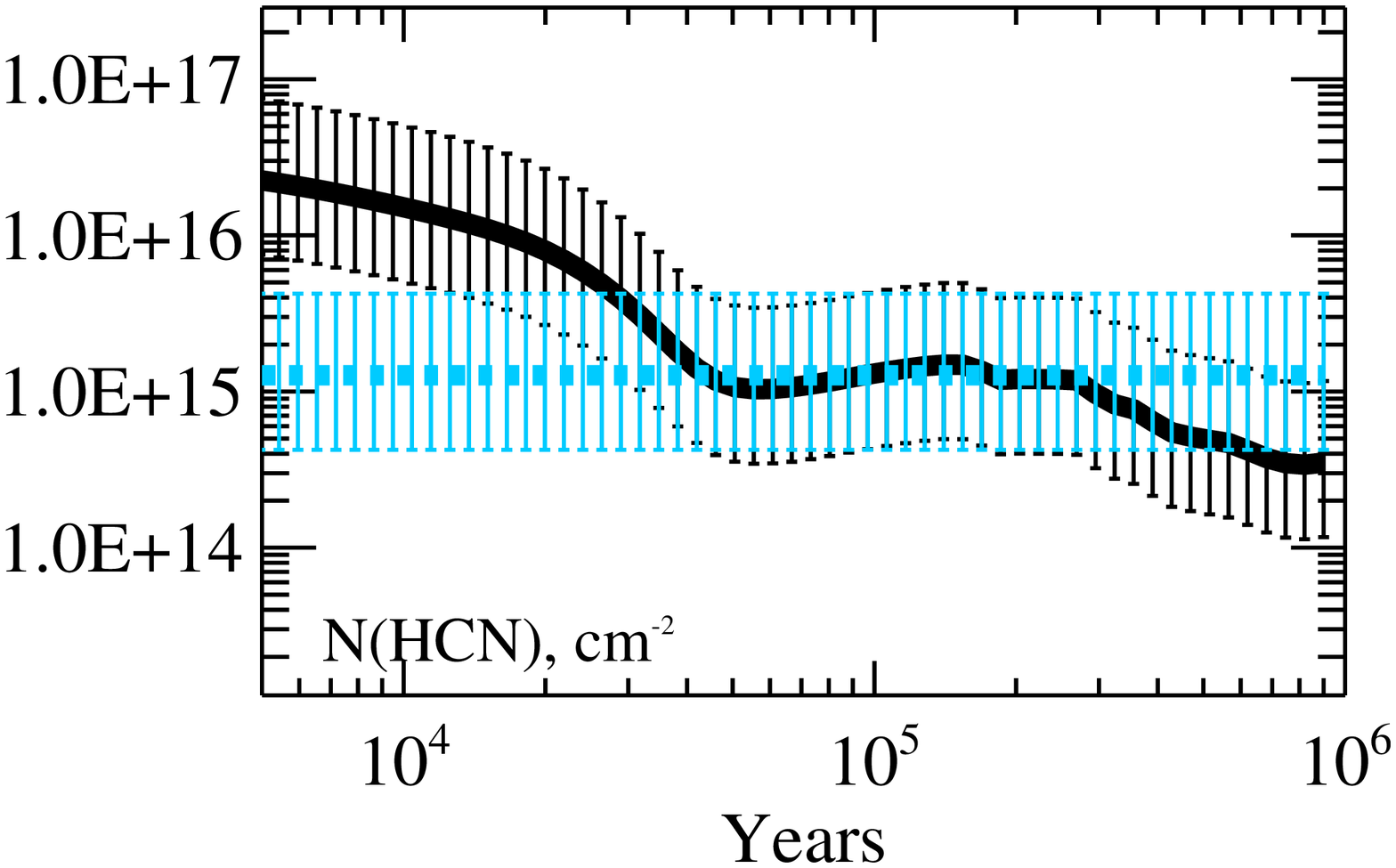}\\
\includegraphics[width=0.32\textwidth]{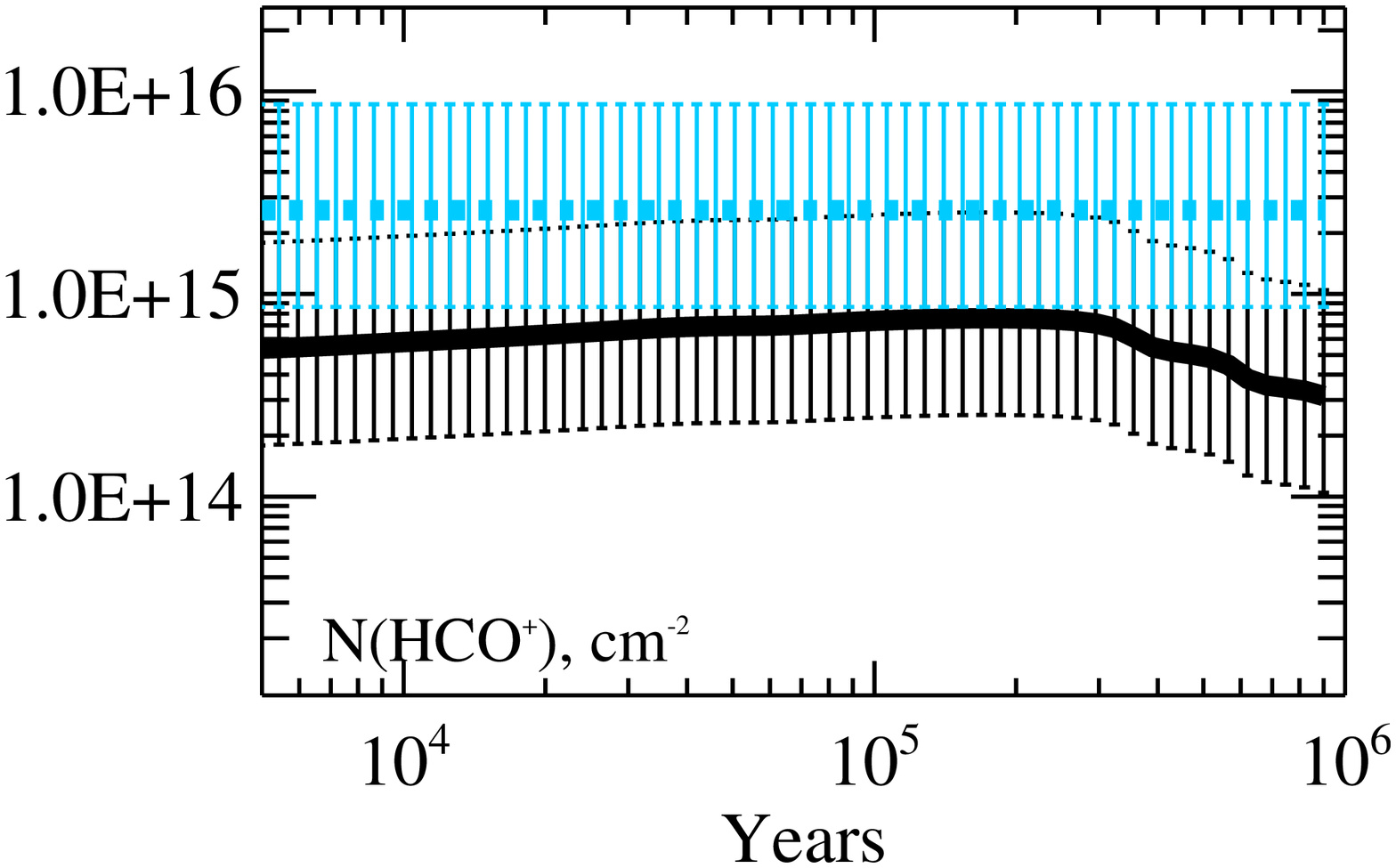}
\includegraphics[width=0.32\textwidth]{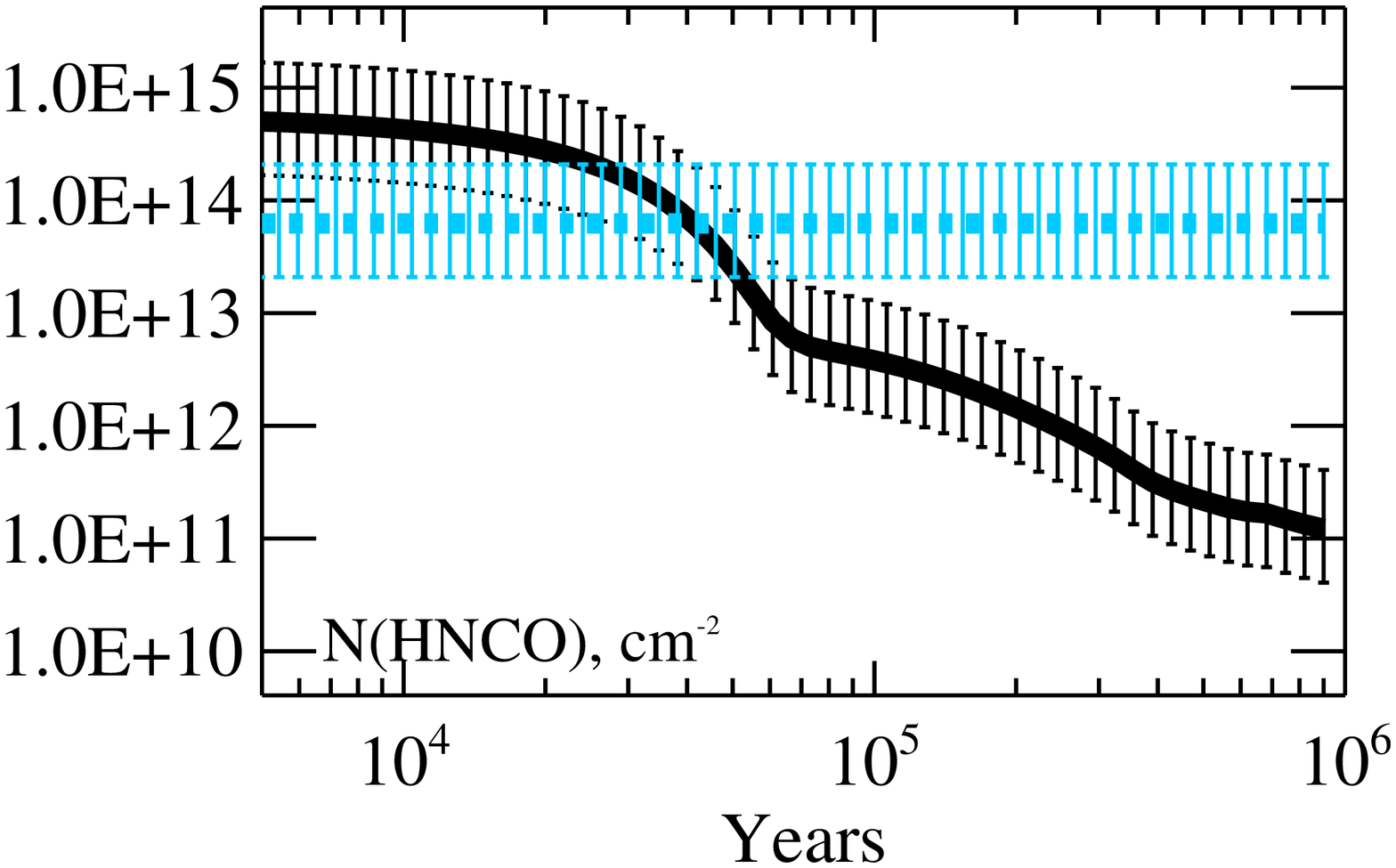}
\includegraphics[width=0.32\textwidth]{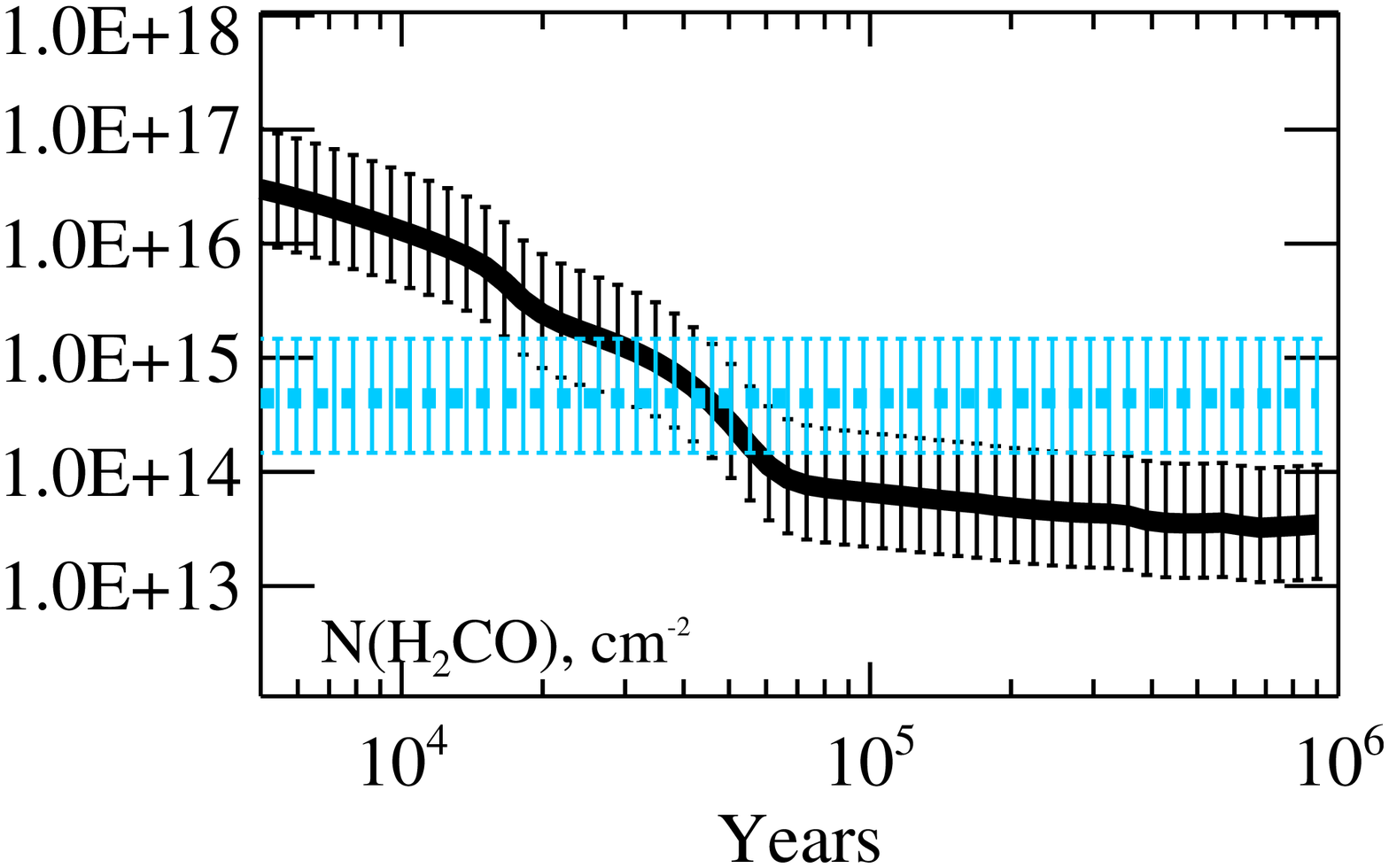}\\
\includegraphics[width=0.32\textwidth]{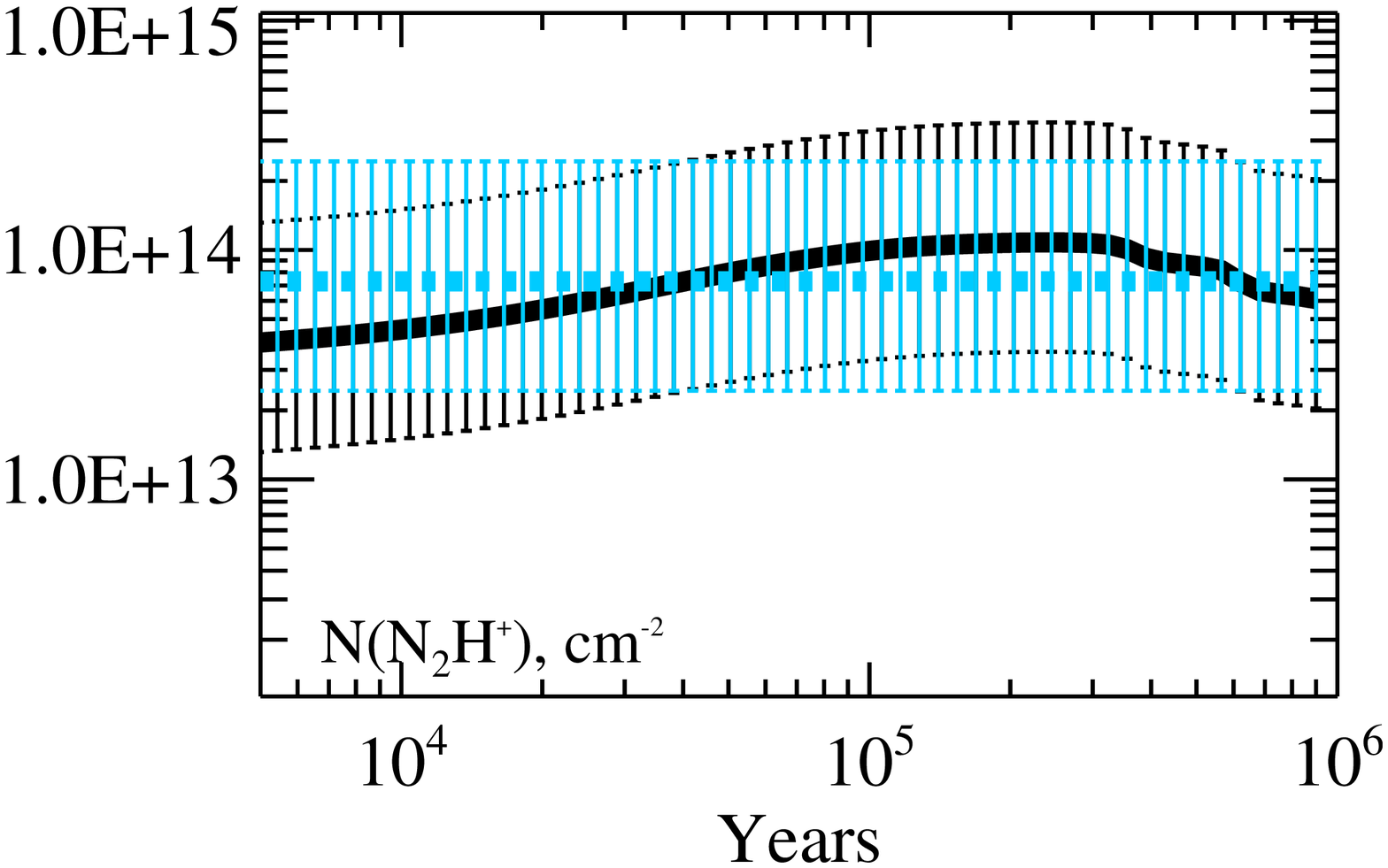}
\includegraphics[width=0.32\textwidth]{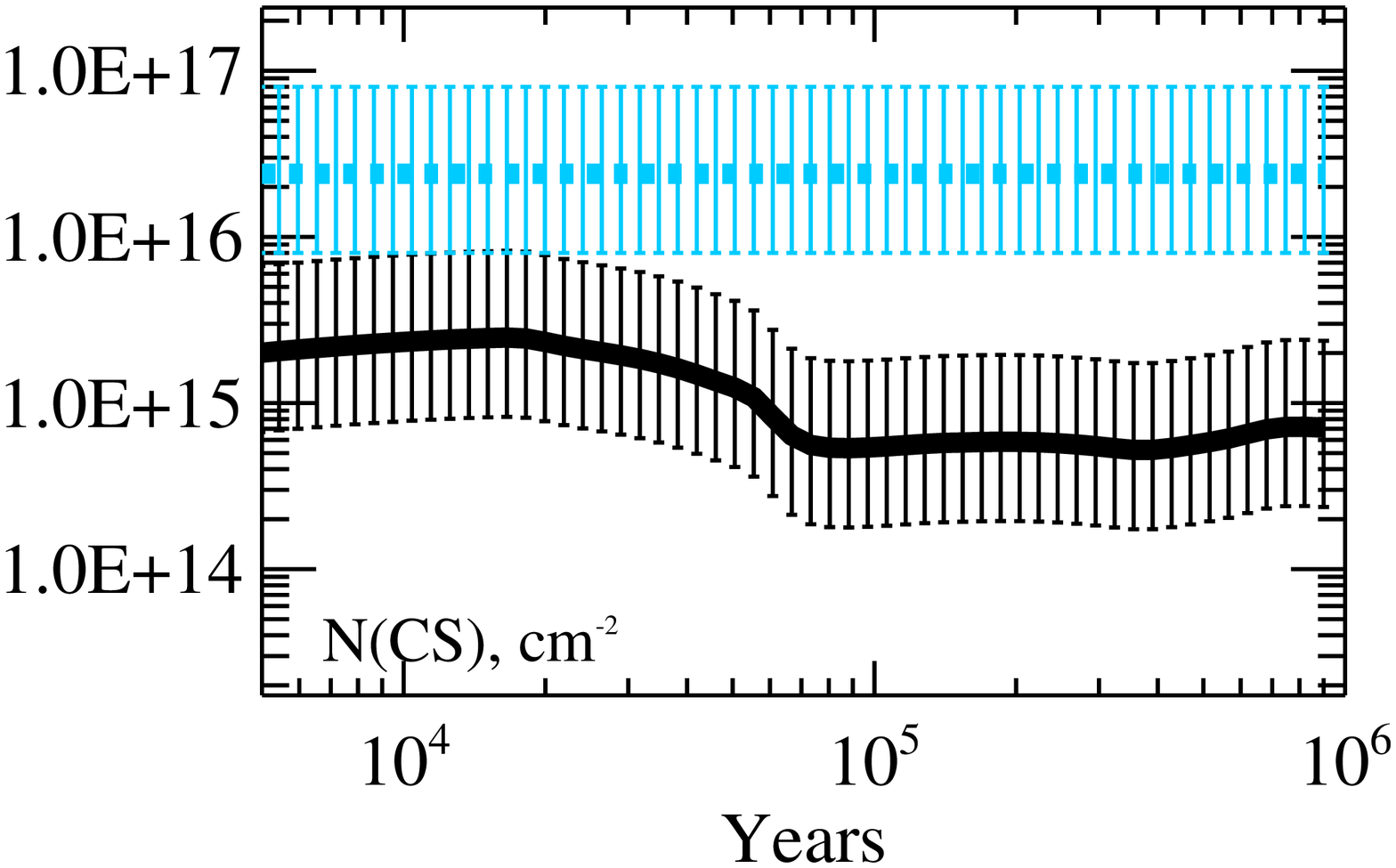}
\includegraphics[width=0.32\textwidth]{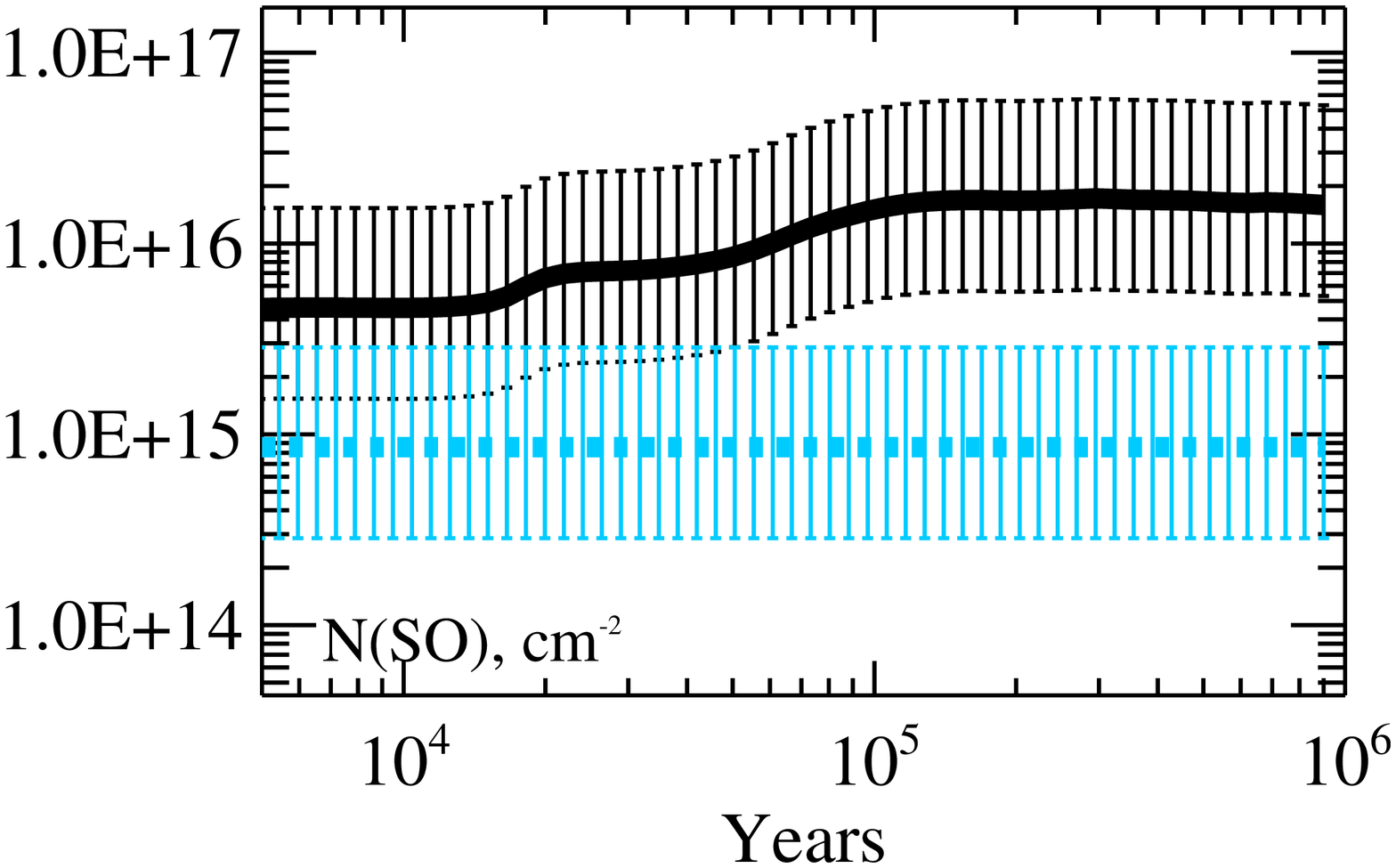}\\
\includegraphics[width=0.32\textwidth]{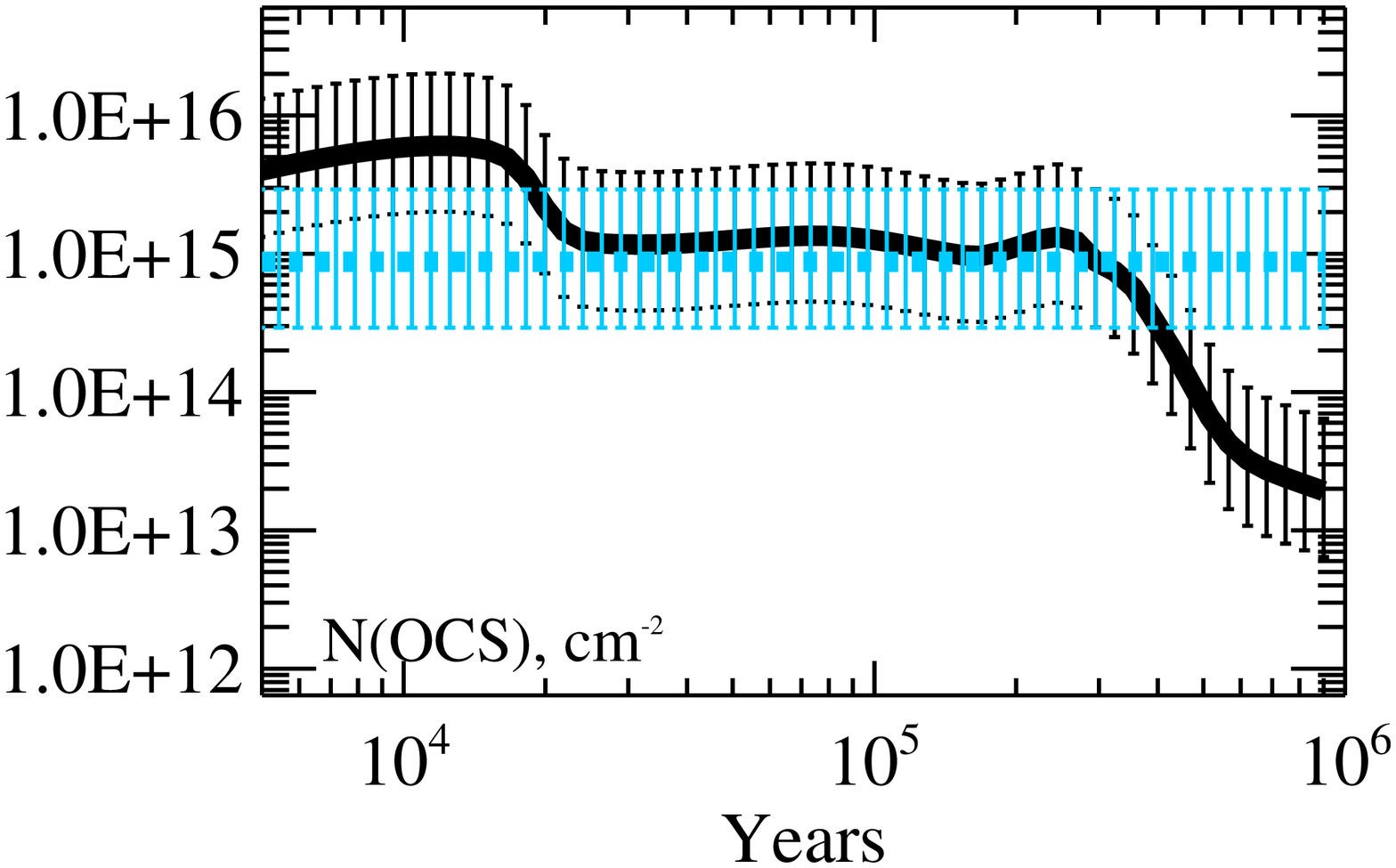}
\includegraphics[width=0.32\textwidth]{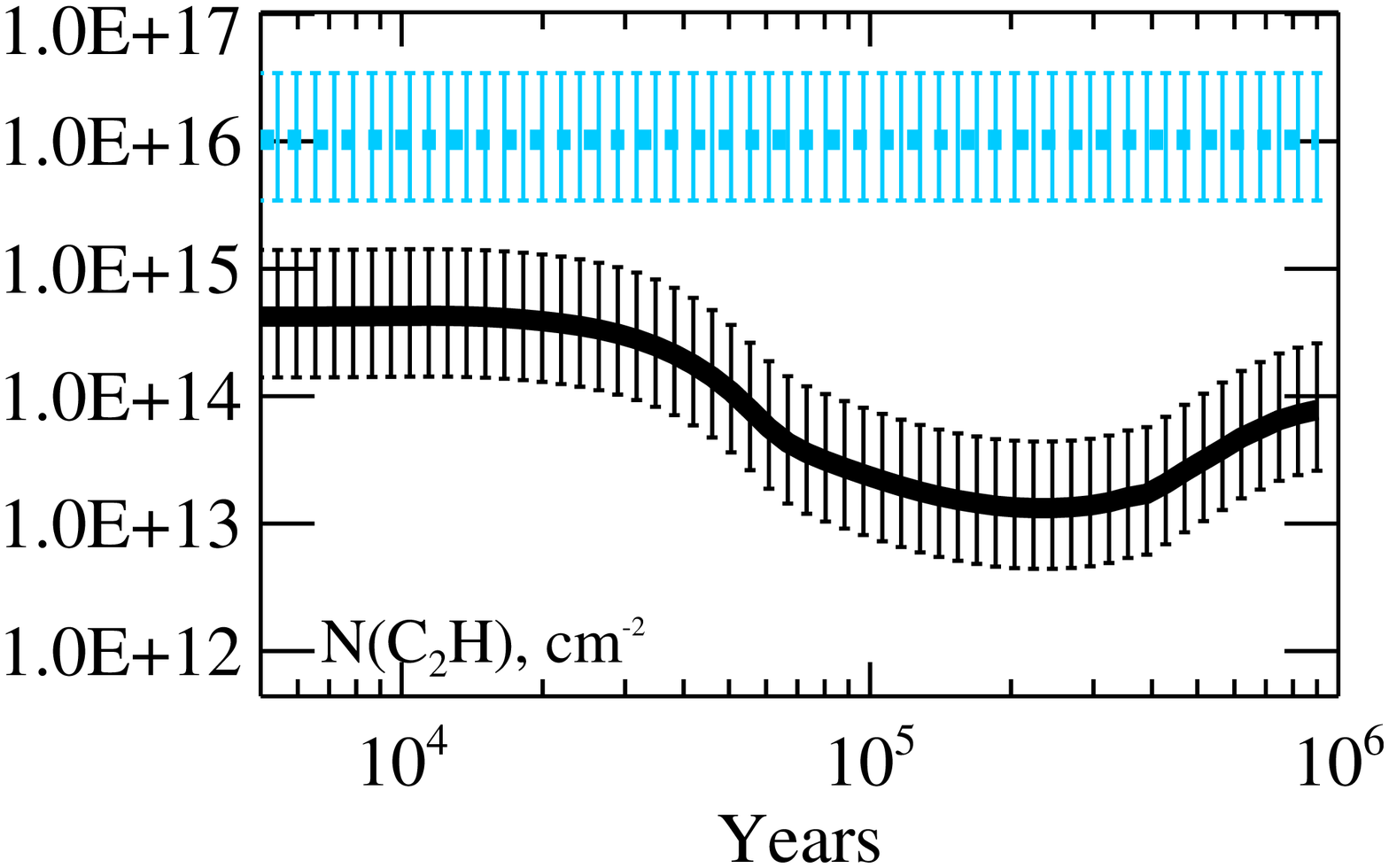}
\includegraphics[width=0.32\textwidth]{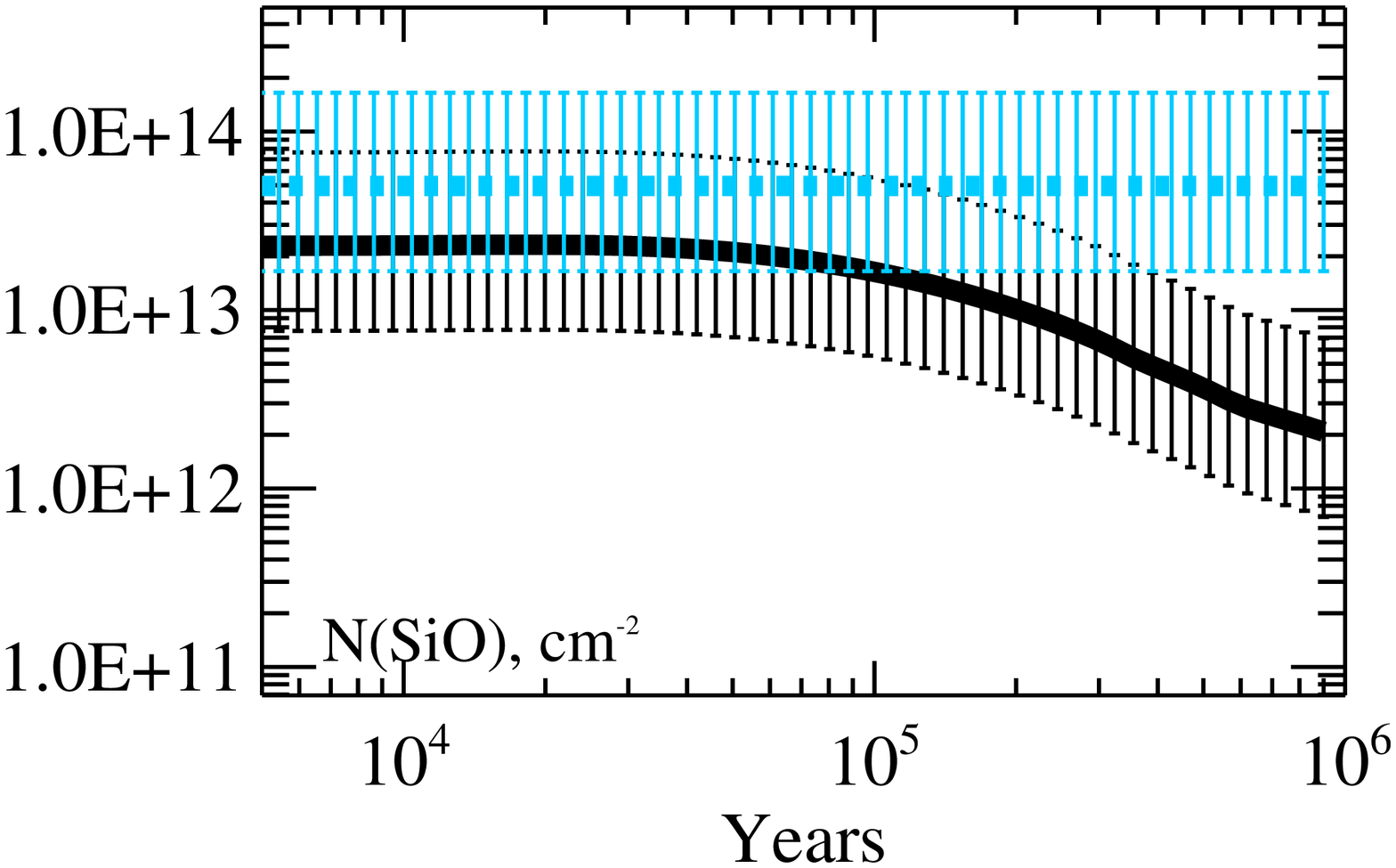}\\
\includegraphics[width=0.32\textwidth]{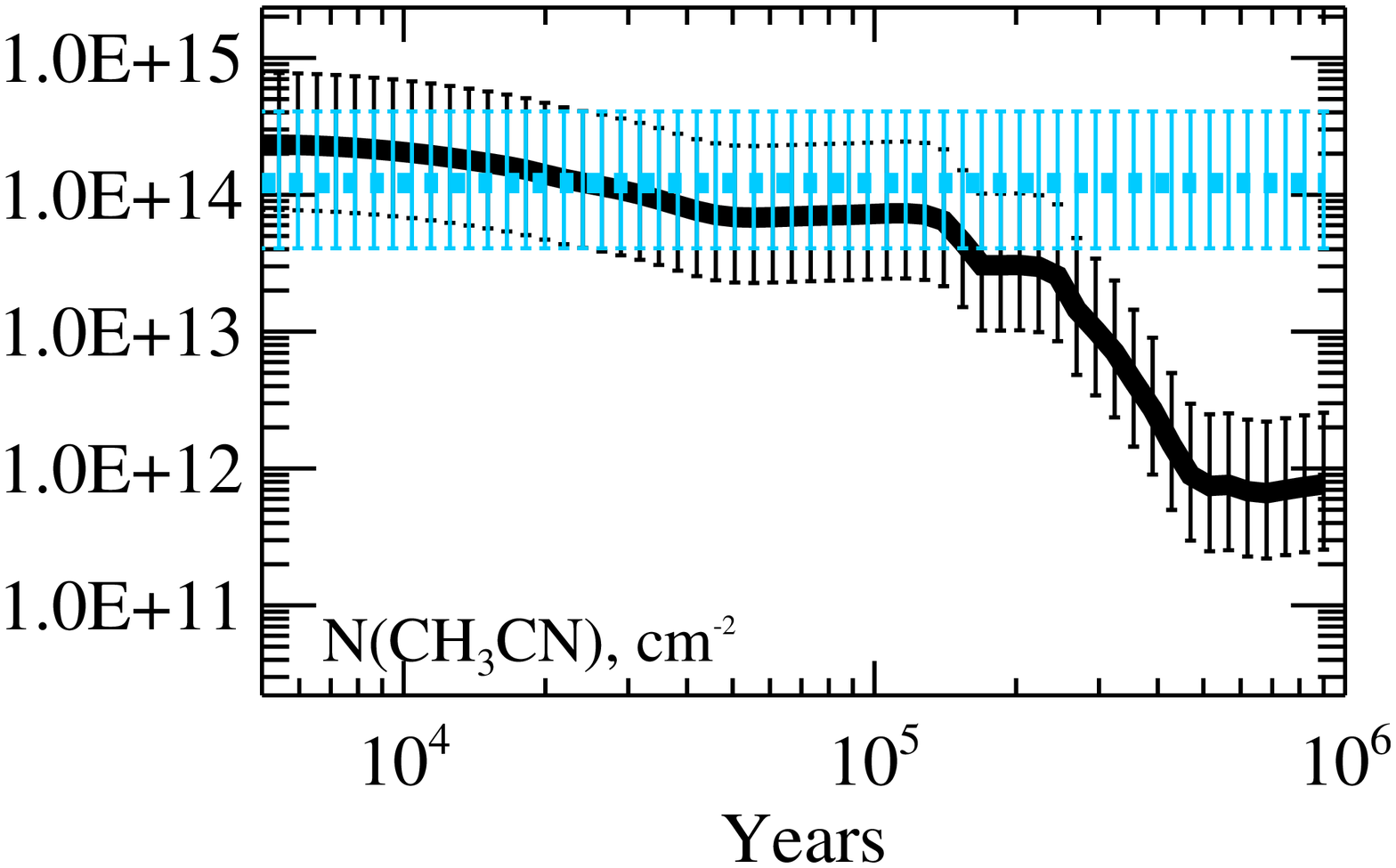}
\includegraphics[width=0.32\textwidth]{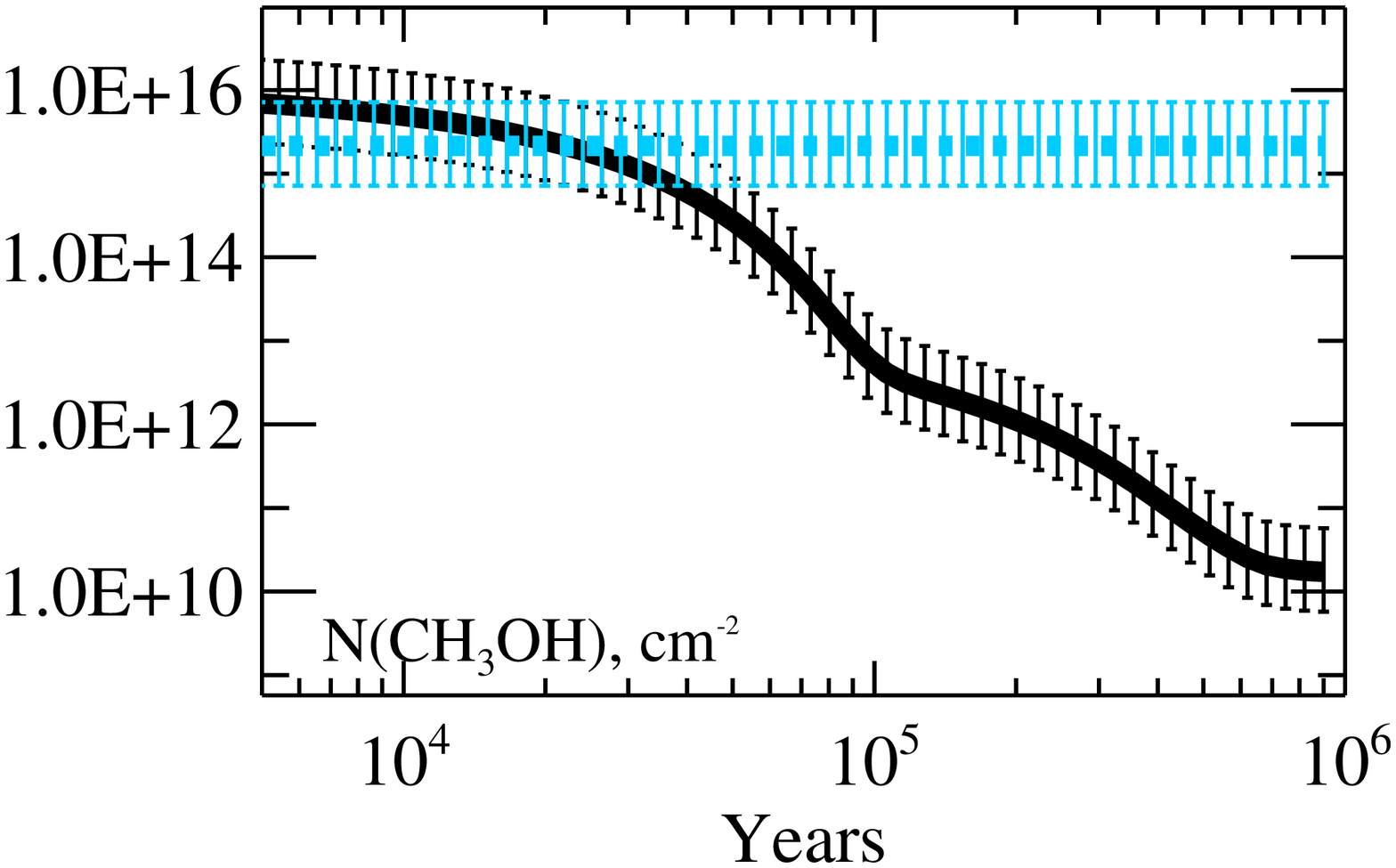}
\caption{Modeled and observed median column densities in cm$^{-2}$ in the HMC stage. The modeled values are shown in black and the observed values in blue. The error bars are given by the vertical marks. The molecule name is labeled in the plot.}
\label{fig:coldens_hmc}
\end{figure*}
}
\clearpage
\onlfig{
\begin{figure*}
\includegraphics[width=0.32\textwidth]{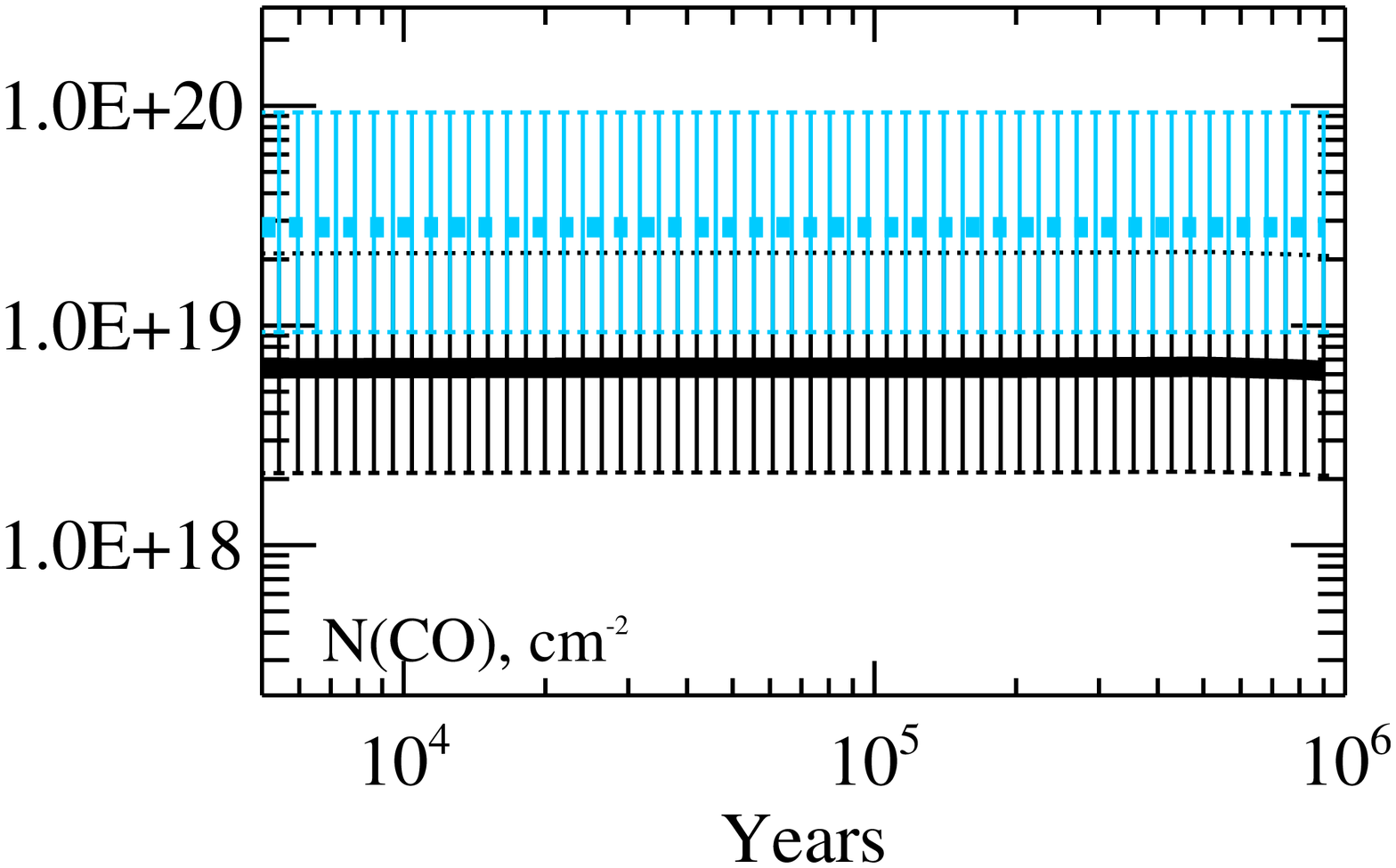}
\includegraphics[width=0.32\textwidth]{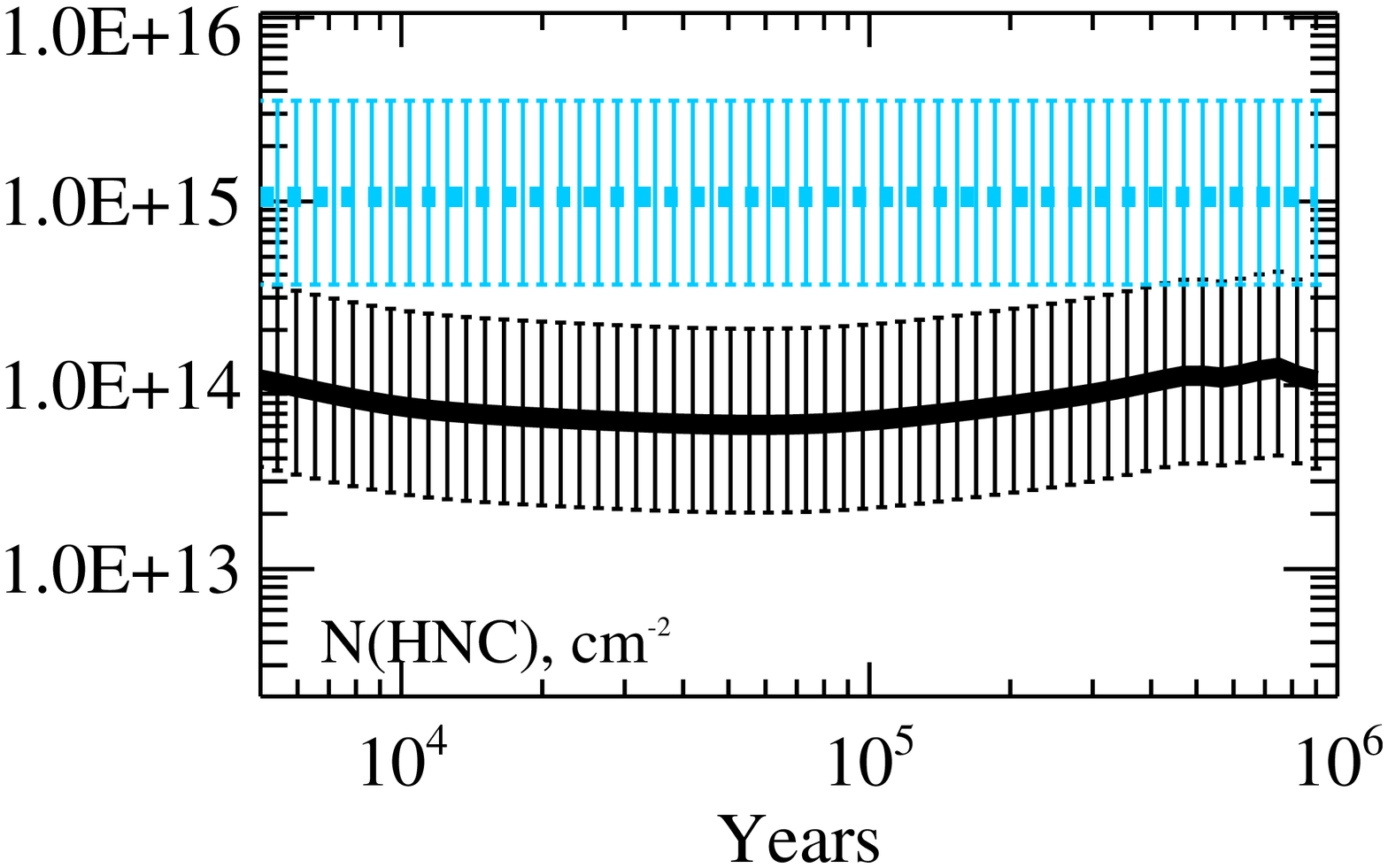}
\includegraphics[width=0.32\textwidth]{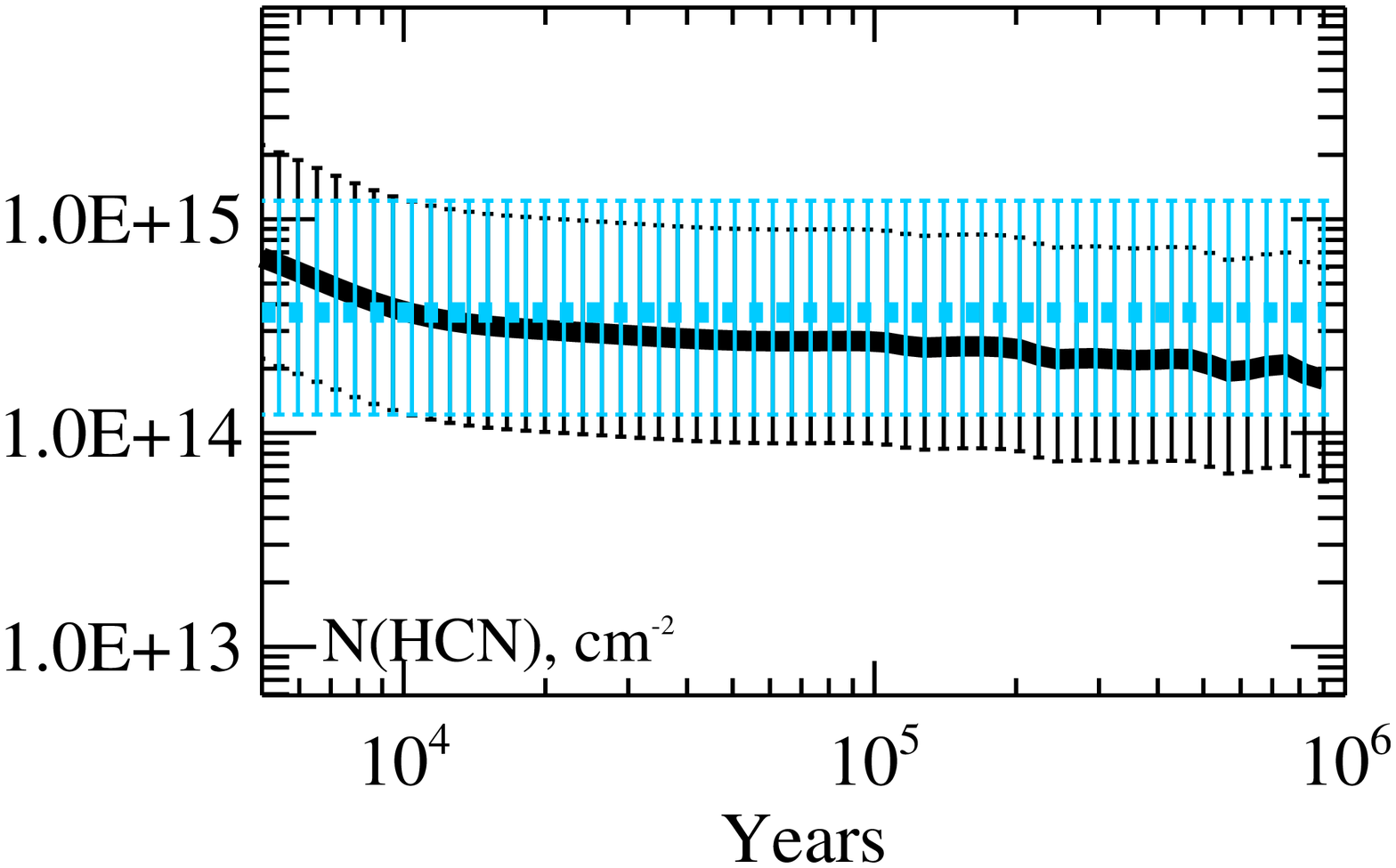}\\
\includegraphics[width=0.32\textwidth]{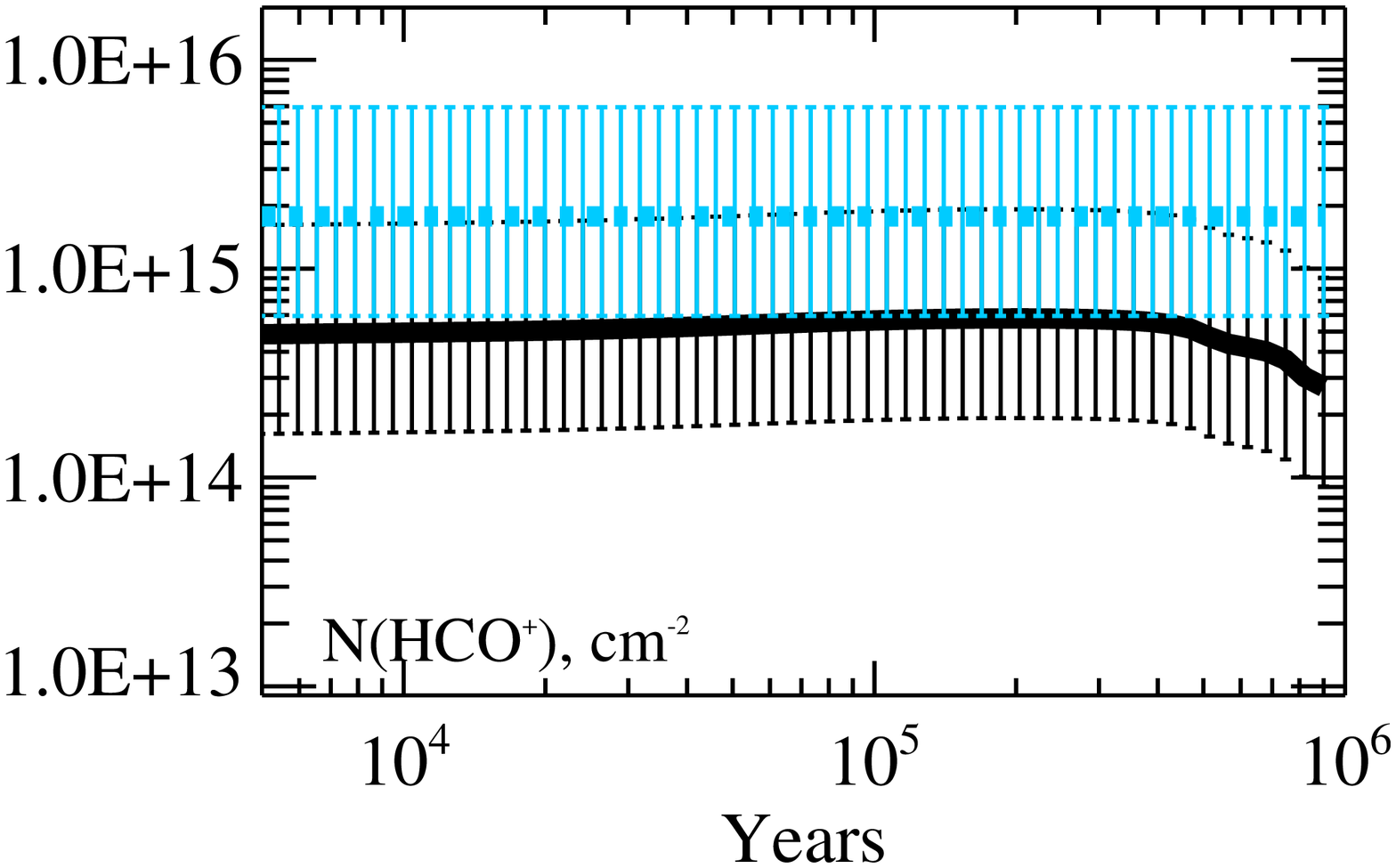}
\includegraphics[width=0.32\textwidth]{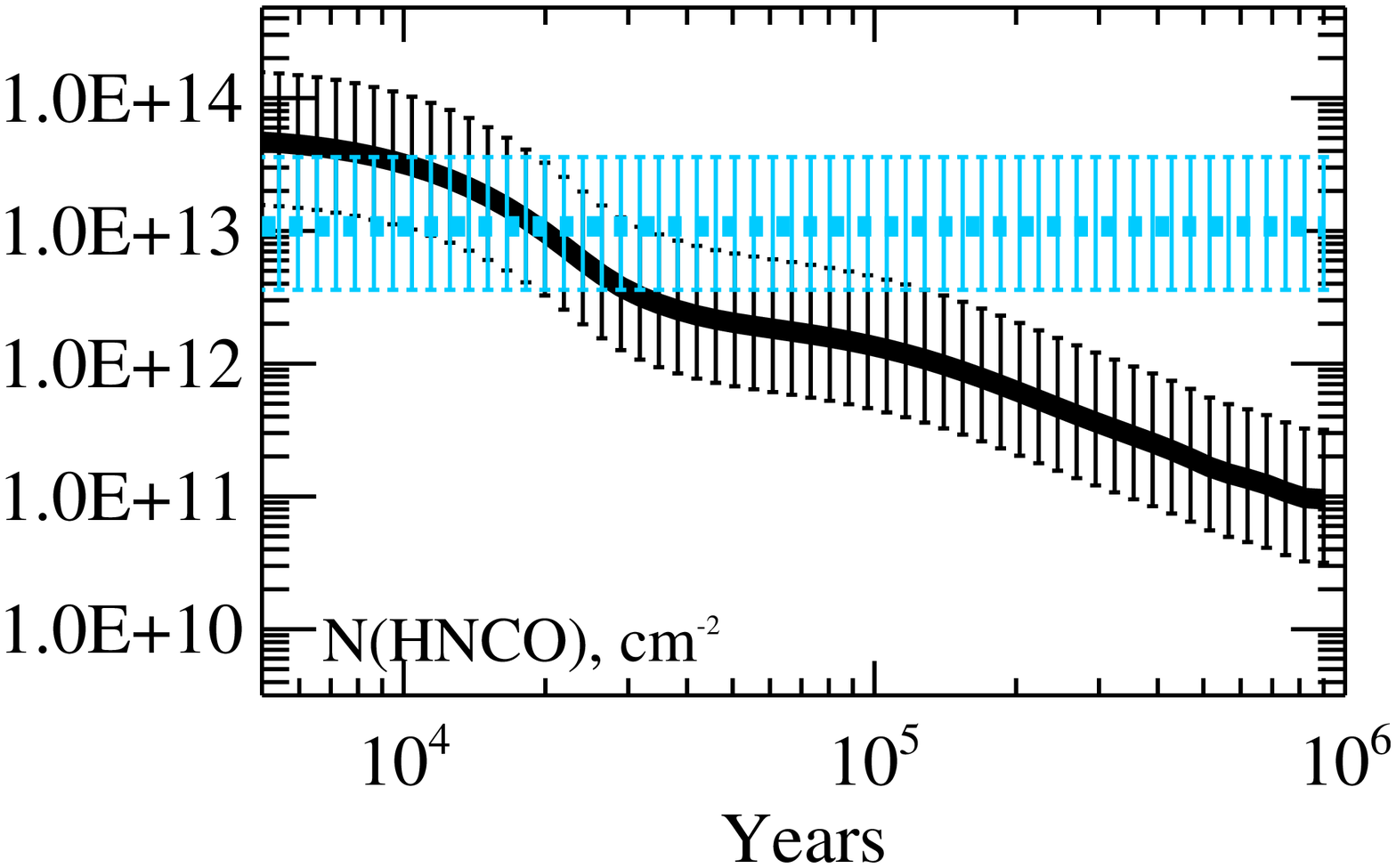}
\includegraphics[width=0.32\textwidth]{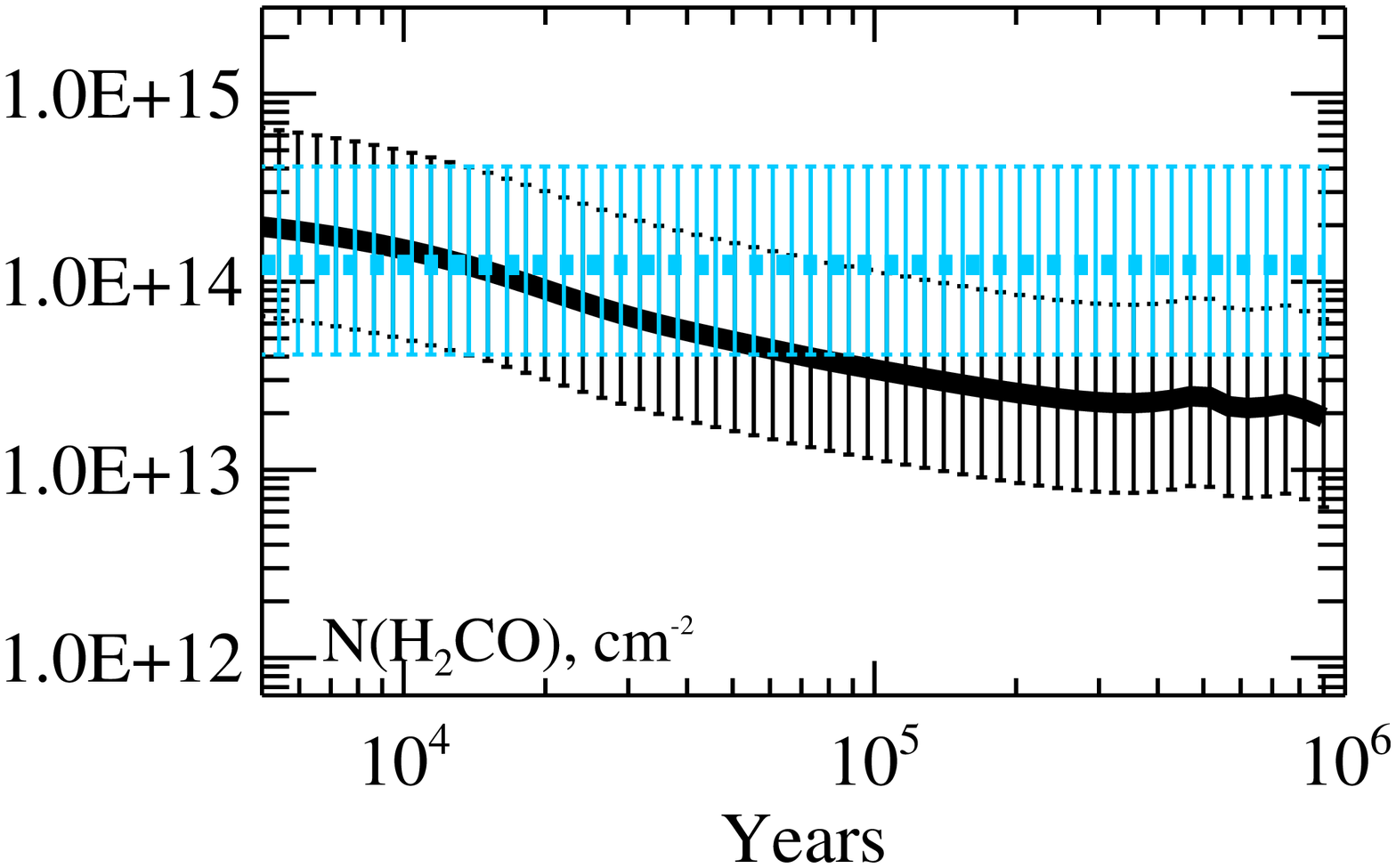}\\
\includegraphics[width=0.32\textwidth]{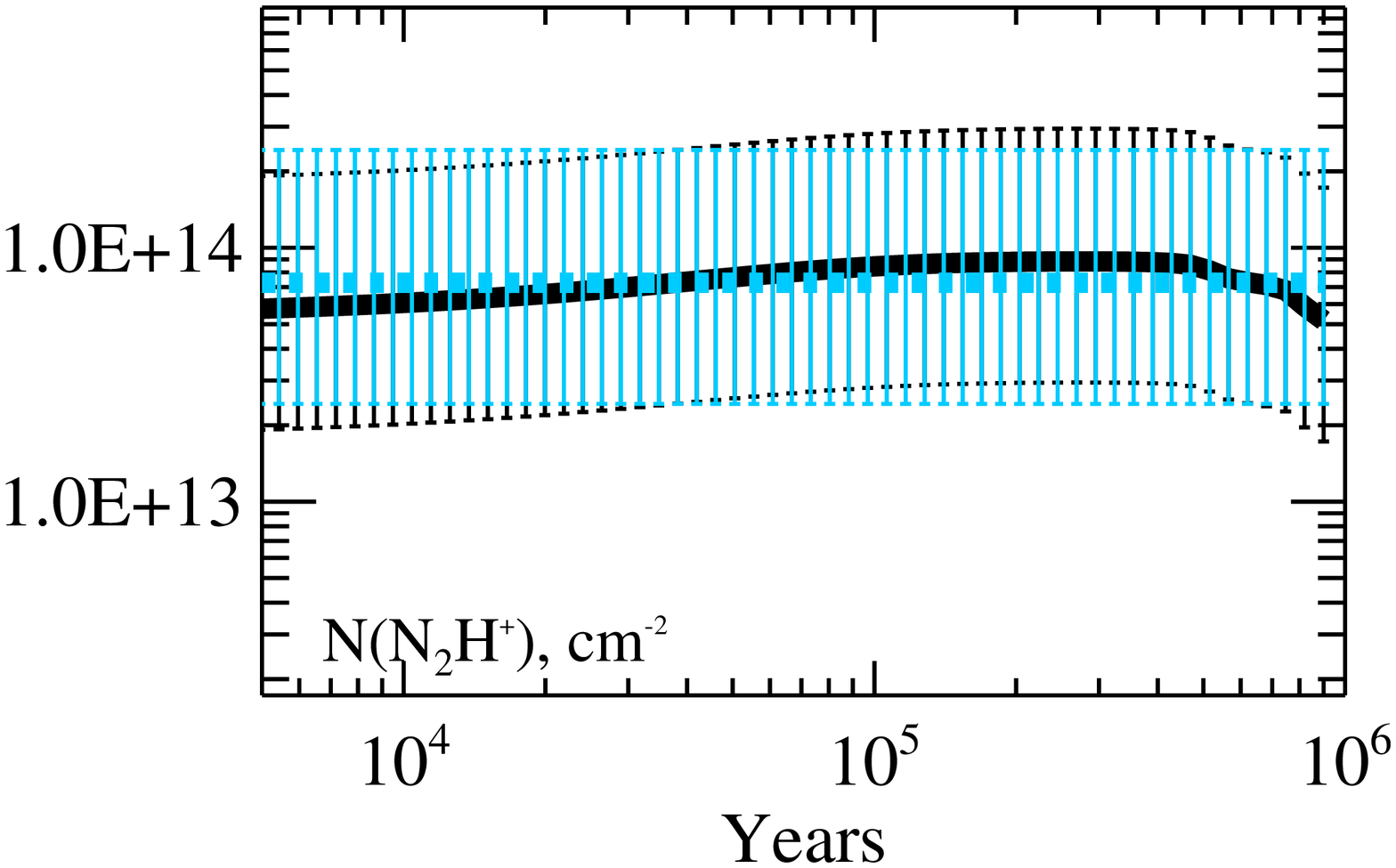}
\includegraphics[width=0.32\textwidth]{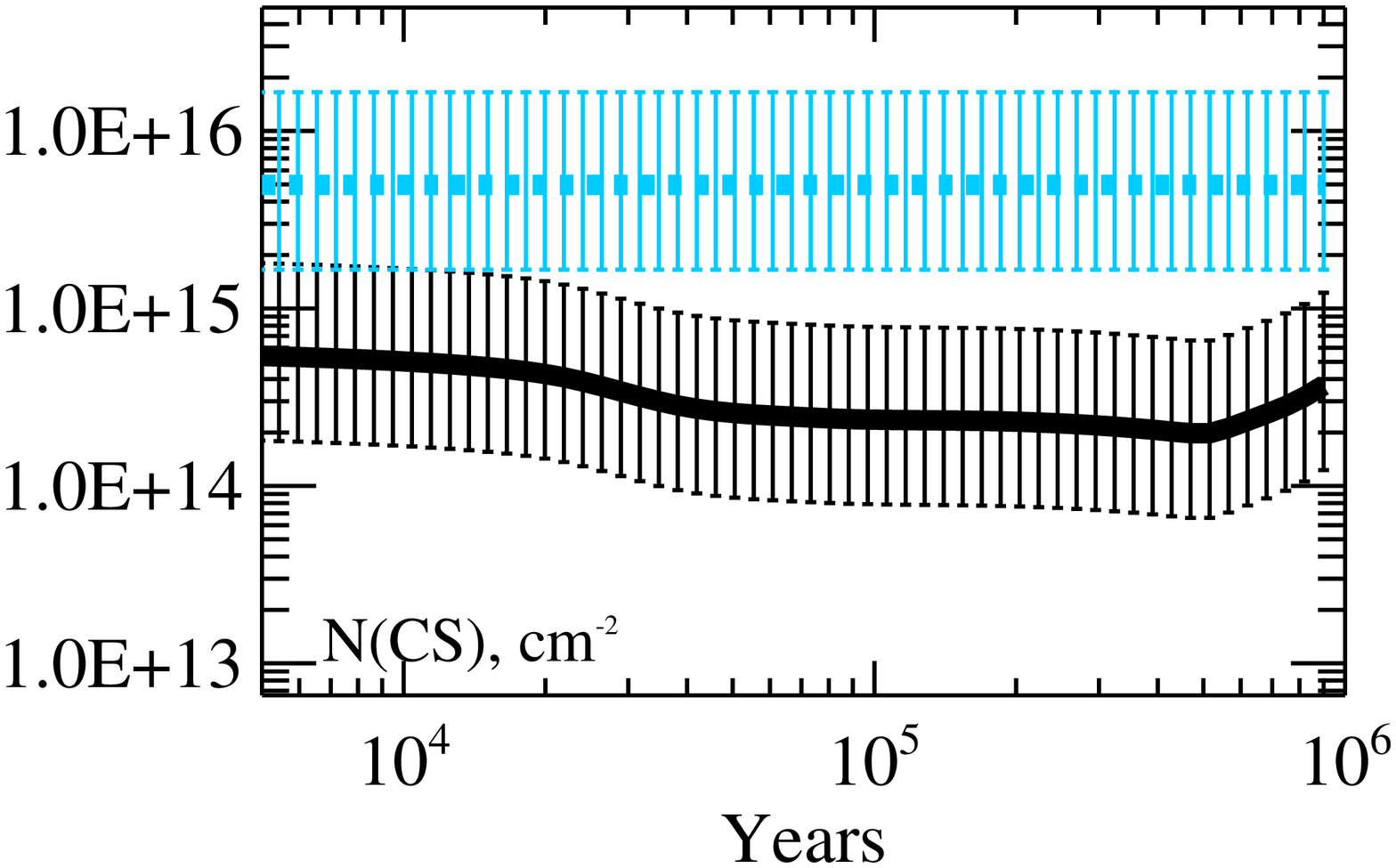}
\includegraphics[width=0.32\textwidth]{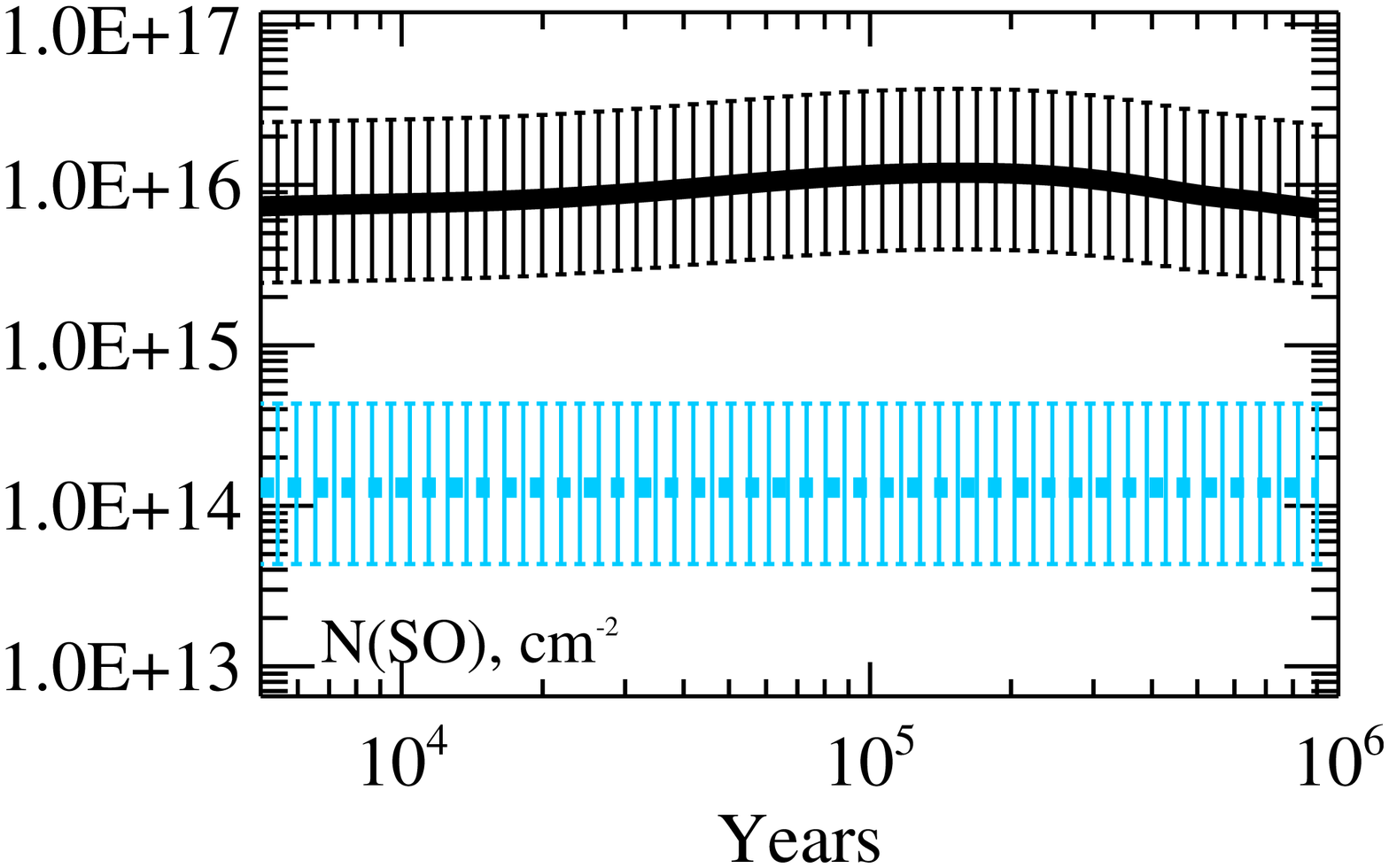}\\
\includegraphics[width=0.32\textwidth]{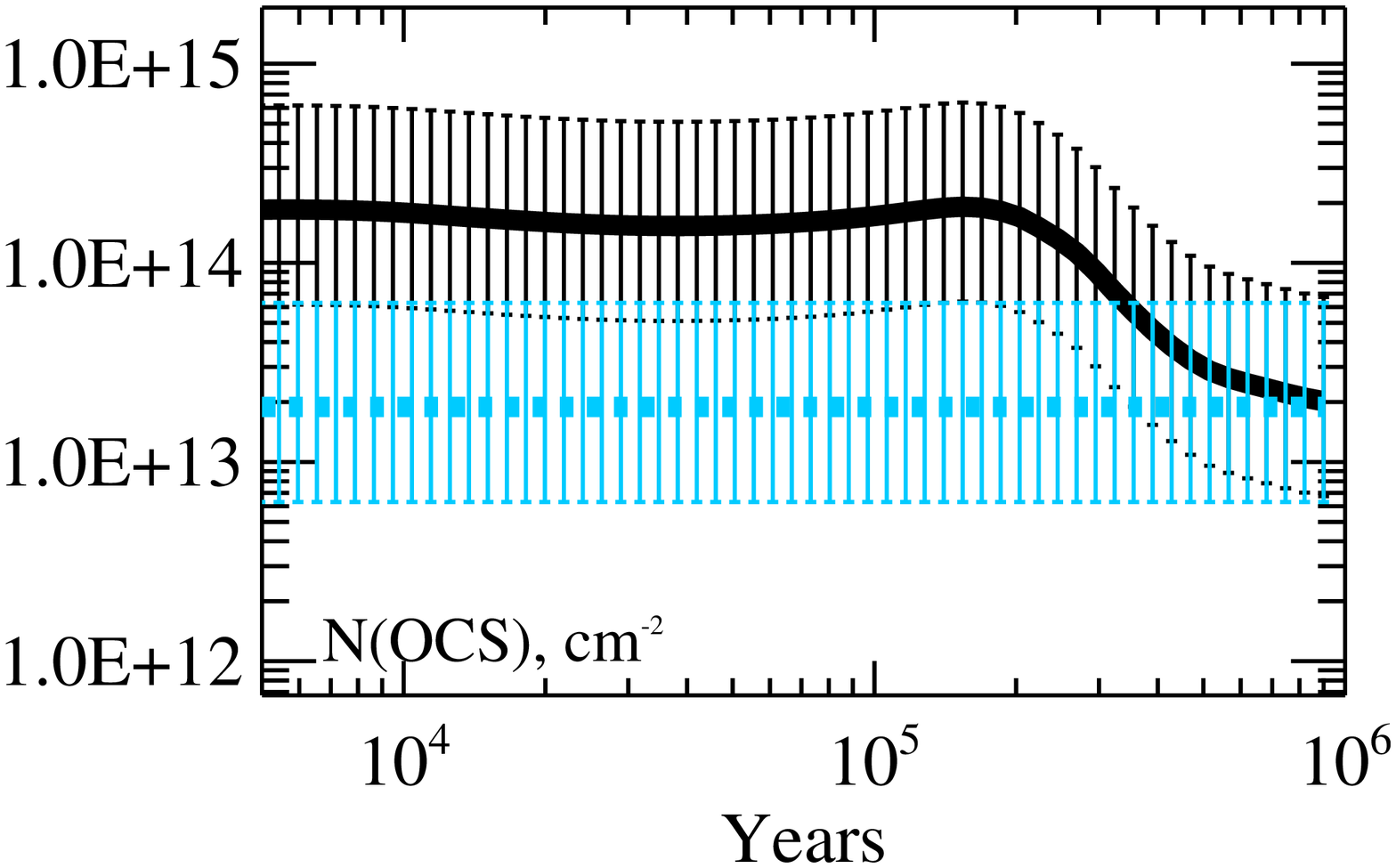}
\includegraphics[width=0.32\textwidth]{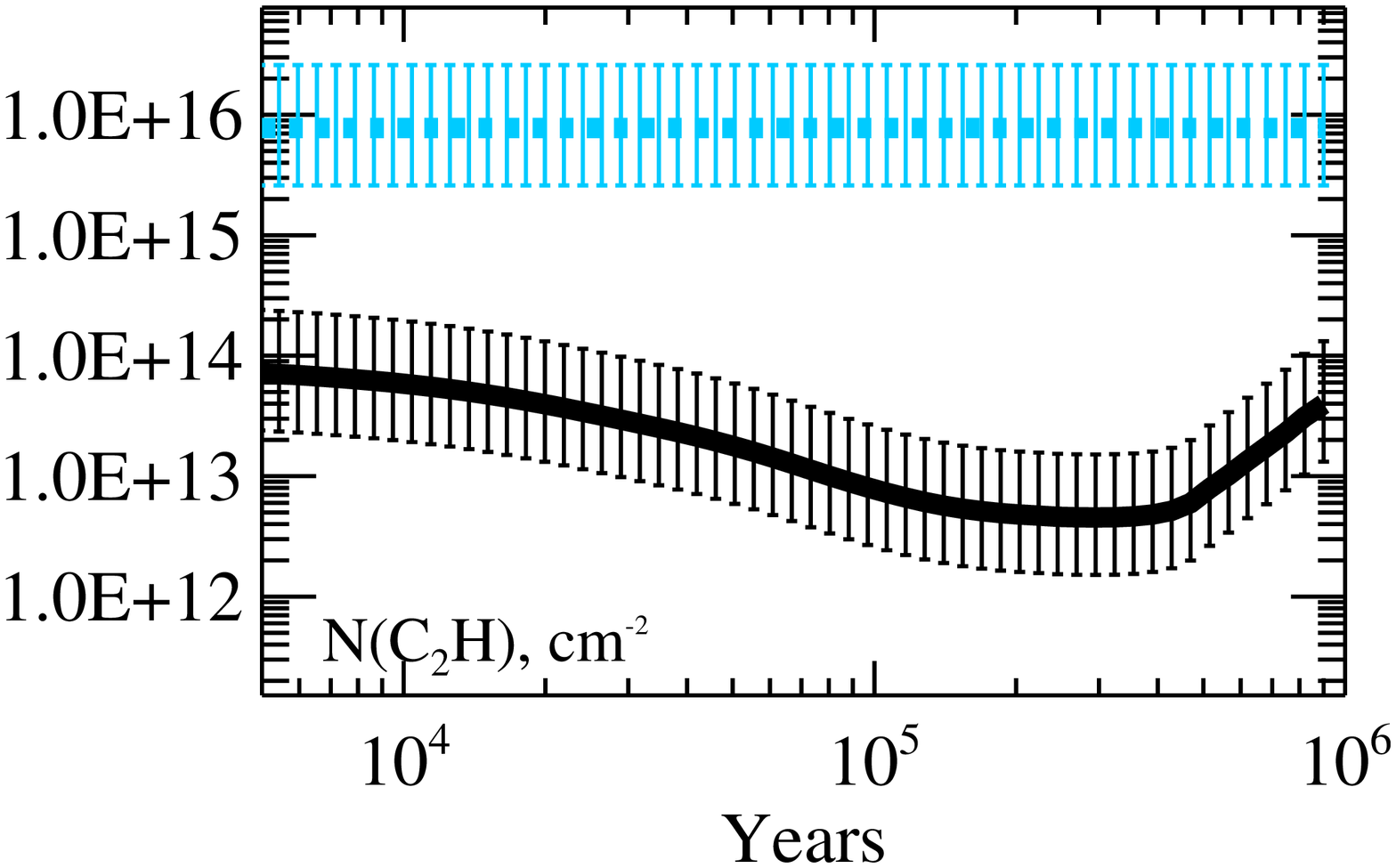}
\includegraphics[width=0.32\textwidth]{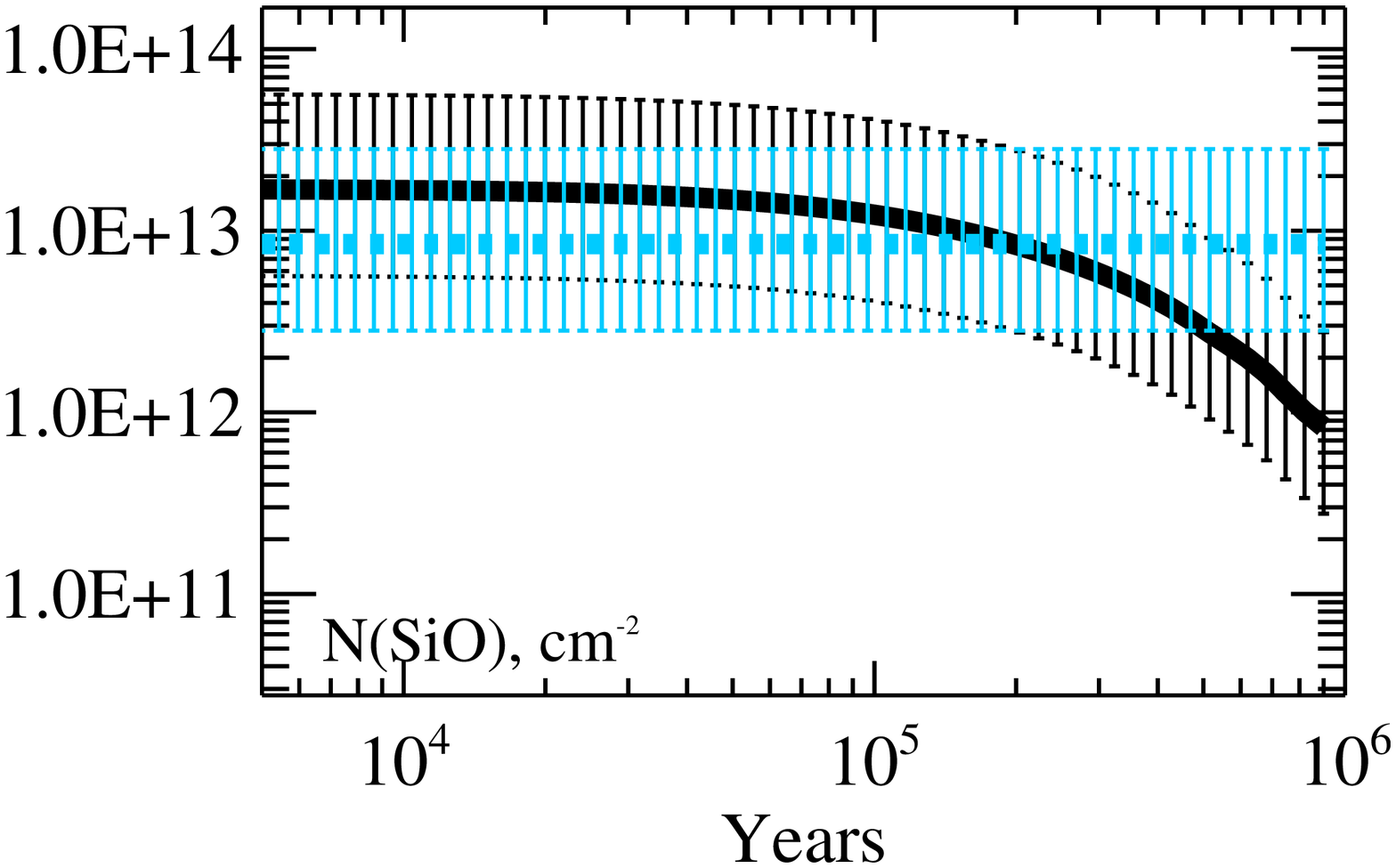}\\
\includegraphics[width=0.32\textwidth]{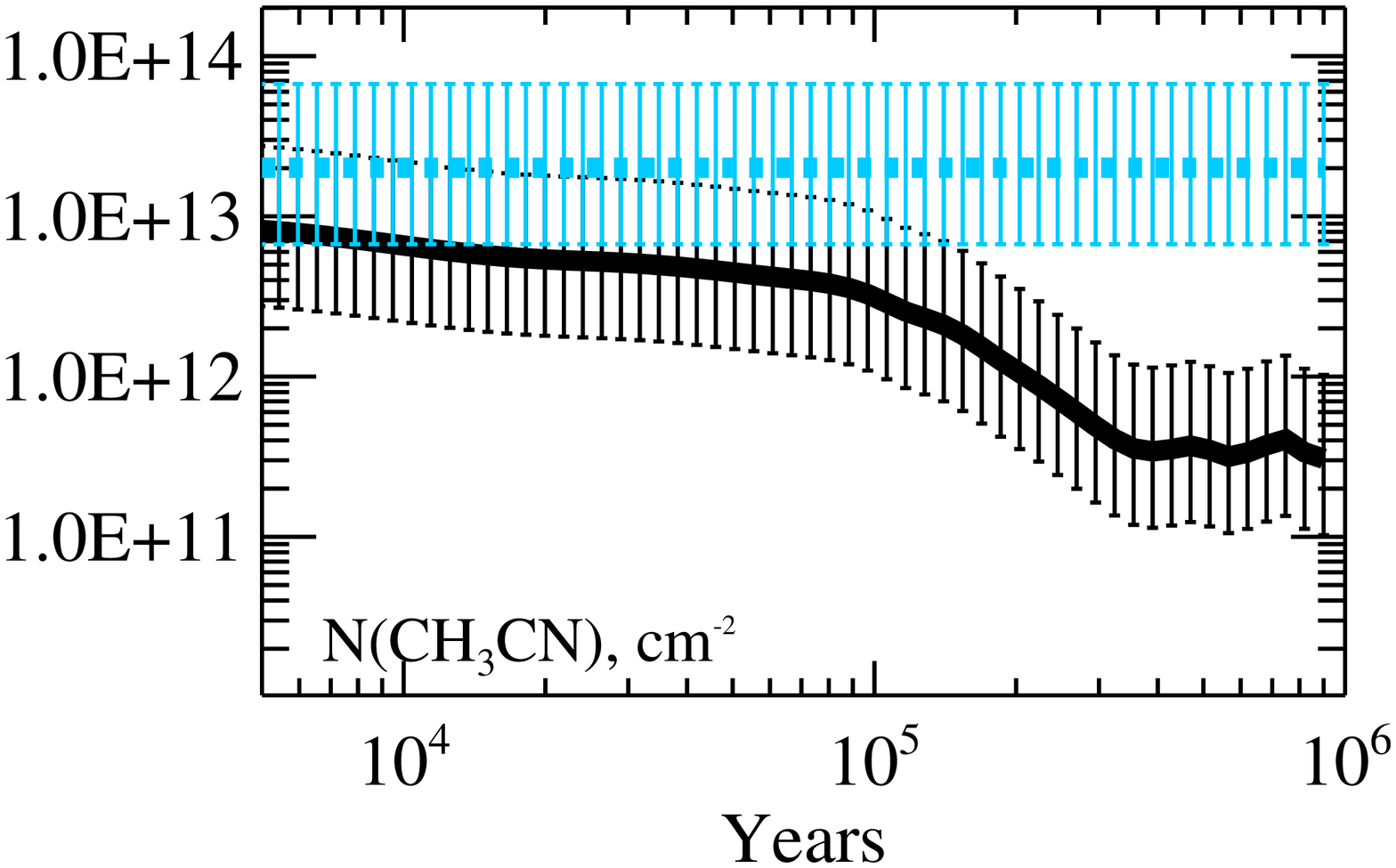}
\includegraphics[width=0.32\textwidth]{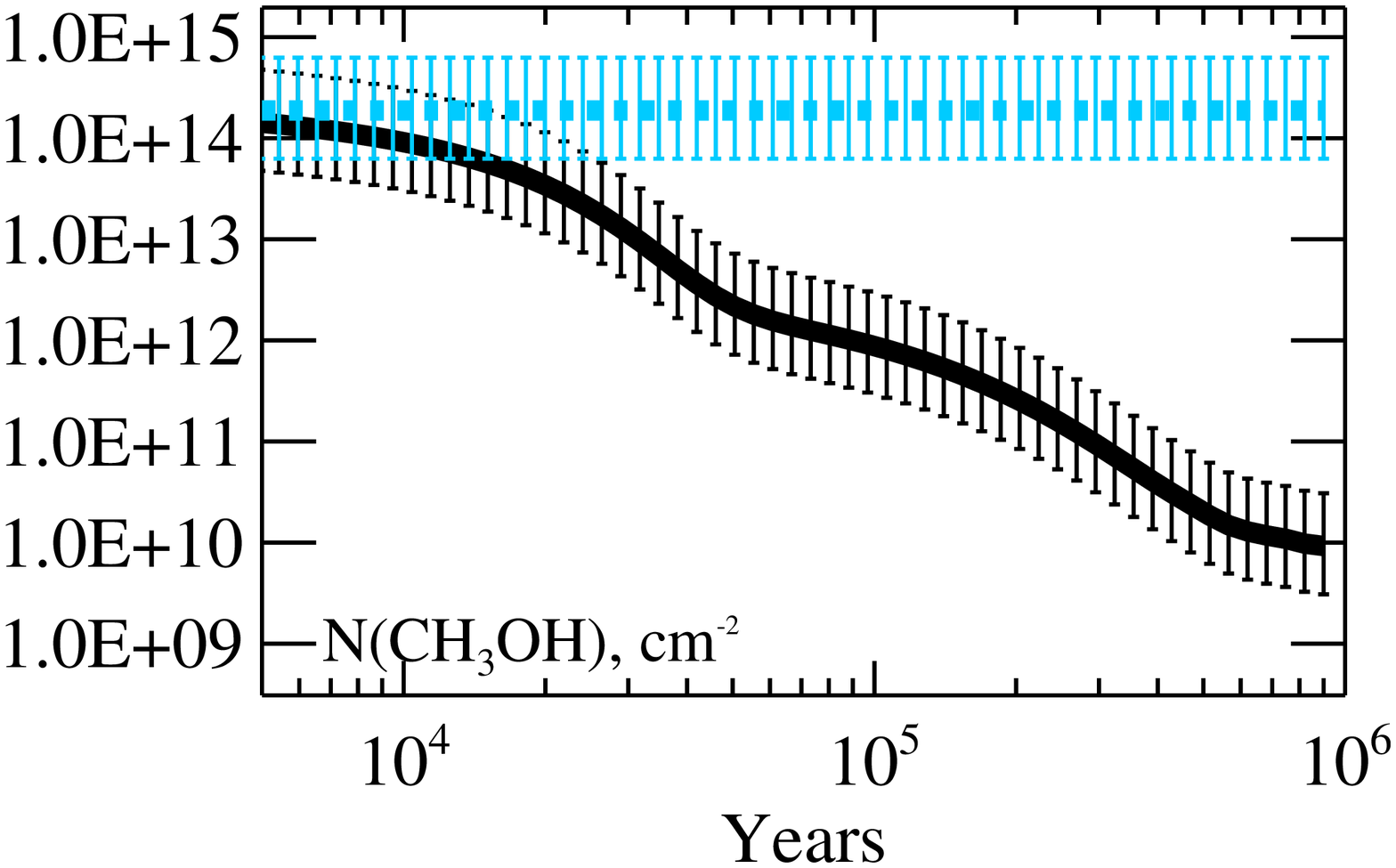}
\caption{Modeled and observed median column densities in cm$^{-2}$ in the UCH{\sc ii} stage. The modeled values are shown in black and the observed values in blue. The error bars are given by the vertical marks. The molecule name is labeled in the plot.}
\label{fig:coldens_uch}
\end{figure*}
}


\end{document}